\documentclass[final,1p,11pt,times]{elsarticle}

\usepackage{graphics}

\usepackage{amssymb}
\usepackage{amsmath}

\renewcommand{\Im}{\mathop{\rm Im}}

\newcommand{\sign}{\mathop{\rm sgn}}
\newcommand{\vep}{\varepsilon}

\newcommand{\Itot}{I}
\newcommand{\Idif}{\check{I}}
\newcommand{\phitot}{\phi}
\newcommand{\phidif}{\check{\phi}}
\newcommand{\Hdif}{\check{H}}
\newcommand{\Thetadif}{\check{\Theta}}
\newcommand{\pdif}{\check{p}}

\newcommand{\curI}{J^\mathcal{I}}
\newcommand{\curHwI}{J^{\mathcal{H}}}

\newcommand{\resfr}{R}

\begin{document}

\begin{frontmatter}



\title{Weak chaos in the disordered nonlinear Schr\"odinger chain:
destruction of Anderson localization by Arnold diffusion}


\author{D. M. Basko}
\ead{denis.basko@grenoble.cnrs.fr}
\address{UJF-Grenoble 1/CNRS,
Laboratoire de Physique et Mod\'elisation des Milieux Condens\'es UMR~5493,
BP166, 38042 Grenoble, France}

\begin{abstract}
The subject of this study is the long-time equilibration dynamics
of a strongly disordered one-dimensional chain of coupled weakly
anharmonic classical oscillators. It is shown that
chaos in this system has a very particular spatial
structure: it can be viewed as a dilute gas of chaotic spots.
Each chaotic spot corresponds to a stochastic pump which drives the
Arnold diffusion of the oscillators surrounding it, thus leading
to their relaxation and thermalization.
The most important mechanism of relaxation at long distances is
provided by random migration of the chaotic spots along the chain,
which bears analogy with variable-range hopping of electrons in
strongly disordered solids. The corresonding macroscopic transport
equations are obtained.
\end{abstract}


\end{frontmatter}




\section{Introduction}
\label{sec:Intro}
\setcounter{equation}{0}

Anderson localization~\cite{Anderson1958} is a general phenomenon 
occurring in many linear wave-like systems subject to a disordered
background (see Ref.~\cite{Lagendijk2009} for a recent review).
It is especially pronounced in one-dimensional systems, where even
an arbitrarily weak disorder localizes all normal modes of the
system~\cite{Gertsenshtein1959,Mott1961}. 
Anderson localization in linear systems (single-particle problems
in quantum mechanics) has been thoroughly studied over the last
50 years; its physical picture is quite clear by now (although some
open questions still remain),
and even rigorous mathematical results have been established.
The situation is much less clear in the presence of
nonlinearities/interactions.

One of the simplest systems where the effect of a classical
nonlinearity on the Anderson localization can be studied, is
the disordered nonlinear Schr\"odinger equation (DNLSE) in one
dimension, discrete or continuous. It has attracted a lot of
interest in the last few years, including both stationary and
non-stationary~problems, as well as the problem of the dynamic
stability of stationary solutions~%
\cite{Frohlich1986,Gredeskul1992,Shepelyansky1993,Molina1994,%
Molina1998,%
Paul2005,Paul2007,Iomin2007,Bourgain2008,FishmanIomin2008,%
Kopidakis2008,Pikovsky2008,FishmanKrivolapov2008,Albert2008,%
Tietsche2008,FlachKrimer2009,Skokos2009,Fishman2009,Iomin2009,%
Veksler2009,Paul2009,Mulansky2009,Albert2010,Bodyfelt2010,%
Veksler2010,Krivolapov2010,Iomin2010,Skokos2010,Flach2010,%
MulanskyPikovsky2010,Laptyeva2010,Pikovsky2010,Johansson2010}.
This interest was especially motivated by the experimental
observation of the Anderson localization of light in disordered
photonic lattices%
~\cite{Schwartz2007,Lahini2008,Lahini2009}, where DNLSE
describes light propagation in the paraxial approximation,
and of cold atoms in disordered optical lattices%
~\cite{Billy2008,Roati2008,Deissler2010}, for which DNLSE
provides the mean-field description.
Other systems have also been studied, such as a
chain of anharmonic oscillators~%
\cite{Frohlich1986,
Kopidakis2008,DharLebowitz2008,DharSaito2008,FlachKrimer2009,%
Skokos2009,Skokos2010,Flach2010,Laptyeva2010,MulanskyAhnert2010},
a chain of classical spins~\cite{Frohlich1986,Oganesyan2009},
a nonlinear Stark ladder~\cite{GarciaMata2009,Krimer2009}.

In contrast to linear problems which can always be reduced to
finding the normal modes and the corresponding eigenvalues, the
problem of localization in a nonlinear system can be stated in
different inequivalent ways.
For example, one can study solutions of the stationary nonlinear
Schr\"odinger equation with disorder~%
\cite{Iomin2007,FishmanIomin2008,Bodyfelt2010}; however, in a
nonlinear system they are not directly related to the dynamics.
Another possible setting is the system subject to an external
perturbation or probe; studies of the transmission of a
finite-length sample where an external flow is imposed~%
\cite{Gredeskul1992,Molina1994,Molina1998,Paul2005,%
Paul2007,Tietsche2008,Paul2009,Albert2010},
or dipolar oscillations in an external trap potential%
~\cite{Albert2008}, fall into this category.
Most attention has been paid to the problem of spreading of an
initially localized wave packet of a finite norm~%
\cite{Shepelyansky1993,Molina1994,Molina1998,Bourgain2008,%
Kopidakis2008,Pikovsky2008,FishmanKrivolapov2008,Wang2009,%
FlachKrimer2009,Skokos2009,Fishman2009,%
Iomin2009,GarciaMata2009,Krimer2009,Veksler2009,Veksler2010,%
Krivolapov2010,Iomin2010,Skokos2010,Flach2010,MulanskyPikovsky2010}
(in linear systems it remains exponentially suppressed at long
distances for all times).
This setting corresponds directly to experiments~%
\cite{Schwartz2007,Lahini2008,Lahini2009,Billy2008,Roati2008}.
A closely related but not equivalent problem is that of
thermalization of an initially out-of-equilibrium system
(in a linear localized system thermalization does not occur).
The difference between these two settings is that in the latter
the total energy stored in the system is proportional to its
(infinite) size, while in the former the infinite system
initially receives a finite amount of energy.
The problem of thermalization has also received attention~%
\cite{DharLebowitz2008,DharSaito2008,Oganesyan2009,Mulansky2009},
and it is the main subject of the present work (although the
problem of wave packet spreading will also be briefly discussed).
%
Still, despite a large body of work, detailed understanding of the
equilibration dynamics in one-dimensional disordered nonlinear
systems is still lacking.

%
%
On the one hand, the direct numerical integration of the
differential equation shows that an initially localized wave
packet of a finite norm does spread indefinitely, its size
growing with time as a sub-diffusive power law~%
\cite{Shepelyansky1993,Molina1998,Kopidakis2008,Pikovsky2008,%
FlachKrimer2009,Skokos2009,Skokos2010,Flach2010,Laptyeva2010}.
%
Existence of different regimes of spreading, depending on whether
the system is in the regime of strong or weak chaos, has been
discussed~\cite{Flach2010,Laptyeva2010}.
%
%
Several authors suggested a macroscopic description of the
long-time dynamics in terms of a nonlinear diffusion-type
equation resulting in sub-diffusive spreading of the wave
packet~\cite{Flach2010,Iomin2010,MulanskyPikovsky2010};
however, arguments used to justify this equations are based on
an \emph{a priori} assumption of spatially uniform chaos.

On the other hand, it was argued that among different initial
conditions with a finite norm, at least some should exhibit
regular quasi-periodic dynamics, and thus would not spread
indefinitely~\cite{Frohlich1986,Bourgain2008,Johansson2010}.
For finite-size systems, scaling of the probability for the
system to be in the chaotic or regular regime has been studied
numerically~\cite{Pikovsky2010}. When scaling arguments are
applied to the results of the numerical integration of the
equation of motion, they indicate slowing down of the power-law
spreading~\cite{MulanskyPikovsky2010}. Rigorous mathematical
arguments support the conclusion that at long times the wave
packet should spread (if spread at all) slower than any power
law~\cite{Wang2009}.

The aim of the present work is the analysis 
of statistical properties of weak chaos in the discrete
one-dimensional disordered nonlinear Schr\"odinger equation
with the initial conditions corresponding to finite norm and
energy \emph{densities}. The main focus is on long-time dynamics
and relaxation at large distances, when no stationary superfluid
flow can exist in the system~\cite{Paul2007,Albert2010} and the
dynamics is chaotic.
It is often assumed that upon thermalization chaos has no spatial
structure, and all sites of the chain are more or less equally
chaotic; here it is argued not to be the case.\footnote{
We will call the motion of an oscillator the more chaotic the
faster it loses the phase memory.}
Namely, in the regime of strong disorder and weak nonlinearity
chaos is concentrated on a small number of rare chaotic spots
(essentially the same picture was also proposed in
Ref.~\cite{Oganesyan2009} for a disordered chain of coupled
classical spins, but it was not put on a quantitative basis).
A chaotic spot is a collection of resonantly coupled oscillators,
in which one can separate a collective degree of freedom (namely,
their relative phase whose dynamics is slow because of the
resonance), which performs chaotic motion. Under the conditions
of weak coupling between neighboring oscillators and weak
nonlinearity (i.~e., Anderson localization being the strongest
effect), the distance between different chaotic spots is much
greater than the typical size of the spot.
The stochastic motion of the collective degree of freedom of
each chaotic spot acts as a stochastic pump, i.~e. drives the
exchange of energy between other oscillators, non-resonantly
coupled to the spot, which corresponds to the Arnold diffusion.
This represents the main mechanism for relaxation and
thermalization of the oscillators, as well as for the transport
of conserved quantities (energy and norm). An important role is
played by the fact that chaotic spots can migrate along the
chain, as the collective degree of freedom may get in and out of
the chaotic region of its phase space. A similar phenomenon has
recently been proven to exist in a chain of weakly coupled
pendula~\cite{Kaloshin2010}. This migration also bears analogy
with the variable-range hopping of electrons in strongly disordered
solids~\cite{Mott1969,Kurkijarvi1973}. Still, there is an important
difference that the chaotic spot does not carry any energy or norm;
it only drives relaxation of oscillators surrounding it.

The main technical challenge in this work is the analysis of high
orders of the perturbation theory, which is necessary both to
separate the collective degree of freedom performing the chaotic
motion, and to couple other oscillators to this degree of freedom.
%
The perturbation theory is divergent, as chaos is a
non-perturbative phenomenon. The divergence occurs because of
overlapping resonances, according to Chirikov's criterion of chaos.
In this work, the probability for resonances to occur is estimated
in each order of  perturbation theory. The subsequent treatment of
each resonance and description of the associated chaotic motion
is based on the stochastic pump model of the Arnold diffusion,
which was proposed for systems with few degrees of freedom%
~\cite{Chirikov1979,Lichtenberg1983}. Thus, the present work is not
more rigorous than the stochastic pump model; it does not add
anything new to the current understanding of how chaos is generated
in nonlinear systems; it rather deals with the statistics of chaos
in a spatially extended system with an infinite number of degrees of
freedom and local coupling between them.



The main quantitative result of the present work is the system
of macroscopic equations which describe transport of the conserved
quantities (energy and norm) along the chain. Such tansport
determines equilibration of the chain at large distances, which 
occurs at long time scales. These equations are of
nonlinear-diffusion type, and explicit expressions for the transport
coefficients are derived. The macroscopic equations are valid at
distances exceeding a certain length scale, which is also explicitly
estimated.

The paper is organized as follows. In Sec.~\ref{sec:Model} the
model is formulated and the main assumptions are discussed.
Sec.~\ref{sec:Results} summarizes the main results.
In Sec.~\ref{sec:Picture} we qualitatively describe the physical
picture and the main ingredients of the solution; various
implications and some related issues are discussed. 
In Sec.~\ref{sec:Spots} we analyze the chaotic phase space in
few-oscillator configurations.
In Sec.~\ref{sec:Perturbation} perturbation theory is developed
and statistics of different terms is analyzed.
In Sec.~\ref{sec:Density} this perturbation theory is used to
study the statistics of the chaotic phase volume for an arbitrary
number of oscillators.
Sec.~\ref{sec:Pump} is dedicated to the analysis of the coupling
between chaotic spots and other oscillators, and to the description
of Arnold diffusion.
In Sec.~\ref{sec:Breaks} the macroscopic transport coefficients
are found.

\section{Model and assumptions}
\label{sec:Model}
\setcounter{equation}{0}

The main subject of this study is the one-dimensional discrete nonlinear
Schr\"odinger equation with diagonal disorder:
\begin{equation}\label{DDNLS=}
i\,\frac{d\psi_n}{d{t}}=
\omega_n\psi_n-\tau\Delta\left(\psi_{n-1}+\psi_{n+1}\right)
+g\psi_n^*\psi_n^2.
\end{equation}
Here the integer $n$ runs from $-\infty$ to $+\infty$ and labels sites
of a one-dimensional lattice. To each site a pair of complex conjugate
variables $\psi_n,\psi_n^*$ is associated.

Eq.~(\ref{DDNLS=}) together with its complex conjugate are the Hamilton
equations,
\begin{equation}
\frac{d\psi_n}{dt}=\frac{\partial{H}}{\partial(i\psi_n^*)},\quad
\frac{d(i\psi_n^*)}{dt}=-\frac{\partial{H}}{\partial\psi_n},
\end{equation}
corresponding to the classical Hamiltonian
\begin{equation}\label{Hpsipsi=}
H(\{i\psi_n^*\},\{\psi_n\})=
\sum_n\left[\omega_n\psi_n^*\psi_n
-\tau\Delta\left(\psi_n^*\psi_{n+1}+\psi_{n+1}^*\psi_n\right)
+\frac{g}2\,(\psi_n^*)^2\psi_n^2\right],
\end{equation}
where $i\psi_n^*$ is the canonical momentum conjugate to the
coordinate~$\psi_n$. For this $\psi_n$ must have a dimensionality
of $(\mathrm{action})^{1/2}$.
One can pass to another set of
canonical variables, the actions $I_n$ and the phases $\phi_n$, as
\begin{equation}
\psi_n=\sqrt{I_n}\,e^{-i\phi_n},
\end{equation}
so that the Hamiltonian becomes
\begin{equation}\label{HIphi=}
H(\{I_n\},\{\phi_n\})=
\sum_n\left[\omega_nI_n
-2\tau\Delta\sqrt{I_nI_{n+1}}\cos(\phi_n-\phi_{n+1})
+\frac{g}2\,I_n^2\right],
\end{equation}
and the equations of motion are
\begin{equation}
\frac{d\phi_n}{dt}=\frac{\partial{H}}{\partial{I}_n},\quad
\frac{d{I}_n}{dt}=-\frac{\partial{H}}{\partial\phi_n}.
\end{equation}
It is convenient to introduce the action-dependent frequency
of each oscillator,
\begin{equation}
\tilde\omega_n(I_n)=\omega_n+gI_n,
\end{equation}
to be distinguished from the bare frequencies $\omega_n$.

The first term on the right-hand side of Eq.~(\ref{DDNLS=}) represents
the diagonal disorder. The bare frequencies~$\omega_n$ are assumed to 
be independent random variables uniformly distributed over the interval
\begin{equation}
-\frac\Delta{2}\leqslant\omega_n\leqslant\frac\Delta{2}.
\end{equation}
The second term on the right-hand side of Eq.~(\ref{DDNLS=}) will be
referred to as tunneling.%
\footnote{May the reader not be confused by this ``quantum-mechanical''
term: here we are always dealing with a classical Hamiltonian system.
This term is the standard one in studies of Anderson localization on
a lattice.}
It is convenient to measure its strength with respect to the disorder,
so the dimensionless parameter~$\tau$ is introduced. Together with the
disorder term, the tunneling term constitutes the linear part of the
problem, whose normal modes are determined from the eigenvalue equation
\begin{equation}
(\omega-\omega_n)\psi_n+\tau\Delta(\psi_{n-1}+\psi_{n+1})=0.
\end{equation}
As we are dealing with a one-dimensional system, any eigenfunction has
an envelope, exponentially localized around some site~$\bar{n}$,
$\psi_n\sim{e}^{-\kappa|n-\bar{n}|/2}$. The smaller~$\tau$, the stronger
the localization. We assume
\begin{equation}\label{vepless1=}
\tau\ll{1}.
\end{equation}
In this case $\kappa\approx\ln(1/2e\tau)\gg{1}$,\footnote{This
estimate can be obtained, e.~g. from the Thouless relation between
the localization length and the density of states~\cite{Thouless1972}, 
approximating the latter by the box of the width~$\Delta$.}
so each eigenstate is essentially localized on a single site, with
weak perturbative tails on the neighboring sites.

The third term on the right-hand side of Eq.~(\ref{DDNLS=}) represents
the nonlinearity whose strength is governed by the coupling constant
$g>0$.\footnote{Generally speaking, the sign of~$g$ is not important
for the dynamics, since the change $g\to-g$ is equivalent to
$\psi_n\to(-1)^n\psi_n^*$, $\omega_n\to-\omega_n$; the latter has no
effect since the disorder distribution is symmetric. However, later
we will be interested in the thermodynamics of the chain, so it is
convenient to have the Hamiltonian bounded from below.}
As mentioned above, we are interested in the case when the nonlinearity
is weak. However, it would be meaningless to simply say that $g$~is
small. Indeed, Eq.~(\ref{DDNLS=}) is invariant under the change
\begin{equation}
\psi_n\to{C}\psi_n,\quad g\to\frac{g}{C^2},
\end{equation}
where $C$ is an arbitrary constant.
Thus, it is the product $g|\psi_n|^2$, the nonlinear frequency shift
of the $n$th oscillator, which can be compared to other frequency
scales.  This nonlinear shift is determined by~$g$, as well as by
the initial conditions for Eq.~(\ref{DDNLS=}). We will assume that
it is much smaller than the disorder and denote the corresponding
dimensionless small parameter by~$\rho$:
\begin{equation}\label{tauless1=}
\frac{g|\psi_n|^2}\Delta\sim\rho\ll{1}.
\end{equation}

Eq.~(\ref{tauless1=}) also expresses another assumption of the
present work, that the typical scale of $|\psi_n|^2$ is about
the same for all oscillators; in other words, there is no
significant correlation between $|\psi_n|^2$ and~$\omega_n$.
As an example of the opposite one could prepare an initial
condition where most of the norm $|\psi_n|^2$ is put on the
oscillators with $\omega_n$'s near the bottom of the band. This
case is more complicated (e.~g., it includes all physics related
to superfluidity, such as Bose condensate being the ground state
of the classical Hamiltonian~(\ref{Hpsipsi=}), as well as large
localization length for low-frequency phonons~\cite{Bilas2006});
it is beyond the scope of the present work. The criterion can be
conveniently formulated in terms of the two integrals of motion
of Eq.~(\ref{DDNLS=}):
the total energy, given by Eq.~(\ref{Hpsipsi=}), and the total
action (norm),
\begin{equation}
I_{tot}\equiv\sum_{n=-\infty}^\infty|\psi_n|^2.
\end{equation}
In the situation favoring oscillators with frequencies near the
bottom of the band one obtains $H\approx-(\Delta/2){I}_{tot}$,
since the dominant contribution to both $H$ and $I_{tot}$ comes
from the oscillators with $\omega_n$ close to $-\Delta/2$
(the coupling and the nonlinear terms in~$H$ are assumed to be
small). 
In the case with $|\psi_n|^2$ independent of~$\omega_n$ there is
a significant  cancellation between contributions
$\omega_n|\psi_n|^2$ to the Hamiltonian~(\ref{Hpsipsi=}) with
$\omega_n>0$ and $\omega_n<0$, so assuming all oscillators to be
excited on the same scale is equivalent to assuming
$|H|/{I}_{tot}\lesssim\Delta/2$. In the following, we will assume
the strong inequality,
\begin{equation}\label{HllIDelta=}
\frac{|H|}{I_{tot}}\ll\Delta,
\end{equation}
which is made for pure technical convenience only. It does not
affect the qualitative conclusions of the work, but simplifies
the calculations since it effectively decouples different
realizations of disorder and different initial conditions.

Finally, we will assume
\begin{equation}\label{Hless0}
H<0.
\end{equation}
It turns out that for certain values of $H$ and $I_{tot}$ the
system cannot thermalize, which can be seen from very basic 
thermodynamic arguments (see Ref.~\cite{Rasmussen2000} and
\ref{app:thermodynamics} for details). Condition~(\ref{Hless0})
in combination with Eqs.~(\ref{tauless1=}),~(\ref{HllIDelta=})
ensures that the system is not in that non-thermal region.

When all three terms on the right-hand side of Eq.~(\ref{DDNLS=})
are present, the equation has only two integrals of motion: the
total energy, $H$, and the total action, $I_{tot}=\sum_nI_n$.
However, if
$\tau=0$ or $g=0$,
the system is integrable. Thus, one
is free to choose any of the two as the perturbation destroying the
integrability. The most natural choice is the tunneling. Then, each
$I_n$ being an integral of motion at $\tau=0$,
Hamiltonian~(\ref{HIphi=}) is already written in the correct variables.
It may seem that this implies the condition $\tau\ll\rho$ (since
$\rho$ is the effective strength of the nonlinearity). Indeed,
as will be seen below (Sec.~\ref{sec:Spots}), some elements of the
physical picture admit a simpler quantitative description at
$\tau\ll\rho$. However, the main results of this paper are obtained
using the perturbation theory both in $\tau$ and $g$, and are
valid for any relation between $\tau$ and $\rho$, as long as both
are small (on the issue of smallness, see also Sec.~\ref{sec:smallness}).

\section{Main results}
\label{sec:Results}
\setcounter{equation}{0}

The first statement is that for general initial conditions,
satisfying the assumptions of the previous section, the system
locally thermalizes with a finite relaxation time.
Local thermalization means that on a finite, sufficiently long
segment of the chain [the corresponding length scale~$L_*$
is given explicitly below, Eq.~(\ref{Lstar=})], the actions
and the phases are distributed according to the grand canonical
distribution, $e^{-\beta(H-\mu{I}_{tot})}$. To respect the two
conserved quantities, the total energy~$H$ and action~$I_{tot}$
of the chosen segment, two Lagrange multipliers are introduced:
$\beta\equiv{1}/T$ is the inverse temperature, and $\mu$ has the
meaning of the chemical potential, although it has the
dimensionality of a frequency. The issue of thermalization is
discussed in more detail in Sec.~\ref{sec:thermalization}.

The values of $\mu$~and~$T$, which define the thermal
distribution on each segment, are determined by the total norm
and energy initially contained on this segment, that is, by the
initial conditions. It turns out that for some values of energy
and norm (namely, when the energy is too high for a given norm),
the corresponding $\mu$~and~$T$ do not exist, as discussed in
Ref.~\cite{Rasmussen2000} and \ref{app:thermodynamics}, so the
system may equilibrate in the microcanonical sense, but still
not be thermal. The assumption of negative total energy, made
in the previous section [Eq.~(\ref{Hless0})], ensures that
such problem does not arise, and the system indeed thermalizes.
However, even when the local thermalization has occurred,
generally speaking, $\mu$~and~$T$ still have different values
for different segments, depending on the initial conditions,
so the system is still not in the global equilibrium. The
results described in the following, concern precisely the global
equilibration of the system.

To account for the dependence of $\mu$ and $T$ on the position,
we introduce a continuous variable, $n\to{x}$, and assume a
dependence on the position, $T=T(x)$, $\mu=\mu(x)$, smooth on
the scale $L_*$. Let us define the smoothed macroscopic energy
and action densities,
\begin{subequations}\begin{eqnarray}
&&\mathcal{I}(x)=\frac{1}{L_*}\sum_{n=x-L_*/2}^{x+L_*/2}|\psi_n|^2,\\
&&\mathcal{H}(x)=\frac{1}{L_*}\sum_{n=x-L_*/2}^{x+L_*/2}
\left[\omega_n|\psi_n|^2
-\tau\Delta\left(\psi_n^*\psi_{n+1}+\psi_{n+1}^*\psi_n\right)
+\frac{g}2\,|\psi_n|^4\right].
\end{eqnarray}\end{subequations}
At each point~$x$ they are related to the local values of
$T(x),\mu(x)$ by the standard thermodynamics (see
\ref{app:thermodynamics}). It is convenient to define
dimensionless variables
\begin{equation}\label{rho=h=}
\rho\equiv\frac{g\mathcal{I}}{\Delta},\quad
h\equiv\frac{g\mathcal{H}}{\Delta^2}.
\end{equation}
Under the assumptions of the previous section, namely,
$\tau\ll{1}$, $\rho\ll{1}$, $|h|\ll\rho$, the thermodynamic
relations take a very simple form (the condition $|h|\ll\rho$
turns out to be equivalent to $\mu<0$, $|\mu|\gg\Delta$):
\begin{equation}\label{rhoh=Tmu}
\rho=\frac{gT}{|\mu|\Delta},\quad
h=\left(\frac{gT}{|\mu|\Delta}\right)^2
-\frac{1}{12}\frac{gT}{\mu^2}.
\end{equation}

In the global equilibrium, $T$~and~$\mu$ must be constant
over the entire chain. As the values of $\mu(x)$ and $T(x)$
are determined by the densities $\mathcal{I}(x)$~and
$\mathcal{H}(x)$, during the global equilibration norm and
energy must be transported along the chain. Thus, the
corresponding currents, $\curI$ and $\curHwI$, must be
flowing during the equilibration. The conservation laws for
the total energy and norm impose the corresponding continuity
equations:
\begin{eqnarray}\label{continuity=}
\frac{\partial}{\partial{t}}\left[\begin{array}{c}
\mathcal{I} \\ \mathcal{H}\end{array}\right]=
-\frac{\partial}{\partial{x}}
\left[\begin{array}{c} \curI \\ \curHwI \end{array}\right].
\end{eqnarray}
The currents must vanish for $T(x),\mu(x)=\mathrm{const}$.
Thus, they can be expanded in the gradients,
and only the first term is retained:
\begin{equation}\label{currents=}
\left[\begin{array}{c} \curI \\ \curHwI \end{array}\right]
=-\sigma(T,\mu)\,\frac{\partial}{\partial{x}}
\left[\begin{array}{c} \mu/T \\ -1/T \end{array}\right].
\end{equation}
$\sigma(T,\mu)$ is a $2\times{2}$ matrix of the transport
coefficients which will be called conductivity.
Since $-\mu/T$ and $1/T$ are thermodynamically conjugate to
$\mathcal{I}$ and $\mathcal{H}$, the matrix $\sigma$ is
symmetric by virtue of the Onsager relations.
The statement about the finiteness of~$\sigma$, equivalent
to the validity of the gradient expansion, and to the normal
character of diffusion in the system, is already not trivial.
Indeed, $\sigma$ could be zero or infinite, leading to anomalous
diffusion; in Ref.~\cite{Iomin2010} $\sigma=0$ was suggested.
Our result is that $\sigma$ is finite, and given by 
\begin{subequations}
\begin{equation}\label{sigmamuT=}
\begin{split}
&\sigma(T,\mu)\sim\frac{\Delta^3}{g^2}\,
\left[\begin{array}{cc} 1 &  O(\rho\Delta) \\
O(\rho\Delta) & O(\Delta^2) \end{array}\right]
\exp\left\{-\mathcal{C}'\left(\frac{\ln(1/\rho)}{\ln(1/\tau)}\right)
\ln^2\frac{1}{\tau^p\rho}\,\ln\frac{1}{\rho}\right\}.
\end{split}
\end{equation}
It depends on $T$ and $\mu$ only through the combination
$\rho=gT/(|\mu|\Delta)$, which is the consequence of the
condition $|\mu|\gg\Delta$, or, equivalently, of
Eq.~(\ref{HllIDelta=}).
The argument of the exponential is written with the logarithmic
accuracy.
The function $\mathcal{C}'(\ln(1/\rho)/\ln(1/\tau))$
is bounded from above and below,
\begin{equation}
\frac{1/3}{[1+\ln(1+x)]^2}\leqslant\mathcal{C}'(x)\leqslant
\frac{8}{[1+\ln(1+x)]^2}.
\end{equation}
Finally, the power~$p$ is bounded by
\begin{equation}\label{pest=}
\frac{1}2\leqslant{p}\leqslant{3}.
\end{equation}
The author has not been able to calculate $\mathcal{C}'(x)$ and
$p$ explicitly. Given this uncertainty, it would make little sense
to calculate slow prefactors. The prefactor $\Delta^3/g^2$ is put
to keep the correct dimensionality. Also, an order-of-magnitude
estimate for the relative magnitude of different matrix elements
of~$\sigma$ could be established.
\end{subequations}

Equations (\ref{rho=h=})
--(\ref{pest=}) form a
closed system of macroscopic equations describing action and
energy transport along the chain and constitute the main
quantitative result of the present work.
Eq.~(\ref{sigmamuT=}) is already sufficient to conclude that the
dependence of~$\sigma$ on the integrability-breaking parameters
$\tau$ and $\rho$ is stronger than any power law, but weaker than
a stretched exponential which would be the expected dependence of
the Arnold diffusion rate on the integrability-breaking
perturbation~\cite{Chirikov1979,Lichtenberg1983,Zaslavsky1985}.
This phenomenon is known as fast Arnold diffusion~\cite{Chirikov1993},
and arises because of a large number of degrees of freedom in the
system.

The above expression for $\sigma(T,\mu)$, Eq.~(\ref{sigmamuT=}),
does not imply averaging over many disorder realizations. In fact,
its inverse, $\sigma^{-1}$, is a self-averaging quantity. The length
scale $L_*$ at which this self-averaging occurs can be viewed as
boundary separating macroscopic and microscopic distances. It can be
estimated as
\begin{equation}\label{Lstar=}
\ln{L}_*\sim\mathcal{C}'\ln^2\frac{1}{\tau^p\rho}.
\end{equation}
Segments much longer than~$L_*$ contain a sufficient number of sites
to effectively replace the average over realizations. Transport
properties of a segment shorter than~$L_*$ are strongly
realization-dependent, and fluctuations are of the same order as
the average.
This situation is analogous to that with hopping conduction in
one dimension~\cite{Raikh1989}.

The macroscopic transport equations given above describe
equilibration of a chain whose average action $\mathcal{I}$ and
energy $\mathcal{H}$ per oscillator are finite.
Still, having written these macroscopic equations, one can always 
prepare an initial condition $\mathcal{I}(x),\mathcal{H}(x)$ in
the form of a finite-size packet, and see how it spreads according
to the equations. In \ref{app:spreading} the equations are shown
to give the size of the packet, $L(t)$, to grow as
\begin{equation}\label{Ltsim}
L(t)\sim{e}^{[(\mathcal{C}')^{-1}\ln{t}]^{1/3}}.
\end{equation}
This is slower than any power law, in agreement with
Ref.~\cite{Wang2009}, and in disagreement with Refs.~%
\cite{Shepelyansky1993,Molina1998,Kopidakis2008,Pikovsky2008,%
FlachKrimer2009,Skokos2009,Skokos2010,Flach2010,Laptyeva2010}.
Possible reasons for this disagreement will be discussed in
Sec.~\ref{sec:smallness}. Note, however, that as the packet
size $L(t)$ increases, the corresponding density inside the
packet, $\rho\propto{1}/L(t)$, decreases. Thus, the length
scale~$L_*$, given by Eq.~(\ref{Lstar=}), very quickly diverges
so after some time we will inevitably have $L(t)<L_*$. Since the
macroscopic equations are only valid at distances exceeding~$L_*$,
they lose their applicability as the packet spreads.

\section{Qualitative picture}
\label{sec:Picture}
\setcounter{equation}{0}

\subsection{Search for chaos}
\label{sec:SearchChaos}

Appearance of chaos in systems with a few degrees of freedom has
been thoroughly studied in the past%
~\cite{Chirikov1979,Lichtenberg1983,Zaslavsky1985}; the main
steps will be reviewed in Sec.~\ref{sec:Spots}.
The essential ingredients are the following.
\begin{enumerate}
\item
When an integer linear combination of several frequencies
$\tilde\omega_n(I_n)$ of the integrable system vanishes for some
values of~$I_n$, that is
$m_1\tilde\omega_{n_1}+\ldots+m_N\tilde\omega_{n_N}=0$
(guiding resonance, in Chirikov's terminology~\cite{Chirikov1979}),
one can separate a slow degree of freedom~$\tilde\phi$,
which is the corresponding linear combination of the phases:
$\tilde\phi=m_1\phi_{n_1}+\ldots+m_N\phi_{n_N}$.
Exactly at resonance this degree of freedom has no dynamics at
all (in the unperturbed system).
\item
If the perturbation has a term resonant with~$\tilde\phi$, the
latter acquires a non-trivial dynamics whose characteristic
frequency $\Omega$ is proportional to the square root of the
perturbation strength. The corresponding phase space has a
separatrix.
\item
Chaos arises upon destruction of this separatrix by another term
of the perturbation which oscillates at some characteristic
frequency~$\varpi$. The volume of the resulting chaotic region
around the separatrix in the phase space of the slow degree of
freedom (the so-called stochastic layer)
(i)~is proportional to the strength of the separatrix-destroying
perturbation (up to a logarithmic factor), and
(ii)~is exponentially small, $\sim{e}^{-\varpi/\Omega}$, when
$\varpi\gg\Omega$ (which is typically the case). Destruction of
the separatrix will be also reviewed in more detail in
Sec.~\ref{sec:2oscstoch}. 
\end{enumerate}
In the particular case of Eq.~(\ref{DDNLS=}) the perturbation
conserves the total action and energy, so the minimal number of
oscillators, necessary to generate chaos, is three.

To determine the main contribution to the chaotic phase
volume, we need (i)~to identify the guiding resonances which have
the largest resonant perturbation strength and thus the highest
frequency~$\Omega$; (ii)~to find the separatrix-destroying terms
of the perturbation which oscillate at the smallest possible
frequency~$\varpi$ and are not too weak (layer resonances, in
Chirikov's terminology~\cite{Chirikov1979}). It is important that
the mentioned perturbation terms include not only those directly
appearing in the Hamiltonian (\ref{Hpsipsi=}) or (\ref{HIphi=}),
but also terms of the kind
\begin{equation}\begin{split}
V&=\mathcal{K}\left(\psi_{n_1}^*\ldots\psi_{n_N}^*
\psi_{\bar{n}_1}\ldots\psi_{\bar{n}_N}
+\psi_{\bar{n}_1}^*\ldots\psi_{\bar{n}_N}^*\psi_{n_1}\ldots\psi_{n_N}\right)\\
&=2\mathcal{K}\sqrt{I_{n_1}\ldots{I}_{n_N}I_{\bar{n}_1}\ldots{I}_{\bar{n}_N}}\,
\cos\left(\phi_{n_1}+\ldots+\phi_{n_N}-\phi_{\bar{n}_1}
-\ldots-\phi_{\bar{n}_N}\right).
\label{Nperturbation=}
\end{split}\end{equation}
Such ``$N$-particle'' terms are effectively generated in high
orders of the perturbation theory in $\tau$~and~$g$, which is
developed in Sec.~\ref{sec:Perturbation}. They oscillate at
the frequency
$\varpi=\tilde{\omega}_{n_1}+\ldots+\tilde{\omega}_{n_N}%
-\tilde{\omega}_{\bar{n}_1}-\ldots-\tilde{\omega}_{\bar{n}_N}$,
which is easier to make small for larger~$N$, as discussed in
Sec.~\ref{sec:numres}. However, the corresponding
strength~$\mathcal{K}$ contains a high power of the coupling
constants.

Consider the simplest resonance, that of two oscillators:
\begin{equation}\label{twooscres=}
\omega_n+gI_n=\omega_{n'}+gI_{n'}.
\end{equation}
Since the typical values of actions $I_n,I_{n'}\sim{T}/|\mu|$,
this equation can be satisfied only if
\begin{equation}\label{twooscmismatch=}
|\omega_n-\omega_{n'}|\sim\frac{gT}{|\mu|}.
\end{equation}
Two arbitrary
sites have $|\omega_n-\omega_{n'}|\sim\Delta$, so for weak
nonlinearity, $gT/|\mu|\ll\Delta$ [Eq.~(\ref{tauless1=})],
one has to look for special pairs. If we fix~$n$, the appropriate
partner~$n'$ will be found typically at distances
$|n-n'|\sim|\mu|\Delta/(gT)\equiv{1}/\rho$. This distance gives
also the minimal order of the perturbation theory in the tunneling
which couples two such oscillators and produces a separatrix, so
that $\Omega\propto\sqrt{\tau^{1/\rho}}$. The thickness of the
corresponding stochastic layer, $\propto{e}^{-\varpi/\Omega}$, will
be very small. The same occurs if one tries an $N$-oscillator
resonance: even though the equation
\begin{equation}\label{Noscres=}
m_1(\omega_{n_1}+gI_{n_1})+\ldots+m_N(\omega_{n_N}+gI_{n_N})=0
\end{equation}
is easier to satisfy than Eq.~(\ref{twooscres=}), a high order of
the perturbation theory is required to couple these oscillators
altogether, and thus $\Omega$ is again proportional to a high power
of~$\tau$, which then enters as $e^{-\varpi/\Omega}$.

The ``cheapest'' way to produce a moderately
small~$\Omega$ is to find a pair of nearest neighbors, $n'=n\pm{1}$,
such that $|\omega_n-\omega_{n+1}|\sim{g}T/|\mu|$. Such pairs
are rare, they occur every $1/\rho$ sites on the average, but they have
$\Omega\propto\sqrt\tau$. The low density of these pairs gives a small
factor $\sim\rho$ in the total chaotic phase volume, but it is much
less dramatic than $\tau^{1/\rho}$ exponentiated. It is here
that the assumption $\tau\ll\rho$ is necessary. Indeed, in the
opposite case one would have to solve the
linear part of the problem first and find the two eigenfrequencies,
whose difference cannot be smaller than $\tau\Delta$ (the so-called
level repulsion). These eigenfrequencies should then be substituted
into Eq.~(\ref{twooscmismatch=}) which cannot be satisfied if
$\tau\gg\rho$. In this latter case the resonant pair should be made
out of oscillators separated by some distance~$l$, so that their
effective coupling is $\sim\tau^l\lesssim\rho$, which can always be
satisfied for a sufficiently large~$l$ (see also
Sec.~\ref{sec:Motttriples}).

Next, one has to identify the separatrix-destroying perturbation which
produces the thickest stochastic layer. If we simply consider coupling
between one of the sites of the resonant pair $n,n+1$ and another
oscillator~$n'$, the frequency of this perturbation is
$\varpi=|\tilde\omega_n-\tilde\omega_{n'}|$. Typically, this frequency
is $\sim\Delta$, so the effect of the perturbation is exponentially
small. Still, out of many available oscillators, it is possible to find
one with a sufficiently close frequency, so that $\varpi$ is also small.
However, this oscillator will be far away, so the stochastic layer
thickness will be proportional to a high power of~$\tau$. In high orders
of perturbation theory many-oscillators terms can also be generated.
Thus, one has to find an optimum between decreasing $\varpi$ in
$e^{-\varpi/\Omega}$ and multiplying it by a large power of~$\tau$.
An analogous optimization procedure has been considered by Chirikov
and Vecheslavov~\cite{Chirikov1993}.
For our problem the resulting width of the stochastic layer is
proportional to $e^{-\ln^2(1/\tau)}$.

Again, the way to bypass this smallness is to notice that the above
arguments concern a resonant pair in a typical surrounding. There is,
however, a small fraction of pairs which have a neighbor with a close
frequency, i.~e. \emph{resonant triples}. Clearly, it makes no sense
to tune the frequency of the third oscillator more precisely than the
frequency mismatch of the resonant pair, so the density of such
resonant triples is $\sim\rho^2$. Moreover, when all three frequencies
are close, the chaotic region of the phase space does not have to be
just a thin layer around the destroyed separatrix, but can be of the
size of the separatrix itself.
Thus, the chaotic region of the $\{I_1,I_2,I_3\}$ space is concentrated
around the straight line $\omega_1+gI_1=\omega_2+gI_2=\omega_3+gI_3$.
A resonant triple will be called a \emph{chaotic spot} if its dynamical
variables fall in the chaotic region. The probability of this latter
event,  which turns out to be $\sim\tau/\rho$ if $\tau\ll\rho$ and
$\sim{1}$ if $\tau\gg\rho$, multiplied by the density of resonant 
triples, $\sim\rho^2$, gives the density of the chaotic spots,
$\bar{w}\sim\min\{\tau\rho,\rho^2\}$. This is discussed in detail for
$\tau\ll\rho$ in Sec.~\ref{sec:3oscchaotic}.

\subsection{Arnold diffusion}

The key property of a chaotic spot is that
the Fourier transform
of $\psi_n(t)$ has a continuous frequency spectrum, as discussed in
Sec.~\ref{sec:spectrum} (see also Sec.~\ref{sec:correlations}).
When the stochastic layer around the separatrix is thin, the spectrum
can be evaluated explicitly via Melnikov-Arnold
integral~\cite{Melnikov1962,Arnold1964,Chirikov1979}.
The spectrum is peaked around $\tilde\omega_n$ with the characteristic
width~$\Omega$, and has an exponential tail,
$e^{-|\omega-\tilde\omega_n|/\Omega}$ away from~ $\tilde\omega_n$.
This should be contrasted with the case of an integrable system
where the Fourier spectrum of the motion is made of $\delta$-peaks with
zero width.

The chaotic spot exerts a random force on other oscillators, thereby
allowing them to exchange action and energy with the spot and among
themselves.
This exchange occurs by small random increments, so the relevant
picture is that of random diffusion in the multi-dimensional space of
actions $\{I_n\}$.  
This corresponds to the standard picture of stochastic pump driving
Arnold diffusion~\cite{Chirikov1979, Lichtenberg1983}.
This diffusion (i)~must respect the total action and
energy conservation laws, and (ii)~cannot ``turn off'' the chaotic
spot, since the dynamic variables of the spot oscillators cannot cross
the stochastic layer boundary~\cite{Chirikov1979}, so that the system
stays on the guiding resonance $\sum_nm_n\tilde\omega_n=0$.
The corresponding
diffusion equation is analyzed in Secs.~\ref{sec:Pumpdiffeq}
and~\ref{sec:relaxremote}.

To study the diffusion of an oscillator, located at large distance from
the nearest chaotic spot, one should find a suitable perturbation term
of the form~(\ref{Nperturbation=}) involving both this oscillator and
an oscillator belonging to the spot (and, typically, many others).
Again, one seeks (i)~to be close to resonance with the spot oscillation
frequency in order to avoid suppression $e^{-\varpi/\Omega}$, and
(ii)~to avoid too high orders of the perturbation theory,
since the diffusion coefficient is proportional to the perturbation
strength squared. Still, if the distance between the chaotic spot and
the oscillator is~$L$, the smallness $\tau^{2L}$ in the diffusion
coefficient is already guaranteed. This is discussed in
Sec.~\ref{sec:couplingremote}.

For a given number $N$ of oscillators, the conditions
\begin{equation}
\sum_{n=1}^NI_n=N\mathcal{I}\,,\quad
\sum_{n=1}^N\left(\omega_nI_n+\frac{g}2\,I_n^2\right)=N\mathcal{H}\,,
\quad\sum_{n=1}^Nm_n(\omega_n+gI_n)=0,
\quad I_1,\ldots,I_N>0\,,
\end{equation}
define a region on an $(N-3)$-dimensional hypersurface whose
$(N-3)$-dimensional volume is finite. Hence, it takes a finite
time for the system to explore this region and to establish the
micro-canonical distribution. This time, however, grows
exponentially with~$N$, as long as only one chaotic spot is
present. 

For a longer segment, such that it hosts two chaotic spots, the
oscillators which are close to one spot equilibrate among
themselves, and those close to the other spot -- among themselves,
much faster than equilibration between the two groups of
oscillators occurs. Indeed, in order to transfer action and energy
between the two groups, one has to find a perturbation term which 
couples oscillators from both groups to one of the spots. Since
the coupling decays exponentially with distance, the best way to
do this would be to couple oscillators which are near the boundary
between the two groups, i.~e., in the middle between the two spots.
As mentioned in the previous subsection, the typical distance
between  chaotic spots is $\sim{1}/\min\{\tau\rho,\rho^2\}$, so
the  characteristic time scale of equilibration is of the order of
$(1/\tau)^{1/\min\{\tau\rho,\rho^2\}}$, which is enormously long.
For the long chain with many spots it may become even worse if one
considers anomalously long segments of the chain, not containing
chaotic spots: even though such segments are rare, they efficiently
block the transport along the chain.

The above picture assumes the positions of chaotic spots to be fixed.
There is, however, a mechanism which provides a faster way of 
equilibration at long distances. Consider first a chaotic spot and the
closest resonant triple which is not chaotic, assuming $\tau\ll\rho$.
As discussed in Sec.~\ref{sec:SearchChaos}, the typical distance to
this triple, $\sim{1}/\rho^2$, is smaller than the distance to the next
chaotic spot, $\sim{1}/(\tau\rho)$. Thus, the diffusion coefficient
for the actions of the triple is small as $\tau^{1/\rho^2}$. As the 
triple explores its phase space in the course of Arnold diffusion, it
may hit its own chaotic region. Importantly, this requires the same 
amount of time as the equilibration itself.\footnote{
This is a property of the two-dimensional diffusion.
Indeed, as the chaotic region of the three-dimensional action
space of the triple is concentrated around the straight line
$\omega_1+gI_1=\omega_2+gI_2=\omega_3+gI_3$, it is sufficient
to determine the time it takes for the system to touch the
corresponding region in the two-dimensional section, perpendicular
to this line. Consider diffusion with the coefficient~$D$ on a
square $L\times{L}$, divided into small cells $a\times{a}$. It
takes the time $\sim{a^2}/D$ to move to an adjacent cell, and the
time $\sim{L}^2/D$ to cross the whole square. Thus, while crossing
the square, the system visits $\sim{L^2/a^2}$ cells, i.~e., a
fraction of the order of unity. In dimensionality higher than two,
only a small fraction of cells would be visited.}
Thus, a new chaotic spot is created, relatively close to the old one.
However, now each of them can diffuse out of its stochastic layer,
driven by the other one. If it happens to the old one, this means
that the system has effectively switched from one guiding resonance
to another one at the point of their intersection in the phase space. 
This switching is also accompanied by a jump of the chaotic spot in
the real space.

It is important to realize that this jump does not correspond to a
transfer of any action or energy: the chaotic spot does not carry
any action or energy by itself; these are stored in the numerous
oscillators surrounding the spot and transferred from one oscillator
to another by Arnold diffusion, while the chaotic spot only drives
this diffusion. Thus, to transfer action and energy between two
groups of oscillators, which already have equilibrated among
themselves, it is quicker to wait until the chaotic spot moves closer
to the boundary between the two groups and drives the Arnold diffusion
there, rather than doing it by diffusion driven from a remote spot.

If now the consideration is extended on triples which are not well
in resonance, but have a noticeable frequency mismatch, this mismatch
will translate into an activation barrier~$E$, so the time required
for a triple to reach its chaotic region will be longer than the
equilibration time by a factor $e^{E/T}$. However, since the typical 
activation barrier of an arbitrary triple is $\sim|\mu|\Delta/g$,
hopping to a typical nearest neighbor will only take the time
$e^{1/\rho}$, much shorter than $(1/\tau)^{1/\rho^2}$. This means
that resonant triples, on the one hand, host chaotic spots for most
of the time, but on the other, are not the ones that determine the
transport.
Thus, one has to return to the beginning of Sec.~\ref{sec:SearchChaos} 
and look  for guiding and layer resonances in a typical environment.
For this,  many-oscillator resonances have to be considered, and thus, 
high orders of the perturbation theory have to be analyzed.

\subsection{Variable-range hopping of chaotic spots}
\label{sec:spotVRH}

Even when the guiding resonance involves several oscillators, it is
still convenient to assign it to a single site of the chain (defined, 
e.~g., as the leftmost site participating in the resonance).
Then, each site~$n$ can be characterized by~$w_n$, the chaotic fraction
of the thermally-weighted phase volume, summed over all possible guiding
resonances involving the site~$n$ and sites to the right of it (see
Section~\ref{sec:Densitydefinitions} for details).
It is determined by the configuration of the static disorder
$\{\omega_{n'}\}$ around the site, but not by the values of the
dynamical variables (over the latter, the thermal average is performed). 
It represents the probability that at a given instant of time the
site~$n$ is hosting a chaotic spot, and plays the same role in the
migration of a chaotic spot as the thermal weight $e^{-E(x)/T}$ in the
Brownian motion of a particle in an energy landscape $E(x)$.%
\footnote{We emphasize again that the sum $\sum_nw_n$, equal to the
average total number of chaotic spots on the chain, is dominated by
rare ``good'' sites corresponding to resonant triples for $\tau\ll\rho$,
as discussed in Sec.~\ref{sec:SearchChaos}. Here we are interested in
migration of chaotic spots between these triples.}
Let us estimate the typical value of~$w$ in the typical environment.
Let $N_g$~and~$N_\ell$ be the number of oscillators involved in the
guiding and layer resonances, respectively. Roughly, 
\begin{equation}
w\sim\sum_{N_g,N_\ell=0}^\infty
\exp\left(-\frac{e^{-N_g}}\rho\right)
\exp\left(-\frac{e^{-N_\ell}}{(\tau^p\rho)^{N_g}}\right)
(\tau^p\rho)^{N_\ell}.
\end{equation}
The first exponential is just the thermal activation probability
$e^{-(H-\mu{I}_{tot})/T}$ for
a resonance like that in Eq.~(\ref{Noscres=}), where the smallest 
frequency mismatch that can be obtained out of $\sim{e}^{N_g}$ 
combinations of $N_g$~frequencies (we do not distinguish between
$2^{N_g}$, $e^{N_g}$, or any other $a^{N_g}$ with $a\sim{1}$),
is $\sim\Delta{e}^{-N_g}$.
The characteristic frequency of the separatrix motion is then
$\Omega\propto(\tau^p\rho)^{N_g}$, where $pN_g$ is the typical
order in~$\tau$, needed to couple all the oscillators. The value
of~$p$ is estimated in Sec.~\ref{sec:Density} as
$1/2\leqslant{p}\leqslant{3}$.
The second exponential corresponds to the exponential smallness of
the Melnikov-Arnold integral $e^{-\varpi/\Omega}$, where the smallest
frequency mismatch of the layer resonance is
$\varpi\sim\Delta{e}^{-N_\ell}$. Finally,
$(\tau^p\rho)^{N_\ell}$ is the strength of the layer resonance
perturbation, determining the width of the stochastic layer. The
leading exponential asymptotics of a sum of competing strong
exponentials is determined by the largest term of the sum, or by
the maximum of the exponent:
\begin{equation}\label{lambda1typ=}
\ln\frac{1}w\sim\min_{N_g,N_\ell}
\left(\frac{e^{-N_g}}\rho
+\frac{e^{-N_\ell}}{(\tau^p\rho)^{N_g}}
+N_\ell\ln\frac{1}{\tau^p\rho}\right)
\sim\ln^2\frac{1}{\tau^p\rho}\ln\frac{1}\rho\,,
\end{equation}
the optimal values being
$N_\ell\sim{N}_g\ln(1/\tau^p\rho)\sim\ln(1/\rho)\ln(1/\tau^p\rho)$.
Note that this already reproduces the main exponential part of the
$\tau,\rho$-dependence of $\sigma$ in Eq.~(\ref{sigmamuT=}).

Thus, to each site is associated a random quantity $w_n$. For different
sites, $n\neq{n}'$, $w_n$ and $w_{n'}$ are assumed to be independent%
\footnote{This is not obvious \emph{a priori}, given the extended nature
of the guiding and layer resonances. Note, however, that replacing
$\omega_{n_1}$ by $\omega_{n_1\pm{1}}$ in Eq.~(\ref{Noscres=}) already
changes the left-hand side by an amount $\sim\Delta$. Thus, the optimal 
guiding resonances corresponding to sites $n$~and~$n+1$ are likely to be
completely different.}.
Then, it is natural to apply the variable-range hopping picture,
proposed for electron transport in strongly disordered solids by
Mott~\cite{Mott1969}: upon changing the guiding resonance, the
chaotic spot typically shifts by a distance~$L$ which optimizes
the compromise between the possibility to find a site with a
large~$w_n$ (which is typically far away) and the exponential
suppression of the hopping rate by the factor $\tau^L$. It is also
known that in one dimension the transport is determined not by
typical hops, but by rare strong obstacles~\cite{Kurkijarvi1973}.
Such obstacles occur with a small probability, but in one dimension
they block the transport efficiently as they cannot be bypassed, in
contrast to higher dimensions. Following Ref.~\cite{Raikh1989}, we
will use the term ``breaks'' to denote such obstacles.\footnote{
Other terms have also been used in the literature for these objects,
such as ``blockades''~\cite{Shante1973} or ``weak
links''~\cite{Lee1984}. The term ``break''~\cite{Raikh1989,Rodin2009}
may not be the most precise, since the chain is not fully broken.
Still, since the approach of the present work is analogous to that
of Ref.~\cite{Raikh1989}, the same terminology is used.}
In the present problem, a break is a region of the chain where the
chaotic fraction~$w$ assumes anomalously small values on many
consecutive sites, i.~e., a region of the chain rarely visited by
chaotic spots.

To implement these arguments, one has to know the distribution of
the random quantities $w_n$. The standard variable-range-hopping
picture~\cite{Mott1969} assumes that $\ln(1/w)\equiv\lambda$ is
uniformly distributed. Here it is definitely not the case, so the
estimate of the typical value~(\ref{lambda1typ=}) is not sufficient;
one has to calculate the whole distribution function $p(\lambda)$.
This implies calculating the statistics of different terms of the
perturbation theory of an arbitrarily large order. The structure of
the perturbation theory for Eq.~(\ref{DDNLS=}) is discussed in
Sec.~\ref{sec:Perturbation}, and $p(\lambda)$ is calculated in
Sec.~\ref{sec:Density}. The result is given by the Gumbel
distribution:
\begin{equation}\label{plambda=}
p(\lambda)=-\frac{d}{d\lambda}\exp\left(-\mathcal{C}_1\rho
e^{[\mathcal{C}\ln^2(1/\tau^p\rho)]^{-1}\lambda}\right).
\end{equation}
Here $\mathcal{C}$ is again a bounded function of $\ln\rho/\ln\tau$.
It differs from $\mathcal{C}'$, which appears in Eq.~(\ref{sigmamuT=})
for~$\sigma$, by a factor of 2 at most. $\mathcal{C}_1$~is another
logarithmic function of $\tau$~and~$\rho$; it enters
Eq.~(\ref{sigmamuT=}) in the combination $\ln(1/\mathcal{C}_1\rho)$,
so it is beyond the logarithmic accuracy and has been omitted from
Eq.~(\ref{sigmamuT=}). Eq.~(\ref{plambda=}) is valid only for
sufficiently large~$\lambda\gg\ln^2(1/\tau^p\rho)$, so it cannot be
used to find the average value of $w=e^{-\lambda}$. Indeed, the
average is determined by smaller $\lambda\sim\ln(1/\tau\rho)$,
corresponding to resonant triples, which should be treated separately.
Given $p(\lambda)$ in the form~(\ref{plambda=}), it is clear that
the quantity which is uniformly distributed is the exponential under
the derivative. Thus, $\lambda_L^\mathrm{typ}$, the typical smallest
value of $\lambda$ which can be found within a segment of a
length~$L$, is found as
\begin{equation}\label{lambdaLtyp=}
\frac{1}L\sim{1}-\exp\left[-\mathcal{C}_1\rho
\left(e^{[\mathcal{C}\ln^2(1/\tau^p\rho)]^{-1}\lambda_L^\mathrm{typ}}
-1\right)\right],
\quad
\lambda_L^\mathrm{typ}=\mathcal{C}\ln^2(1/\tau^p\rho)
\ln\frac{1}{\mathcal{C}_1\rho{L}}\,.
\end{equation}
For $L\sim{1}$ this agrees with the estimate~(\ref{lambda1typ=}).

It is crucial that all the discussion of this subsection
is valid regardless of the relation between $\tau$~and~$\rho$.
Indeed, the typical order of the perturbation theory needed to form
the guiding resonance is $N_g\sim\ln(1/\rho)$. Then the effective
coupling between the oscillators of the guiding resonance,
$\sim(\tau^p\rho)^{\ln(1/\rho)}$, is always smaller than $\rho$,
regardless of~$\tau$.

\subsection{Macroscopic transport coefficients}
\label{sec:QualitativeMacroscopic}

To relate the chaotic fraction $w$ and the macroscopic
conductivity~$\sigma$, we use the fact that each segment between
neighboring breaks contains many chaotic spots (this will be
verified in the end of this subsection). Then, equilibration within
each segment occurs much faster than between different segments, so
each segment~$k$ of the length $L_k$ can be characterized by its
own values of the temperature and the chemical potential,
$\mu_k,T_k$, as well as the total action $I_{tot,k}$ and energy
$H_k$. These latter ones slowly change due to the currents through
the breaks $\curI_k,\curHwI_k$
(which can be viewed as the definition of the currents).
It is convenient to denote the corresponding 2-columns by
\begin{equation}
Q_k=\left[\begin{array}{c} I_{tot,k} \\ H_k \end{array}\right],\quad
J_k=\left[\begin{array}{c} \curI_k \\ \curHwI_k \end{array}\right],\quad
V_k=\left[\begin{array}{c} \mu_k/T_k \\ -1/T_k \end{array}\right],
\end{equation}
where the $k$th break is assumed to separate the $k$th and the
$(k+1)$st segments. Then energy and action conservation can be
written as
\begin{subequations}
\begin{equation}\label{dQkdt=}
\frac{dQ_k}{dt}=J_{k-1}-J_{k}.
\end{equation}
The current through the $k$th break can be related to
$\mu_k,T_k,\mu_{k+1},T_{k+1}$ with the help of the diffusion
equation in the space of actions $\{I_n\}$ (it is done in
Sec.~\ref{sec:Breaks}), and can be written as
\begin{equation}\label{Jk=RkVk}
J_k=R_k^{-1}(V_k-V_{k+1}),
\end{equation}
to the linear order in the potential difference $V_k-V_{k+1}$.
The validity of the linear response theory follows quite
generally from the diffusive character of the system dynamics
in the space of actions. This is equivalent to finiteness of
the coefficient $R_k^{-1}$, which is a symmetric $2\times{2}$
matrix and can be viewed as the conductance of the break. It
is determined by the whole profile $\{\lambda_n\}$ of the break,
$R_k=R(\{\lambda_n\})$, and is found in Sec.~\ref{sec:Breaks}.
Finally, we have a thermodynamic relation
between $I_{tot,k},H_k$ and $T_k,\mu_k$, discussed in
Sec.~\ref{sec:Results} and \ref{app:thermodynamics}
[see Eqs.~(\ref{ImuT=}),~(\ref{HmuT=})], which we write here as
\begin{equation}\label{Qk=LkqVk}
Q_k=L_k\,\mathcal{Q}(V_k).
\end{equation}
The explicit form of the function $\mathcal{Q}(V)$ follows from the
discussion of Sec.~\ref{sec:Results} and \ref{app:thermodynamics},
but is not important here.
\end{subequations}

\begin{figure}
\begin{center}
\includegraphics[width=10cm]{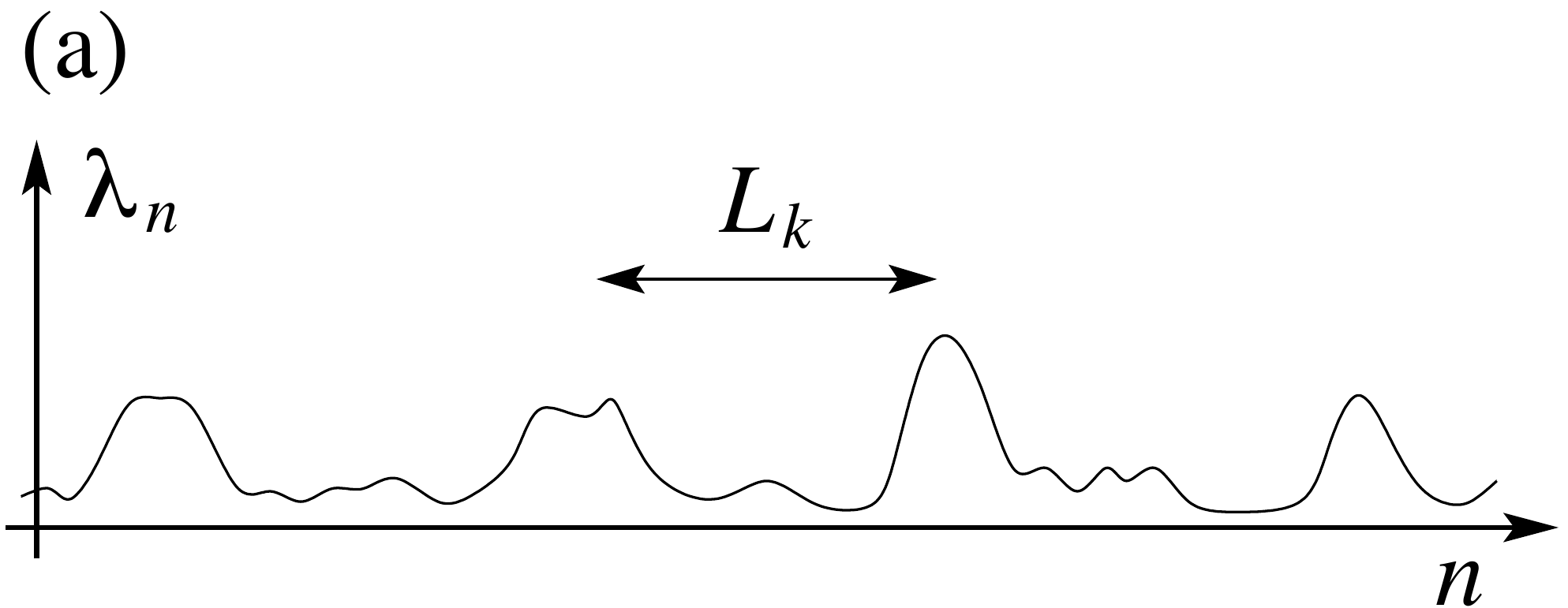}
\includegraphics[width=10cm]{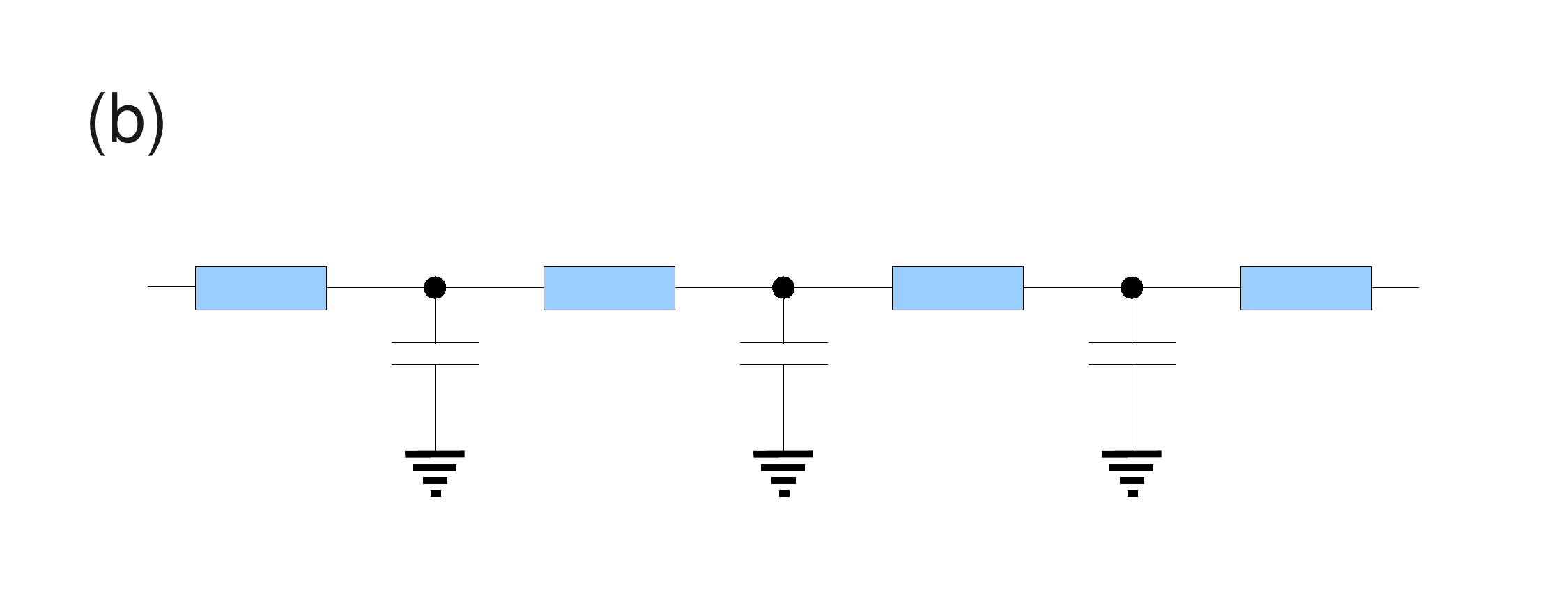}
\end{center}
\caption{\label{fig:circuit}
(a)~A~realization of $\lambda_n\equiv\ln(1/w_n)$, where $w_n$
is the chaotic fraction of the thermally weighted phase volume,
associated with the site~$n$ (see details in the text,
Sec.~\ref{sec:spotVRH}). Four breaks are shown.
(b)~The equivalent electric circuit with resistors and
nonlinear capacitors (the bottom plate of each capacitor
is grounded).}
\end{figure}

Now we pass to macroscopic equations, valid at large distances
$L\gg{L}_k$. Eqs.~(\ref{dQkdt=}) and (\ref{Qk=LkqVk}) lead to
the continuity equation,
$\partial\mathcal{Q}(V)/\partial{t}=-\partial{J}/\partial{x}$.
To relate the current~$J$ to the voltage gradient, we note that
in a stationary situation, when a constant uniform current~$J$
flows through the chain, the potentials can be found from 
Eq.~(\ref{Jk=RkVk}) as $V_k=-\sum_{k'<k}R_{k'}J$. The potential
drop over a distance~$L$ including many segments determines the
macroscopic gradient:
\begin{equation}\label{dVdx=}
\frac{\partial{V}}{\partial{x}}\approx\frac{V(x+L/2)-V(x-L/2)}{L}=
-\frac{\sum_kR_k}{L}\,J =-\sigma^{-1}J.
\end{equation}
Substituting the expression for the current in the continuity
equation, we obtain the macroscopic transport equation:
\begin{equation}\label{dQdVdVdt=}
\frac{d\mathcal{Q}(V)}{dV}\,\frac{\partial{V}}{\partial{t}}=
\frac\partial{\partial{x}}\,\sigma(V)\,\frac{\partial{V}}{\partial{x}}.
\end{equation}

Eqs.~(\ref{dQkdt=})--(\ref{Qk=LkqVk}) are formally equivalent to
those which describe an electric circuit, shown in
Fig.~\ref{fig:circuit}(b).
Namely, each segment corresponds to a capacitor, each break to a
resistor, while $Q_k$ and $V_k$ correspond to electric charge and
voltage. The only difference is that instead of the usual linear
relation $Q_k=C_kV_k$, where $C_k$ is the capacitance, we have a
nonlinear relation~(\ref{Qk=LkqVk}). In Eq.~(\ref{dQdVdVdt=})
$d\mathcal{Q}/dV$ plays the role of the macroscopic capacitance per
unit length, and $\sigma$~is the macroscopic conductivity.
The physical reason for this analogy is that the theory of electric
circuits describes nothing else but the transport of a conserved
quantity (electric charge) driven by the gradient of the
corresponding thermodynamically conjugate variable (voltage).
In our problem the phenomenology is the same.

Eq.~(\ref{dVdx=}) expresses the usual addition rule for resistors
in series. Since $\sigma^{-1}$ is proportional to the sum of
resistances of individual breaks~$R_k$, and these resistances are
independent due to large spatial separation between breaks, then
by the law of large numbers $\sigma^{-1}$ is self-averaging at long
distances. Thus, we can write
\begin{equation}\label{sigmaRdP=}
\sigma^{-1}=\int{R}(\{\lambda_n\})\,dP(\{\lambda_n\}),
\end{equation}
where $dP(\{\lambda_n\})$ is the probability measure (per unit length)
of the break configurations $\{\lambda_n\}$, determined by product of
independent measures~(\ref{plambda=}).
The integral in Eq.~(\ref{sigmaRdP=}) is mostly determined by the
vicinity of a certain configuration which optimizes the strong
competition between the exponentially growing $R(\{\lambda_n\})$
and quickly decreasing $dP(\{\lambda_n\})$ (Fig.~\ref{fig:optbreak}).
$R(\{\lambda_n\})$ is determined by the diffusion coefficient for
oscillators in the break region. This
diffusion coefficient contains the same small factors as the chaotic
fraction~$w_n$, so they are strongly correlated. We take this into
account by assuming the diffusion coefficient to be proportional to
${e}^{-\mathcal{C}_2\lambda_L^\mathrm{typ}}$,
$1\leqslant\mathcal{C}_2\leqslant{2}$
(see Sec.~\ref{sec:couplingremote}). Calculation of the integral in
Eq.~(\ref{sigmaRdP=}) is performed in Sec.~\ref{sec:Breaks}, and is
quite analogous to the optimization procedure of Ref.~\cite{Raikh1989}.
The result is given by Eq.~(\ref{sigmamuT=}).
%
The self-averaging of $\sigma^{-1}$ occurs at distances exceeding
the typical distance between the optimal breaks,
$L_*\sim{e}^{\mathcal{C}\ln^2(1/\tau^p\rho)}$.
Thus, it is $L_*$ that determines the border
between the microscopic and macroscopic scales, Eq.~(\ref{Lstar=}).
Note that this distance is indeed much greater than the typical
distance between chaotic spots (a power of $1/\tau,1/\rho$).

\subsection{Various remarks}

\subsubsection{On the decay of correlations}
\label{sec:correlations}

Let us see how the difference between oscillators which belong to a
chaotic spot and the rest of oscillators manifests itself in their
dynamics. One can study the two-time correlator:
\begin{equation}
\tilde{A}_{nn}(t)=
\lim_{t_0\to\infty}\int_0^{t_0}\frac{dt'}{t_0}\,\psi_n(t'+t)\,\psi_n^*(t'),
\end{equation}
or its Fourier transform, $A_{nn}(\omega)$, which can be called the
spectral function.
For uncoupled oscillators (at $\tau=0$) the correlator simply
oscillates as
$\tilde{A}_{nn}(t)=I_n{e}^{-i(\omega_n+gI_n)t}$, without any decay,
so $A_{nn}(\omega)$ is proportional to a $\delta$-function. 

For a chaotic spot residing on a resonant triple, the phase of each
of its three oscillators is randomized on the time scale of the destroyed
separatrix, $t_c\sim{1}/\Omega$, which should give the time scale of
decay of correlations (we neglect the small residual correlations
found in Ref.~\cite{Khodas2000} for the standard map).
Let us now consider an oscillator which is far away from a chaotic spot.
Its phase is randomized due to fluctuations of the action~$I_n$, which
translates into fluctuations of the frequency
$\tilde\omega_n(I_n)=\omega_n+gI_n$. Namely,
\[
\psi_n(t'+t)\,\psi_n^*(t')\propto
\exp\left\{-i\int\limits^t_{t'}\left[\omega_n+gI_n(t')+g\,\delta{I}_n(t'')
\right]dt''\right\},
\]
where $\delta{I}_n(t'')=I_n(t'')-I_n(t')$ is the random increment
of the action. For the diffusive dynamics
$\delta{I}_n(t'')\sim\sqrt{D_n|t''-t'|}$, where
$D_n$~is the diffusion coefficient. The correlation is suppressed
when the random part of the phase becomes $\sim{1}$, so it is natural
to estimate the correlation time as $t_c\sim(g^2D_n)^{-1/3}$.
The diffusion coefficient is exponentially suppressed at least as
$\tau^{2L}$, where $L$~is the distance to the nearest chaotic spot,
so the correlation time is exponentially longer than the time
$1/\Omega$ for the resonant triple.

However, it would be wrong to associate thus obtained~$t_c$ with the
decay time of $\tilde{A}_{nn}(t)$ and $1/t_c$ with the width of the
spectral peak in $A_{nn}(\omega)$. Indeed, the peak is supposed to be
centered at the oscillator frequency; but as the latter depends on
$I_n$, contributions from different times~$t'$ correspond to $I_n$'s
differing on the thermal scale $T/|\mu|$, so the peak is smeared over
$\sim{g}T/|\mu|=\rho\Delta$. This smearing is at least of the same order
as $\Omega\sim\Delta\sqrt{\min\{\tau\rho,\rho^2\}}$. Thus, on the one
hand, the spectrum of $A_{nn}(\omega)$ is broadened for all oscillators,
and has no infinitely sharp $\delta$-peaks, so the motion of all
oscillators is chaotic. On the other hand, $A_{nn}(\omega)$ does not
distinguish between oscillators belonging to chaotic spots and other
oscillators.

To eliminate this ``inhomogeneous broadening'' effect, one can use
the conditional average
\begin{equation}
\tilde{A}_{nn}(t|I)=
\lim_{t_0\to\infty}\frac{\int_0^{t_0}dt'\,\psi_n(t'+t)\,\psi_n^*(t')\,%
\delta(I_n(t')-I)}
{\int_0^{t_0}dt'\,\delta(I_n(t')-I)}.
\end{equation}
Then the main peak in $A_{nn}(\omega|I)$ would be positioned at
$\omega=\omega_n+gI$, and broadened by $1/\Omega$ for oscillators
belonging to resonant triples, or by an exponentially small value
$(g^2D_n)^{1/3}$ for other oscillators. Note that the spectrum
$A_{nn}(\omega|I)$ also contains weak satellite peaks at different
frequencies due to perturbative coupling to other oscillators.

Another way to elimitate the ``inhomogeneous broadening'' effect is
to consider a four-time correlator,
\[
\tilde{C}_{nnnn}(t)=\lim_{t_0\to\infty}\int_0^{t_0}\frac{dt'}{t_0}\,
\psi_n(t'+t)\,\psi_n^*(t')\,\psi_n^*(t')\,\psi_n(t'-t),
\]
where the regular oscillation simply cancels out. For uncoupled
oscillators this quantity is simply given by $I_n^2$, while in the
coupled system it decays on a time scale $t_c$ which is fast or slow
for oscillators belonging or not to a chaotic spot, respectively,
as discussed above.
Accounting for the fact that a given oscillator can belong to
a chaotic spot just for some intervals of time results in the
correlator $\tilde{C}_{nnnn}(t)$ having both fast- and slow-decaying
components with the corresponding weights.

\subsubsection{On the thermalization of oscillators}
\label{sec:thermalization}

Let us define thermalization of oscillators for a given trajectory of
the system in the phase space $\{I_n(t),\phi_n(t)\}$.
Let $N$~oscillators $n_1,\ldots,n_N$ be called thermalized if the
probability to find them in a given point of the phase space in the
course of dynamics is given by the product of independent Gibbs
distributions: 
\begin{equation}\label{thermalization=}
\lim_{\vep\to{0}}\lim_{t_0\to\infty}
\int\limits_0^{t_0}\frac{dt}{t_0}\,
\prod_{k=1}^N\delta_\vep(I_{n_k}(t)-I_{n_k})\,
\delta_\vep(\phi_{n_k}(t)-\phi_{n_k})
=\prod_{k=1}^N
\frac{e^{-\beta(\omega_{n_k}-\mu)I_{n_k}-\beta{g}I_{n_k}^2/2}}%
{2\pi\int{e}^{-\beta(\omega_{n_k}-\mu)I-\beta{g}I^2/2}\,dI}+O(\tau).
\end{equation}
Here $\delta_\vep$ denotes the $\delta$-function of a small but finite
width~$\vep$.

Clearly, this must hold for oscillators which do not belong to a chaotic
spot for most of the time, as their actions diffuse independently. 
However, this is not the case for oscillators which actually provide the
main contribution to the chaotic fraction. Consider, for example, a
resonant triple on the sites $n=1,2,3$ at $\tau\ll\rho$. When it hosts
a chaotic spot, the three actions stay close to the line
$\omega_1+gI_1=\omega_2+gI_2=\omega_3+gI_3$, so their joint probability
distribution does not split into a product. Even though the chaotic spot
eventually leaves the triple, and the three oscillators can diffuse
independently for some time, the relative contribution of such
``non-Gibbsian'' time intervals to the integral in
Eq.~(\ref{thermalization=}) may be significant.

Thus, the statement about thermalization of the chain is approximate.
Namely, there is a small fraction of oscillators for which the Gibbs
distribution is not valid. Nevertheless, it does not affect the 
conclusions of Sec.~\ref{sec:Results} and \ref{app:thermodynamics}. 
Indeed, when one
calculates the total action and energy, their relative contribution
is $\sim\min\{\tau\rho,\rho^2\}$, and thus is already beyond the
precision of Eqs.~(\ref{ImuT=}), (\ref{HmuT=}). Still, the situation
is somewhat counter-intuitive: the ``most chaotic'' oscillators turn
out to be the ``least thermal''.

\subsubsection{On the Kolmogorov-Arnold-Moser theorem}
\label{sec:KAM}

Thermalization of the oscillator chain for general initial conditions,
discussed above, seems to contradict the Kolmogorov-Arnold-Moser (KAM)
theorem~\cite{Kolmogorov1954,Moser1962,Arnold1963}. According to the
theorem, as the strength of the integrability-breaking perturbation
$\tau\to{0}$, the fraction of the chaotic part of the phase space must
vanish. At sufficiently small~$\tau$ most of the phase space must be
occupied by the invariant tori, where no thermalization is possible,
while here thermalization is argued to be the case for general initial
conditions. There is, however, an essential difference between systems
with finite and infinite number of degrees of freedom.

As discussed in the previous subsections, the typical spatial density
of the chaotic spots is $\bar{w}\sim\min\{\tau\rho,\rho^2\}$. This
means that the probability for a chain of length~$L$ to contain no
chaotic spots at all, is given by $e^{-\bar{w}L}$. This is just the
fraction of the total phase volume, occupied by the KAM tori. Even 
though it goes to unity as $\tau\to{0}$ or $g\to{0}$ for any
fixed~$L$, if one lets $L\to\infty$ first, it vanishes. Equivalently, 
even though one can prepare the initial conditions such that the
chaotic dynamics is absent, they form a set of zero measure.

Infinite number of degrees of freedom by itself is not sufficient to
destroy the regular regions of the phase space. There are two more
conditions: (i)~disorder and (ii)~extensive norm, and both are crucial.
Indeed, early studies on one-dimensional systems of interacting
particles have shown that even though the proportion of the regular
regions tended to decrease as the number of particles~$N$
increased~\cite{Froeschle1975}, it remained finite even at
$N\to\infty$ as energy per particle was kept finite, provided
that interactions were sufficiently
short-ranged~\cite{Diana1976,Casartelli1976}. At the same time,
for disordered systems invariant tori have been shown to exist%
~\cite{Frohlich1986,Poschel1990,Johansson2010}, but they corresponded
to a finite norm.

One can view the collective effect of the infinite length, randomness,
and extensive norm in the following way. Let us consider finite segments
of~$N$ neighboring oscillators: $n=n_0+1,\ldots,n_0+N$. Then, by simply
varying $n_0$ one will always find a combination of frequencies
$\omega_{n_0+1},\ldots,\omega_{n_0+N}$, arbitrarily close to some
given one, and the norm density on that segment will not be vanishingly
small (i.~e., bounded from below with a bound independent of~$n_0$). 

\subsubsection{On the effects of quantization}
\label{sec:quantum}

The most essential element of the picture developed in this work
is the chaotic spot -- a collection of a \emph{finite} number of
oscillators, whose frequency spectrum has a continuous component.
Because of its the continuous spectrum, the chaotic spot can
redistribute arbitrarily large amounts of energy between the
oscillators (limited only by the total amount of the system's
energy, which is infinite for the infinite chain), given enough
time.
In quantum mechanics, the energy spectrum of a system of any
finite number of oscillators is always discrete, so the amount
of energy which oscillators can exchange is finite even for
infinite time. In such a system, irreversibility can arise only
when the system has a possibility to visit an infinite number of
discrete quantum states in an infinite sequence of tunneling events.
A necessary (but not sufficient) condition for this is that the
coupling between different states should be strong enough as
compared to the energy mismatch between these states (this mismatch
is always finite when the levels are discrete). The appropriate
physical picture is that of Anderson localization-delocalization
transition~\cite{Anderson1958}. However, now it should be considered
not in the one-dimensional space of the chain, but in the space of
excitations of the many-oscillator system.

The quantum version of Arnold diffusion for a few oscillators was
considered in Refs.~\cite{Leitner1997,Demikhovskii2002}. The minimal
number of oscillators giving the Arnold diffusion in each model was
taken, so the quantum system was effectively one-dimensional
(moreover, of a finite length, since the number of oscillators was
finite). Since in one-dimensional quantum problems with disorder all 
eigenstates are inevitably localized, the conclusion was that Arnold 
diffusion in a quantum system is always stopped by Anderson
localization after a sufficiently long time. 

In the present problem the Arnold diffusion occurs in a system with
many degrees of freedom, so one may expect the effective
dimensionality of the quantum localization problem to be very high.
In this situation, Anderson transition can take place. The control
parameter for this transition would be the measure of ``quantumness''
of the system, i.~e. the Plank constant~$\hbar$ (of course, the
measure of ``quantumness'' can also be expressed in terms of the
physical parameters of the system at fixed $\hbar$, as discussed
below). Namely, for $\hbar$ not small enough thermalization and
transport in the system can be completely blocked by Anderson
localization. This transition would then be of the same nature as
the many-body localization transition in a one-dimensional system
of interacting bosons subject to disorder~\cite{Aleiner2010}.

Indeed, quantization of the Hamiltonian~(\ref{Hpsipsi=}) corresponds
to replacing the canonical variables $\psi_n,i\psi^*_n$ by operators
$\hat\psi_n,i\hat\psi_n^\dagger$ with the standard commutation 
relation, $[\hat\psi_n,i\hat\psi_n^\dagger]=i\hbar$. The resulting
quantum Hamiltonian corresponds to the disordered Bose-Hubbard model.
This model describes quantum bosonic particles, which interact with
each other via contact potential, and whose motion is confined to an
array of potential wells coupled by tunnelling.
Then $\hat\psi_n^\dagger/\sqrt\hbar$ and $\hat\psi_n/\sqrt\hbar$
are creation and annihilation operators for the particle in the
$n$th potential well, $\hbar\omega_n$ is the energy of the bound
state in the $n$th well, $\hbar\tau\Delta$ is the tunnelling matrix
element between bound states in the neighboring wells,
and $\hbar^2g$ corresponds to the Hubbard~$U$, the interaction
energy between two particles occupying the same potential well.
The classical limit, formally corresponding to
$\hbar\to{0}$, can be also expressed in terms of physical quantities.
If one starts from the Bose-Hubbard model with a given disorder
energy bandwidth~$W$ (corresponding to $\hbar\Delta$) at a
temperature~$T$, the classical limit at fixed~$W$ corresponds to
$T/W\to\infty$, $U/W\to{0}$, $UT/W^2=\mathrm{const}\ll{1}$.

The quantum bosonic system considered in Ref.~\cite{Aleiner2010}
corresponded to the continuum limit of the Bose-Hubbard model:
the lattice spacing $a\to{0}$, the site number $|n|\to\infty$,
maintaining $x=na=\mathrm{const}$, and also $\tau\to\infty$,
while keeping $\tau{a}^2=\mathrm{const}$.
Thus, no direct correspondence between the systems studied in present
work and that of Ref.~\cite{Aleiner2010} can be established.

\subsubsection{On the Mott's law for variable-range hopping}
\label{sec:onMott}

Taking the typical value, $\lambda_L^\mathrm{typ}$,
from Eq.~(\ref{lambdaLtyp=}), and maximizing
$\tau^Le^{-(1+\mathcal{C}_2)\lambda_L^\mathrm{typ}}$ with respect
to~$L$ gives the optimal distance for the hopping of the chaotic
spot $L_{VRH}\sim\ln^2(1/\tau^p\rho)/\ln(1/\tau)$, and the
variable-range-hopping (VRH) estimate for $\sigma$ in the spirit
of Mott~\cite{Mott1969}. This estimate is not supposed to be valid
in one dimension because of rare breaks~\cite{Kurkijarvi1973}.
In our case, however, the only
difference of this estimate from expression~(\ref{sigmamuT=}) is in
the factor $\mathcal{C}'$, while the main functional form is the
same (which still means that the resistance of a break is much
greater than that of a long segment between breaks).
This is due to the dramatic suppression (double exponential) of
the probability $p(\lambda)$, Eq.~(\ref{plambda=}), at very
large~$\lambda$, which makes too strong breaks very improbable
(in the standard Mott's picture $\lambda$~has a uniform
distribution, corresponding to a constant density of states).

It may also seem surprising that the picture of hopping transport,
which usually gives the temperature dependence in the form of a
stretched exponential ($e^{-1/T^{d+1}}$ in the dimensionality $d>1$
\cite{Mott1969} and $e^{-1/T}$ in $d=1$~\cite{Kurkijarvi1973}),
translates here into $e^{-\ln^3(1/T)}$. The reason for this is that
the hopping object, the chaotic spot, can be viewed as having many
internal degrees of freedom. Indeed, on every site optimization
is performed over a large number of available guiding resonances
to minimize the frequency mismatch.
As a result, the usual stretched exponential is stretched so much in
this case, that its argument is logarithmic rather than power-law.
This large number of internal degrees of freedom is also the reason
for the strong suppression of $p(\lambda)$ at very large~$\lambda$.

%
%

\subsubsection{On the smallness of small parameters}
\label{sec:smallness}

As mentioned in Sec.~\ref{sec:Results}, the results of the numerical
integration of Eq.~(\ref{DDNLS=})
\cite{Shepelyansky1993,Molina1998,Kopidakis2008,Pikovsky2008,%
FlachKrimer2009,Skokos2009,Skokos2010,Flach2010,Laptyeva2010}
correspond to much faster dynamics than predicted in the present work. 
Moreover, the observed behavior could be understood assuming that upon
thermalization of the expanding cloud, the chaos uniformly spreads
over all oscillators of the cloud~\cite{Skokos2009}. Again, this is 
quite different from the picture of rare chaotic spots. The average
density of the chaotic spots was estimated in the present work as
$\bar{w}\sim\min\{\tau\rho,\rho^2\}$, and the numerical coefficient
was not obtained. Thus, the chaotic spot picture should be valid when
$\tau$ and~$\rho$ are \emph{sufficiently} small.
The smallest values of $\tau$ and $\rho$, for which the direct
numerical integration of Eq.~(\ref{DDNLS=}) produced reliable
results on the packet spreading,
are $\tau=1/40$ and $\rho$ about~1~\cite{FlachTBP}.
So the question arises, starting from which values of $\tau$
and~$\rho$ will the chaotic spot picture really work?

One of the main physical assumptions used in the present work is that
most of the normal modes of the linear problem are localized on one
site of the lattice.
Numerical studies of the linear part of Eq.~(\ref{DDNLS=}) have shown
that at $\tau=1$ the localization length is about 100, and it becomes
$\sim{1}$ only at $\tau\approx{0}.1$~\cite{Czycholl1981,Kappus1981}. 
This means that the disorder can be considered strong only for
$\tau<0.1$ at most.

In Ref.~\cite{Pikovsky2010}, chains of finite length~$L$ were
considered, and the probability for the chain to be in the regular
KAM regime was analyzed numerically. At low densities, this
probability was found to scale as
$\exp[-10^6(2\rho)^{9/4}(2\tau)^{35/16}(L/16)]$.
In the present work, it is interpreted as the probability to have
no chaotic spots on the segment, $e^{-\bar{w}L}$, as discussed in
Sec.~\ref{sec:KAM}. Thus, the results of Ref.~\cite{Pikovsky2010}
approximately correspond to $\bar{w}\approx{10}^6\tau^2\rho^2$.
Apart from the difference from both $\bar{w}\propto\tau\rho$ and
$\bar{w}\propto\rho^2$, obtained in the present work (the reasons
for this discrepancy are unclear at the moment, so this issue
deserves to be studied in the future), one should
note the presence of the large numerical coefficient $10^6$.
This means that for the chaotic spots to be rare very small values
$\tau\rho<10^{-3}$ are needed.

These observations suggest that the transport regime, described in
the present work, should set in at very small values of $\tau$
and~$\rho$, which probably have not been accessed in the numerical
works up to the present date.

\section{Resonant triples}
\label{sec:Spots}
\setcounter{equation}{0}

This section represents a detailed analysis of chaos in a system
of three oscillators whose frequencies happen to be close to
each other.
Although such resonant triples do not contribute to the main result
for the conductivity, Eq.~(\ref{sigmamuT=}), they have the thickest
stochastic layers.
Estimation of their chaotic phase volume is the main task of the
present section.

First, we discuss a resonant pair of oscillators where chaos is
generated by an external perturbation, which can be studied by the
standard methods~\cite{Chirikov1979,Lichtenberg1983,Zaslavsky1985}.
Next, the closed system of three oscillators is considered, for
which no analytical solution is available. Still, an estimate
for the chaotic phase volume can be obtained by analyzing various
limiting cases and using the results for a perturbed resonant pair.

The results of this section are relevant in the limit $\tau\ll\rho$.
The opposite limiting case, $\tau\gg\rho$, requires analysis of high
orders of perturbation theory, and will be discussed in 
Sec.~\ref{sec:Motttriples}.

\subsection{Two oscillators: separatrix}
\label{sec:2oscsep}

The Hamiltonian of two oscillators has the form
\begin{equation}\label{H12=}
H=\omega_1I_1+\frac{g}2\,I_1^2+\omega_2I_2+\frac{g}2\,I_2^2
-2\tau\Delta\sqrt{I_1I_2}\cos(\phi_1-\phi_2).
\end{equation}
Let us perform a canonical change of variables:
\begin{subequations}\begin{eqnarray}
&&\phidif=\phi_1-\phi_2,\quad \Idif=\frac{I_1-I_2}2,\quad
\phitot=\frac{\phi_1+\phi_2}{2},\quad\Itot=I_1+I_2,\\
&&H=\frac{\omega_1+\omega_2}2\,\Itot+\frac{g}4\,\Itot^2
+(\omega_1-\omega_2)\Idif+g\Idif^2
-2\tau\Delta\sqrt{\Itot^2/4-\Idif^2}\cos\phidif.\label{Hpairnew=}
\end{eqnarray}\end{subequations}
The total action $\Itot$ is conserved, so only $\Idif,\phidif$
have a non-trivial dynamics:
\begin{subequations}\begin{eqnarray}
&&\frac{d\phidif}{dt}=\frac{\partial{H}}{\partial\Idif}
=\omega_1-\omega_2+2g\Idif
+2\tau\Delta\,\frac{\Idif\cos\phidif}{\sqrt{\Itot^2/4-\Idif^2}}\,,\\
&&\frac{d\Idif}{dt}=-\frac{\partial{H}}{\partial\phidif}
=-2\tau\Delta\sqrt{\Itot^2/4-\Idif^2}\sin\phidif\,.
\label{twooscdyn2=}
\end{eqnarray}\end{subequations}
A separatrix in the phase space $(\Idif,\phidif)$ appears when
these dynamical equations have a hyperbolic stationary point.

\begin{figure}
\begin{center}
\includegraphics[width=5cm]{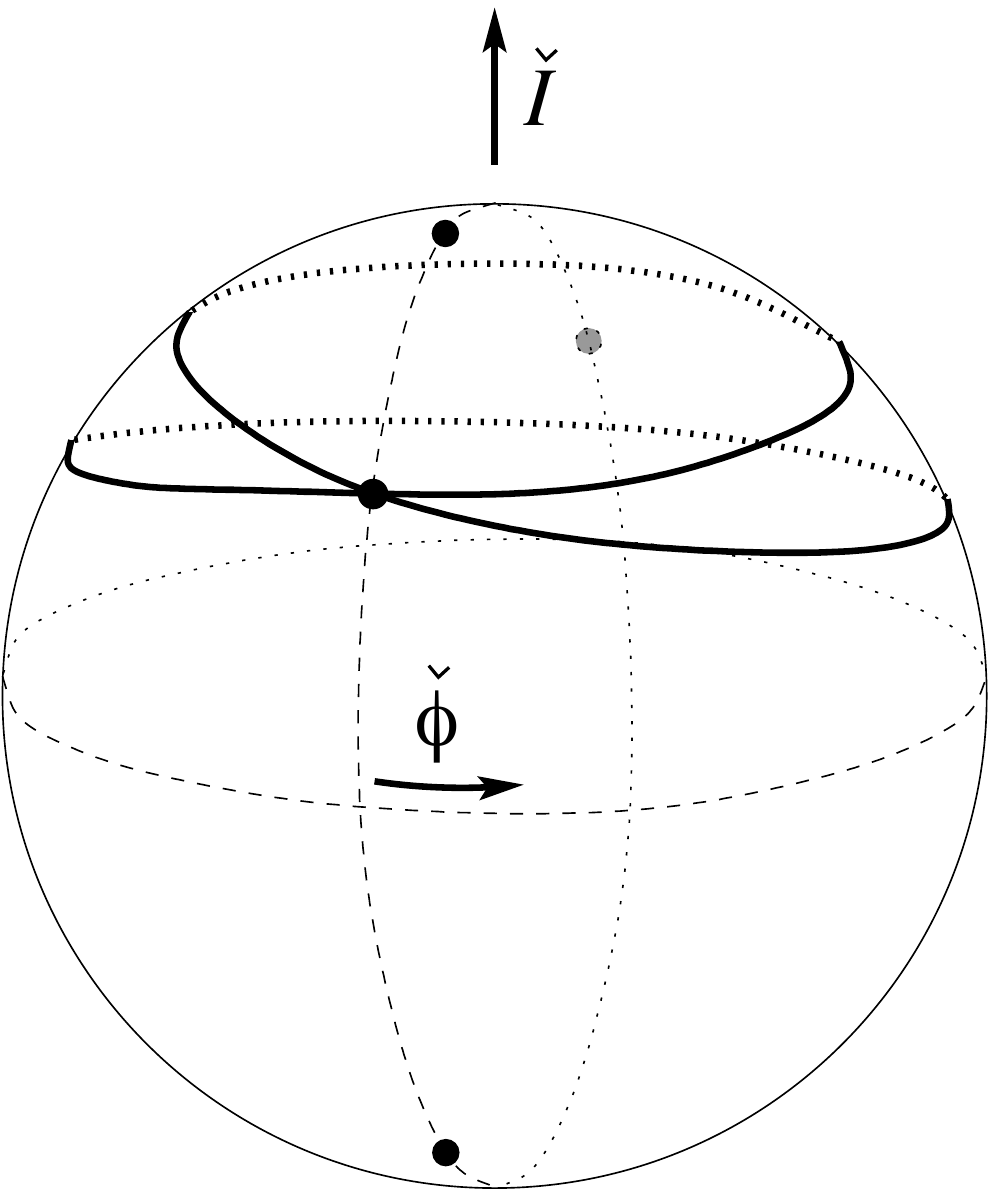}
\end{center}
\caption{\label{fig:sphere}
A schematic view of the  $(\Idif,\phidif)$ phase space, represented
by the sphere.
The circles $\Thetadif=\pi/2$, $0\leqslant\phidif<2\pi$ (``equator'')
and $\phidif=0,\pi$, $0\leqslant\Thetadif\leqslant\pi$  (``Greenwich meridian'')
are shown by dashed lines. The thick solid line represents the
separatrix. The stationary points (one hyperbolic, at the separatrix
self-crossing, and three elliptic) are shown by dark points.}
\end{figure}

To find the stationary points, we make a change of variables
$\Idif=(\Itot/2)\cos\Thetadif$, $0\leqslant\Thetadif\leqslant\pi$, thereby
mapping the phase space onto a sphere (Fig.~\ref{fig:sphere}).
From Eq.~(\ref{twooscdyn2=}) it is seen that the stationary points
require $\phidif=0$ or~$\pi$. It is convenient to fix $\phidif=0$,
and extend $0\leqslant\Thetadif<2\pi$, parametrizing the circle by
$\Thetadif$ only. Then we obtain an equation for stationary points:
\begin{equation}\label{twooscst=}
\frac{\omega_1-\omega_2}{2\Delta}\,\sin\Thetadif
+\tau\cos\Thetadif+\frac{g\Itot}{4\Delta}\,\sin2\Thetadif=0.
\end{equation}
It has four solutions if and only if
\begin{equation}\label{twooscboundary=}
|2\tau\Delta|^{2/3}+|\omega_1-\omega_2|^{2/3}\leqslant(g\Itot)^{2/3}.
\end{equation}
Otherwise, it has two solutions. To see this, we introduce
\begin{equation}
r=2\sqrt{\frac{(\omega_1-\omega_2)^2+(2\tau\Delta)^2}{(g\Itot)^2}},
\quad
\Theta_0=\arctan\frac{\omega_1-\omega_2}{2\tau\Delta},
\end{equation}
so Eq.~(\ref{twooscst=}) becomes
$\sin{2}\Thetadif+r\cos(\Thetadif-\Theta_0)=0$.
At the borderline, it has three solutions, one of which is degenerate.
Equating to zero $\sin{2}\Thetadif+r\cos(\Thetadif-\Theta_0)$ together
with its derivative, we obtain
\[
r^2=1+3\cos^2{2}\Thetadif=4\,\frac{1+\tan^6\Thetadif}{(1+\tan^2\Thetadif)^3},
\quad \tan(\Thetadif-\Theta_0)=-2\cot{2}\Thetadif.
\]
From the second equation we obtain $\tan\Thetadif=(\cot\Theta_0)^{1/3}$.
Substituting it into the first equation, we obtain Eq.~(\ref{twooscboundary=}).

Thus, for the separatrix to exist, it is absolutely necessary to have
$2\tau\Delta<g\Itot$. This is the reason for making the assumption
$\tau\ll\rho$, as the typical value of $\Itot\sim{g}T/|\mu|$. Given
$\tau\Delta\ll{g}\Itot$, the
condition~(\ref{twooscboundary=}) can be taken in zero approximation,
$|\omega_1-\omega_2|<g\Itot$. This is simply the condition for existence
of $I_1,I_2>0$ satisfying $\omega_1+gI_1=\omega_2+gI_2$,
see Eq.~(\ref{twooscres=}). 
Also, at  $\tau\ll{g}\Itot$ the separatrix is concentrated in a narrow
region of~$\Idif$ around the resonant value $\Idif=(\omega_2-\omega_1)/(2g)$,
so in the last term of Hamiltonian~(\ref{Hpairnew=}) one can simply set
$\Idif=(\omega_2-\omega_1)/(2g)$.
Then the Hamiltonian becomes equivalent to that of a pendulum,
\begin{equation}\label{Hpendulum=}
H=\frac{\omega_1+\omega_2}2\,\Itot+\frac{g}4\,\Itot^2
-\frac{(\omega_1-\omega_2)^2}{4g}
+g\left(\Idif-\frac{\omega_2-\omega_1}{2g}\right)^2
-\frac{\Omega^2\sign\tau}{2g}\,\cos\phidif,
\end{equation}
and the pendulum frequency is given by
\begin{equation}\label{Omega2=}
\Omega^2=2|\tau|\Delta\sqrt{(g\Itot)^2-(\omega_1-\omega_2)^2}.
\end{equation}

\subsection{Two oscillators: stochastic layer}
\label{sec:2oscstoch}

Suppose now that a time-dependent perturbation is added to the
Hamiltonian~(\ref{H12=}):
\begin{equation}\label{Hprime=}
H'=2V(I_1,I_2)\cos(m_1\phi_1+m_2\phi_2-\omega{t})
\end{equation}
The frequency $\omega$ is a linear combination of frequencies of other
oscillators. Their motion is taken in zero approximation, so effectively
they exert an external time-dependent force on the resonant pair.
This force produces a stochastic layer around the separatrix, as is
discussed in detail in
Refs.~\cite{Chirikov1979,Lichtenberg1983,Zaslavsky1985}.

\begin{figure}
\begin{center}
\includegraphics[width=8cm]{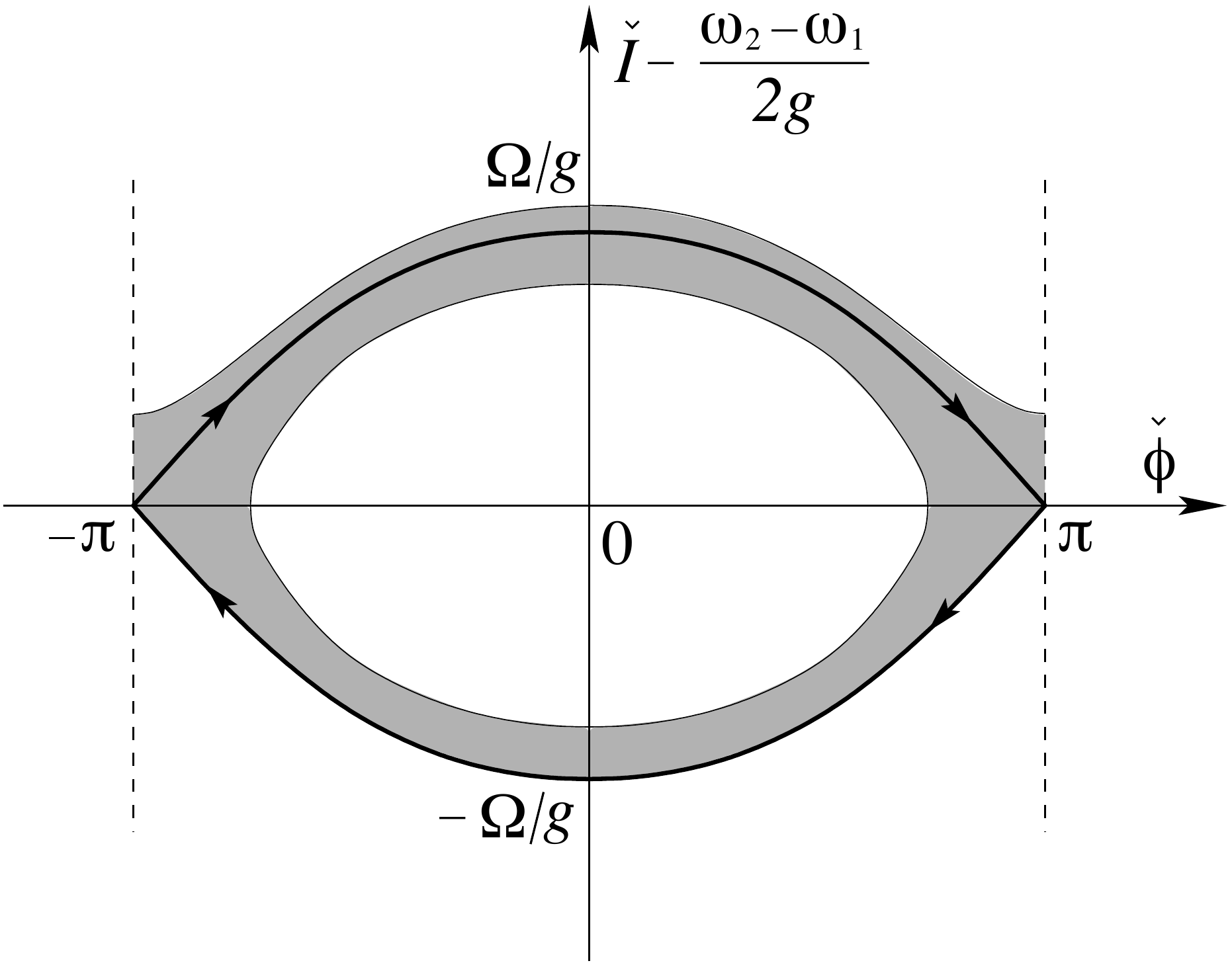}
\end{center}
\caption{\label{fig:separatrix}
Stochastic layer around the separatrix in the $(\Idif,\phidif)$
phase space, as shown by the shaded area ($\tau>0$, $m_1-m_2>0$,
and $\Lambda>0$ are assumed).}
\end{figure}

Perturbation~(\ref{Hprime=}) leads to a change in the old integrals of
motion, $\Itot$ and $H$,
\begin{subequations}\begin{eqnarray}
\label{dIdtpair=}
&&\frac{d\Itot}{dt}=-\frac{\partial{H}'}{\partial\phitot},\\
&&\frac{dH}{dt}=-\frac{\partial{H}}{\partial\Itot}
\frac{\partial{H}'}{\partial\phitot}
-\frac{\partial{H}}{\partial\Idif}\frac{\partial{H}'}{\partial\phidif}
+\frac{\partial{H}}{\partial\phidif}\frac{\partial{H}'}{\partial\Idif}.
\label{dHdtpair=}
\end{eqnarray}\end{subequations}
Out of the three terms on the right-hand side of Eq.~(\ref{dHdtpair=}),
the first one is essentially the repetition of Eq.~(\ref{dIdtpair=}),
while in the third term $\partial{H}/\partial\phidif=O(\tau)$, so it
can be neglected. 
Being interested in the perturbation of the pendulum motion in the
vicinity of the separatrix, one can take the values of $I_1,I_2$ at
the point of resonance, the unperturbed dependence of
$\phitot=(\partial{H}/\partial\Itot)t$, and $\phidif(t)$ right on the
separatrix:
\begin{subequations}\begin{eqnarray}
&&V(I_1,I_2)\to
V\left(\frac\Itot{2}+\frac{\omega_2-\omega_1}{2g},
\frac\Itot{2}-\frac{\omega_2-\omega_1}{2g}\right),\\
&&\phitot(t)\to\frac{\omega_1+\omega_2+g\Itot}{2}\,t,\\
&&\phidif(t)\to\pm\left[4\arctan{e}^{\Omega(t-t_0)}-\pi\right].
\end{eqnarray}\end{subequations}
The two signs of $\phidif(t)$ correspond to the two branches of the
separatrix (Fig.~\ref{fig:sphere}).
The changes in $\Itot$ and $H$ during one passage of the separatrix
are given by
\begin{subequations}\begin{eqnarray}
\delta\Itot&=&2(m_1+m_2)V\int\limits_{-\infty}^\infty
\sin\left[\frac{m_1-m_2}2\,\phidif(t)-\Lambda\Omega{t}\right]dt=\nonumber\\
&=&-\frac{2V}\Omega\,(m_1+m_2)\,
\mathcal{A}_{\pm(m_1-m_2)}(\Lambda)\sin(\Lambda\Omega{t}_0),
\label{changeinI=}\\
\delta{H}&=&\frac{\partial{H}}{\partial\Itot}\,\delta\Itot
+(m_1-m_2)V\int\limits_{-\infty}^\infty\frac{d\phidif(t)}{dt}\,
\sin\left[\frac{m_1-m_2}2\,\phidif(t)-\Lambda\Omega{t}\right]dt=\nonumber\\
&=&
-\frac{2V\omega}\Omega\,\mathcal{A}_{\pm(m_1-m_2)}(\Lambda)\sin(\Lambda\Omega{t}_0),
\label{changeinH=}
\end{eqnarray}\end{subequations}
where the Melnikov-Arnold integral $\mathcal{A}_m(\Lambda)$ is defined as
\begin{equation}\begin{split}
&\mathcal{A}_m(\Lambda)\equiv\int\limits_{-\infty}^\infty
\exp\left[\frac{im}2(4\arctan{e}^s-\pi)\right]e^{-i\Lambda{s}}\,ds,\\
&\Lambda\equiv\frac{\omega-(m_1+m_2)(\omega_1+\omega_2+g\Itot)/2}\Omega.
\label{MelnikovArnold=}
\end{split}\end{equation}
Since $\mathcal{A}_{-m}(\Lambda)=\mathcal{A}_m(-\Lambda)$, and
$\mathcal{A}_{m>0}(\Lambda<0)=(-1)^m\mathcal{A}_{m}(|\Lambda|)\,e^{-\pi|\Lambda|}$,
for $|\Lambda|\gg{1}$ and for a given perturbation~(\ref{Hprime=}),
one of the branches of the separatrix is much more efficient than
the other one.
If we fix $\tau>0$, $m_1-m_2>0$, and $\Lambda>0$ for definiteness,
the main contribution comes from the upper branch of the separatrix,
shown in Fig.~\ref{fig:separatrix}.
At $\Lambda\gg{m}$ the Melnikov-Arnold integral is exponentially
suppressed,
\begin{equation}\label{MAasymp=}
\mathcal{A}_{m>0}(\Lambda\gg{m})\approx
4\pi\,\frac{(2\Lambda)^{m-1}}{\Gamma(m)}\,e^{-\pi\Lambda/2},
\end{equation}
where $\Gamma(m)$ is the Euler's gamma function.

Let us represent the pendulum Hamiltonian [the last two terms of
Eq.~(\ref{Hpendulum=})] as $(\Omega^2/2g)(h+1)$, where $h$~is the
dimensionless energy of the pendulum, counted from the separatrix.
Then $h$~measures the distance of a trajectory from the separatrix.
Denoting
\begin{equation}\label{hs=}
h_s\equiv\frac{4gV}{\Omega^2}\,\Lambda^2\,\mathcal{A}_{m_1-m_2}(\Lambda),\quad
\theta\equiv\Lambda\Omega{t}_0,
\end{equation}
and using Eqs.~(\ref{changeinI=}),~(\ref{changeinH=}), we write
the change in~$h$ upon passing the upper branch of the separatrix as
\begin{subequations}
\begin{equation}\label{sepmaph=}
h\mapsto h+\frac{h_s}{\Lambda}\,\sin\theta,
\end{equation}
For trajectories close to separatrix, $|h|\ll{1}$, the time spent in
the vicinity of the hyperbolic point diverges as $(1/\Omega)\ln(32/|h|)$.
This gives the value of the phase~$\theta$ for the next passage of the
separatrix:
\begin{equation}\label{sepmaptheta=}
\theta\mapsto \theta+(2)\Lambda\ln\frac{32}{|h|}.
\end{equation}
\end{subequations}
The factor of 2 should be taken for trajectories inside the separatrix
(pendulum oscillations), since the hyperbolic point is passed twice
before the system returns to the upper branch. For trajectories just
above the upper branch (pendulum rotations) there is no factor of 2.
Trajectories below the lower branch are excluded from consideration
since they never get to the upper branch.
Eqs.~(\ref{sepmaph=}), (\ref{sepmaptheta=}) define a mapping which
is often called separatrix or whisker mapping. If the mapping is
iterated starting with a sufficiently small value of~$h$, the
phase~$\theta$ tends to randomize.
For $|\Lambda\gg{1}|$ the mapping can be linearized in $h$ around
one of the resonant values $h_n$ by introducing a new variable~$z$:
\begin{equation}
z=(2)\Lambda\,\frac{h-h_n}{h_n}\,,\quad
h_n=32\exp\left[\-\frac{2\pi{n}}{(2)\Lambda}\right].
\end{equation}
The separatrix mapping then reduces to the so-called standard mapping,
also called Chirikov-Taylor mapping:
\begin{equation}\label{stmap=}
z\mapsto{z}+K\sin\theta,\quad
\theta\mapsto \theta-z,
\end{equation}
where the stochasticity parameter is defined as $K(h_n)=(2)h_s/h_n$.
As discussed in Refs.~\cite{Chirikov1979,Lichtenberg1983,Zaslavsky1985},
the standard mapping becomes chaotic when $K>K_c$. Numerically,
$K_c$ is very close to~1, and we neglect $K_c-1$ in what follows.
This determines the stochastic region for the separatrix mapping
(\ref{sepmaph=})--(\ref{sepmaptheta=}) as $|h|<(2)h_s$.
Using this result, one can calculate the total phase volume of the
stochastic layer (the shaded area in Fig.~\ref{fig:separatrix}):
\begin{equation}\label{pendulumWs=}
W_s\equiv\int\limits_\mathrm{layer}\frac{d\Idif\,d\phidif}{2\pi}
=\frac{\Omega}{2\pi{g}}\left(2h_s\ln\frac{32e}{2h_s}
+\frac{h_s}2\ln\frac{32e}{h_s}\right).
\end{equation}

For $|\Lambda|\gg{1}$ the change in $h$ upon one iteration is much 
smaller than the layer width, so the dynamics of $h$ is diffusive.
If the phases at successive iterations are assumed uncorrelated,
the average square of displacement in $h$ after $N$~iterations is
$(N/2)(h_s/\Lambda)^2$. It takes the time $N\delta{t}$, where
$\delta{t}=(2)\Omega^{-1}\ln(32/(2)h)$ is the time step, which
gives the following estimate for the diffusion coefficient in~$h$:
\begin{equation}\label{Dh=}
D_h(h)=\frac{\Omega}{2}\,\frac{h_s^2}{\Lambda^2}
\left[(2)\ln\frac{32}{(2)h}\right]^{-1}.
\end{equation}
In fact, the assumption of completely random phases is never
fully correct. For the standard mapping, Eq.~(\ref{stmap=}),
even at relatively large $K\sim{10}-20$ small residual
correlations of the phase (at the level of $10^{-3}-10^{-4}$)
have been found~\cite{Khodas2000}. These residual correlations
are usually explained by the system sticking to various structures
inside the chaotic region, such as stability islands where the
diffusion is anomalously slow, or accelerator modes, where the
diffusion is anomalously fast. Still, in spite of the residual
correlations, the diffusion coefficient in $z$ for the standard
mapping for $K>2$ was found in Ref.~\cite{Khodas2000} to agree
quite well with its random-phase value $K^2/2$ together with
the leading in $1/\sqrt{K}$ correction~\cite{Rechester1980};
the largest relative deviations from $K^2/2$ were less than
$100\%$ and occurred for special values of $K$ corresponding to
accelerator modes. The accelerator modes should be much less
important for the separatrix mapping than for the standard
mapping; indeed, since $K=K(h)$, the fast diffusion in~$h$ will
quickly move the system away from the special values of~$K$.

The random phase assumption is completely wrong close to
the stochastic layer boundary. Indeed, the system cannot
leave the stochastic layer and end up on a stable periodic
orbit, so the diffusion coefficient must vanish at the
boundary. According to numerics, it vanishes as a power law,
$D_h(h)\propto((2)h_s-|h|)^{2.55\ldots}$~\cite{Chirikov1979}.
As will be seen later (see Eq.~(\ref{ArnoldD=}) and
\ref{app:diffusion}), what will matter for the present problem
is a certain average of the diffusion coefficient over the
stochastic layer, so we estimate the relative error of
Eq.~(\ref{Dh=}) to be $\sim{1}$, which is beyond the
accuracy of our final result, Eq.~(\ref{sigmamuT=}).

As the perturbation causes changes in~$\Itot$, the latter also
undergoes diffusion.
However, Eqs.~(\ref{changeinI=}),~(\ref{changeinH=}), rigidly
relate the change in~$\Itot$ to the change in~$h$ on each step:
\begin{equation}
\delta\Itot=\frac{(m_1+m_2)\Omega}{2g\Lambda}\,\delta{h},
\end{equation}
so, as $h$~is limited to a thin stochastic layer, a single perturbation
term~(\ref{Hprime=}) cannot produce a large excursion in $\Itot$.

\subsection{Three oscillators: resonances}
\label{sec:3oscres}

\begin{figure}
\begin{center}
\includegraphics[width=8cm]{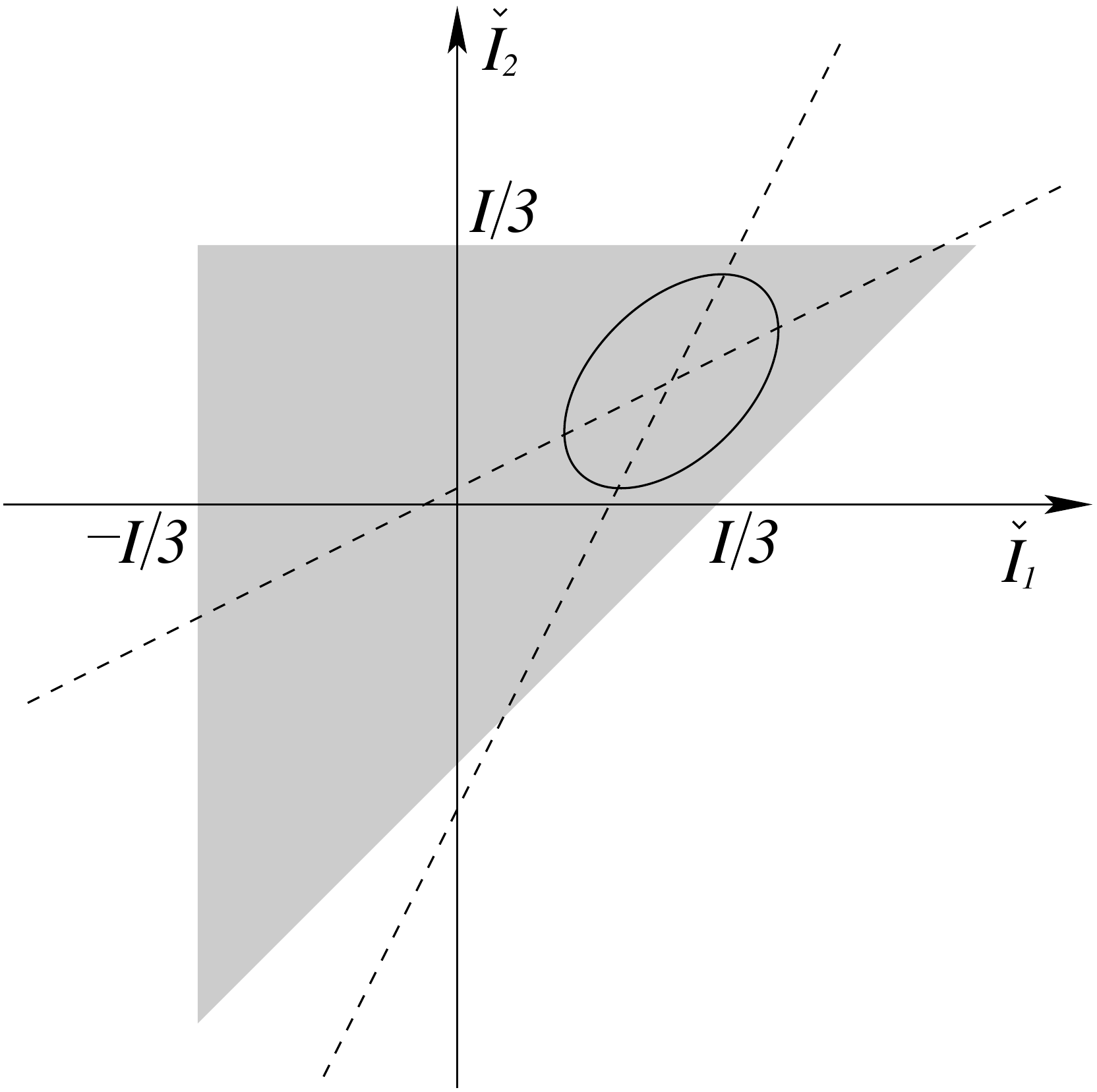}
\end{center}
\caption{\label{fig:triangle}
The allowed region of $\Idif_1,\Idif_2$, shown by the shaded area.
The resonance manifolds defined by Eqs.~(\ref{res12=}), (\ref{res23=})
are shown by dashed lines. The ellipse represents a constant-energy
surface.}
\end{figure}

The Hamiltonian of the triple is given by
\begin{equation}
H=\sum_{n=1}^3\left[\omega_nI_n+\frac{g}2\,I_n^2\right]
-2\tau\Delta\sum_{n=1}^2\sqrt{I_nI_{n+1}}\cos(\phi_n-\phi_{n+1}).
\end{equation}
We perform the following canonical change of variables:
\begin{subequations}\begin{eqnarray}
&&\phitot=\frac{\phi_1+\phi_2+\phi_3}{3}\,,\quad
\phidif_1=\phi_1-\phi_2\,,\quad\phidif_2=\phi_2-\phi_3\,,\\
&&\Itot=I_1+I_2+I_3\,,\quad
\Idif_1=\frac{2I_1-I_2-I_3}3\,,\quad\Idif_2=\frac{I_1+I_2-2I_3}3\,,\\
&&H=\frac{\omega_1+\omega_2+\omega_3}{3}\,\Itot+\frac{g\Itot^2}6+\nonumber\\
&&\qquad{}+(\omega_1-\omega_2)\Idif_1+g\Idif_1^2
+(\omega_2-\omega_3)\Idif_2+g\Idif_2^2-g\Idif_1\Idif_2-\nonumber\\
&&\qquad{}-2\tau\Delta\sqrt{I_1I_2}\cos\phidif_1
-2\tau\Delta\sqrt{I_2I_3}\cos\phidif_2\,.\label{Htriple=}
\end{eqnarray}\end{subequations}
The allowed region of $\Idif_1,\Idif_2$, determined by the
conditions $I_1,I_2,I_3>0$, is a triangle
$-\Itot/3<\Idif_1<2\Itot/3$, $\Idif_1-\Itot/3<\Idif_2<\Itot/3$,
shown in Fig.~\ref{fig:triangle}.

We consider two possibilities for the guiding resonance: either
$\omega_1+gI_1=\omega_2+gI_2$ or $\omega_2+gI_2=\omega_3+gI_3$.
The resonance between the oscillators 1 and 3 has the effective
coupling $\sim\tau^2$, so its stochastic layer is negligibly thin.
The resonance conditions are
\begin{subequations}\begin{eqnarray}
&&\frac{d\phidif_1}{dt}=0\;\;\;\Rightarrow\;\;\;
\Idif_2-2\Idif_1=\frac{\omega_1-\omega_2}g\,,\label{res12=}\\
&&\frac{d\phidif_2}{dt}=0\;\;\;\Rightarrow\;\;\;
\Idif_1-2\Idif_2=\frac{\omega_2-\omega_3}g\,.\label{res23=}
\end{eqnarray}\end{subequations}
They intersect at the point
\begin{equation}\label{respoint=}
\Idif_{1r}
=\frac{-2\omega_1+\omega_2+\omega_3}{3g}\,,\quad
\Idif_{2r}
=\frac{-\omega_1-\omega_2+2\omega_3}{3g}\,.
\end{equation}
This point lies inside the allowed triangle if
$gI>2\omega_i-\omega_j-\omega_k$ for any
$\{i,j,k\}$ being a permutation of $\{1,2,3\}$,
or, equivalently,
\begin{equation}\label{intriangle=}
g\Itot>3\max\limits_{1\leqslant{n}\leqslant{3}}\{\omega_n\}
-\sum_{n=1}^3\omega_n\equiv{g}\Itot_{min}.
\end{equation}

The importance of the intersection point can be seen by considering
the result of the previous subsection for the 1-2 resonance, where
the separatrix-destroying perturbation comes from the coupling to the
third oscillator. Let us set $m_1=0$, $m_2=1$, $\omega=\omega_3+gI_3$,
$V=-\tau\Delta\sqrt{I_2I_3}$ in Eq.~(\ref{Hprime=}), and take the
large-$\Lambda$ asymptotics~(\ref{MAasymp=}) of the Melnikov-Arnold
integral appearing in Eq.~(\ref{hs=}) for $h_s$, which, in turn, 
enters Eq.~(\ref{pendulumWs=}). This gives the following estimate for
the $I_3$-dependent volume of the stochastic layer around the 1-2
separatrix (numerical and logarithmic factors are omitted):
\begin{equation}
{W}_s^{(12)}\sim\frac\Delta\Omega\sqrt{I_2I_3}\,\tau\Lambda^2{e}^{-\pi\Lambda/2},
\quad
\Lambda=\frac{|\omega_2-\omega_3+g(I_2-I_3)|}{\Omega}\,.
\end{equation}
If we now integrate $W_s^{(12)}$ over $I_3$ to find the total volume
of the stochastic region in the space $(I_1,I_2,I_3)$, the exponential
factor will ensure that the integral is dominated by values of $I_3$
close to the resonance intersection point. 

\subsection{Three oscillators: chaotic phase space}
\label{sec:3oscchaotic}

Let us expand the Hamiltonian, Eq.~(\ref{Htriple=}), around the
resonance intersection point, Eq.~(\ref{respoint=}), writing
$\Idif_{1,2}=\Idif_{1,2}^r+p_{1,2}$:
\begin{eqnarray}
H&=&H_r(\Itot)+\Hdif(p_1,p_2;\phidif_1,\phidif_2)=\nonumber\\
&=&
-\frac{(\omega_1-\omega_2)^2+(\omega_2-\omega_3)^2+(\omega_1-\omega_3)^2}{6g}
+\frac{\omega_1+\omega_2+\omega_3}{3}\,\Itot+\frac{g\Itot^2}6
+\nonumber\\
&&{}+gp_1^2-gp_1p_2+gp_1^2-2V_1\cos\phidif_1-2V_2\cos\phidif_2\,.
\label{H3expand=}
\end{eqnarray}
Taking $V_{1,2}$ at the intersection point is allowed when
the resonance intersection point $(\Idif_1^r,\Idif_2^r)$ is not too
close to the sides of the allowed triangle in the $(\Idif_1,\Idif_2)$
plane. This is possible when $g\Itot\gg\tau\Delta$.

The main task of the present section is to estimate the fraction
of the total (thermally weighted) phase volume, occupied by the
chaotic part:
\begin{equation}\label{wtripledef=}
w=\frac{\int\limits_\mathrm{chaotic}e^{-(H-\mu\Itot)/T}\,
\prod_{n=1}^3dI_nd\phi_n/(2\pi)}
{\int{e}^{-(H-\mu\Itot)/T}\,\prod_{n=1}^3dI_nd\phi_n/(2\pi)}.
\end{equation}
At $gT/|\mu|\ll\Delta\ll|\mu|$, the denominator can be
simply taken to be $(T/|\mu|)^3$. Inside the allowed triangle
the typical value of $\Hdif\sim{g}\Itot^2\ll{T}$, so its
contribution to the thermal weight can be neglected. Then the
calculation of~$w$ can be done in two steps:
\begin{subequations}\begin{eqnarray}\label{Wssdef=}
&&W_{ss}(\Itot)=\int\limits_\mathrm{chaotic}
\frac{d\Idif_1d\phidif_1}{2\pi}\frac{d\Idif_2d\phidif_2}{2\pi}\,,\\
&&w=\left(\frac{|\mu|}T\right)^3\int\limits_0^\infty
e^{-[H_r(\Itot)-\mu\Itot]/T}\,W_{ss}(\Itot)\,d\Itot.\label{wtriple=}
\end{eqnarray}\end{subequations}

Not being able to calculate $W_{ss}(\Itot)$ at arbitrary $V_1,V_2$, we
first consider the case when one of the couplings is much stronger
than the other (e.~g., $V_1\gg{V}_2$), and then extrapolate
to $V_1\sim{V}_2$.
At $V_2=0$ we write $\Hdif$ as
\begin{equation}
\left.\Hdif\right|_{V_2=0}
=g\left(p_1-\frac{p_2}2\right)^2-2V_1\cos\phidif_1+\frac{3}4\,gp_2^2
\equiv\Hdif_1+\frac{3}4\,gp_2^2\,,
\end{equation}
and find the separatrix solution for $\phidif_1$, as in
Sec.~\ref{sec:2oscstoch}:
\begin{equation}
\phidif_1(t)=\pm\left[4\arctan{e}^{\Omega_1(t-t_0)}-\pi\right],
\quad\phidif_2(t)=-\frac{\phidif_1(t)}{2}+\frac{3}2\,gp_2t,
\quad\Omega_1=\sqrt{4gV_1}.
\end{equation}
Again, we calculate the changes in the action $p_2$ and the
pendulum energy~$\Hdif_1$:
\begin{subequations}\begin{eqnarray}
&&\delta{p}_2=-\int{2}V_2\sin\phidif_2(t)\,dt
=-\frac{V_2}{\sqrt{gV_1}}\,\mathcal{A}_{\pm{1}}\!
\left(\frac{3gp_2}{4\sqrt{gV_1}}\right)
\sin\frac{3gp_2t_0}{4\sqrt{gV_1}}\,,\\
&&\delta\Hdif_1=-\int\frac{\partial\Hdif_1}{\partial{p}_2}\,
2V_2\sin\phidif_2(t)\,dt
=\frac{3gp_2V_2}{2\sqrt{gV_1}}\,\mathcal{A}_{\pm{1}}\!
\left(\frac{3gp_2}{4\sqrt{gV_1}}\right)
\sin\frac{3gp_2t_0}{4\sqrt{gV_1}}\,,\qquad
\end{eqnarray}\end{subequations}
where 
$\mathcal{A}_{\pm{1}}(\Lambda)=\pm{2}\pi{e}^{\pm\pi\Lambda/2}/\sinh(\pi\Lambda)$.

Again, one arrives at the separatrix mapping, analogous to
Eqs.~(\ref{sepmaph=})--(\ref{sepmaptheta=}). 
The main contribution to the stochastic phase volume comes
from $g|p_2|\sim\sqrt{gV_1}$, i.~e. $|\Lambda|\sim{1}$. This
is still much larger than the inverse time of the passage of
the separatrix due to the logarithmic factor, so the phase
is randomized after a few iterations.
However, at $|\Lambda|\sim{1}$~the change in the Hamiltonian
per iteration is not small compared to the stochastic layer
width, so the standard approach for determining the layer width
(linearization of Eq.~(\ref{sepmaph=}) around some value of~$h$
and reduction to the standard mapping) does not apply in this
case. The diffusive picture for the motion across the stichastic
layer is not valid either.
Also, for the pendulum rotation, $\Hdif_1>2V_1$, the
contribution $\propto{e}^{-3\pi|\Lambda|/2}$ from the
``inefficient'' rotation direction should also be kept.
Thus, the stochastic layer is determined by the inequalities
\begin{eqnarray}
-4V_2\Lambda^2\tilde{\mathcal{A}}_o(\Lambda)<\Hdif_1-2V_1<
2V_2\Lambda^2\tilde{\mathcal{A}}_\pm(\Lambda),
\end{eqnarray}
where $\tilde{\mathcal{A}}_{o,\pm}(\Lambda)$ are some functions,
analogous to Melnikov-Arnold integral~(\ref{MelnikovArnold=}),
about which we only know that
\begin{subequations}\begin{eqnarray}
&&\tilde{\mathcal{A}}_o(-\Lambda)
=\tilde{\mathcal{A}}_o(\Lambda)\,,\quad
\tilde{\mathcal{A}}_\pm(-\Lambda)
=\tilde{\mathcal{A}}_\mp(\Lambda)\,,\\
&&|\Lambda|\sim{1}:\quad\tilde{\mathcal{A}}_{o,\pm}(\Lambda)\sim{1},\\
&&|\Lambda|\gg{1}:\quad
\tilde{\mathcal{A}}_o(\Lambda)\approx{2}\pi{e}^{-\pi|\Lambda|/2},\quad
\tilde{\mathcal{A}}_\pm(\Lambda)\approx{2}\pi{e}^{-\pi|\Lambda|\pm\pi\Lambda/2}.
\end{eqnarray}\end{subequations}
Now we can calculate the volume of the stochastic layer in the
pendulum phase space analogously to Eq.~(\ref{pendulumWs=}) for
each $p_2$, and then integrate over $p_2$ [equivalently,
over~$\Lambda=3gp_2/(4\sqrt{gV_1})$]:
\begin{equation}
W_{ss}=\frac{2}{3\pi}\frac{V_2}{g}\int\limits_{-\infty}^\infty{d}\Lambda
\left[4\Lambda^2\tilde{\mathcal{A}}_o(\Lambda)\ln\frac{16eV_1/V_2}%
{\Lambda^2\tilde{\mathcal{A}}_o(\Lambda)}
+\sum_\pm\Lambda^2\tilde{\mathcal{A}}_\pm(\Lambda)\ln\frac{32eV_1/V_2}%
{\Lambda^2\tilde{\mathcal{A}}_\pm(\Lambda)}\right].
\end{equation}
The integral is dominated by $|\Lambda|\sim{1}$ where the functions
$\tilde{\mathcal{A}}_{o,\pm}(\Lambda)$ are not known.
Thus, one can only conclude that
\begin{equation}\label{Wssasymm=}
W_{ss}=C_{ss}\,\frac{V_2}{g}\ln\frac{V_1}{V_2}\,,
\end{equation}
with some unknown constant $C_{ss}\sim{1}$.

At $V_2\ll{V}_1$, the main region of integration over $p_2$,
$|p_2|\sim\sqrt{V_1/g}$, is far away from the separatrix of the
$(p_2,\phidif_2)$ motion, which occurs at $|p_2|\sim\sqrt{V_2/g}$.
When $V_2\sim{V}_1$, the two separatrices should be treated on
equal footing, and the author is not aware of any accurate method
for description of the stochastic layer.
A reasonable extrapolation of Eq.~(\ref{Wssasymm=}) is
\begin{equation}\label{Wsstriangle=}
W_{ss}=C_{ss}\,\frac{V_{min}}{g}
\ln\frac{C_{ss}'V_{max}}{V_{min}}\,,
\end{equation}
where $V_{max}=\max\{V_1,V_2\}$, $V_{min}=\min\{V_1,V_2\}$,
and $C_{ss},C_{ss}'\sim{1}$ are some bounded functions of
$V_{max}/V_{min}$. It is worth reminding that this expression is
valid when the resonance intersection point is inside the allowed
triangle and not too close to its boundaries, i.~e., $\Itot$
satisfies inequality~(\ref{intriangle=}) with a positive correction
$\sim\tau\Delta$ to its right-hand side.

Now let us turn to the integral in Eq.~(\ref{wtriple=}). The whole
$\Itot$~integration range consists of three parts: (i)~the smallest
$\Itot$, $0<\Itot<\Itot_{min}^{(line)}$, when the allowed triangle
is so small that none of the dashed lines in Fig.~\ref{fig:triangle}
enters it,
(ii)~intermediate~$\Itot$, $\Itot_{min}^{(line)}<\Itot<\Itot_{min}$
when one or both dashed lines enter the triangle, but not their 
intersection point, and
(iii)~sufficiently large~$\Itot>\Itot_{min}$, when the triangle
contains the resonance intersection point [Eq.~(\ref{intriangle=})].
For a typical configuration of the random oscillator frequencies,
$\Itot_{min}^{(line)},\Itot_{min},\Itot_{min}-\Itot_{min}^{(line)}%
\sim\Delta/g$. Hence, even though $W_{ss}$ in the intermediate region 
has an additional smallness $\sim{e}^{-{1}/\sqrt\tau}$ from the 
Melnikov-Arnold integral, a thermal factor $\sim{e}^{1/\rho}$ is
gained with respect to $\Itot>\Itot_{min}^{(point)}$ due to the higher
activation barrier of the latter. However, as will be seen below, the 
main contribution to the disorder average of~$w$ comes not from the 
typical configurations, but from those where
$\Itot_{min}\sim{T}/|\mu|$. In this case the contribution from
the region $\Itot_{min}^{(line)}<\Itot<\Itot_{min}$ can be neglected,
and one can focus on $\Itot$ satisfying inequality~(\ref{intriangle=})
and use Eq.~(\ref{Wsstriangle=}) for~$W_{ss}$.

As $\Idif$ increases, the intersection point enters the triangle
through one of its sides, depending on which of
$\omega_1,\omega_2,\omega_3$ is the largest. It is convenient to
rewrite $H_r(\Itot)$ as
\begin{subequations}\begin{eqnarray}
&&H_r(\Itot)=E_a
+\max\limits_{1\leqslant{n}\leqslant{3}}\{\omega_n-\mu\}\,(\Itot-\Itot_{min})
+\frac{g}6\,(\Itot-\Itot_{min})^2,\label{Hrtriple=}\\
&&E_a=3\max\limits_{1\leqslant{n}\leqslant{3}}\frac{(\omega_n-\mu)^2}{2g}
-\sum_{n=1}^3\frac{(\omega_n-\mu)^2}{2g}\,.
\end{eqnarray}\end{subequations}
The contribution of the last term in Eq.~(\ref{Hrtriple=}) to the
thermal weight can be neglected.
$E_a$ can be viewed as the activation barrier for the triple.
Its probability distribution is calculated as (using $\Delta\ll|\mu|$)
\begin{equation}\begin{split}
&p_\omega(E_a)=\int\limits_{-\Delta/2}^{\Delta/2}
\frac{d\omega_1}{\Delta}\frac{d\omega_2}{\Delta}\frac{d\omega_3}{\Delta}\,
\delta\left(3\max\limits_{1\leqslant{n}\leqslant{3}}\frac{(\omega_n-\mu)^2}{2g}
-\sum_{n=1}^3\frac{(\omega_n-\mu)^2}{2g}-E_a\right)=\\
&=\frac{3g}{|\mu|\Delta}\left\{\begin{array}{ll}
\displaystyle\frac{gE_a}{|\mu|\Delta}
\left(1-\frac{3}4\frac{gE_a}{|\mu|\Delta}\right),
&\displaystyle 0<\frac{gE_a}{|\mu|\Delta}<1,\\
{}\displaystyle\left(1-\frac{1}{2}\frac{gE_a}{|\mu|\Delta}\right)^2,
&\displaystyle 1<\frac{gE_a}{|\mu|\Delta}<2,\\
0,&\mathrm{otherwise}.
\end{array}\right.
\end{split}\end{equation}
Let us assume that $\omega_1>\omega_2,\omega_3$, which occurs with
probability 1/3. Denoting $g(\Itot-\Itot_{min})=\omega'$,
$\omega_1-\omega_2=\omega_{12}$, $\omega_1-\omega_3=\omega_{13}$,
we obtain
\begin{equation}
w=\frac{2C_{ss}\tau\Delta}{3(gT/|\mu|)^3}\,
e^{-E_a/T}
\int\limits_0^\infty\sqrt{\omega'(\omega'+3\omega_{12})}\,
e^{-\omega_1\omega'/(gT)}
\ln\frac{C_{ss}'\sqrt{\omega'+3\omega_{13}}}{\sqrt{\omega'}}\,d\omega'.
\end{equation}
The integral can be calculated analytically in two limiting
cases:
\begin{subequations}\begin{equation}\begin{split}\label{wtriple1=}
&w=\frac{2C_{ss}\tau}{3}\frac{|\mu|\Delta}{gT}\,e^{-E_a/T}
\left\{\begin{array}{ll}
\ln{C}_{ss}',& \omega_{12},\omega_{13}\ll{g}T/|\mu|,\\
\displaystyle\sqrt{\frac{3\pi}{16}\frac{\omega_1\omega_{12}}{gT}}
\ln\frac{C_{ss}''\omega_{13}\omega_1}{gT}\,,&
\omega_{12},\omega_{13}\gg{g}T/|\mu|,
\end{array}\right.\\
&C_{ss}''=3(C_{ss}')^2e^{\ln{2}-1+\gamma/2},
\end{split}\end{equation}
where $\gamma=0.577\ldots$ is the Euler-Mascheroni constant.
For $\omega_3>\omega_1,\omega_2$ one obtains the same result with
$\omega_1$ and $\omega_3$ interchanged.
For $\omega_2>\omega_1>\omega_3$ the calculation is fully analogous,
and one obtains
\begin{equation}\label{wtriple2=}
w=\frac{2C_{ss}\tau}{3}\frac{|\mu|\Delta}{gT}\,e^{-E_a/T}
\left\{\begin{array}{ll}
\ln{C}_{ss}',& \omega_{21},\omega_{23}\ll{g}T/|\mu|,\\
\displaystyle\sqrt{\frac{3\pi}{16}\frac{\omega_2\omega_{21}}{gT}}
\ln\frac{(C_{ss}')^2\omega_{23}}{\omega_{21}}\,,&
\omega_{21},\omega_{23}\gg{g}T/|\mu|.
\end{array}\right.
\end{equation}\end{subequations}
Again, for $\omega_2>\omega_3>\omega_1$ it is sufficient to interchange
$\omega_1$ and $\omega_3$.

Now one can average $w$ over the disorder, which gives the average
spatial density of the chaotic spots:
\begin{equation}\label{wav=}
\overline{w}\equiv\int\limits_{-\Delta/2}^{\Delta/2}
\frac{d\omega_1}{\Delta}\frac{d\omega_2}{\Delta}
\frac{d\omega_3}{\Delta}\,w
=C_w\,\tau\,\frac{gT}{|\mu|\Delta}\equiv{C}_w\tau\rho\,,
\end{equation}
where $C_w$ is a constant. The average is dominated by rare triples
which have $E_a\sim{T}$, that is
$|\omega_{12}|\sim|\omega_{23}|\sim{g}T/|\mu|$. Since expressions
(\ref{wtriple1=})--(\ref{wtriple2=}) cover only limiting cases, no
analytical expression for the constant~$C_w$ is available.

The estimate~(\ref{wav=}) is not very much sensitive to the
fact that in Eq.~(\ref{wtripledef=}) the chaotic fraction is
defined using the thermal weight. The latter only fixes the
typical value of the action $I_n\sim{g}T/|\mu|$. Let us
repeat the estimate simply fixing $I_1=I_2=I_3=\Itot/3$, so
instead of the thermal exponential in Eq.~(\ref{wtripledef=})
we put
$\delta(I_1-\Itot/3)\,\delta(I_2-\Itot/3)\,\delta(I_3-\Itot/3)$.
Then $V_1=V_2=(2/3)\tau\Delta\Itot$, and
$\overline{w}$ is simply given by the probability that the
point $\Idif_1=\Idif_2=0$ lies inside the area
$W_{ss}(\Itot)=C_{ss}'''\tau\Delta\Itot/g$ (we denote
$C_{ss}'''=(2/3)C_{ss}\ln{C}_{ss}'$) around the resonance
point~(\ref{respoint=}), which gives
$\overline{w}=3C_{ss}'''\tau{g}\Itot/\Delta$.
Defining $\rho=g\Itot/(3\Delta)$, we obtain an estimate fully
analogous to Eq.~(\ref{wav=}). 

\section{Perturbation theory}
\label{sec:Perturbation}
\setcounter{equation}{0}

\subsection{General remarks}

One could formulate the peturbation theory for Eq.~(\ref{DDNLS=})
by taking independent oscillations,
$\psi_n^{(0)}(t)=\sqrt{I_n}\,e^{-i(\omega_n+gI_n)t+i\phi_n^0}$,
as the zero approximation and solving Eq.~(\ref{DDNLS=}) for
$\psi_n(t)$ by iterations.
Clearly, such perturbation theory is divergent. Indeed, if it were
convergent, there would be no chaos. Formally, its divergent
character can be seen already from the exact solution for two
oscillators, given in Sec.~\ref{sec:2oscsep}. The pendulum frequency
depends on $\tau$ as $\Omega\propto\sqrt{|\tau|}$
[Eq.~(\ref{Omega2=})], while the dependence on~$g$ has a threshold
[Eq.~(\ref{twooscboundary=})]. As the dependence on $\tau$ and $g$
is non-analytic at $\tau,g\to{0}$, the perturbation series has
zero radius of convergence.


The crucial point of the present work is that instead of studying
the whole perturbation series, the most dangerous terms of this
series should be identified
(more, presicely, the probability for them to occur), and the subsequent
analysis of their effect should be performed non-perturbatively, using
an effective model, such as driven pendulum. In other words, once a
small frequency difference in the denominator has occurred in the
pertubation series, one step back should be taken, the combination
of the oscillator phases corresponding to this frequency combination
should be called the phase $\phidif$ of the effective pendulum, while
the prefactor multiplying the inverse of this frequency difference
should be called the effective coupling, which multiplies $\cos\phidif$
in the effective pendulum Hamiltonian.

Divergence of the perturbation theory due to resonances corresponds
to an instability at long times. Indeed, perturbation theory in the
time domain in the nonlinearity has been shown to converge only for
not too long times~\cite{FishmanKrivolapov2008,Fishman2009}.
Its divergence at long times signals the chaotic instability.
In fact, the perturbation theory developed here is closely related
to that of Refs.~\cite{FishmanKrivolapov2008,Fishman2009}, as will
be discussed in more detail the end of Sec.~\ref{sec:derivdiag}.

The perturbation theory developed here is designed to generate the
effective many-oscillator coupling in high orders in $\tau$ and $g$.
We define the effective coupling between $2N$ oscillators
$n_1,\ldots,n_N,\bar{n}_1,\ldots,\bar{n}_N$ (these are, generally
speaking, $2N$ arbitrary integers) as the coupling term in the
Hamiltonian which involves
no other oscillators and leads to the same equations of motion for
these $2N$ oscillators, as the equations obtained by eliminating all
other oscillators from the full equations of motion,
Eq.~(\ref{DDNLS=}). Thus, the effective coupling is the sum of
terms of the kind
\begin{equation}\label{VN=}
V^{(N)}=\frac{1}N
\,\mathcal{K}^{(N)}_{n_1\ldots{n}_N\bar{n}_1\ldots\bar{n}_N}
\psi_{n_1}^*\ldots\psi_{n_N}^*\psi_{\bar{n}_1}\ldots\psi_{\bar{n}_N},
\end{equation}
Strictly speaking, the above definition of the effective coupling
is meaningful only exactly at resonance,
$\omega_{n_1}+\ldots+\omega_{n_N}=
\omega_{\bar{n}_1}+\ldots+\omega_{\bar{n}_N}$, as will be seen in
Sec.~\ref{sec:derivdiag}. This does not pose a problem, since it
is just in the vicinity of the resonance that this coupling is
needed, where it determines the parameters of the effective
pendulum. Thus, the frequency difference
$|\omega_{n_1}+\ldots+\omega_{n_N}-\omega_{\bar{n}_1}-\ldots-\omega_{\bar{n}_N}|$
is assumed to be smaller than any of the intermediate frequency
denominators involved in the coefficient~$\mathcal{K}$. In other
words, the $2N$-oscillator perturbation term is generated under
the assumption of no resonances for $2N'<2N$ oscillators, involved
in the construction of this $2N$-oscillator term. Indeed, if a
$2N'$-oscillator resonance occurs, it would be more efficient in 
producing chaos due to larger effective coupling obtained in a
lower order of the perturbation theory, and then $V^{(N)}$ is
not needed at all.

Since we are interested in the leading approximation to the
effective pendulum frequency, we will take the lowest order in~$g$
which couples the $2N$ oscillators altogether (that is, $g^{N-1}$). 
Similarly, we will take the lowest possible order in~$\tau$, which
is determined by the spatial arrangement of the oscillators.
According to this argument, one should ignore terms of the
kind~(\ref{VN=}) where some of $n$'s coincide with some of
$\bar{n}$'s. Indeed, the coinciding ones do not enter the slow phase
$\phi_{n_1}+\ldots+\phi_{n_N}-\phi_{\bar{n}_1}-\ldots-\phi_{\bar{n}_N}$,
so the same effective pendulum can be constructed in a lower order
of the perturbation theory. Thereby we neglect all secular terms
of the perturbation theory, and, consequently, the nonlinear
frequency shifts in the frequency denominators involved in the
coefficient~$\mathcal{K}$. 
Due to the condition $g|\psi_n|^2/\Delta\ll{1}$, Eq.~(\ref{tauless1=}),
these shifts are small compared to the dispersion of the random
bare frequencies~$\omega_n$, and thus give small corrections to the
statistics of the denominators. Such corrections are beyond the
accuracy of the present work.

\subsection{Diagrammatic rules}
\label{sec:diagRules}

In this subsection we describe the formal procedure for generating
$2N$-oscillator couplings~(\ref{VN=}). Its derivation will be given
in Sec.~\ref{sec:derivdiag}.

For two given oscillators $n_1,n_2$, the coupling by tunneling is
taken in the lowest order in $\tau$ (that is, $\tau^{|n_1-n_2|}$).
This is done by defining the Green's function (propagator) as
\begin{subequations}
\begin{equation}
G_{n_1n_2}(\omega)=\frac{1}{\omega-\omega_{n_{min}}}\,
(-\tau\Delta)\,\frac{1}{\omega-\omega_{n_{min}+1}}\,(-\tau\Delta)
\ldots(-\tau\Delta)\,\frac{1}{\omega-\omega_{n_{max}}}\,,
\end{equation}
where $n_{min}=\min\{n_1,n_2\}$, $n_{max}=\max\{n_1,n_2\}$.
We also define the shortened propagators:
\begin{equation}
G_{n_1n_2}'(\omega)=(\omega-\omega_{n_1})\,G_{n_1n_2}(\omega),\quad
G_{n_1n_2}''(\omega)=G_{n_1n_2}(\omega)\,(\omega-\omega_{n_2}).
\end{equation}
\end{subequations}
This is nothing but the Green's function of the linear problem,
taken in the leading order of the locator expansion. In contrast
to other works where the solutions of the linear problem are
assumed to be known and their properties are postulated
\cite{FishmanKrivolapov2008,FlachKrimer2009,Skokos2009,%
Fishman2009,Iomin2009,Iomin2010,Flach2010,Johansson2010},
here these solutions are
constructed explicitly using the perturbation theory in~$\tau$.
This is the reason why the Green's function of the linear
problem appears.

\begin{figure}
\begin{center}
\includegraphics[width=12cm]{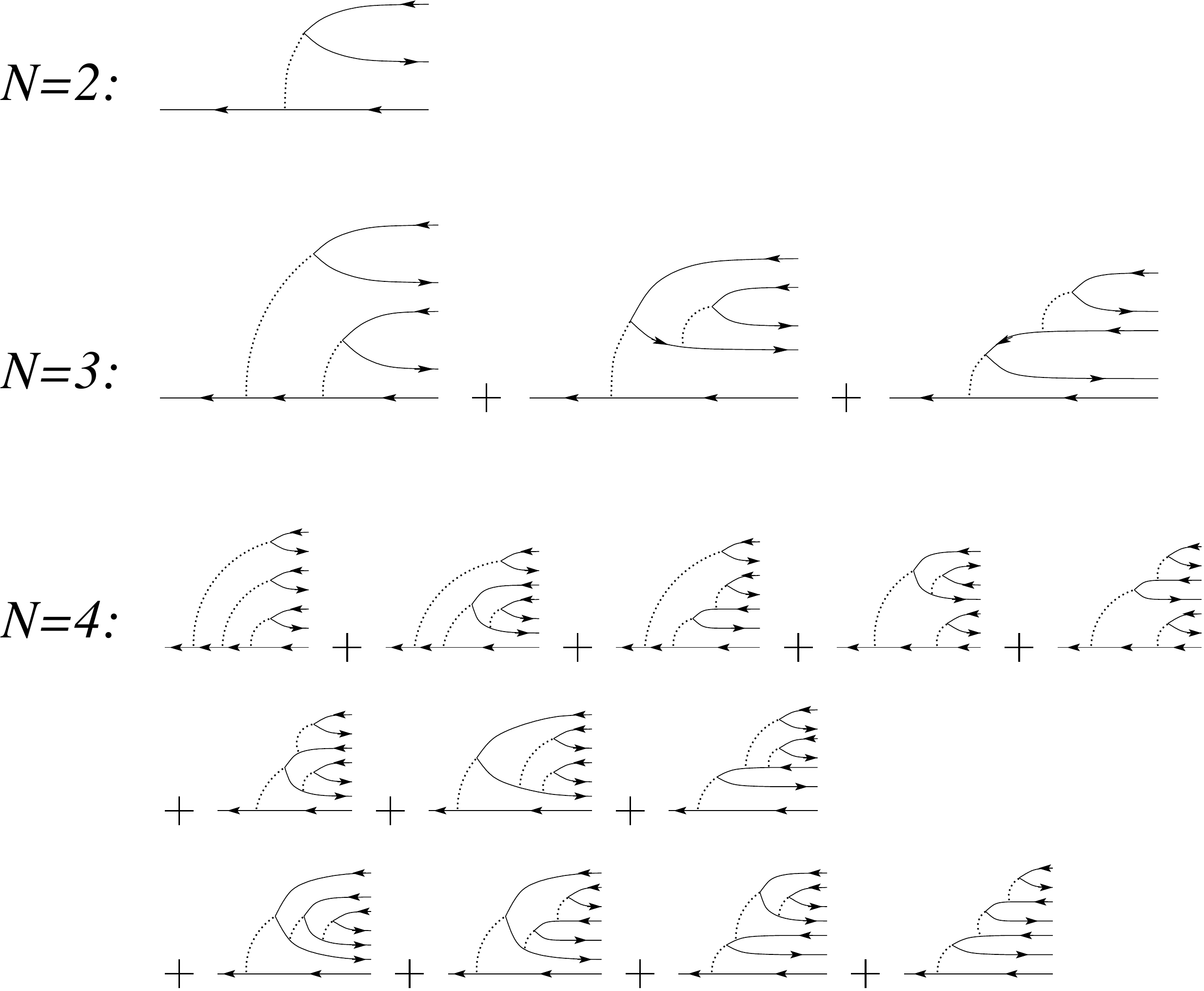}
\end{center}
\caption{\label{fig:SCBAiter}
All distinct diagrams generating effective couplings for
$N=2,3,4$ oscillators.}
\end{figure}

\begin{figure}
\begin{center}
\includegraphics[width=12cm]{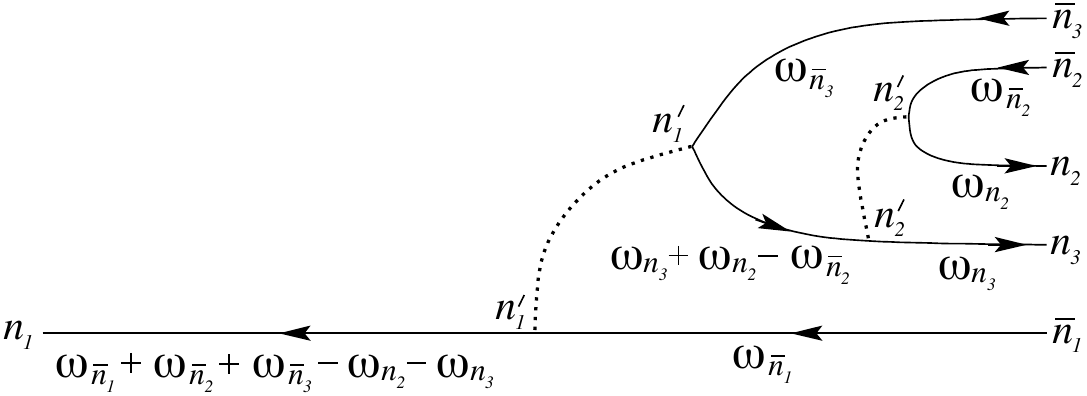}
\end{center}
\caption{\label{fig:SCBAexpr}
Labeling the lines and vertices of a diagram.}
\end{figure}

The effective $2N$-oscillator coupling is obtained from the
following procedure.
\begin{enumerate}
\item
Draw all distinct diagrams for the left-going solid line obtained
after $N-1$ iterations of the diagrammatic equation
\begin{subequations}
\begin{equation}\label{SCBAleft=}
\mbox{\includegraphics[width=8cm]{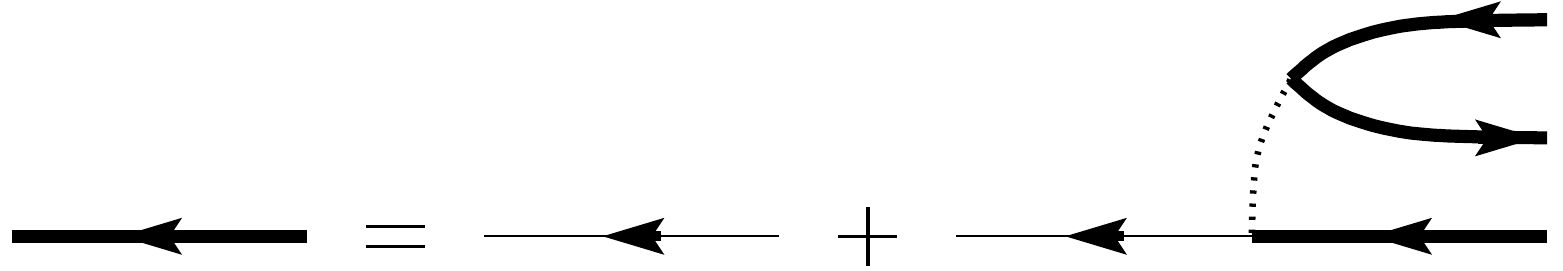}},
\end{equation}
supplemented by an analogous equation for the right-going line
\begin{equation}\label{SCBAright=}
\mbox{\includegraphics[width=8cm]{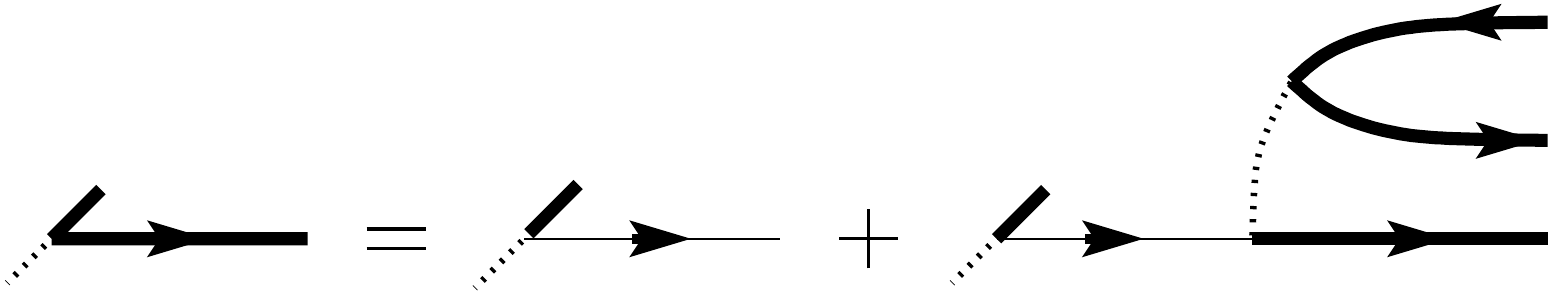}}.
\end{equation}
\end{subequations}
To illustrate the diagram generation, all distinct diagrams for
$N=2,3,4$ are shown explicitly in Fig.~\ref{fig:SCBAiter}.
\item
Each of the free ends of solid lines corresponds to some of the
sites $n_1,\ldots$, $n_N$, $\bar{n}_1,\ldots,\bar{n}_N$. Each
dotted-solid-solid vertex also has a site index, the two ends
of the same dotted line corresponding to the same site.
\item
Each solid line has a frequency argument. For external lines
it is the frequency of the site at its free end. For internal
lines it is determined from the conservation law holding for
each dotted line: the sum of the two frequencies of the entering
solid lines is equal to the sum of the two outgoing frequencies.
An example of labeling a diagram is shown in
Fig.~\ref{fig:SCBAexpr}.
Note that this procedure is consistent only when
$\omega_{n_1}+\ldots+\omega_{n_N}=
\omega_{\bar{n}_1}+\ldots+\omega_{\bar{n}_N}$.
\item
One obtains an analytical expression corresponding to each
diagram by
(i)~associating a factor~$g$ to each dotted line;
(ii)~associating a factor $G_{nn'}(\omega)$ to each internal
solid line which starts on site~$n$, ends on site~$n'$, and
carries frequency~$\omega$;
(iii)~associating a factor $G_{n_kn}'(\omega_{n_k})$ or
$G_{n\bar{n}_k}''(\omega_{\bar{n}_k})$ to each external
outgoing or incoming line, respectively, with free end on site
$n_k$ or $\bar{n}_k$;
(iv)~summing over internal site indices.
For example, the diagram in Fig.~\ref{fig:SCBAexpr} produces
an expression
\[\begin{split}
&g^2\sum_{n_1',n_2'}G'_{n_1n_1'}
(\omega_{\bar{n}_1}+\omega_{\bar{n}_2}+\omega_{\bar{n}_3}
-\omega_{n_2}-\omega_{n_3})\,
G_{n_1'\bar{n}_1}''(\omega_{\bar{n}_1})\,
G_{n_2n_2'}'(\omega_{n_2})\,
G_{n_2'\bar{n}_2}''(\omega_{\bar{n}_2})\\
&\qquad{}\times G_{n_3n_2'}'(\omega_{n_3})\,
G_{n_2'n_1'}(\omega_{n_3}+\omega_{n_2}-\omega_{\bar{n}_2})\,
G_{n_1'\bar{n}_3}''(\omega_{\bar{n}_3}).
\end{split}\]
\item
Summation over all diagrams gives the coefficient
$\mathcal{K}^{(N)}_{n_1\ldots{n}_N\bar{n}_1\ldots\bar{n}_N}$
in Eq.~(\ref{VN=}).
\end{enumerate}

\subsection{Derivation of the diagrammatic rules}
\label{sec:derivdiag}

Let us write the equations of motion, Eq.~(\ref{DDNLS=})
in the form
\begin{equation}\label{DDNLSpt=}
(i\partial_{t}-\omega_n)\psi_n
=-\tau\Delta(\psi_{n+1}+\psi_{n-1})+\psi_n\sum_{n'}g_{nn'}\psi_{n'}^*\psi_{n'}\,,
\end{equation}
where $g_{nn'}=g\delta_{nn'}$ is introduced to simplify the diagrammatic
rules (otherwise, additional combinatorial coefficients would have
to be assigned to each diagram).
The right-hand side of Eq.~(\ref{DDNLSpt=})
can be viewed as an external force acting on the oscillator~$n$. Then,
the response of the oscillator to an external periodic force is
determined as
\begin{equation}\label{response=}
(i\partial_{t}-\omega_n)\psi_n=
\sum_\alpha{f}_\alpha{e}^{-i\omega_\alpha{t}}
\;\;\;\Rightarrow\;\;\;
\psi_n=\psi_n^{(0)}e^{-i\omega_nt}
+\sum_\alpha\frac{{f}_\alpha{e}^{-i\omega_\alpha{t}}}%
{\omega_\alpha-\omega_n}\,.
\end{equation}

Let us start with two oscillators, $n$~and~$\bar{n}$ (assume $n<\bar{n}$
for concreteness). Suppose the $\bar{n}$th oscillator performs some
given motion
\begin{equation}
\psi_{\bar{n}}(t)=\int\frac{d\omega}{2\pi}\,
\tilde\psi_{\bar{n}}(\omega)\,e^{-i\omega{t}},
\end{equation}
and let us find the corresponding force acting on oscillator~$n$.
To the leading order in $\tau$ (which is $\bar{n}-n$) and~$g$ (which is
zero), this force is found by writing Eq.~(\ref{DDNLSpt=}) in the Fourier
space, and eliminating oscillators $n+1,\ldots,\bar{n}-1$ perturbatively:
\begin{equation}\begin{split}
f_{n\leftarrow\bar{n}}(t)&=\int\frac{d\omega}{2\pi}\,
(-\tau\Delta)\,\frac{1}{\omega-\omega_{n+1}}\,(-\tau\Delta)\ldots
(-\tau\Delta)\,\frac{1}{\omega-\omega_{\bar{n}-1}}\,(-\tau\Delta)\,
\tilde\psi_{\bar{n}}(\omega)\,e^{-i\omega{t}}\equiv\\
&\equiv\int\frac{d\omega}{2\pi}\,G_{n\bar{n}}'''(\omega)\,
\tilde\psi_{\bar{n}}(\omega)\,e^{-i\omega{t}}.
\end{split}\end{equation}
We are interested in the case when the spectrum of
$\tilde\psi_{\bar{n}}$
is concentrated around $\omega\approx\omega_{\bar{n}}\approx\omega_n$,
i.~e. slow modulations of the periodic oscillator motion. Then one can
replace $G_{n\bar{n}}'''(\omega)\to{G}_{n\bar{n}}'''(\omega_{\bar{n}})$
and take it out of the integral. Then the effective equation of motion
for the $n$th oscillator is
\begin{equation}
(i\partial_{t}-\omega_n)\,\psi_n(t)
={G}_{n\bar{n}}'''(\omega_{\bar{n}})\,\psi_{\bar{n}}(t)\,,
\end{equation}
which can be obtained from the effective coupling Hamiltonian
\begin{equation}\label{Vtwoosc=}
V={G}_{n\bar{n}}'''(\omega_{\bar{n}})\,\psi_n^*\psi_{\bar{n}}.
\end{equation}
This corresponds to the diagram consisting of a single solid line
with two free ends, so we have introduced the Green's function
$G'''$, shortened on both ends. To the coupling Hamiltonian in
Eq.~(\ref{Vtwoosc=}) another term should be added, which has the
same form, but with $n$ and $\bar{n}$ interchanged. The resulting
Hamiltonian is real only when the expression
${G}_{n\bar{n}}'''(\omega_{\bar{n}})$ is symmetric with respect
to interchange $n\leftrightarrow\bar{n}$, that is only when
$\omega_{\bar{n}}=\omega_n$. 

To see how the procedure works for a larger number of oscillators,
let us calculate the force acting on oscillator $n_1$. First, we write
\begin{equation}
f_{n_1}(t)=\sum_n\int\frac{d\omega}{2\pi}\,
G_{n_1n}'''(\omega)\,\tilde\psi_n(\omega)\,e^{-i\omega{t}},
\end{equation}
as before. Now, we find $\tilde\psi_n(\omega)$ from
Eq.~(\ref{response=}) where the force comes from the nonlinear term:
\begin{equation}\begin{split}
\tilde\psi_n(\omega)=&\frac{1}{\omega-\omega_n}\sum_{n'}
\int{dt}\,e^{i\omega{t}}\left[\psi_n(t)\,g_{nn'}\,
\psi_{n'}(t)\,\psi_{n'}^*(t)\right]=\\
&=\sum_{n'}\frac{g_{nn'}}{\omega-\omega_n}
\int\frac{d\omega'}{2\pi}\,\frac{d\omega''}{2\pi}\,
\tilde\psi_n(\omega')\,\tilde\psi_{n'}(\omega'')\,
\tilde\psi_{n'}^*(\omega'+\omega''-\omega)
\end{split}\end{equation}
Expressing $\tilde\psi_{n},\tilde\psi_{n'},\tilde\psi_{n'}^*$ in terms
of some other oscillators $n_2,\bar{n}_1,\bar{n}_2$,
we obtain
\begin{equation}\begin{split}
f_{n_1}(t)&=\int\frac{d\omega}{2\pi}\,e^{-i\omega{t}}
\int\frac{d\omega'}{2\pi}\,\frac{d\omega''}{2\pi}\,
\sum_{n,\bar{n}_1} G_{n_1n}'(\omega)\,G_{n\bar{n}_1}''(\omega')\,
\tilde\psi_{\bar{n}_1}(\omega')\sum_{n'}g_{nn'}
\times\\ &{}\times
\sum_{n_2,\bar{n}_2}\tilde\psi_{n_2}^*(\omega'+\omega''-\omega)\,
G_{n_2n'}'(\omega'+\omega''-\omega)\,G_{n'\bar{n}_2}''(\omega'')\,
\tilde\psi_{\bar{n}_2}(\omega'')\,.
\end{split}\end{equation}
Again, inside the Green's functions one can replace
$\omega'\to\omega_{\bar{n}_1}$, $\omega''\to\omega_{\bar{n}_2}$,
$\omega'+\omega''-\omega\to\omega_{n_2}$. When they are taken out
of the triple frequency integral, the latter evaluates to
$\psi_{\bar{n}_1}(t)\,\psi_{\bar{n}_2}(t)\,\psi_{n_2}^*(t)$.
As a result,
\begin{equation}
(\partial_t-\omega_{n_1})\psi_{n_1}=\sum_{n_2,\bar{n}_1,\bar{n}_2}
\mathcal{K}^{(2)}_{n_1n_2\bar{n}_1\bar{n}_2}
\psi_{n_2}^*\psi_{\bar{n}_1}\psi_{\bar{n}_2},
\end{equation}
where $\mathcal{K}^{(2)}_{n_1n_2\bar{n}_1\bar{n}_2}$ is the expression
corresponding to the diagram for $N=2$ in Fig.~\ref{fig:SCBAiter},
according to the rules of the previous subsection. Finally, the factor
1/2 has to be added when deducing $V^{(2)}$ form $\mathcal{K}_2$,
in order to avoid double counting upon differentiation
$\partial{V}^{(2)}/\partial\psi_{n_1}^*$.

Generalization to the $N$-oscillator case is straightforward:
calculation of the force by iterative solution of Eq.~(\ref{DDNLSpt=})
corresponds to iterations of diagrammatic equations 
(\ref{SCBAleft=}), (\ref{SCBAright=}). The factor $1/N$ in $V^{(N)}$
takes care of double counting upon differentiation.

To conclude this subsection, we note that the perturbation theory,
developed here, is closely related to that of
Refs.~\cite{FishmanKrivolapov2008,Fishman2009}. One obvious difference
is that the linear part of the problem was assumed to be diagonalized 
in Refs.~\cite{FishmanKrivolapov2008,Fishman2009}, while here the
linear coupling is treated perturbatively. This difference is purely
technical; the present calculation corresponds just to the perturbative
evaluation of the overlap sums which dress the nonlinear coupling
in the approach of Refs.~\cite{FishmanKrivolapov2008,Fishman2009}.
Second, in Refs.~\cite{FishmanKrivolapov2008,Fishman2009} one starts
from a single excited normal mode as zero approximation and then
calculates the force acting on other oscillators. Here, the
independently excited oscillators are taken as the zero approximation,
and the force acting on some given oscillator~$n$ is calculated.
Clearly, the two approaches are closely related, so the structure
of the perturbation theory is very similar in the two cases. It is
the subsequent use of the information obtained from the perturbation
theory which is quite different.

\subsection{On the number of diagrams}
\label{sec:Ndiag}

It turns out that the total number of diagrams of the order $g^{N-1}$,
generated according to the rules given in Sec.~\ref{sec:diagRules},
is given by the following exact expression:
\begin{equation}\label{Ndiag=}
\mathcal{D}_N=\sum_{k=1}^{N-1}\frac{2^{N-1}}{2^kk!}
\sum_{0\leqslant{m}_1\leqslant\ldots\leqslant{m}_{k-1}\leqslant{N}-1-k}
\,(3m_1+2)(3m_2+3)\ldots(3m_{k-1}+k),
\end{equation}
to be derived below.
Even though explicit, this expression is not straightforward
to evaluate. Still, it can be shown that $\mathcal{D}_N$
grows exponentially with~$N$. Let us estimate it from below
and from above:
\begin{equation}
\sum_{k=1}^{N-1}\frac{2^{N-1}}{2^kk!}\,\frac{(3N-3-2k)!}{(3N-2-3k)!}
<\mathcal{D}_N<
\sum_{k=1}^{N-1}\frac{2^{N-1}}{2^kk!}\,
\frac{(3N-3-2k)!}{(3N-2-3k)!}\,\frac{(N-2)!}{(k-1)!(N-1-k)!}.
\end{equation}
The lower bound is obtained by simply taking one term
of the $m$-sum with all $m_i=N-1-k$; this is the largest term,
so the upper bound is obtained by multiplying it by the
total number of terms in the sum, where we use the formula%
\footnote{This formula can be obtained by noting that the sum,
denoted by $s_{k,n}$, satisfies the recurrence relation
$s_{k,n}=s_{k-1,n}+s_{k,n-1}$. Then one can (i)~check directly
that it is satisfied by $(n+k-1)!/[k!(n-1)!]$, or (ii)~introduce
the characteristic function,
$\chi(\phi,\theta)=\sum_{k,n}e^{ik\phi+i(n-1)\theta}S_{k,n}$,
for which the recurrence relation reduces to a simple algebraic
equation, so that $\chi(\phi,\theta)=(1-e^{i\phi}-e^{i\theta})^{-1}$,
from which $s_{k,n}$ is easily reconstructed.}
\begin{equation}\label{sumformula=}
\sum_{1\leqslant{m}_1\leqslant\ldots\leqslant{m}_k\leqslant{n}}1
=\frac{(n-1+k)!}{k!\,(n-1)!}.
\end{equation}
At $N\gg{1}$ we approximate the factorials by Stirling's formula,
denote $k/N=x$, and pass to the integral:
\begin{equation}\begin{split}
&\int\limits_0^1\frac{dx}{\sqrt{8\pi{N}^3}}
\sqrt{\frac{(3-3x)^3}{(3-2x)^5}}
\left[\frac{2^{1-x}(3-2x)^{3-2x}}{x^x(3-3x)^{3-3x}}\right]^N
<\\&<\mathcal{D}_N<
\int\limits_0^1\frac{dx}{4\pi{N}^2}
\sqrt{\frac{(3-3x)^3}{x(1-x)(3-2x)^5}}
\left[\frac{2^{1-x}(3-2x)^{3-2x}}
{(1-x)^(1-x)x^{2x}(3-3x)^{3-3x}}\right]^N,
\end{split}\end{equation}
The maxima of the expressions in the square brackets on the left and
on the right are 4.365\ldots and 8.722\ldots, respectively. 
For the numerical evaluation of $\mathcal{D}_N$ it is convenient
to cast Eq.~(\ref{Ndiag=}) into a recursive form:
\begin{subequations}
\begin{eqnarray}
&&\mathcal{D}_N=\sum_{k=1}^{N-1}\frac{2^{N-1}}{2^k}
{S}_{k,N-1-k},\\
&&{S}_{k+1,n}=\sum_{m=0}^{n}\left(1+\frac{3m}{k+1}\right){S}_{k,m},
\quad S_{1,n}=1.
\end{eqnarray}
\end{subequations}
Numerical evaluation of the leading exponential gives
\begin{equation}
\frac{\mathcal{D}_{N+1}}{\mathcal{D}_N}\mathop{\sim}\limits_{N\to\infty}
(6.75\ldots)\times\left(1+\frac{1}{N}\right)^{-1.5\ldots}.
\end{equation}

\begin{figure}
\begin{center}
\includegraphics[width=12cm]{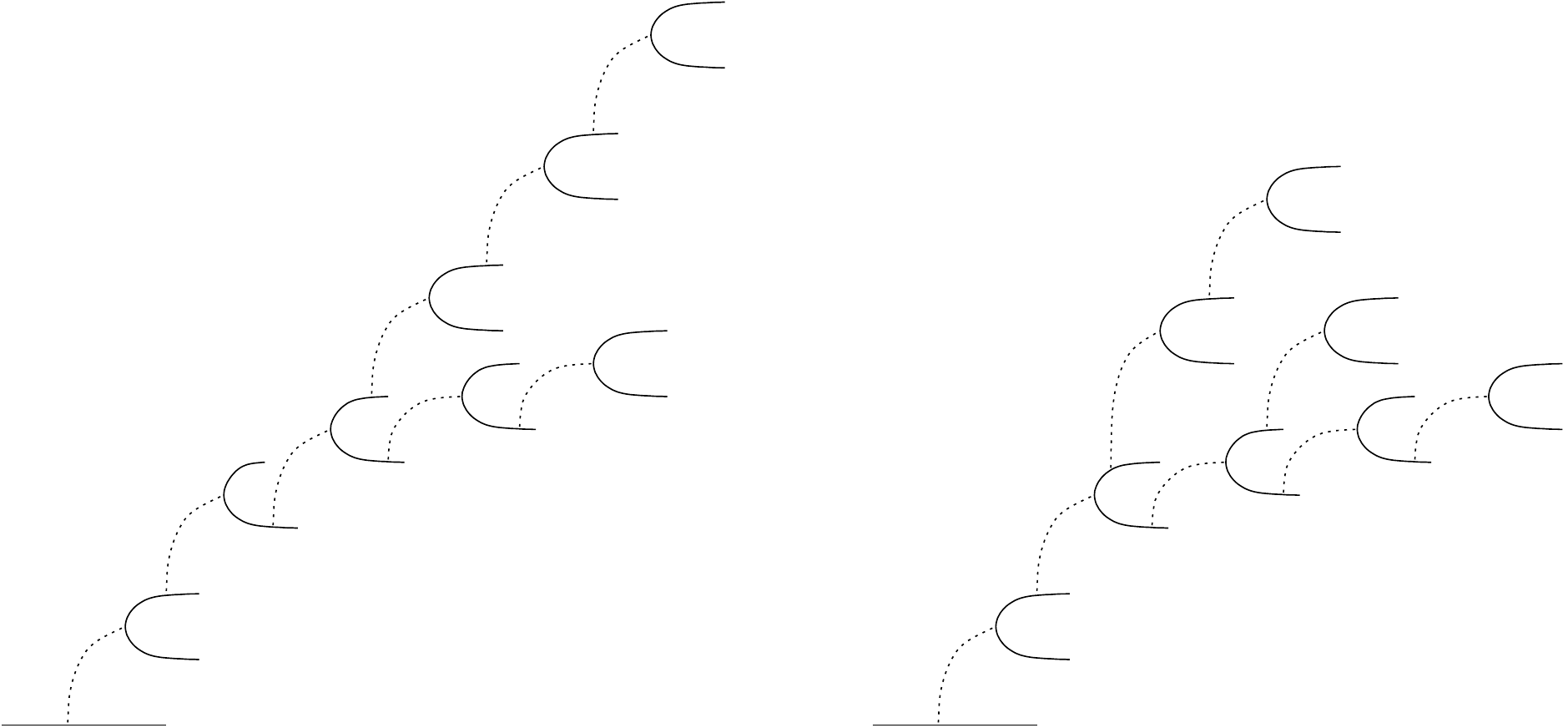}
\end{center}
\caption{\label{fig:branch}
Examples of two- and three-branch diagrams for effective 18-oscillator coupling.}
\end{figure}

To derive Eq.~(\ref{Ndiag=}) let us view the diagrams as trees
formed by the dotted lines (Fig.~\ref{fig:branch}).
The simplest one is a single-branch tree: in the order $g^{N-1}$
there are $2^{N-2}$ such diagrams. If there is one lateral branch,
the number of such diagrams is
\begin{equation}
\mathcal{D}_{N}^{(2)}=\frac{1}{2}\sum_{l_1,l_2=1}^{N-1}[3(l_1-1)+2]
2^{l_1-1}2^{l_2-1}\,\delta_{l_1+l_2,N-1}\,,
\end{equation}
because $2^{l_1-1}$ is the number of ways to build the main branch,
$2^{l_2-1}$ is the number of ways to build the lateral branch,
$3(l_1-1)+2$ is the number of electron lines where the lateral branch
can be attached to the main one, and 1/2 stands because of the
arbitrary choice of which branch is main and which one is lateral.
Summing over the number of branches, we obtain
\begin{eqnarray}\nonumber
\mathcal{D}_{N}&=&\sum_{k=1}^{N-1}\sum_{l_1,\ldots,l_k\geqslant{0}}
\frac{1}{k!}\,(3l_1+2)(3l_1+3l_2+3)
\ldots(3l_1+\ldots+3l_{k-1}+k)\times\\
\label{NSCBA=} && \hspace*{2cm}\times
2^{l_1+\ldots+l_k}\,\delta_{l_1+\ldots+l_k,N-1-k}\,,
\end{eqnarray}
which is equivalent to Eq.~(\ref{Ndiag=}).

It turns out that Eq.~(\ref{FishmanDN=}) is equivalent to the
following recursive relation, derived in Ref.~\cite{Fishman2009}:%
\footnote{Strictly speaking, the quantity, for which
Eq.~(\ref{FishmanDN=}) was derived in Ref.~\cite{Fishman2009}, is
not defined in exactly the same way as $\mathcal{D}_N$, since the
perturbation theory developed in Ref.~\cite{Fishman2009} is not
strictly identical to the present one. Still, the two perturbation
theories are closely related, as discussed in the end of
Sec.~\ref{sec:derivdiag}, and the equivalence between
Eq.~(\ref{FishmanDN=}) and Eq.~(\ref{Ndiag=}) is a manifestation of
this relation.}
\begin{equation}\label{FishmanDN=}
\mathcal{D}_N=\sum_{N_1,N_2,N_3\geqslant{1}}
\mathcal{D}_{N_1}\mathcal{D}_{N_2}\mathcal{D}_{N_3}
\delta_{N_1+N_2+N_3,N+1},\quad
\mathcal{D}_1=\mathcal{D}_2=1.
\end{equation}
The author was not able to prove the equivalence between
Eq.~(\ref{FishmanDN=}) and Eq.~(\ref{Ndiag=}) algebraically,
however, it can be checked numerically for quite large~$N$.
For example, both equations give
$\mathcal{D}_{50}=30426054945480277365983787382745806500$.

\subsection{On the statistics of diagrams}
\label{sec:statistics}

Let us define a resonance $r$ as a sequence
$n_1^r\leqslant{n}_2^r\leqslant\ldots\leqslant{n}^r_{N_r}$,
$\bar{n}_1^r\leqslant\bar{n}_2^r\leqslant\ldots\leqslant\bar{n}^r_{N_r}$, where
none of the $n$'s coincides with any of the $\bar{n}$'s. The
freedom of interchanging $n\leftrightarrow\bar{n}$ can be fixed
by requiring
\begin{equation}\label{nbarnfix=}
\omega_{n_1^r}+\ldots+\omega_{n^r_{N_r}}<
\omega_{\bar{n}_1^r}+\ldots+\omega_{\bar{n}^r_{N_r}}.
\end{equation}
The frequency mismatch of the resonance is thus defined to be
positive:
\begin{equation}
\varpi_r=\omega_{\bar{n}_1^r}+\ldots+\omega_{\bar{n}^r_{N_r}}
-(\omega_{n_1^r}+\ldots+\omega_{n^r_{N_r}}).
\end{equation}
To associate the resonance with a single site~$n$, one can fix
$n_1^r=n$. In the case when some of the $n$'s or some of the
$\bar{n}$'s coincide, one can introduce the multiplicity of each
site $n$, $m^r_n$, defined as the number of times it occurs in
the sequence of $n^r$'s or minus the number of times it occurs
in the sequence of $\bar{n}^r$'s.
For sites~$n$ that do not participate in the resonance, $m_n^r=0$.
Due to action conservation,
\begin{equation}\label{summn=0}
\sum_{n}m^r_n=0,\quad \sum_{n}|m^r_n|=2N_r.
\end{equation}
Viewing the sequence $\{m_n^r\}$ as a vector, we define the scalar
product and the norm in the usual way:
\begin{equation}
(\underline{m}^r,\underline{m}^{r'})=
\sum_{n=-\infty}^\infty{m}^r_nm^{r'}_n,
\quad|\underline{m}^r|^2=(\underline{m}^r,\underline{m}^r).
\end{equation}
The oscillators are coupled by an effective Hamiltonian
\begin{equation}\begin{split}
&V=2V_r
\cos\left(\phi_{n_1^r}+\ldots+\phi_{n^r_{N_r}}-
\phi_{\bar{n}_1^r}-\ldots-\phi_{\bar{n}^r_{N_r}}\right),
\\
&V_r=\mathcal{K}_r
\sqrt{I_{n_1^r}\ldots{I}_{n^r_{N_r}}I_{\bar{n}_1^r}\ldots{I}_{\bar{n}^r_{N_r}}},\\
&\mathcal{K}_r=\frac{1}N\sum
\mathcal{K}^{(N)}_{n_1\ldots{n}_{N_r}\bar{n}_1\ldots\bar{n}_{N_r}}.
\label{V=Vr}
\end{split}\end{equation}
The summation is performed over $(N_r!)^2/\prod_n|m_n^r|!$
distinct permutations of $n_1^r,\ldots,n^r_{N_r}$ and of
$\bar{n}_1^r,\ldots,\bar{n}^r_{N_r}$.
Besides the summation over permutations, $\mathcal{K}_r$ represents
a sum over different diagrams whose number was estimated in
Sec.~\ref{sec:Ndiag}, and for each diagram the summation over the
positions of $N_r-1$ nonlinear vertices should be performed.
Thus, $\mathcal{K}_r$ is a sum of many terms, whose number will be
denoted by $\mathcal{N}$. Each
term contains a product of $L_r+N_r-2$ frequency denominators,
where $L_r\geqslant{N}_r$ is the minimal order in $\tau$ which couples
all oscillators together.

It is convenient to explicitly separate all trivial deterministic
factors and introduce the dimensionless random quantity
\begin{equation}\label{tildeKr=}
\tilde{\mathcal{K}}_r=
\frac{2^{L_r+N_r-2}\Delta^{N_r-2}}{\tau^{L_r}g^{N_r-1}}\,
\mathcal{K}_r.
\end{equation}
Our primary interest will be the probability 
that $\tilde{\mathcal{K}}_r$ exceeds a certain value $e^{-\lambda}$:
\begin{equation}
\mathcal{P}\left\{\tilde{\mathcal{K}}_r<e^{-\lambda}\right\}
=\overline{\theta\left(\ln\frac{1}{\tilde{\mathcal{K}}_r}-\lambda\right)}\,,
\end{equation}
where $\theta(x)$ is the Heaviside step function, and
the overline denotes the statistical averaging
over the static frequency disorder.

\begin{figure}
\begin{center}
\includegraphics[width=8cm]{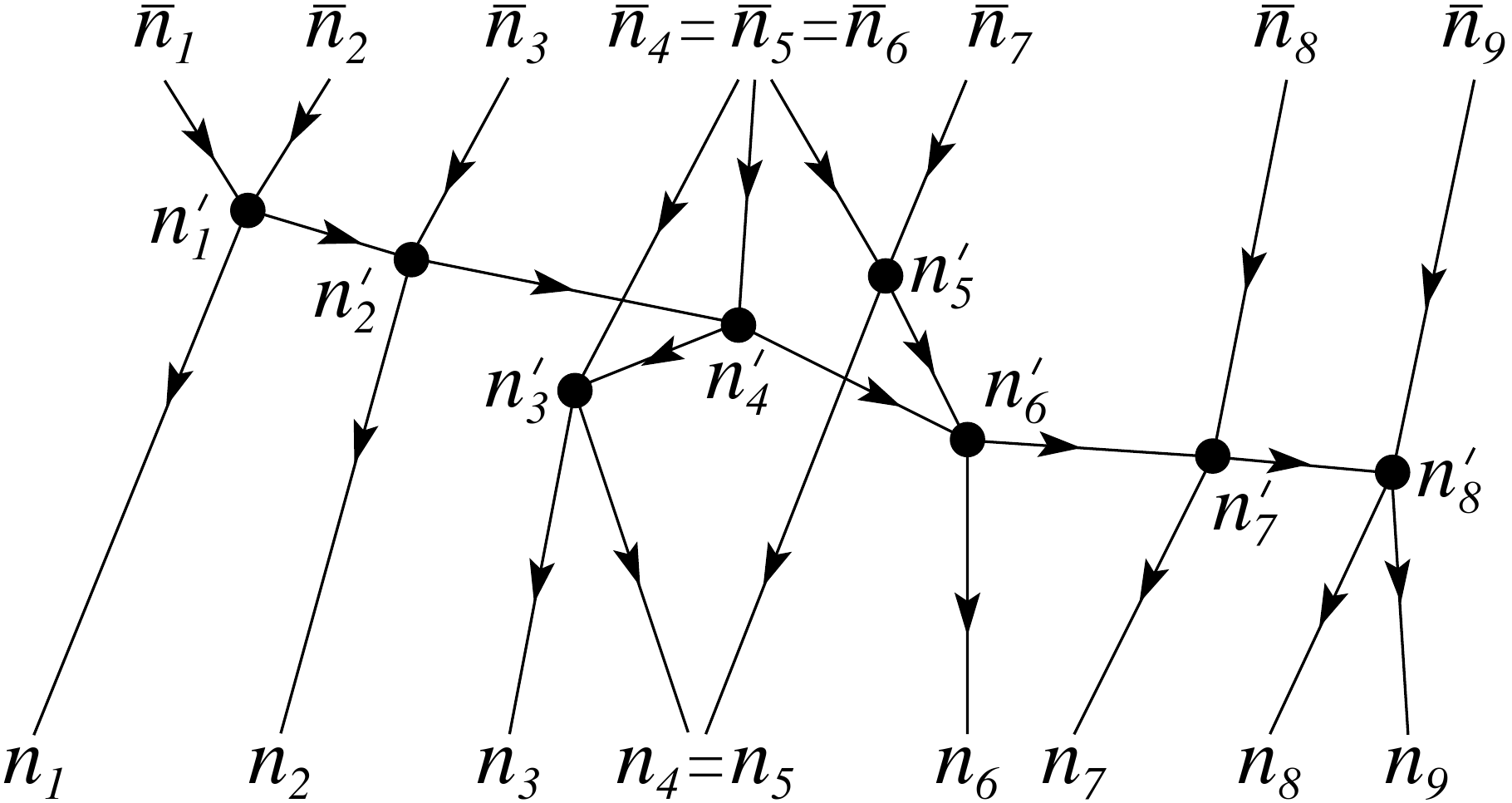}
\end{center}
\caption{\label{fig:spacediag}
An example of the spatial structure for an $N=9$ diagram.}
\end{figure}

First, let us argue that the number of terms $\mathcal{N}$ is
at most exponential in ${N}_r,L_r$. It is convenient
to introduce a pictorial representation of each term, which
illustrates its spatial structure. Namely, let the direction
from left to right represent the position on the chain. Then,
each term of the perturbation theory, contributing to
$\mathcal{K}$, corresponds to definite locations of the free
ends of the solid lines and definite locations of the dotted
lines of the corresponding diagram. For the latter reason it
is more convenient to represent the nonlinear vertex by a smal
circle:
\begin{equation}\label{dotdash=}
\includegraphics[width=6cm]{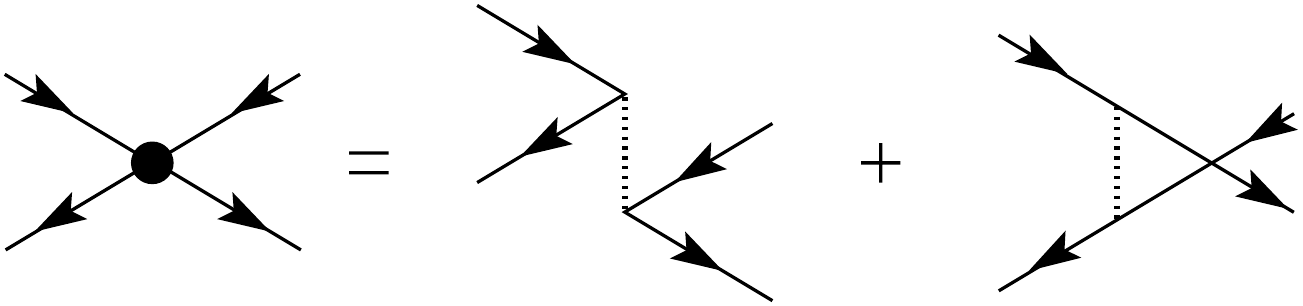}
\end{equation}
An example of a diagram with a spatial structure is shown in
Fig.~\ref{fig:spacediag}.
The vertical direction on the picture has no particular meaning
and is used just for the convenience of presentation. What really
counts is the length of the projection of each solid line on the
horizontal axis, as the sum of these projections for all lines
determines the order in~$\tau$.

Now it is quite easy to see that out of a factorially large number
of permutations of $n^r$'s and $\bar{n}^r$'s, very few actually
contribute. Indeed, while permutation
$\bar{n}_1\leftrightarrow\bar{n}_2$ in Fig.~\ref{fig:spacediag}
does not change the order in $\tau$, the permutation
$\bar{n}_2\leftrightarrow\bar{n}_3$ already increases the order
in~$\tau$, and $\bar{n}_1\leftrightarrow\bar{n}_9$ almost doubles
the order. In fact, a typical permutation contributing to the
factorial number, has a smallness $\tau^{\sim{L}_r^2}$, which
suppresses even the large factorial factor. In fact, the number
of permutations which preserve the order in~$\tau$ cannot exceed
the number of vertices, i.~e., $N_r-1$.

\begin{figure}
\begin{center}
\includegraphics[width=8cm]{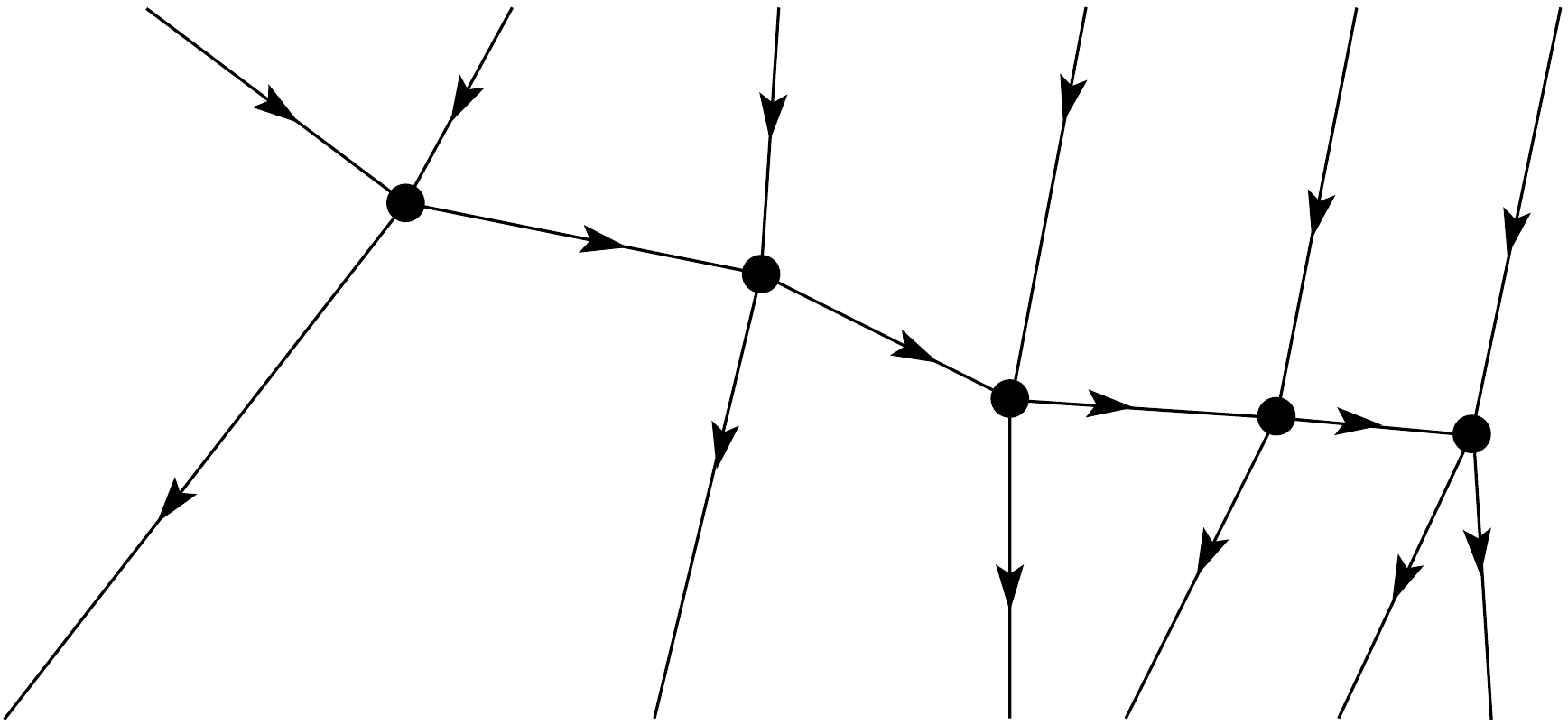}
\end{center}
\caption{\label{fig:sparsediag}
The structure of the diagram for a sparse resonance.}
\end{figure}

Another source of different terms is the summation over the
positions of the nonlinear vertices. In Fig.~\ref{fig:spacediag},
for example, $n_1'$ can be varied between $\bar{n}_1$ and
$\bar{n}_2$ without changing the order in~$\tau$. Clearly, this
freedom is most important when the sites $n_1^r,\ldots{n}_{N_r}^r$
and $\bar{n}_1^r,\ldots\bar{n}_{N_r}^r$ are sparse, i.~e.,
$L_r\gg{N}_r$ (which will be called a sparse resonance).
In this case the most important diagrams have the structure shown
in Fig.~\ref{fig:sparsediag}: the solid lines should be drawn as
vertically as possible, but at least one should go across the whole
spatial region in order to connect all the oscillators. Moreover, for
sparse resonances having $|m_n|>1$ clearly represents a
disadvantage since imposing constraints on the oscillator positions
decreases the freedom in making $\varpi_r$ small, without any gain
in~$\tau$. Then, each nonlinear vertex can be moved within a
segment $\sim{L}_r/N_r$, so all vertices together give a factor
$\sim(L_r/N_r)^{N_r}$. Using the fact that
$\lim(\mathcal{D}_{N+1}/\mathcal{D}_N)=6.75\ldots<e^2$, we arrive
at the following bounds:
\begin{equation}\label{Nterms<}
1\leqslant\mathcal{N}<\left(\frac{L_r}{N_r}\right)^{N_r}e^{2N_r}.
\end{equation}

Let us now turn to the statistics of of each individual term
contributing to $\tilde{\mathcal{K}}_r$.
The denominators of the $2N_r$ external Green's functions contain
differences of two frequencies only, whose probability distribution is
\begin{equation}
\varpi=\left|\omega_n-\omega_{n'}\right|\;\;\;\Rightarrow\;\;\;
p_{1,-1}(\varpi)=\frac{2}{\Delta}
\left\lfloor{1}-\frac{\varpi}{\Delta}\right\rfloor,
\end{equation} 
where we denoted $\lfloor{x}\rfloor\equiv{x}\,\theta(x)$, i.~e.,
$\lfloor{x}\rfloor=x$ for $x>0$ and $\lfloor{x}\rfloor=0$ otherwise.
The $N_r-2$ internal Green's functions contain linear combinations of
many frequencies, so their probability distribution can be estimated
from the central limit theorem:
\begin{equation}\label{CLT=}
\varpi=\left|\sum_{n=-\infty}^\infty{m_n}\omega_n\right|
\;\;\;\Rightarrow\;\;\;
p_{\underline{m}}(\varpi)=\frac{2}\Delta\sqrt{\frac{6}{\pi|\underline{m}|^2}}
\exp\left(-\frac{6\varpi^2}{|\underline{m}|^2\Delta^2}\right),
\end{equation} 
since the dispersion of a single frequency is
\begin{equation}
\int\limits_{-\Delta/2}^{\Delta/2}\frac{d\omega}\Delta\,\omega^2
=\frac{\Delta^2}{12}.
\end{equation}
When the number of terms $\mathcal{N}$ is very large, the distribution
at $\varpi\to{0}$ is the most important (the main contribution to
$\tilde{\mathcal{K}}_r$ coming from largest terms, thus small denominators),
so we replace $p(\varpi)$ by a box distribution of the height
$p(\varpi\to{0})$. The factor $(2/\Delta)$ has been included explicitly
in Eq.~(\ref{tildeKr=}), and for external Green's functions nothing
else is needed. For internal ones, however, $\sqrt{6/(\pi|\underline{m}|^2)}$
should be included.
Since the contribution of the internal Green's functions to $L_r$ can
vary from 0 to 100\%, and internal Green's functions can involve
up to $2N_r$ frequencies, so that $|\underline{m}|^2$ can approach $2N_r^2$,
we can represent each term contributing to $\tilde{\mathcal{K}}_r$ as
\begin{equation}
\tilde{\mathcal{K}}_r^{(\mathrm{1\:term})}=a\zeta,\quad ,
\end{equation}
where the non-random coefficient $a$ is bounded by
\begin{equation}\label{factorialsuppression=}
\frac{1}{(\sqrt{3/\pi}\,N_r)^{L_r}}<a<1,
\end{equation}
and $\zeta$ is a random variable constructed as
\begin{equation}\label{zetadef=}
\zeta=\frac{1}{x_1\ldots{x}_{L_r+N_r-2}},
\end{equation}
$x_1,\ldots,{x}_{L_r+N_r-2}$ being random variables uniformly
distributed between $0$ and $1$.

Strictly speaking, $x_i$'s are not statistically independent,
since the same oscillator frequency can enter several factors
in the denominator.
The resulting correlations can lead to both bunching (meaning
that if one of $x_i$'s is small, than some others are also
likely to be small) and antibunching (if one of $x_i$'s is small,
than some others are less likely to be small) of $x_i$'s.
First, we will estimate $\tilde{\mathcal{K}}_r$ assuming all
$x_i$'s to be
statistically independent. After that, we will study the effect
of both type of correlations on the obtained estimate.

The probability distribution for $\zeta$, defined by
Eq~(\ref{zetadef=}) with independent $x_i$'s,
\begin{equation}\label{pzeta=}
p_{N_r+L_r-2}(\zeta)=\frac{\theta(\zeta-1)}{\zeta^2}\,
\frac{\ln^{N_r+L_r-3}\zeta}{(N_r+L_r-3)!},
\end{equation}
is discussed in detail in \ref{app:productdist}. It is crucial that all
moments of this distribution diverge, so for the sum of $\mathcal{N}$
such random variables with random signs,
$\pm\zeta_1\pm\ldots\pm\zeta_\mathcal{N}$,
the central limit theorem does not hold. Unable to calculate explicitly
the probability distribution of this sum, we estimate the sum from
above and from below. This corresponds to estimating the probability
for the absolute value of the sum to be smaller than some value $z$,
$\mathcal{P}\left\{|\pm\zeta_1\pm\ldots\pm\zeta_\mathcal{N}|<z\right\}$,
from below and from above, respectively.

As discussed in \ref{app:productdist}, the lower bound for the
probability is obtained if one replaces the sum by its largest term:
\begin{equation}
\mathcal{P}\left\{|\pm\zeta_1\pm\ldots\pm\zeta_\mathcal{N}|<z\right\}
\geqslant\mathcal{P}\left\{\zeta_1,\ldots,\zeta_\mathcal{N}<z\right\}
\approx\exp\left[-\mathcal{N}\int\limits_z^\infty
{p}_{N_r+L_r-2}(\zeta)\,d\zeta\right].
\end{equation}
When $\mathcal{N}$ is sufficiently large (exponential in $N_r,L_r$),
it can be approximated as 
\begin{equation}\label{Pz=}
\mathcal{P}\left\{\zeta_1,\ldots,\zeta_\mathcal{N}<z\right\}
\approx\exp\left[-\frac{1}z\,\frac{\mathcal{N}}{(N_r+L_r-3)!}
\frac{\ln^{N_r+L_r-2}z}{\ln(z/e^{N_r+L_r-3})}\right].
\end{equation}
This probability is (i)~very strongly suppressed for $z$~smaller than a
typical value at which the argument of the exponential is of the order
of unity, and (ii)~slowly approaches unity for larger~$z$, so that all
moments of~$z$ diverge.
This typical value of~$z$ is found from
\begin{equation}
\frac{1}{z\,(N_r+L_r-3)!}\frac{\ln^{N_r+L_r-2}z}{\ln(z/e^{N_r+L_r-3})}
=\frac{1}{\mathcal{N}}.
\end{equation}
At large $N_r,L_r$ the solution of this equation is given by
\begin{equation}
z\sim\left[e\,f(\mathcal{N}^{1/(N_r+L_r)})\right]^{N_r+L_r},
\end{equation}
where we omitted the prefactor, a power of $N_r+L_r$, and
the function $f(x)$ is defined as the solution of the equation
\begin{equation}
\frac{f}{1+\ln{f}}=x.
\end{equation}
In particular, if $\mathcal{N}\sim{c}^{N_r+L_r}$, then
the typical $z\sim(ef(c))^{N_r+L_r}$.
To give an idea of the behavior of $f(x)$, we note that
$f(1)=1$, and $x\ln(ex)\leqslant{f}(x)<2x\ln(ex)$ for $x>1$.
To obtain the approximate expression in the region around the
typical value of $z$, we linearize the logarithm of the argument
of the exponential in Eq.~(\ref{Pz=}) with respect to $\ln{z}$.
Using the right inequality~(\ref{Nterms<}) the inequality
\begin{equation}
f\left((L/N)^{N/(L+N)}e^{2N/(L+N)}\right)<e^2,\quad L\geqslant{N},
\end{equation}
and the right inequality~(\ref{factorialsuppression=}) we arrive
at the following bound:
\begin{equation}\label{PK<lambda>}
\mathcal{P}\left\{\tilde{\mathcal{K}}_r<e^{-\lambda}\right\}
>\exp\left[-e^{\lambda+3(L_r+N_r)}\right].
\end{equation}

A lower bound for $|\pm\zeta_1\pm\ldots\pm\zeta_\mathcal{N}|$
is obtained by ``squeezing'' its probability distribution towards
smaller values. As discussed in \ref{app:productdist}, this is
done by first replacing distribution~(\ref{pzeta=}) for each
term~$\zeta$ by the log-normal distribution, applying the central
limit theorem to the latter, and squeezing the resulting Gaussian
into a box of the same height. This corresponds to assuming the sum
to be uniformly distributed in the range
\begin{equation}
0<|\pm\zeta_1\pm\ldots\pm\zeta_\mathcal{N}|
<\sqrt{\frac\pi{2}}\,e^{2(L_r+N_r-3)},
\end{equation}
where the left inequality~(\ref{Nterms<}) was used.
Using the left inequality~(\ref{factorialsuppression=}), we
translate it into the uniform distribution of
$\tilde{\mathcal{K}}_r$ in the range
\begin{equation}\label{0<Kr<}
0<\tilde{\mathcal{K}}_r<\sqrt{\frac\pi{2}}\,
\frac{e^{2(L_r+N_r-3)}}{(\sqrt{3/\pi}\,N_r)^{L_r}}.
\end{equation}
Squeezing this range further, down to $1/N_r^{L_r}$,
for pure analytical convenience,
we obtain 
\begin{equation}\label{PK<lambda<}
\mathcal{P}\left\{\tilde{\mathcal{K}}_r<e^{-\lambda}\right\}
<e^{L_r\ln{N}_r-\lambda}\,\theta(\lambda-L_r\ln{N}_r)
+\theta(L_r\ln{N}_r-\lambda).
\end{equation}

Having found the bounds for the statistics of
$\tilde{\mathcal{K}}_r$,
let us obtain bounds for the following average:
\begin{equation}
\overline{\lfloor\tilde\lambda+\ln\tilde{\mathcal{K}}_r\rfloor^N}
=-\int\limits_{-\infty}^{\tilde\lambda}
(\tilde\lambda-\lambda)^N\,\frac{d}{d\lambda}
\mathcal{P}\left\{\tilde{\mathcal{K}}_r<e^{-\lambda}\right\}
d\lambda,\quad
\lfloor{x}\rfloor\equiv{x}\,\theta(x).
\end{equation}
For the bound~(\ref{PK<lambda>}) the integral can be calculated as
\begin{equation}\begin{split}
&-\int\limits_{-\infty}^{\tilde\lambda}
(\tilde\lambda-\lambda)^N\,\frac{d}{d\lambda}\,
\exp\left(-e^{\lambda-\lambda_r}\right)\,
d\lambda\approx\int\limits_{\max\{0,\tilde\lambda-\lambda_r\}}^\infty
\lambda^Ne^{\tilde\lambda-\lambda_r-\lambda}\,d\lambda=\\
&=\left\{\begin{array}{ll}
(\tilde\lambda-\lambda_r)^{N+1}/(\tilde\lambda-\lambda_r-N),
&\tilde\lambda-\lambda_r-N\gg\sqrt{N},\\
N!\,e^{\tilde\lambda-\lambda_r},&
N-(\tilde\lambda-\lambda_r)\gg\sqrt{N},
\end{array}\right.
\label{thetalnAv12=}
\end{split}
\end{equation}
where the abrupt cutoff of $\exp(-e^{\lambda-\lambda_r})$
at $\lambda>\lambda_r$ was simply replaced by a step function.
For the bound~(\ref{PK<lambda<}) we obtain the integral
\begin{equation}
\theta(\tilde\lambda-\lambda_r)\int\limits_{\lambda_r}^{\tilde\lambda}
(\tilde\lambda-\lambda)^Ne^{\lambda_r-\lambda}\,d\lambda
=\left\{\begin{array}{ll}
(\tilde\lambda-\lambda_r)^{N+1}/(\tilde\lambda-\lambda_r-N),
&N\gg{1},\\
(\tilde\lambda-\lambda_r)^N,& N\ll\tilde\lambda-\lambda_r.
\end{array}\right.
\end{equation}
In the following, we will be interested in the situation
$\tilde\lambda-\lambda_r\gg{N}$, so our working estimate will be
\begin{equation}\label{thetalnAv=}
\lfloor\tilde\lambda-L_r\ln{N}_r\rfloor^N
<\overline{\lfloor\tilde\lambda+\ln\tilde{\mathcal{K}}_r\rfloor^N}
<\lfloor\tilde\lambda+3(L_r+{N}_r)\rfloor^N.
\end{equation}

Let us now see how the above estimates are modified by correlations
between different $x_i$'s in Eq.~(\ref{zetadef=}). To estimate the
effect of bunching, we consider its extreme case by requiring that
variables in groups of~$m$ are identical:
$x_1=x_2=\ldots=x_m$, $x_{m+1}=\ldots=x_{2m}$, etc., 
This corresponds to replacement $\zeta\to\zeta^{1/m}$,
$L_r+N_r-2\to(L_r+N_r-2)/m$ is Eq.~(\ref{zetadef=}). Repeating the
same steps, we obtain Eq.~(\ref{PK<lambda>}) with
$e^{\lambda+3(L_r+N_r)}\to{e}^{\lambda/m+3(L_r+N_r)/m}$,
and Eq.~(\ref{0<Kr<}) with $e^{2(L_r+N_r-3)}\to{e}^{2m(L_r+N_r-3)}$.
Note that $m$~cannot be too large, as this imposes strong restrictions
on the structure of the resonance, so we assume $m\ll{L}_r,N_r$.
Thus, the modification of Eq.~(\ref{PK<lambda>}) has no effect on
the estmate (\ref{thetalnAv=}) since one still can replace
$\exp(-e^{(\lambda-\lambda_r)/m})$ by a step function in
Eq.~(\ref{thetalnAv12=}). As for modification of Eq.~(\ref{0<Kr<}),
it is completely absorbed in the passage from Eq.~(\ref{0<Kr<})
to Eq.~(\ref{PK<lambda<}). To estimate the effect of antibunching,
we set some of the variables in Eq.~(\ref{zetadef=}) to $x_i=1$,
thereby replacing $L_r+N_r-2$ by a smaller number $N'$. This
replacement can only decrease the value of $\zeta$, so the
probabilistic upper bound on $\zeta$, expressed by
Eq.~(\ref{PK<lambda>}), remains valid. In Eq.~(\ref{0<Kr<}) we
have to replace $e^{2(L_r+N_r-3)}\to{e}^{2(N'-1)}$, so this
modification is also absorbed in the passage to
Eq.~(\ref{PK<lambda<}). To summarize, the bounds established
previously for uncorrelated $x_i$'s, are loose enough, so that the
effects of correlations fit in these bounds as well.

\subsection{On the number of resonances}
\label{sec:numres}

Here we estimate the number of different coupling terms of the
form (\ref{VN=}), $\mathcal{R}(N_r,L_r)$, corresponding to given
values of $N_r$ and $L_r$. One of the
oscillators 
is assumed to be fixed; otherwise, $\mathcal{R}(N_r,L_r)$ would be
infinite due to the translational degree of freedom.
As will be seen shortly, $\mathcal{R}(N_r,L_r)$ at large
$N_r,L_r$ can be represented as
\begin{equation}\label{RNrLr=}
\mathcal{R}(N_r,L_r)=\mathcal{R}_0(N_r,L_r)
\left[\resfr(L_r/N_r)\right]^{N_r},
\end{equation}
where $\mathcal{R}_0(N_r,L_r)$ is a power-law prefactor.
Thus, most interesting for us is the quantity
\begin{equation}
\resfr(p)=\lim_{N_r\to\infty}\left[\mathcal{R}(N_r,pN_r)\right]^{1/N_r}.
\end{equation}
Finiteness of this quantity (i.~e., the fact that it is
neither zero, nor infinity) is equivalent to the validity of
representation~(\ref{RNrLr=}). The quantity $p=L_r/N_r$ is
nothing but the ratio between the order of the perturbation
theory in~$\tau$ and the order in~$g$.
Note that $L_r\geqslant{N_r}$, since $n^r_k\neq\bar{n}_{k'}^r$ for
any $k,k'$, and the oscillators must be coupled. Thus,
\begin{equation}
\resfr(p<1)=0.
\end{equation}

To obtain an upper bound on $R(p)$, let us choose some
$2N_r$-oscillator
diagram and fix one of its free ends to a given site. Let us now
count all possible positions of all other free ends and nonlinear
vertices which produce the required~$L_r$. In total, the diagram
has $2N_r-2$ Green's functions, each corresponding to a certain
displacement along the chain. Thus, we simply have to sum over all
possible displacements, keeping their sum equal to~$L_r$. As
mentioned above, $N_r$~displacements cannot be smaller than unity,
others can be zero:
\begin{equation}\begin{split}
&\sum_{k_1,\ldots,k_{N_r}\geqslant{1}}
\sum_{k_{N_r+1},\ldots,k_{3N_r-2}\geqslant{0}}
\delta_{k_1+\ldots+k_{3N_r-2},L_r}=
\sum_{k_1,\ldots,k_{3N_r-2}\geqslant{0}}
\delta_{k_1+\ldots+k_{3N_r-2},L_r-N_r}=\\
&=\int\limits_0^{2\pi}\frac{d\phi}{2\pi}
\frac{e^{-i(L_r-N_r)\phi}}{(1-e^{i\phi})^{3N_r-2}}
=\frac{(L_r+2N_r-3)!}{(L_r-N_r)!\,(3N_r-3)!}.
\end{split}\end{equation}
This expression should be multiplied by $2^{3N_r-2}$, since
each $k_j$ can correspond to a step to the left or to the right.
This procedure produces all resonances with the given~$L_r$ which
involve the starting site. Moreover, each resonance is counted
several times, because the spatial arrangements of the diagram,
differing by the positions of the nonlinear vertices, but with 
the same positions of the ends, are counted as different
configurations. Thus, it indeed produces an upper bound,
and we have
\begin{equation}
\resfr(p)\leqslant\frac{8(p+2)^{p+2}}{27(p-1)^{p-1}}\leqslant{8}(2p-1)^3.
\end{equation}

To obtain a lower bound, consider a resonance with $n_k=2k-1$,
$\bar{n}_k=2k$,
$k=1,\ldots,N_r$. Coupling the oscillators by a diagram similar
to the one shown in Fig.~\ref{fig:sparsediag} gives $L_r=3N_r-2$.
If one exchanges $\bar{n}_k\leftrightarrow{n}_{k+1}$ for some~$k$,
$L_r$ is not changed, but a different resonance is obtained.
Thus, we can construct explicitly $2^{N_r-1}$ resonances with
$L_r=3N_r-2$, which means that $\resfr(p=3)\geqslant{2}$.
One can generalize this estimate by considering resonances with
$\bar{n}_{N_r}-n_1=L-1$, and fixed $|\bar{n}_k-n_{k+1}|=1$.
Then $L_r=L+N_r-2$, while the number of such resonances is at
least $2^{N_r-1}(L/2-1)!/[(N_r-1)!(L/2-N_r)!]$ ($L$~is divided
by two, since we are counting the ways to distribute
nearest-neighbor pairs). This gives
\begin{equation}
\resfr(p\geqslant{3})\geqslant\sqrt{\frac{(p-1)^{p-1}}{(p-3)^{p-3}}}
\geqslant{e}(p-3).
\end{equation}

Taking into account the power-law prefactors, we obtain the
upper and lower bounds as
\begin{equation}\begin{split}\label{<RNrLr<}
\frac{e^{N_r}}{\sqrt{2\pi{N}_r}}
\left\lfloor\frac{L_r}{N_r}-3\right\rfloor^{N_r-1}
\leqslant\mathcal{R}(N_r,L_r)
\leqslant{64}^{N_r}
\left\lfloor\frac{L_r}{N_r}-\frac{1}2\right\rfloor^{3N_r},
\end{split}\end{equation}
where $\lfloor{x}\rfloor\equiv{x}\,\theta(x)$, i.~e.,
$\lfloor{x}\rfloor=x$ for $x>0$ and $\lfloor{x}\rfloor=0$
otherwise.
We will also need an upper bound for the contribution to
$\mathcal{R}(N_r,L_r)$ from resonances with typical values
of multiplicities $|m_n^r|\sim{m}_r$. If we
constrain $|m_n^r|\geqslant{m}_r$, the number of
oscillators is $m^r$ times smaller. Thus, in the number of
combinations we replace $N_r\to{N}_r/m_r$ (but not in the
ratio $L_r/N_r$, since it cannot be too small even for high
multiplicities): 
\begin{equation}\label{RNrLrm=}
\mathcal{R}(N_r,L_r)\leqslant\sum_{m_r=1}^{N_r}{64}^{N_r/m_r}
\left\lfloor\frac{L_r}{N_r}-\frac{1}2\right\rfloor^{3N_r/m_r}.
\end{equation}



\subsection{On resonant triples for not too weak disorder}
\label{sec:Motttriples}

The analysis of the stochastic phase volume of nearest-neighbor
triples was performed in Sec.~\ref{sec:Spots} under the condition
$\tau\ll\rho$. This condition guaranteed that for two neighboring 
oscillators the phase space had a separatrix when the difference
between their bare frequencies $|\omega_n-\omega_{n+1}|$ was 
sufficiently small. In the opposite case, $\tau\gg\rho$, the minimal
splitting of frequencies of the neighboring oscillators is
$2\tau\Delta$ because of the coupling (also known as level repulsion,
in the quantum-mechanical language), so the nonlinearity is too
weak to produce the separatrix, as seen from Eq.~({twooscboundary=}).

Two normal modes of the disordered linear chain can have 
eigenfrequencies much closer than $\tau\Delta$ only if they are
well separated in space. Indeed, the effective coupling between
two oscillators separated by a distance~$l$ can be obtained in
the $(l-1)$st order of the perturbation theory, and is given
by Eq.~(\ref{Vtwoosc=}). Strictly speaking, it is a random quantity,
so its statistics should be analyzed. However, we note that the
typical value of the splitting scales as $\tau^l\Delta$, rapidly
falling off with the distance. As soon as this splitting becomes
$\sim\rho\Delta$ or smaller, the separatrix can appear. On the one
hand, this can always be satisfied when $l$ is large. On the other,
if $l$ is too large, the coupling will be very weak, so the
effective pendulum frequency will be very small, and the stochastic
layer will be very thin due to Melnikov-Arnold exponential. Thus,
the optimal value of~$l$ is such that $\tau^l\sim\rho$. The same
arguments can be used for the choice the third oscillator, needed to
destroy the separatrix and to produce the stochastic layer. Thus,
we face the situation similar to that of Sec.~\ref{sec:Spots}, but
the coupling constant~$\tau$ should be replaced by
$\tau\to\tau^l\sim\rho$. Hence, the average density of the chaotic
spots for $\tau\gg\rho$ is $\bar{w}\sim\rho^2$.

Finally, we note that the situation described above (effective
tunnelling coupling between two oscillators being of the same
order as the nonlinearity and their frequency mismatch) corresponds
to nonlinear coupling between two double-humped eigenstates of the
linear problem. The distance between the two humps is given by~$l$,
and the two eigenstates correspond to different mutual signs of the
humps of the wavefunction. The importance of double-humped states
for the wave-packet spreading was discussed in Ref.~\cite{Veksler2010}.

\section{Probability distribution of the chaotic fraction}
\label{sec:Density}
\setcounter{equation}{0}

\subsection{General remarks on the procedure}
\label{sec:remarkstat}

Each site~$n$ of the chain can be characterized by a variable~$w_n$,
the chaotic fraction of the thermally-weighted phase volume, summed
over all possible guiding resonances involving the site~$n$ and sites
to the right of it [to be formally defined below, Eq.~(\ref{wgdef=})].
For a resonant nearest-neighbor triple at $\tau\ll\rho$ this quantity
was defined in Eq.~(\ref{wtripledef=}) and estimated in
Sec.~\ref{sec:3oscchaotic},
the result being [see Eqs.~(\ref{wtriple1=}),~(\ref{wtriple2=})]
\begin{equation}\label{wtrip=}
w_n\sim\frac\tau\rho\,{e}^{-E_n/T},\quad 2gE_n=
3\max_{k=n,n+1,n+2}(\omega_k-\mu)^2-\sum_{k=n}^{n+2}(\omega_k-\mu)^2.
\end{equation}

Note that the definition of $w_n$ involves the thermal averaging
but not the disorder averaging. Thus, $w_n$ is still a random
quantity, determined by random frequencies of the oscillators in
the vicinity of the site~$n$. Our goal in this section is to
find its probability distribution function for values $w$ much
smaller than its disorder-averaged value,
$w\ll\min\{\tau\rho,\rho^2\}$. For this
purpose Eq.~(\ref{wtrip=}) is of no use, as it greatly
underestimates the typical value of~$w$ (the value where the
probability distribution has a maximum), giving
$w\sim(\tau/\rho)e^{-1/\rho}$. In fact, one can typically find
chaotic phase volume at lower energies by taking into consideration
guiding and layer resonances which involve many oscillators.

Typically, for a given guiding resonance~$g$ and a given layer
resonance~$\ell$, the corresponding contribution to the chaotic
fraction, $w^{(g,\ell)}$, is very small (much smaller than
for a nearest-neighbor triple, as high orders of perturbation
theory have to be involved). That is, the probability that it
exceeds some value $w_0=e^{-\lambda}$,
$\mathcal{P}\left\{w^{(g,\ell)}>w_0\right\}$, is small
unless $w_0$~itself is very small. However, $w<w_0$
means that $w^{(g,\ell)}<w_0$ for all of the numerous candidates
for guiding
and layer resonances. The probability of this event can be
written as (assuming statistical independence of the contributions
from different resonances)
\begin{equation}\label{sumwl=}
\mathcal{P}\{w<w_0\}=\prod_{g,\ell}
\left(1-\mathcal{P}\left\{w^{(g,\ell)}>w_0\right\}\right)
\approx\exp\left(-\sum_{g,\ell}\mathcal{P}\left\{w^{(g,\ell)}>w_0\right\}\right)
\equiv{e}^{-S}.
\end{equation}
Thus, $S$ is determined by the competition of the smallness
of each individual probability
$\mathcal{P}\left\{w^{(g,\ell)}>w_0\right\}$
and the large number of terms in the sum.

It is more convenient technically to perform the summation
in Eq.~(\ref{sumwl=}) in two steps. First, we calculate
\begin{subequations}
\begin{equation}
S_g(\lambda)=\sum_\ell\overline{\theta(w^{(g,\ell)}-e^{-\lambda})},
\end{equation}
where $\theta(x)$ is the Heaviside step function, and
the overline denotes the statistical averaging
over the frequencies involved in the layer resonance,
while the guiding resonance is kept fixed.
Thus, $e^{-S_g}$ is the probability of $w^{(g)}<e^{-\lambda}$
for a given guiding resonance, but after the optimization over
the layer resonances. On the second step, we calculate
\begin{equation}
S(\lambda)=\sum_g\mathcal{P}\left\{w^{(g)}>e^{-\lambda}\right\}
=\sum_g\left(1-\overline{e^{-S_g(\lambda)}}\right),
\end{equation}
where the averaging is performed over the frequencies of the
guiding resonance. This latter averaging is assumed to be
independent from the former one.
\end{subequations}
Note that expanding the exponential and simply averaging
$\overline{S_g}$ would produce a wrong result, since at
large~$\lambda$ (the ones we are interested in) the main
contribution to the average would come from configurations
for which the expansion $1-e^{-S_g}\approx{S}_g$ is not valid.
This point will be discussed in more detail in 
Sec.~\ref{sec:Densityguidinggeneral}.

The main body of the bulky calculations is presented in
Secs.~\ref{sec:Densitythermal}--\ref{sec:Densityguiding},
and the result is summarized in Sec.~\ref{sec:DensitySummary}.

\subsection{Definitions}
\label{sec:Densitydefinitions}

Consider a guiding resonance~$g$
(defined as in Sec.~\ref{sec:statistics}, $r=g$)
with the effective coupling as in Eq.~(\ref{V=Vr}):
\begin{equation}
V=-2V_g
\cos\left(\phi_{n_1^g}+\ldots+\phi_{n^g_{N_g}}-
\phi_{\bar{n}_1^g}-\ldots-\phi_{\bar{n}^g_{N_g}}\right).
\end{equation}
This coupling determines the effective pendulum Hamiltonian~$\Hdif_g$
and the frequency~$\Omega_g$,
\begin{equation}\label{Hpendulumg=}
\Hdif_g=\frac{g}2\,|\underline{m}^g|^2\pdif^2-2V_g\cos\phidif,\quad
\Omega_g=\sqrt{2|\underline{m}^g|^2g|V_g|},
\end{equation}
where $\phidif=\phi_{n_1^g}+\ldots+\phi_{n^g_{N_g}}-
\phi_{\bar{n}_1^g}-\ldots-\phi_{\bar{n}^g_{N_g}}$ is the slow
phase (pendulum coordinate), and $\pdif$~is the pendulum momentum
(see \ref{app:change} for the discussion of the kinetic energy
term).

The resonance condition reads as
\begin{equation}\label{Igmin=}
I_{n^g_1}+\ldots+I_{n^g_{N_g}}
-I_{\bar{n}^g_1}-\ldots-I_{\bar{n}^g_{N_g}}=
-\frac{\omega_{n_1^g}+\ldots+\omega_{n^g_{N_g}}
-\omega_{\bar{n}_1^g}-\ldots-\omega_{\bar{n}^g_{N_g}}}g
\equiv{I}^\mathrm{min}_g.
\end{equation}
The right-hand side is positive, as we have imposed
condition~(\ref{nbarnfix=}). Then the resonant configuration with
the lowest activation barrier is obtained by setting to zero all
$I_{\bar{n}}$'s, and all $I_{n}$'s except one, $I_{n^g_*}$, where
$n^g_*$ is the site having the largest multiplicity~$m^g_n$, and
if there are several such sites, the
one with the smallest $\omega_n$ should be chosen. For this site
one should set $I_{n^g_*}={I}^\mathrm{min}_g/m^g_{n^g_*}$.
This lowest activation barrier is given by
\begin{equation}\label{Egmindef=}
E_g=
(\omega_{n^g_*}-\mu)\frac{I^\mathrm{min}_g}{m^g_{n^g_*}}
+\frac{g}2\left(\frac{I^\mathrm{min}_g}{m^g_{n^g_*}}\right)^2.
\end{equation}
The chaotic fraction $w_n^{(g)}$ corresponding to a given guiding
resonance~$g$ is defined as
\begin{equation}\label{wgdef=}
w^{(g)}_{n_1^g}=
\left\langle|\underline{m}^g|^2{W}_s\,\delta\left(I_{n_1^g}+\ldots+I_{n_{N_g}^g}
-I_{\bar{n}_1^g}-\ldots-I_{\bar{n}_{N_g}^g}-I^\mathrm{min}_g\right)
\right\rangle_T,
\end{equation}
where the thermal average is defined in the usual way,
\begin{equation}
\langle\ldots\rangle_T=
\frac{\int(\ldots){e}^{(H-\mu{I}_{tot})/T}
\prod_{n=-\infty}^\infty{d}I_nd\phi_n/2\pi}
{\int{e}^{(H-\mu{I}_{tot})/T}\prod_{n=-\infty}^\infty{d}I_nd\phi_n/2\pi}.
\end{equation}
The prefactor $|\underline{m}^g|^2{W}_s$ in front of the
$\delta$-function in Eq.~(\ref{wgdef=}) represents the effective
thickness of the resonant hyperplane defined by the
$\delta$-function, and is discussed in detail
in \ref{app:change}. $W_s$~is the volume of the stochastic
layer in the phase space of the effective pendulum. It is
determined by the strength and the frequency mismatch of the
layer resonance, so it depends on the actions of all oscillators
involved in the layer resonance. The definitions and notations
for the layer resonance are again as in Sec.~\ref{sec:statistics},
$r=\ell$. Then $W_s$ can be found similarly to the case of a
resonant pair, Eq.~(\ref{pendulumWs=}):
\begin{subequations}\begin{equation}\label{Wsgl=}
W_s=\frac{\Omega_g}{\pi|\underline{m}^g|^2g}
\frac{5}2\frac{|V_\ell|}{|V_g|}\,\Lambda^2\mathcal{A}_m(\Lambda)
\ln\frac{2^{21/5}e|V_g|/|V_\ell|}{\Lambda^2\mathcal{A}_m(\Lambda)},
\end{equation}
where the Melnikov-Arnold integral is defined in
Eq.~(\ref{MelnikovArnold=}), and its index and the argument
are given by
\begin{equation}
m=2\,\frac{|(\underline{m}^g,\underline{m}^\ell)|}{|\underline{m}^g|^2},
\quad
\Lambda=
\frac{\tilde\omega_{n_1^\ell}+\ldots+\tilde\omega_{n^\ell_{N_\ell}}
-\tilde\omega_{\bar{n}_1^\ell}-\ldots-\tilde\omega_{\bar{n}^\ell_{N_\ell}}}%
{\Omega_g}
\equiv\frac{\varpi_\ell}{\Omega_g}.
\end{equation}\end{subequations}

Definition~(\ref{wgdef=}) and expression~(\ref{Wsgl=})
are meaningful only when $\Lambda\gg{1}$.
When this is not the case, the two resonances should be treated
on equal footing, so the situation is analogous to the case of the
resonant triple. In particular, one arrives at a Hamiltonian
similar to Eq.~(\ref{H3expand=}) with the kinetic energy given by
$(g/2)|\underline{m}^g|^2p_1^2
+g(\underline{m}^g,\underline{m}^\ell)p_1p_2
+(g/2)|\underline{m}^\ell|^2p_2^2$.
In Eq.~(\ref{wgdef=}) the second $\delta$-function should be
introduced, and instead of $|\underline{m}^g|^2W_s$ the prefactor
should be
$(|\underline{m}^g|^2|\underline{m}^\ell|^2
-(\underline{m}^g,\underline{m}^\ell)^2)W_{ss}$, where
$W_{ss}$~is defined analogously to Eq.~(\ref{Wssdef=}).
It will be seen below, however, that for those values of~$w$ which
are important for the transport, $\Lambda\gg{1}$ holds. Also,
$\Lambda\gg{m}$ (indeed, $m$~can grow at most linearly with
$N_g,N_\ell$, while $\Lambda$ is exponential in ${N}_g$),
so that the asymptotic expression~(\ref{MAasymp=}) for the
Melnikov-Arnold integral can be used.

Expression~(\ref{Wsgl=}) for $W_s$ is already quite cumbersome
for explicit analytical calculations. Hence, we will work with
its upper and lower bounds,
\begin{equation}\label{<m2Ws<}
10N_g\,\frac\Delta{g}\left(\frac{g|V_\ell|}{\Delta^2}\right)
\frac{1}{(2N_\ell)!}\,e^{-\pi\Lambda/2}
<|\underline{m}^g|^2W_s
<20N_g^2\,\frac\Delta{g}\left(\frac{g|V_\ell|}{\Delta^2}\right)^{1/\pi}
\frac{(2\Lambda)^{(m+1)/\pi}}{[\Gamma(m)]^{1/\pi}}\,
e^{-\Lambda/2}.
\end{equation}
They can be obtained using the following inequalities,
which are easy to check:
\begin{eqnarray*}
&&x\ln\frac{2^{21/5}e}{x}<8x^{1/\pi},\quad
2N_g\leqslant|\underline{m}^g|^2<2N_g^2,\quad
\frac{1}{2N_g^2}\leqslant
\frac{|(\underline{m}^g,\underline{m}^\ell)|}{|\underline{m}^g|^2}<N_\ell,\\
&&\Omega_g<\Delta\;\;\;\Rightarrow\;\;\;|V_g|<\frac{\Delta^2}{4gN_g}.
\end{eqnarray*}
The upper bound for $|\underline{m}^g|^2W_s$ can be further
simplified by noting that large values of $m$ are extremely
inefficient. Indeed, if the guiding and
the layer resonances have just one site in common, $m\sim{1}/N_g$.
To increase $m$ to unity and beyond, about $mN_g$ oscillators of
the layer resonance should be restricted to overlap with the guiding
resonance. According to the estimate~(\ref{<RNrLr<}), such a
restriction would decrease the number of available layer resonances
by a factor $\sim{e}^{mN_g}$, which would increase the typical
smallest $\varpi_\ell$ by the same factor. Thus, we can estimate
$\Lambda_m\sim\Lambda_0e^{mN_g}$. The minimum of the function
\[
\frac{(\Lambda_0e^{mN_g})^{(m+1)/\pi}}{[\Gamma(m)]^{1/\pi}}
\exp\left(-\Lambda_0e^{mN_g}/2\right)
\]
for $\Lambda_0\gg{1}$ is reached at small $m\to{0}$.
Using also the fact that $\Lambda^{1/\pi}e^{-\Lambda/2}<e^{-\Lambda/3}$,
we obtain the following upper bound:
\begin{equation}
|\underline{m}^g|^2W_s
<25N_g^3\,\frac\Delta{g}\left(\frac{g|V_\ell|}{\Delta^2}\right)^{1/\pi}
e^{-\Lambda/3}.
\end{equation}

\subsection{Thermal averaging}
\label{sec:Densitythermal}

To integrate over the actions, it is convenient to resolve the
$\delta$-function in Eq.~(\ref{wgdef=}) with respect to $I_{n_*^g}$.
As will be seen later, the typical value of action for oscillators
with $n^g,\bar{n}^g,n^\ell,\bar{n}^\ell\neq{n}^g_*$, as well as the
typical thermal value of the action, $T/|\mu|$, are much smaller
than the typical value of $I_g^\mathrm{min}$:
\begin{equation}\label{ITwllIgmin=}
I_n,\frac{T}{|\mu|}\ll{I}_g^\mathrm{min}.
\end{equation}
Thus, we simply set $I_{n_*^g}=I_g^\mathrm{min}$.

\subsubsection{Lower bound}

In order to obtain the lower bound for~$w$ we write
\begin{equation}
\frac{\pi\Lambda}2<
\frac{\pi\varpi_\ell}{4\sqrt{N_gg\mathcal{K}_g(I_g^\mathrm{min})^{m_{n_*^g}/2}}}
\prod_{n\neq{n^g_*}}\frac{1}{I_n^{|m_n^g|/4}}
<\frac{\pi\varpi_\ell}{4\sqrt{N_gg\mathcal{K}_g(T/|\mu|)^{m_{n_*^g}/2}}}
\prod_{n\neq{n^g_*}}\frac{1}{I_n^{|m_n^g|/4}},
\end{equation}
while $|V_\ell|$ can be averaged over the thermal distribution
independently.
For the integral over the oscillators of the guiding resonance
we use the inequality, derived in \ref{app:intGibbs}
for $A\gg{1}$:
\begin{equation}\begin{split}
&\int\limits_0^\infty
\exp\left[-A\prod_{n\neq{n^g_*}}\frac{1}{x_n^{|m_n^g|/4}}\right]
\prod_{n\neq{n^g_*}} e^{-x_n}dx_n
>\exp
\left[-(2N_g-m_{n_*^g}+4)\left(\frac{A}{4}\right)^{4/(2N_g-m_{n_*^g}+4)}\right],
\label{lowerestint=}
\end{split}\end{equation}
and simplify the argument of the exponential using
\begin{equation}
(2N_g-m_{n_*^g}+4)
\left(\frac{\pi}{16}\frac{1}{\sqrt{N}_g}\right)^{4/(2N_g-m_{n_*^g}+4)}<2N_g,
\quad A^{{4/(2N_g-m_{n_*^g}+4)}}<A^{4/N_g},
\end{equation}
while for the oscillators of the layer resonance we use the
inequality
\begin{equation}
\int\limits_0^\infty\prod_nx_n^{|m_n^\ell|/2}\,e^{-x_n}dx_n>e^{-N_\ell},
\end{equation}
following from the fact that $\Gamma(x+1)>e^{-x}$.
%
As a result, the lower bound for $w$ is
\begin{equation}
w^{(g,\ell)}>\frac{\mathcal{K}_\ell(T/e|\mu|)^{N_\ell-1}}{\Delta(2N_\ell)!}\,
e^{-E_g/T}e^{-2N_g\Lambda_T^{4/N_g}},
\quad\Lambda_T\equiv
\frac{\varpi_\ell}{\sqrt{g\mathcal{K}_g(T/|\mu|)^{N_g}}}.
\end{equation}

\subsubsection{Upper bound}

Let us first average the actions of the oscillators belonging to the
layer resonance, assuming it to have only one oscillator in common
with the guiding resonance 
(see the discussion in the end of Sec.~\ref{sec:Densitydefinitions}).
\begin{equation}\begin{split}
w^{(g,\ell)}<{}
\frac{25N_g^3}{\Gamma(1/(2\pi)+1)}\,\frac{|\mu|\Delta}{gT}
\left[\sqrt{\frac{E_g}{T}}\frac{g\mathcal{K}_\ell}{\Delta^2}
\left(\frac{T}{|\mu|}\right)^{N_\ell}\right]^{1/\pi}
{e}^{-E_g/T}\left\langle{e}^{-\Lambda/3}\right\rangle_T
\prod_{n=-\infty}^\infty\Gamma\left(\frac{|m_n^\ell|}{2\pi}+1\right).
\end{split}\end{equation}
As in Sec.~\ref{sec:Densitydefinitions}, this estimate can be
simplified using the fact that large values of
$|m_n^\ell|$ are inefficient. Indeed, suppose all
$|m_n^\ell|$ are either equal to $m$ or 0
($1\ll{m}<N_\ell$). Then the product of the gamma functions can be
estimated as
\[
[\Gamma(m/2\pi+1)]^{2N_\ell/m}\sim
\left(\frac{m}{2\pi{e}}\right)^{N_\ell/\pi}.
\]
At the same time, this leads to an increase of $\Lambda$,
$\Lambda\sim\Lambda_{m=1}e^{N_\ell}e^{-N_\ell/m}$. Clearly,
the maximum of the function
$m^{N_\ell/(2\pi)}e^{-\Lambda_1e^{N_\ell}e^{-N_\ell/m}}$
is reached at the smallest $m=1$.
Analogously, one can show that restriction of one of the sites
of the layer resonance to $n_*^g$ again does more harm by
increasing $\Lambda$ by a factor of the order of~1, than
help by introducing a factor $E_g/T$. Using
$\Gamma(1/(2\pi)+1)<1$, we can simply replace all gamma
functions by unity.

The average $\left\langle{e}^{-\Lambda/3}\right\rangle_T$ is
sensitive to the multiplicities $m_n^g$. Again,
large multiplicities are not efficient in lowering
$E_g$, however, this fact is not obvious at this
stage of the calculation, and it will be proven later, in
Sec.~\ref{sec:Densityguidingupper}. For the moment, we keep
the typical multiplicity $m^g$ as an additional degree of
freedom and optimize with respect to it afterwards. The
thermal average is estimated using the inequality
\begin{equation}\label{uglyineq=}
\int\limits_0^\infty 
\exp\left[-x_1-\ldots-x_N-\frac{A}{(x_1\ldots{x}_N)^\alpha}\right]
dx_1\ldots{d}x_N
\leqslant{N}\left(\frac{8\pi}{e^2}\right)^{N/2}
\exp\left[-\frac{N}{2}\,(\alpha{A})^{1/(N\alpha+1)}\right],
\end{equation}
proven in \ref{app:intGibbs}. Let set $\alpha=m^g/4$,
$N=(2N_g-m^g_{n_*^g})/m^g$. In fact,
\[\begin{split}
&\frac{2N_g-m^g_*}{m^g}\left(\frac{8\pi}{e^2}\right)^{(2N_g-m_*^g)/(2m^g)}
\exp\left[-\frac{2N_g-m_*^g}{2m^g}\,
(\alpha{A})^{4/(2N_g-m_g^*+4)}\right]\leqslant\\
&\leqslant{2}N_g\left(\frac{8\pi}{e^2}\right)^{N_g/m^g}
\exp\left[-\frac{N_g}{2m^g}\,
(\alpha{A})^{2/N_g}\right],
\end{split}\]
where we denoted $m^g_*\equiv{m}^g_{n_g^*}$ for compactness.
Thus, we arrive at the following upper bound:
\begin{equation}\begin{split}
&w^{(g,\ell)}<
50N_g^4\,\frac{|\mu|\Delta}{gT}
\left[\sqrt{\frac{E_g}{T}}\frac{g\mathcal{K}_\ell}{\Delta^2}
\left(\frac{T}{|\mu|}\right)^{N_\ell}\right]^{1/\pi}
\left(\frac{8\pi}{e^2}\right)^{N_g/m^g}{e}^{-E_g/T}\times\\
&\qquad{}\times\exp\left[-\left(\frac{N_g}{2m^g}\right)^{1-2/N_g}
\left(\frac{T}{E_g}\right)^{m^g_*/(2N_g)}\,
\left(\frac{\Lambda_T}{12}\right)^{2/N_g}\right],\\
&\Lambda_T\equiv
\frac{\varpi_\ell}{\sqrt{g\mathcal{K}_g(T/|\mu|)^{N_g}}}.
\end{split}
\end{equation}

\subsection{Summation over layer resonances}
\label{sec:Densitylayer}

The previous subsection was dedicated to the estimation of the
contribution to the chaotic fraction $w^{(g,\ell)}$ due to a
given pair of guiding and layer resonances $(g,\ell)$.
Here we are going to perform the summation over~$\ell$ in
Eq.~(\ref{sumwl=}):
\begin{equation}
S_g(\lambda)=\sum_\ell\mathcal{P}\left\{w^{(g,\ell)}>e^{-\lambda}\right\}
\end{equation}
Since afterwards this quantity will be summed over the guiding
resonances, we are interested in those~$\lambda$ for which $S_g\ll{1}$.
In Sec.~\ref{sec:Densitythermal} upper and lower bounds for
$w^{(g,\ell)}$ were given, $w^{(g,\ell)}_<<w^{(g,\ell)}<w^{(g,\ell)}_>$.
Then, $S_g$ can be estimated as
\begin{equation}
\sum_\ell\overline{\theta\left(\lambda-\ln\frac{1}{w^{(g,\ell)}_<}\right)}
<S_g(\lambda)<
\sum_\ell\overline{\theta\left(\lambda-\ln\frac{1}{w^{(g,\ell)}_>}\right)}\,,
\end{equation}
where the averaging is performed over the frequencies involved
in the layer resonance.

\subsubsection{Lower bound}
\label{sec:Densitylayerlower}

The lower bound, $w^{(g,\ell)}_<$, depends on the frequencies
through two quantities: the frequency mismatch~$\varpi_\ell$ and
the coupling $\mathcal{K}_\ell$. For the former we simply use
the central limit theorem, Eq.~(\ref{CLT=}).
Moreover, since we are interested in very small $\varpi_\ell$,
the Gaussian exponential can be omitted. Then we obtain
\begin{equation}\begin{split}
S_g>\sqrt{\frac{g\mathcal{K}_g(T/|\mu|)^{N_g}}{\Delta^2}}
\frac{1}{(2N_g)^{N_g/4}}\sum_\ell\frac{1}{N_\ell}
\overline{\left\lfloor
\lambda-\frac{E_g}{T}
-\ln\frac{(2N_\ell)!}{(\rho/e)^{N_\ell-1}}
-\ln\frac{g^{N_\ell-1}}{\Delta^{N_\ell-2}\mathcal{K}_\ell}
\right\rfloor^{N_g/4}},
\end{split}\end{equation}
where $\rho\equiv{g}T/(|\mu|\Delta)$, and where
$\lfloor{x}\rfloor\equiv{x}\,\theta(x)$, i.~e.,
$\lfloor{x}\rfloor=x$ for $x>0$ and $\lfloor{x}\rfloor=0$
otherwise.

To average over $\mathcal{K}_\ell$, we 
use the left inequality in Eq.~(\ref{thetalnAv=}):
\begin{equation}\begin{split}
S_g>{}&\sqrt{\frac{g\mathcal{K}_g(T/|\mu|)^{N_g}}{\Delta^2}}
\frac{1}{(2N_g)^{N_g/4}}\sum_\ell\frac{1}{N_\ell}
\times\\&{}\times
\left\lfloor\lambda-\ln\frac{\sqrt{2\pi{N}_\ell}\rho}{4e}
-\frac{E_g}{T}
-N_\ell\ln\left(\frac{4N_\ell^2}{e\rho}+2\right)
-L_\ell\ln\frac{2N_\ell}{\tau}\right\rfloor^{N_g/4}.
\label{sumLell=}
\end{split}\end{equation}
We sum over $L_\ell,N_\ell$ using the left inequality~(\ref{<RNrLr<}):
\begin{equation}\begin{split}
S_g>{}&\sqrt{\frac{g\mathcal{K}_g(T/|\mu|)^{N_g}}{\Delta^2}}
\frac{1}{(2N_g)^{N_g/4}}
\sum_{N_\ell}\frac{(e/N_\ell)^{N_\ell}}{\sqrt{2\pi{N}_\ell}}
\times\\&{}\times
\sum_{L_\ell}\left\lfloor\lambda-\ln\frac{\sqrt{2\pi{N}_\ell}\rho}{4e}
-\frac{E_g}{T}
-N_\ell\ln\left(\frac{4N_\ell^2}{e\rho}+2\right)
-L_\ell\ln\frac{2N_\ell}{\tau}\right\rfloor^{N_g/4}
\left\lfloor{L}_\ell-3N_\ell\right\rfloor^{N_\ell}.
\end{split}\end{equation}
The sum over~$L_\ell$ is estimated from below by its
largest term,
\begin{subequations}\begin{eqnarray}
&&L_\ell-3N_\ell=\frac{N_\ell}{N_g/4+N_\ell}
\frac{\lambda-E_g/T-\ln(\sqrt{2\pi{N}_\ell}\rho/4e)
-N_\ell\ln[32N_\ell^5/(e\tau^3\rho)]}{\ln(2N_\ell/\tau)},\nonumber\\ &&\\
&&S_g>\sqrt{\frac{g\mathcal{K}_g(T/|\mu|)^{N_g}}{\Delta^2}}\,
\frac{1}{2^{3N_g/4}}\sum_{N_\ell}\frac{1}{\sqrt{2\pi{N}_\ell}}
\left[\frac{e}{\ln(2N_\ell/\tau)}\right]^{N_\ell}
\frac{1}{(N_\ell+N_g/4)^{N_\ell+N_g/4}}
\nonumber\times\\&&\qquad{}\times
\left\lfloor\lambda-\ln\frac{\sqrt{2\pi{N}_\ell}\rho}{4e}
-\frac{E_g}{T}
-N_\ell\ln\frac{32N_\ell^5}{e\tau^3\rho}\right\rfloor^{N_\ell+N_g/4}.
\end{eqnarray}\end{subequations}
Again, we estimate the sum over $N_\ell$ from below by its largest term.
It is easy to show that
\begin{subequations}
\begin{equation}\label{maxANaN=}
\max_N\left(\frac{A}N-a\right)^N=e^{(A/a)F(a)},\quad
N_{max}=\frac{A/a}{1+1/F(a)},
\end{equation}
where the function $F(a)$ is defined as the solution of the equation
\begin{equation}
\ln\frac{a}{F}=1+F.
\end{equation}
We note that $F(a\to{0})=a/e+O(a^2)$,
$F(e^2)=1$, $F(a\to\infty)\sim\ln{a}$, and
\begin{equation}
\frac{1}2\ln{a}\leqslant{F}(a)<\ln(2a),\quad a\geqslant{1}.
\label{Faleqgeq=}
\end{equation}
\end{subequations}
As will be seen later, the values of $\lambda$ which are
important for the transport correspond to $N_g,N_\ell$
satisfying the inequalities
\begin{equation}\label{lambdaNgNlineq=}
N_g\ll{N}_\ell\ll\frac{1}\tau,\frac{1}\rho.
\end{equation}
%
Thus, with logarithmic precision we approximate
\begin{equation}
\ln\frac{32N_\ell^5}{e\tau^3\rho}\approx
\ln\frac{1}{\tau^3\rho},\quad
\ln\frac{4e}{\sqrt{2\pi{N}_\ell}\rho}\approx\ln\frac{1}\rho
\end{equation}
so we arrive at the following bound
\begin{equation}\begin{split}
\label{Sg>=}
&S_g>\frac{1}{\sqrt{2\pi}}
\sqrt{\frac{g\mathcal{K}_g(T/|\mu|)^{N_g}}{\Delta^2}}
\left(\frac{1}{8e}\,\ln\frac{1}{\tau^3\rho}\right)^{N_g/4}
\left\lfloor\frac{\lambda+\ln(1/\rho)-E_g/T}{\ln(1/\tau^3\rho)}
+\frac{N_g}{4}\right\rfloor^{-1/2}
\times\\&\qquad\times
\exp\left\{\left\lfloor\frac{\lambda+\ln(1/\rho)-E_g/T}{\ln(1/\tau^3\rho)}
+\frac{N_g}4\right\rfloor
F\left(\frac{e\ln(1/\tau^3\rho)}{\ln(1/\tau)}\right)\right\}.
\end{split}\end{equation}


\subsubsection{Upper bound}
\label{sec:Densitylayerupper}

The procedure is not very different from that for the lower
bound.
After using the central limit theorem for $\varpi_\ell$
and averaging over $\mathcal{K}_\ell$ with the help of the
right inequality in Eq.~(\ref{thetalnAv=}), we obtain
\begin{equation}\begin{split}
S_g<{}&24\left(\frac{E_g}{T}\right)^{m_*^g/4}
\sqrt{\frac{g\mathcal{K}_g(T/|\mu|)^{N_g}}{\Delta^2}}
\left(\frac{2m^g}{N_g}\right)^{N_g/2-1}
\times\\ &\times
\sum_\ell\left\lfloor
\lambda-\frac{E_g}{T}
-\ln\left(\frac\rho{50N_g}\left[\frac{T}{16E_g}\right]^{1/2\pi}
\left[\frac{e^2}{8\pi}\right]^{N_g/m^g}\right)
-\frac{N_\ell}\pi\ln\frac{2}{e^3\rho}
-\frac{L_\ell}\pi\ln\frac{2}{e^3\tau}
\right\rfloor^{N_g/2}.
\end{split}\end{equation}
We sum over $L_\ell,N_\ell$ using the right inequality~(\ref{<RNrLr<}).
The sum over~$L_\ell$ is estimated from above by its largest term,
which corresponds to
\[\begin{split}
L_\ell-\frac{N_\ell}{2}{}&=\frac{3N_\ell}{N_g/2+3N_\ell}\,
\frac{\pi}{\ln(2e^{-3}/\tau)}\times\\
&\times\left\lfloor
\lambda-\frac{E_g}{T}
-\ln\left(\frac\rho{50N_g}\left[\frac{T}{16E_g}\right]^{1/2\pi}
\left[\frac{e^2}{8\pi}\right]^{N_g/m^g}\right)
-\frac{N_\ell}{2\pi}\ln\frac{8}{e^9\rho^2\tau}
\right\rfloor,
\end{split}\]
multiplied by the number of terms,
\[
\frac{\pi}{\ln(2e^{-3}/\tau)}
\left\lfloor
\lambda-\frac{E_g}{T}
-\ln\left(\frac\rho{50N_g}\left[\frac{T}{16E_g}\right]^{1/2\pi}
\left[\frac{e^2}{8\pi}\right]^{N_g/m^g}\right)
-\frac{N_\ell}{2\pi}\ln\frac{8}{e^9\rho^2\tau}
\right\rfloor.
\]
This gives
\begin{equation}\begin{split}
S_g<{}&\frac{N_g}{2m^g}
\left(\frac{E_g}{T}\right)^{m_*^g/4}
\sqrt{\frac{g\mathcal{K}_g(T/|\mu|)^{N_g}}{\Delta^2}}
\left(2m^g\right)^{N_g/2}
\times\\ &\times
\sum_{N_\ell}
\left(\frac{24\pi}{\ln(2e^{-3}/\tau)}\right)^{3N_\ell+1}
\frac{1}{(6N_\ell+N_g)^{N_g/2+3N_\ell}}
\times\\ &\times
\left\lfloor\lambda-\frac{E_g}{T}
-\ln\left(\frac\rho{50N_g}\left[\frac{T}{16E_g}\right]^{1/2\pi}
\left[\frac{e^2}{8\pi}\right]^{N_g/m^g}\right)
-\frac{N_\ell}{2\pi}\ln\frac{8}{e^9\tau\rho^2}
\right\rfloor^{N_g/2+3N_\ell+1}.
\end{split}\end{equation}
The sum over $N_\ell$ is again estimated from above by its
largest term multiplied by the number of terms using
Eqs.~(\ref{maxANaN=})--(\ref{Faleqgeq=}) and
Eq.~(\ref{lambdaNgNlineq=}).
Also, we estimate $E_g/T<N_g/(m_*^g\rho)$ in the prefactor
and under the logarithm,
where we replaced the Gaussian distribution~(\ref{CLT=})
for the frequency difference in Eq.~(\ref{Igmin=}) by a
box distribution. This gives
\begin{equation}\begin{split}
&S_g<\frac{3N_g}{m^g}
\left(\frac{N_g}{m_*^g\rho}\right)^{m_*^g/4}
\sqrt{\frac{g\mathcal{K}_g(T/|\mu|)^{N_g}}{\Delta^2}}
\left(\frac{m^g}{12\pi}\ln\frac{1}{\tau}\right)^{N_g/2}
\times\\ &\qquad\times
\left\lfloor\frac{6\pi(\lambda-\lambda_1^>-E_g/T)}{\ln(1/\tau\rho^2)}
+\frac{6\pi\ln(8\pi/e^2)}{\ln(1/\tau\rho^2)}\,\frac{N_g}{m^g}
+\frac{N_g}2\right\rfloor^2
\times\\ &\qquad\times
\exp\left\{\left\lfloor
\frac{6\pi(\lambda-\lambda_1^>-E_g/T)}{\ln(1/\tau\rho^2)}
+\frac{6\pi\ln(8\pi/e^2)}{\ln(1/\tau\rho^2)}\,\frac{N_g}{m^g}
+\frac{N_g}2\right\rfloor
F\left(\frac{2\ln(1/\tau\rho^2)}{\ln(1/\tau)}\right)
\right\},\\
&\lambda_1^>=\ln\left[\frac\rho{50N_g}
\left(\frac{e^9\tau\rho^2}{8}\right)^{1/6\pi}
\left(\frac{m_*^g\rho}{16N_g}\right)^{1/2\pi}\right].
\end{split}\end{equation}

\subsection{Summation over guiding resonances}
\label{sec:Densityguiding}

\subsubsection{Toy calculation}
\label{sec:Densityguidinggeneral}

Before proceeding with the main calculation,
let us illustrate the difference between two ways of calculation
of~$S$ from~$S_g$, mentioned in Sec.~\ref{sec:remarkstat}, namely,
\begin{equation}
S=\sum_g\overline{S_g}
\end{equation}
and
\begin{equation}\label{S=AveSg}
S=\sum_g\overline{(1-e^{-S_g})},
\end{equation}
using a simplified ``toy'' expression, which nevertheless captures
the essential features of the results obtained in
Sec.~\ref{sec:Densitylayer}:
\begin{equation}\label{Sgtoy=}
S_g=(\tau^p\rho)^{N_g}
\exp\left(\frac{\lambda-x_g/\rho}{\ln(1/\tau^p\rho)}\right).
\end{equation}
Here $x_g$ is a random variable uniformly distributed on
$0\leqslant{x}_g\leqslant{1}$, ascribed to each guiding resonance. The
number of resonances corresponding to a given $N_g$ will be
taken to be simply $2^{N_g}$ just for this toy calculation.

The first calculation is straightforward:
\begin{equation}\begin{split}
\sum_g\overline{S_g}{}&=\sum_{N_g=1}^\infty
(2\tau^p\rho)^{N_g}
\left\{\theta(1-\lambda\rho)
\left[\left(e^{\lambda/\ln(1/\tau^p\rho)}-1\right)
\rho\ln\frac{1}{\tau^p\rho}
+1-\lambda\rho\right]\right.+\\
&\qquad+\left.\theta(\lambda\rho-1)\,e^{\lambda/\ln(1/\tau^p\rho)}
\left(1-e^{-1/\rho\ln(1/\tau^p\rho)}\right)
\rho\ln\frac{1}{\tau^p\rho}\right\}\approx\\
&\approx\left(e\tau^p\rho^2\ln\frac{1}{\tau^p\rho}\right)
e^{\lambda/\ln(1/\tau^p\rho)}.
\end{split}\end{equation}
The sum is dominated by $N_g=1$, which corresponds to resonant
pairs. This means that only two resonances (i.~e., $2^{N_g}$ as
mentioned in the previous paragraph, taken at $N_g=1$) effectively
contribute. Thus the whole calculation is meaningless, the
correct one being
\begin{equation}
e^{-S}=\prod_g\mathcal{P}\left\{w^{(g)}<e^{-\lambda}\right\}
=\prod_g\overline{e^{-S_g}}.
\end{equation}
When $1-\overline{e^{-S_g}}\ll{1}$, this is equivalent to
Eq.~(\ref{S=AveSg}).

\begin{figure}
\begin{center}
\includegraphics[width=8cm]{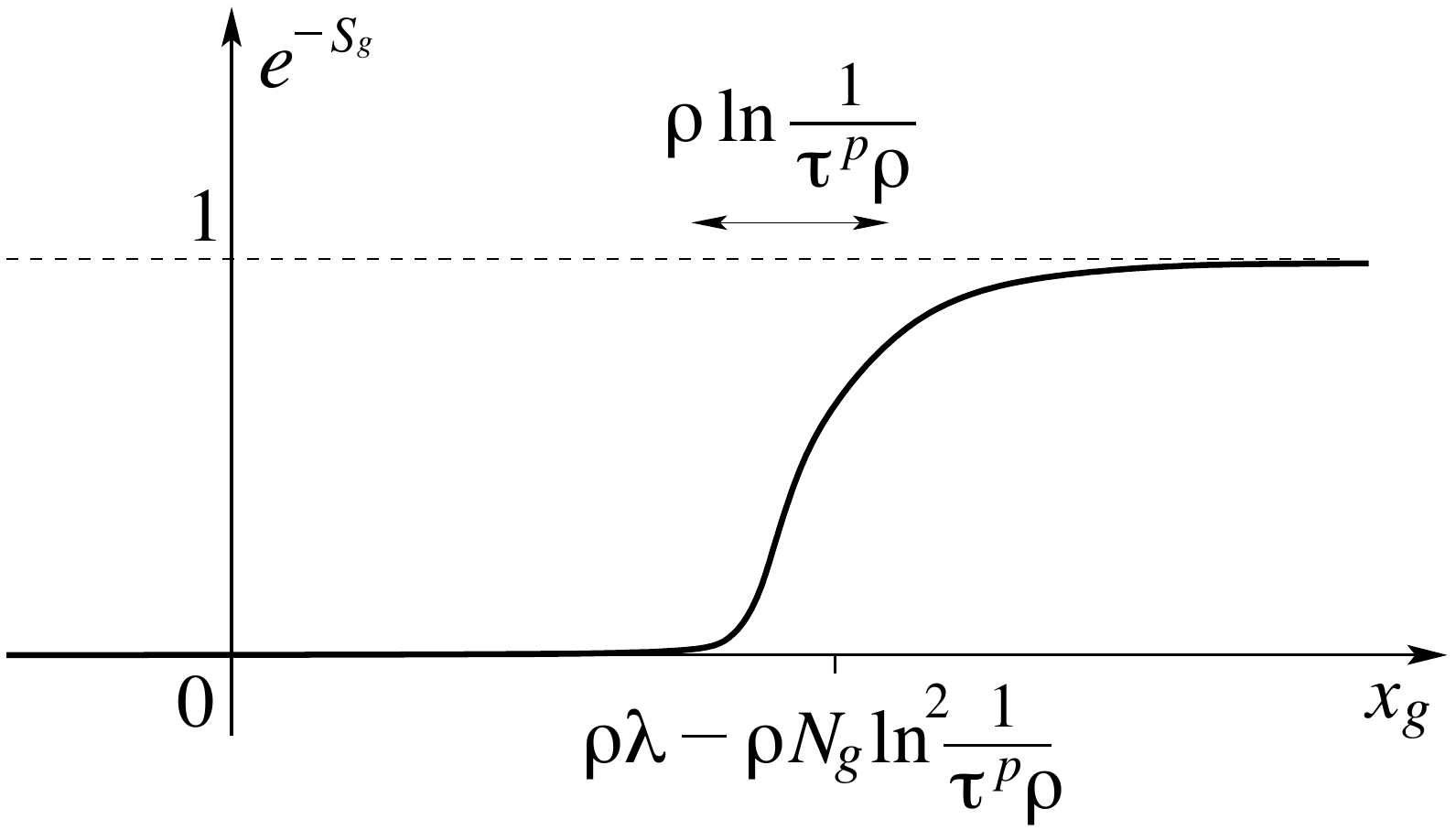}
\end{center}
\caption{\label{fig:Gumbel}
A schematic plot of $e^{-S_g}$ as a function of~$x_g$, as
determined by Eq.~(\ref{Sgtoy=}).}
\end{figure}
The plot of $e^{-S_g}$ as a function of~$x_g$ is schematically
shown in Fig.~\ref{fig:Gumbel}. It can be viewed as a step
function with a width of the order of
$\rho\ln(1/\tau^p\rho)$. As long as this width is smaller
than the offset,
$x_g=\lambda\rho-\rho{N}_g\ln^2(1/\tau^p\rho)$,
defined as the point where $S_g=1$, we can simply approximate
it by the Heaviside function.
Then
\begin{equation}
\overline{e^{-S_g}}=
1-\rho\left\lfloor\lambda-N_g\ln^2\frac{1}{\tau^p\rho}\right\rfloor
\end{equation}
(we remind that $\lfloor{x}\rfloor=x$ for $x>0$ and
$\lfloor{x}\rfloor=0$ otherwise), which gives
\begin{equation}
S=\rho\sum_{N_g=1}^{\lambda/\ln^2(1/\tau^p\rho)}
2^{N_g}\left(\lambda-N_g\ln^2\frac{1}{\tau^p\rho}\right)
=\rho\left(\ln\frac{1}{\tau^p\rho}\right)^2
\left(\sum_{N=1}^\infty 2^{-N}N\right)
e^{\lambda\ln{2}/\ln^2(1/\tau^p\rho)}.
\end{equation}
Note that the typical value of the offset,
$\lambda\rho-\rho{N}_g\ln^2(1/\tau^p\rho)\sim
\rho\ln^2(1/\tau^p\rho)$, which is indeed greater
than the width, $\rho\ln(1/\tau^p\rho)$.


\subsubsection{Lower bound}

As discussed above, from estimate~(\ref{Sg>=})
for $S_g$ we pass to the estimate for $S$:
\begin{equation}\begin{split}
&S>\sum_g\overline{\left[1-\exp\left(-e^{\lambda_g}\right)\right]},\\
&\lambda_g=\left(\frac{\lambda-\ln\rho-E_g/T}{\ln(1/\tau^3\rho)}
+\frac{N_g}4\right)
F\left(\frac{e\ln(1/\tau^3\rho)}{\ln(1/\tau)}\right)
-\frac{1}2\ln\left(\frac{\lambda-\ln\rho-E_g/T}{\ln(1/\tau^3\rho)}
+\frac{N_g}4\right)+\\
&\qquad{}+\frac{N_g}4\ln\left(\frac{1}{8e}\,\ln\frac{1}{\tau^3\rho}\right)
-\frac{1}2\ln\frac{2\pi\Delta^2}{g\mathcal{K}_g(T/|\mu|)^{N_g}}.
\end{split}\end{equation}
To perform the averaging, we replace
$\exp(-e^{\lambda_g})\to\theta(-\lambda_g)$,
due to the abrupt cutoff at $\lambda_g>0$. To transform the
lower bound on $\lambda_g$ into an upper bound on $E_g$, we
note that the solution of the equation $\alpha{x}-\ln\sqrt{x}=A$
is smaller than $(2/\alpha)A$ for $A\gg{1}$, $\alpha\sim{1}$.
This makes the restriction on $E_g$ tighter, so 
using the central limit theorem, Eq.~(\ref{CLT=}), to average over
$E_g$ which for the lower bound can be approximated as
$E_g=|\mu|\varpi_g/g$, and the left inequality~(\ref{thetalnAv=})
to average over $\tilde{\mathcal{K}}_g$, we obtain
\begin{equation}\begin{split}
S>{}&\sum_g\frac{2\rho}{N_g}
\left\lfloor\lambda+\ln\frac{1}\rho\right.-\\
&-\left.\ln\frac{1}{\tau^3\rho}
\left[F\left(\frac{e\ln(1/\tau^3\rho)}{\ln(1/\tau)}\right)\right]^{-1}
\left(\ln\frac\pi{2}
+N_g\ln\frac{4\ln^{1/4}(1/\tau^e)}{\rho\ln^{3/4}(1/\tau^3\rho)}
+L_g\ln\frac{2N_g}\tau\right)\right\rfloor.
\end{split}\end{equation}
%
%
The sum over $L_g,N_g$ is treated in full analogy with
Sec.~\ref{sec:Densitylayerlower}, to give
\begin{equation}\begin{split}
S>{}&\frac{1}{2\sqrt{\pi}e}\,\ln^4\frac{1}{\tau^3\rho}\ln\frac{1}\tau
\left[F\left(\frac{e\ln(1/\tau^3\rho)}{\ln(1/\tau)}\right)\right]^{-5/2}\times\\
&\times\frac{\rho}{\lambda^{3/2}}
\exp\left\{\frac\lambda{\ln^2(1/\tau^3\rho)}
\left[F\left(\frac{e\ln(1/\tau^3\rho)}{\ln(1/\tau)}\right)\right]^2\right\}.
\label{Slower=}
\end{split}\end{equation}

\subsubsection{Upper bound}
\label{sec:Densityguidingupper}

Analogously to the lower bound, we write the upper bound
for~$S$ in the form
\begin{equation}\begin{split}
&S<\sum_g\overline{\left[1-\exp\left(-e^{\lambda_g}\right)\right]},\\
&\lambda_g=\lambda_{gg}F%
\left(\frac{2\ln(1/\tau\rho^2)}{\ln(1/\tau)}\right)
+2\ln\lambda_{gg}
-\frac{1}2\ln\frac{\Delta^2}{g\mathcal{K}_g(T/|\mu|)^{N_g}}
+\\&\qquad
+\frac{N_g}2\ln\left(\frac{m^g}{24\pi}\,\ln\frac{1}{\tau}\right)
+\frac{m^g_*}{4}\ln\frac{N_g}{m_*^g\rho}+\ln\frac{3N_g}{m^g},\\
&\lambda_{gg}=\frac{6\pi(\lambda-\lambda_1^>-E_g/T)}{\ln(1/\tau\rho^2)}
+\frac{6\pi\ln(8\pi/e^2)}{\ln(1/\tau\rho^2)}\,\frac{N_g}{m^g}
+\frac{N_g}2.
\end{split}\end{equation}
To perform the averaging, we again replace
$\exp(-e^{\lambda_g})\to\theta(-\lambda_g)$ and transform the
lower bound on $\lambda_g$ into an upper bound on $E_g$, noting
that the solution of the equation $\alpha{x}+2\ln{x}=A$
is greater than $A/(2\alpha)$ for $A\gg{1}$, $\alpha\sim{1}$.
This makes the restriction on $E_g$ looser, so 
using the central limit theorem, Eq.~(\ref{CLT=}), to average over
$E_g$ which can be approximated as $E_g=|\mu|\varpi_g/(m^g_*g)$,
and the right inequality~(\ref{thetalnAv=})
to average over $\tilde{\mathcal{K}}_g$, we obtain
\begin{equation}\begin{split}
S<{}&\sum_g\frac{2m^g_*\rho}{\sqrt{N_g}}
\left\lfloor\lambda-\lambda_1^>+\frac{N_g}{m^g}\ln\frac{8\pi}{e^2}
+\frac{1}{6\pi}\ln\frac{1}{\tau\rho^2}
\left[F
\left(\frac{2\ln(1/\tau\rho^2)}{\ln(1/\tau)}\right)\right]^{-1}\right.\times\\
&\times\left.\left[-N_g\ln\frac{2}{e^3\rho}
+N_g\ln\left(\frac{m^g}{24\pi}\,\ln\frac{1}{\tau}\right)
+\frac{N_g}{2}\,F
\left(\frac{2\ln(1/\tau\rho^2)}{\ln(1/\tau)}\right)
\right.\right.-\\
&-\left.\left.L_g\ln\frac{2}{e^3\tau}
+\frac{m^g_*}{2}\ln\frac{N_g}{m_*^g\rho}+\ln\frac{12N_g}{m^g}\right]
\right\rfloor.
\end{split}\end{equation}
To sum over the resonances we use a generalized version of
Eq.~(\ref{RNrLrm=}) to include $m^g_*$:
\begin{equation}
\sum_g\to\sum_{m^g,m^g_*<N_g<2L_g}{64}^{(N_g-m^g_*/2)/m_g}
\left\lfloor\frac{L_g}{N_g}-\frac{1}2\right\rfloor^{3(N_g-m^g_*/2)/m_g}.
\end{equation}
The sum over $L_g$ is estimated by the largest term
multiplied by the number of terms:
\begin{equation}\begin{split}
S<{}&\sum_{m^g,m^g_*<N_g}\frac{m^g_*m^g\sqrt{N_g}\,\rho}{72\pi{e}}
\ln\frac{1}{\tau\rho^2}\ln\frac{1}{\tau}
\left[F
\left(\frac{2\ln(1/\tau\rho^2)}{\ln(1/\tau)}\right)\right]^{-1}\times\\
&\times
\left\{\frac{12\pi}{N_g\ln(1/\tau)}
\left\lfloor\frac{\lambda-\lambda_1^>+(N_g/m^g)\ln(8\pi/e^2)}%
{\ln(1/\tau\rho^2)}\,2F
\left(\frac{2\ln(1/\tau\rho^2)}{\ln(1/\tau)}\right)
\right.\right.+\\&\qquad{}+\ln\frac{12N_g}{m^g}
+\frac{m_*^g}{2}\ln\frac{N_g}{m_*^g\rho}{}-{}\\
&\qquad-\left.\left.\frac{N_g}{2}\ln\frac{1}{\tau\rho^2}
+N_g\ln\left(\frac{m^g}{24\pi}\,\ln\frac{1}{\tau}\right)
+\frac{N_g}{2}\,F
\left(\frac{2\ln(1/\tau\rho^2)}{\ln(1/\tau)}\right)
\right\rfloor\right\}^{2+3(N_g-m^g_*/2)/m_g}.
\end{split}\end{equation}
Now we can optimize with respect to $m^g$. The main
competition is between the exponent $\sim{N}_g/m^g$, which
is decreased upon increasing~$m^g$, and the term $N_g\ln{m}^g$
which increases the base. The sum over $N_g$ has a finite range,
$1\leqslant{N}_g\leqslant{N}_g^{max}$. Note that ${N}_g^{max}$ is almost
the same for $m^g=1$ and $m^g=N_g$, since the term $N_g\ln{m}^g$
gives only a small correction to the term $-(N_g/2)\ln(1/\tau\rho^2)$,
due to Eq.~(\ref{lambdaNgNlineq=}). This means that when we fix
$1\leqslant{m}^g\leqslant{N}_g^{max}$, and sum over $N_g$, analogously to
the preceding subsections, the result of the summation has a form
$a^{N_g/m^g}$, where $N_g\gg{1}$ corresponds to the maximum of the
whole expression and $a>{1}$. This maximum is determined
by the balance of different terms which enter~$a$, and thus also
only weakly depends on $m^g$. Thus, already as $m^g$ is increased
from 1 to~2, $a^{N_g}\gg{a}^{N_g/2}$. Hence, the sum is strongly
dominated by $m^g=1$.

The situation with $m_*^g$ is less obvious, as the sum is effectively
contributed by a few $m^g_*\sim{1}$. To see this, let is fix
$m^g_*$ and sum over $N_g$. With logarithmic precision we approximate
\begin{subequations}\begin{eqnarray}
&&{}-\frac{N_g}{2}\ln\frac{1}{\tau\rho^2}
+N_g\ln\left(\frac{m^g}{24\pi}\,\ln\frac{1}{\tau}\right)
+\frac{N_g}{2}\,F\left(\frac{2\ln(1/\tau\rho^2)}{\ln(1/\tau)}\right)
+\frac{m_*^g}{2}\ln\frac{N_g}{m_*^g\rho}\approx\nonumber\\
&&\qquad\approx-\frac{2+3(N_g-m^g_*/2)}6\ln\frac{1}{\tau\rho^2}
-\frac{m^g_*}{2}\ln\frac{1}\tau+\frac{1}3\ln\frac{1}{\tau\rho^2},\\
&&\frac{1}{N_g^{2+3(N_g-m^g_*/2)}}\approx
\left(\frac{3}{2+3(N_g-m^g_*/2)}\right)^{2+3(N_g-m^g_*/2)}e^{2-3m_*^g/2}.
\end{eqnarray}\end{subequations}
The sum over $N_g$ is treated in full analogy with
Sec.~\ref{sec:Densitylayerupper}, to give
\begin{equation}\begin{split}
S<{}&\left(\sum_{m^g_*=1}^\infty
\frac{m^g_*e^{-3m^g_*/2}}{9\pi/e}\right)
\left(\ln\frac{1}{\tau\rho^2}\right)^{-1/2}
\ln\frac{1}\tau
\left[F\left(\frac{2\ln(1/\tau\rho^2)}{\ln(1/\tau)}\right)\right]^{1/2}\times\\
&\times\rho\lambda^{3/2}
\exp\left\{\frac{12\lambda}{\ln^2(1/\tau\rho^2)}\,
F\left(\frac{2\ln(1/\tau\rho^2)}{\ln(1/\tau)}\right)
F\left(\frac{6\pi\ln(1/\tau\rho^2)}{\ln(1/\tau)}\right)\right\},
\label{Supper=}
\end{split}\end{equation}
where the sum over $m^g_*$ converges at $m_*^g\sim{1}$ and can
be estimated as $1/28$.

\subsection{Summary}
\label{sec:DensitySummary}

We have arrived at the following expression for
$S(\lambda)=-\ln\mathcal{P}\left\{w<{e}^{-\lambda}\right\}$:
\begin{subequations}\begin{equation}\begin{split}
\label{Slambda=}
&S(\lambda)=\mathcal{C}_1\lambda^{p_1}\rho\,
\exp\left[\frac{\lambda}{\mathcal{C}\ln^2(1/\tau^p\rho)}\right],\quad
\frac{1}2\leqslant{p}\leqslant{3},\quad-\frac{3}2\leqslant{p}_1\leqslant\frac{3}{2}.
\end{split}\end{equation}
$\mathcal{C}=\mathcal{C}(\ln\rho/\ln\tau)$ is a slow
function of the ratio of two logarithms. It can be bounded 
by estimating the argument of the exponential in Eq.~(\ref{Slower=})
from below, and that in Eq.~(\ref{Supper=}) from above:
\begin{equation}
\frac{1}4\,[1+\ln(1+x)]^2\leqslant\frac{1}{\mathcal{C}(x)}\leqslant{3}[1+\ln(1+x)]^2.
\end{equation}\end{subequations}
$\mathcal{C}_1=\mathcal{C}_1(\ln(1/\tau),\ln(1/\rho))$ is a power-law
function of $\ln(1/\tau)$ and $\ln(1/\rho)$ whose explicit form does
not enter the final result within our precision.
Note that the power $p$ is inherited directly from the value of
the lower cutoff of~$L_r$, $\min(L_r/N_r)$, appearing in the
expression~(\ref{<RNrLr<}) for the number of resonances, while the
exponential asymptotics of this expression contributes
to~$\mathcal{C}$.

From Sec.~\ref{sec:Densityguiding} it is seen that
the sum over the guiding resonances is mostly contributed by
\begin{equation}\label{EgTtyp=}
N_g\sim\frac{\lambda}{\ln^2(1/\tau^p\rho)},\quad
\frac{E_g}{T}\sim\lambda,
\end{equation}
while the values of $L_g$ are those close to the lower cutoff.
Similarly, from Sec.~\ref{sec:Densitylayer} we obtain
\begin{equation}
N_\ell\sim\frac{\lambda}{\ln(1/\tau^p\rho)},\quad
\Lambda_T\sim\left(\frac{\lambda}{N_g}\right)^{N_g/2},\quad
\ln\frac\Delta{\varpi_\ell}\sim
{N}_g\ln\left[\frac{1}{\tau^p\rho\ln^2(1/\tau^p\rho)}\right],
\end{equation}
the values of $L_\ell$ again close to the lower cutoff.
Then, from Sec.~\ref{sec:Densitythermal} we obtain
\begin{equation}\label{Ingtyp=}
\Lambda\sim{N_g}\Lambda_T^{2/N_g}\sim\lambda,\quad
I_{n^g\neq{n}^g_*}\sim
\frac{T}{|\mu|}\left(\frac{\Lambda_T}\Lambda\right)^{2/N_g}
\sim\frac{T}{|\mu|}\ln^2\frac{1}{\tau^p\rho}.
\end{equation}
The inequality
$\lambda-L_\ell\ln{N}_\ell\gg{N}_g$, necessary to use the
first regime of Eq.~(\ref{thetalnAv12=}), becomes evident
at this point.
As a result, the long calculations of Sec.~\ref{sec:Density}
can be summarized in very simple terms: the chaotic fraction
can be written as the product of three factors
\begin{equation}
w\sim\left\langle\frac{g|V_\ell|}{\Delta^2}\right\rangle
\left\langle{e}^{-\varpi_\ell/\Omega_g}\right\rangle
e^{-E_g/T},
\end{equation}
where the prefactors containing $\tau,\rho$ in powers $\sim{1}$
are neglected, and the most important configurations determining
to $\mathcal{P}\{w<{e}^{-\lambda}\}$ correspond to the logarithms
of each of the three factors being of the same order, $-\lambda$.

It may be interesting to compare the contributions to the
stochastic layer width from the layer resonance and from the
next best candidate. For the latter the orders in $\tau,g$
are likely to be the same, $L_r=L_\ell$, $N_r=L_\ell$, while
the frequency mismatch, $\varpi_r$, is greater than that for
the layer resonance, $\varpi_\ell$, by a factor $\sim{1}$.
Thus, we set $\varpi_r=2\varpi_\ell$ for an estimate, so that
$\Lambda_T$ is twice greater for the second resonance, and
focus on the ratio between the values of the
Melnikov-Arnold exponentials,
\begin{equation}\label{secondlayer=}
\frac{\langle{e}^{-\varpi_r/\Omega_g}\rangle}%
{\langle{e}^{-\varpi_\ell/\Omega_g}\rangle}
\sim\exp\left[-(2^{2/N_g}-1)\Lambda\right]
\sim{e}^{-\ln^2(1/\tau^p\rho)}.
\end{equation} 
Thus, the contribution of the next best resonance to the
stochastic layer width is parametrically smaller.

Finally, to justify inequalities~(\ref{lambdaNgNlineq=}),
$N_g\ll{N}_\ell\ll{1}/\tau,1/\rho$, 
as well as assumption~(\ref{ITwllIgmin=}),
$T/|\mu|\ll{I}_{n^g}\ll{I}_g^\mathrm{min}$, we have to
specify~$\lambda$. The typical values of~$\lambda$ are those
for which $S(\lambda)\sim{1}$, i.~e.
$\lambda\sim\ln(1/\rho)\ln^2(1/\tau^p\rho)$, so 
inequalities~(\ref{lambdaNgNlineq=}), (\ref{ITwllIgmin=}) are
indeed satisfied.
As we will see in Sec.~\ref{sec:Breaks}, the transport is
dominated by regions with anomalously small~$w$, i.~e. with
larger~$\lambda$, so that Eq.~(\ref{Slambda=}) works. It
ceases to be valid for $\lambda\lesssim\ln^2(1/\tau^p\rho)$,
so it is not suitable to find the average value of~$w$,
which is dominated by rare resonant configurations (triples).
Moreover, for $\lambda\gg\ln^2(1/\tau^p\rho)$ the typical
strength of the guiding resonance $V_g$, which scales as
$(\tau^p\rho)^{N_g}$, is smaller than $\rho$ even at
$\rho\ll\tau$, so that the range of validity of the pendulum
approximation coincides with that of Eq.~(\ref{Slambda=}).
The relation between $\tau$~and~$\rho$ becomes important
in determining the nature of the rare resonant configurations
(whether they correspond to nearest-neighbor triples, or
spatially separated ones).

\section{Stochastic pump and Arnold diffusion}
\label{sec:Pump}
\setcounter{equation}{0}

The standard description of Arnold diffusion in multidimensional
dynamical systems is based on the stochastic pump model~%
\cite{Chirikov1979,Lichtenberg1983}. Namely, motion of the
oscillators participating in the guiding resonance (the chaotic
spot) is assumed to have a stochastic component with a continuous
frequency spectrum. The continuous spectrum arises because the
motion corresponds to successive passages of the separtrix of
the effective pendulum at random instants of time and is analyzed
in detail in Sec.~\ref{sec:spectrum}.
These oscillators of the chaotic spot are coupled to all other
oscillators in the system (typically, in a high order of the
perturbation theory).
The combination of the two ingredients (coupling and continuous
spectrum) results in energy and action exchange between the
chaotic spot and the surrounding oscillators. This exchange occurs
by small random amounts, which corresponds to diffusion. The
corresponding diffusion equation is written in
Sec.~\ref{sec:Pumpdiffeq}, and its consequences are analyzed
in subsequent subsections.

\subsection{Stochastic pump spectrum}
\label{sec:spectrum}

\begin{figure}
\begin{center}
\includegraphics[width=10cm]{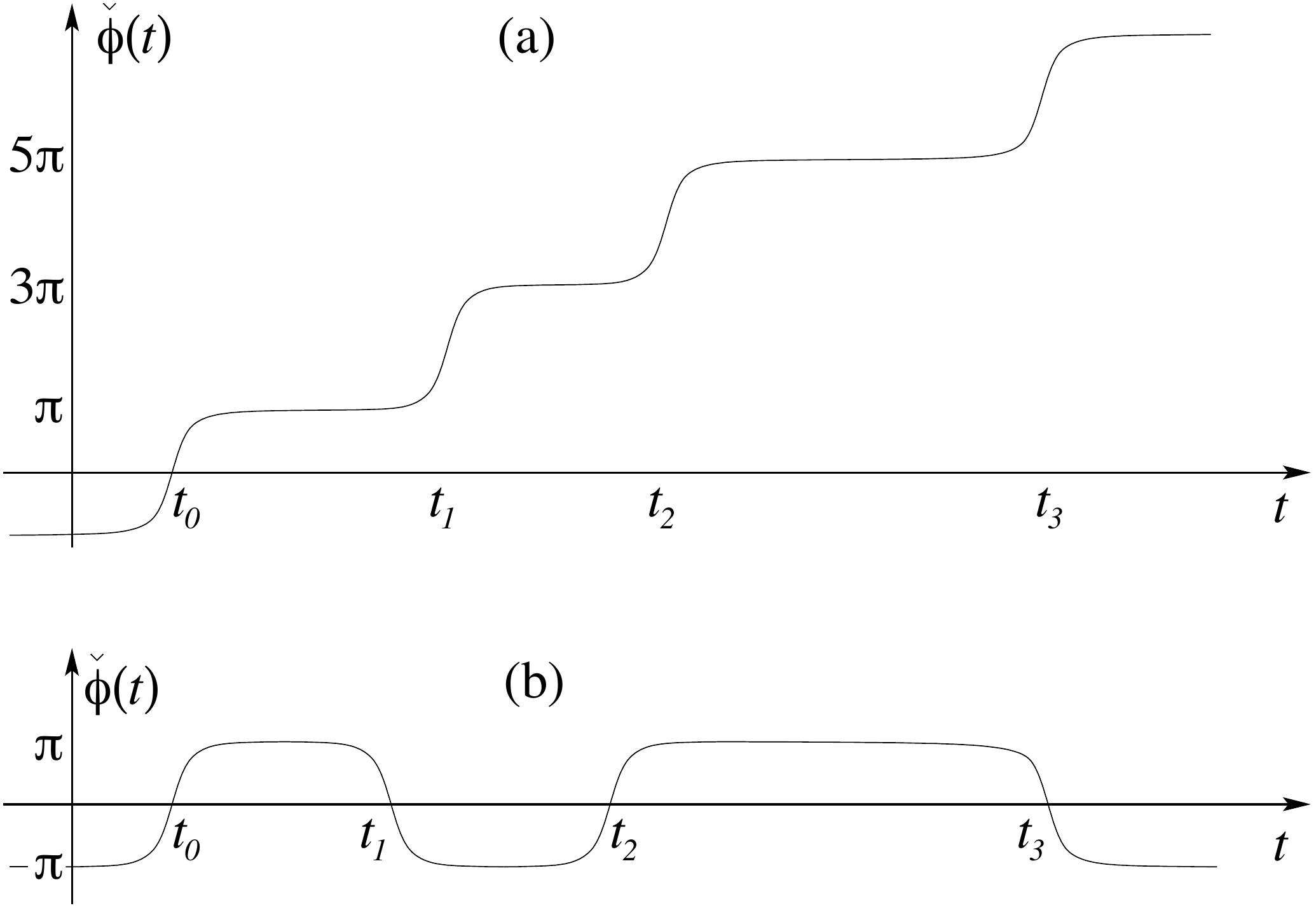}
\end{center}
\caption{\label{fig:phase}
Schematic representation of the behavior of the slow phase
$\phidif(t)$, corresponding to rotation~(a) and oscillation~(b)
of the effective pendulum. The flat regions around odd multiples
of~$\pi$ correspond to long intervals of time spent in the
vicinity of the unstable equilibrium of the pendulum, while the
steps correspond to passages through the stable equilibrium
at random instants of time~$t_j$.}
\end{figure}

Let us assume the general form of the guiding resonance,
as defined in Sec.~\ref{sec:statistics}:
\begin{equation}
\sum_nm_n^g\,\tilde\omega_n=0,
\end{equation}
where $\tilde\omega_n(I_n)=\omega_n+gI_n$, and
the integer numbers $\{m_n^g\}$ satisfy Eq.~(\ref{summn=0}).
If the dynamics of the guiding resonance can be described by
an effective pendulum Hamiltonian, and the stochastic layer
around the separatrix is thin,
for each oscillator participating in the guiding resonance its
phase $\phi_n$ can be written as
\begin{equation}\label{phint=}
\phi_n(t)=\tilde\omega_nt
+\frac{m_n^g}{|\underline{m}^g|^2}\,\phidif(t),
\end{equation}
where
\begin{equation}
\phidif(t)=\sum_nm_n^g\,\phi_n(t)
\end{equation}
is the slow phase which represents the coordinate of the
effective pendulum (see \ref{app:change} for details).
For rotation and
oscillation of the pendulum, corresponding to the pendulum
energy being above or below the separatrix energy, respectively,
$\phidif(t)$ can be represented in the following form (see also
Fig.~\ref{fig:phase}):
\begin{equation}
\phidif(t)=\sum_{j=-\infty}^\infty\left\{\begin{array}{ll}
4\arctan{e}^{\Omega_g(t-t_j)}-\pi,&\mbox{rotation,}\\
{}[4\arctan{e}^{\Omega_g(t-t_j)}-\pi]
-[4\arctan{e}^{\Omega_g(t-t_j/2-t_{j+1}/2)}-\pi],& \mbox{oscillation.}
\end{array}\right.
\end{equation}
Here $\Omega_g$ is the effective pendulum frequency, see
Eq.~(\ref{Hpendulumg=}).
The instants of time $t_j$, which are determined by the separatrix
mapping [Eqs.~(\ref{sepmaph=}), (\ref{sepmaptheta=}) with
$t=\theta/(\Lambda\Omega_g)$], represent the random ingredient of
the problem.
Assuming the typical interval $t_{j+1}-t_j\gg{1}/\Omega_g$,
we can represent the Fourier transform as a sum of independent
contributions:
\begin{equation}
\int\limits_{-\infty}^\infty
{e}^{-i(m/2)\,\phidif(t)}e^{i\omega{t}}\,dt
=\frac{1}{\Omega_g}\sum_{j=-\infty}^\infty
\left\{\begin{array}{l}
\mathcal{A}_m(\omega/\Omega_g)\,{e}^{i\omega{t}_j},\\
\mathcal{A}_m(\omega/\Omega_g)\,{e}^{i\omega{t}_j}
+\mathcal{A}_m(-\omega/\Omega_g)\,{e}^{i\omega(t_j+t_{j+1})/2}.
\end{array}\right.
\end{equation}
The Melnikov-Arnold integral $\mathcal{A}_m(\Lambda)$ is defined
in Eq.~(\ref{MelnikovArnold=}), and here we use the standard
notation for its index $m=2m_n^g/|\underline{m}^g|^2$,
which is not necessarily an integer.
If the times $t_j$ are assumed random and uncorrelated,
the power spectrum is given by
\begin{equation}\begin{split}\label{rhom=Am}
\nu_{mm'}(\omega){}&=\int\limits_{-\infty}^\infty
\left\langle{e}^{-i(m/2)\,\phidif(t)+i(m'/2)\,\phidif(t')}\right\rangle_t
e^{i\omega(t-t')}\,d(t-t')=\\
&=\frac{\langle\delta{t}^{-1}\rangle_t}{2\Omega_g^2}
\left\{\begin{array}{l}
2\,\mathcal{A}_m(\omega/\Omega_g)\,\mathcal{A}_{m'}(\omega/\Omega_g),\\
\mathcal{A}_m(\omega/\Omega_g)\,\mathcal{A}_{m'}(\omega/\Omega_g)
+\mathcal{A}_m(-\omega/\Omega_g)\,\mathcal{A}_{m'}(-\omega/\Omega_g).
\end{array}\right.
\end{split}\end{equation}
Here $\langle\delta{t}^{-1}\rangle_t$ is the average repetition
rate of $t_j$'s for rotation, and the double repetition rate
for oscillation. Note that
$\langle\delta{t}^{-1}\rangle_t\sim\Omega_g$,
up to a logarithmic factor. The angular brackets here correspond
to the average over the dynamics:
\begin{equation}
\langle{f}(t_1,\ldots,t_N)\rangle_t=\lim_{t\to\infty}
\int\limits_0^t\frac{dt'}{t}\,
{f}(t_1+t',\ldots,t_N+t').
\end{equation}

The assumption of random and uncorrelated $t_j$ is not fully
correct. Small residual correlations in the phase found in
Ref.~\cite{Khodas2000}
for the standard mapping translate in correlations between different
$t_j$'s. These residual correlations are usually explained by the
system sticking to various structures inside the chaotic region,
such as stability islands where the diffusion is anomalously slow,
or accelerator modes, where the diffusion is anomalously fast.
When dealing with the standard mapping, one often introduces a
small amount of external noise which prevents the system from
sticking~\cite{Rechester1980,Khodas2000}. The present problem,
however, is more complex than the standard mapping, as we
already discussed in the end of Sec.~\ref{sec:2oscstoch}.
In addition to that discussion, we note that the resonant
triple is not isolated, but it is subject to perturbations
from surrounding oscillators. Even though these perturbations
are not truly random, it is quite plausible to assume that
they are sufficient to prevent the triple from sticking,
and thus effectively fulfill the role of the external noise.

As we have seen in Sec.~\ref{sec:3oscchaotic}, a chaotic spot
is not always described by an effective pendulum with a thin
stochastic layer around the separatrix, and the guiding and
the layer resonances are not always distinguishable, so that
above expressions are not always applicable.
Still, for any oscillator~$n$, belonging to the chaotic spot
($n\in\mathrm{chaotic}$), we can write
\begin{equation}\label{psixi=}
\psi_n(t)=\sqrt{I_n}\,e^{-i\tilde\omega_nt}\xi_n(t),
\end{equation}
where $\xi(t)$ is a certain random process. Its statistical
properties cannot be determined precisely. Still, one can
assume $\langle\xi_n(t)\rangle_t=0$.
Also, for any set of integer numbers $\underline{m}\equiv\{m_n\}$
the two-time correlator and the corresponding spectral density
can be defined:
\begin{equation}\begin{split}
\label{xixi=}
&\left\langle\Xi_{\underline{m}}(t_1)\,\Xi_{\underline{m}}^*(t_2)\right\rangle_t
=\int\frac{d\omega}{2\pi}\,\nu_{\underline{m}}(\omega)\,e^{-i\omega(t_1-t_2)},\\
&\Xi_{\underline{m}}(t)=\prod_{n\in\mathrm{chaotic}}
\left\{\begin{array}{ll}\xi_n^{m_n}(t),&m_n\geqslant{0},\\
{}[\xi_n^*(t)]^{|m_n|}, & m_n<0.\end{array}\right.
\end{split}\end{equation}
When the guiding and the layer resonances are not distinguishable,
the chaotic dynamics has only one characteristic frequency scale,
$\Omega_g$. Then, the leading exponential in the high-frequency
asymptotics $|\omega|\gg\Omega_g$, can be estimated as
\begin{equation}
\nu_{\underline{m}}(\omega)\sim\frac{1}{\Omega_g}\,{e}^{-\pi|\omega|/(2\Omega_g)}.
\end{equation}
For a resonant triple, considered in Sec.~\ref{sec:3oscchaotic},
$\Omega_g\sim\sqrt{\tau\rho}\Delta$, which is valid at $\tau\ll\rho$
As discussed in Sec.~\ref{sec:Motttriples}, in the opposite limiting
case, $\tau\ll\rho$, one should effectively replace $\tau$ by~$\rho$,
so $\Omega_g\sim{\rho}\Delta$.

\subsection{Diffusion equation}
\label{sec:Pumpdiffeq}

Let us now consider the dynamics of oscillators, surrounding
the chaotic spot. The latter is assumed to be described by
the Hamiltonian 
\begin{equation}
H=\sum_n\left(\omega_nI_n+\frac{g}2\,I_n^2\right)
-2V_g\cos\left(\sum_{n}m_n^g\phi_n\right).
\end{equation}
Other oscillators are coupled to those of the guiding
resonance by the Hamiltonian of the form
\begin{equation}
V=-\sum_r2V_r\cos\left(\sum_{n}m_n^r\phi_n\right),
\end{equation}
where the integer vectors $\underline{m}^g$ and
$\underline{m}^r$ have a nonzero overlap. They
satisfy the constraints imposed by the total action
conservation, Eq.~(\ref{summn=0}).
Generally speaking,
$V_g$ and $V_r$ depend on the actions of the participating
oscillators. For the moment we assume the effective pendulum
description of the guiding resonance to be valid, so we take
$V_g$ and $V_r$ exactly at resonance (the opposite case is
considered in the end of this subsection).

Let us restrict our attention to a finite number $N>N_g$ of
oscillators. To separate the slow phase of the guiding
resonance, we perform a canonical transformation (see also
\ref{app:change}):
\begin{equation}\begin{split}\label{ptoI=}
&\phidif=\sum_{n=1}^Nm_n^g\phi_n\equiv
(\underline{m}^g,\underline\phi),\quad
\phi_k'=\sum_{n=1}^Nu_{kn}\phi_k\equiv
(\underline{u}_k,\underline\phi),\quad
k=2,\ldots,N,\\
&I_n=m_n^gp_1+\sum_{k=2}^Np_ku_{kn},\quad
\pdif=p_1+\frac{(\underline{m}^g,\underline{\omega})}{g|\underline{m}^g|^2},
\quad\underline{\omega}=(\omega_1,\ldots,\omega_N).
\end{split}\end{equation}
Here $\pdif$ is the momentum, conjugate to the slow phase,
and counted from the point of the exact resonance.
$u_{kn}$ is some constant matrix satisfying
the appropriate orthogonality relations:
\begin{equation}
(\underline{u}_{k>1},\underline{u}_{k'>1})=\delta_{kk'},\quad
(\underline{u}_{k>1},\underline{m}^g)=0,\quad
(u^{-1})_{nk}=\left\{\begin{array}{ll}
m_n^g/|\underline{m}^g|^2,&k=1,\\ u_{kn},&k>1.
\end{array}\right.
\end{equation}
For oscillators not participating in the guiding resonance
one can set $u_{kn}=\delta_{kn}$, i.~e., $p_k=I_k$. It is
also convenient to introduce the vector $\underline{u}^I$
with all components $u^I_{n}=1$, which corresponds to the
total action:
\begin{equation}
I_{tot}=\sum_{n=1}^NI_n=\sum_{k=2}^Np_k(\underline{u}_k,\underline{u}^I),\quad
(\underline{m}^g,\underline{u}^I)=(\underline{m}^r,\underline{u}^I)=0.
\end{equation}

In the absence of the couplings~$V_r$ the hamiltonian is given by
\begin{equation}
H=\sum_{k=2}^N\left[(\underline{\omega},\underline{u}_k)p_k
+\frac{g}2\,p_k^2\right]
-\frac{(\underline{m}^g,\underline{\omega})^2}{2g|\underline{m}^g|^2}
+\Hdif_g,\quad
\Hdif_g=\frac{g|\underline{m}^g|^2}2\,\pdif^2-2V_g\cos\phidif,
\end{equation}
so the actions
$p_2,\ldots,p_N$, as well as the pendulum Hamiltonian $\Hdif_g$,
are conserved. When the couplings $V_r$ are switched on, the
change in the actions between two moments $t_1<t_2$ is determined
from the
Hamilton's equations, $dp_k/dt=-\partial{V}/\partial\phi_k'$,
and the change in $\Hdif_g$ is found from
$d\Hdif_g/dt=-
(\partial{V}/\partial\phidif)(\partial\Hdif/\partial\pdif)$:
\begin{subequations}\begin{eqnarray}\label{deltaIn=}
&&\delta{p}_{k>1}=-i\sum_rV_r(\underline{m}^r,\underline{u}_k)
\int\limits_{t_1}^{t_2}dt'\exp\left[i\varpi_rt'
+i\,\frac{(\underline{m}^r,\underline{m}^g)}{|\underline{m}^g|^2}\,
\phidif(t')\right] + \mbox{c. c.},\\
&&\delta\Hdif_g=-i\sum_rV_r
\frac{(\underline{m}^r,\underline{m}_g)}{|\underline{m}^g|^2}
\int\limits_{t_1}^{t_2}dt'\,\frac{d\phidif(t')}{dt'}\,
\exp\left[i\varpi_rt'
+i\,\frac{(\underline{m}^r,\underline{m}^g)}{|\underline{m}^g|^2}\,
\phidif(t')\right] + \mbox{c. c.},\nonumber\\
\label{deltaHdif=}
\end{eqnarray}\end{subequations}
where ``c. c.'' stands for the complex conjugate and
$\varpi_r=\sum_nm_n^r\tilde\omega_n=
(\underline{m}^r,\underline{\tilde\omega})$.
Note the relations
\begin{subequations}\begin{eqnarray}
&&\sum_{k=2}^N\delta{p}_k(\underline{u}_k,\underline{u}^I)=0,\\
&&\sum_{k=2}^N\delta{p}_k(\underline{u}_k,\underline{\tilde\omega})
+\delta\Hdif=V(0)-V(t).
\label{dp+dH=0}
\end{eqnarray}\end{subequations}
The first relation follows from the orthogonality relations
and represents the total action conservation.
The second relation is obtained by integrating by parts and
represents the total energy conservation. In fact, these
relations hold separately for each given term~$r$ of the
perturbation.

The integrand in Eqs.~(\ref{deltaIn=}), (\ref{deltaHdif=}) is
randomly oscillating due to the chaotic dynamics of~$\phidif(t)$,
so $\langle\delta{p}_k\rangle$ remains bounded even for large
$t_2-t_1$. In contrast, the average of the product
$\langle\delta{p}_k\delta{p}_{k'}\rangle_t$ has a contribution
growing linearly with time $t_2-t_1$:
\begin{equation}\label{dpkdpk=}
\langle\delta{p}_k\delta{p}_{k'}\rangle_t=
2(t_2-t_1)\sum_rV_r^2\nu_{\underline{m}^r}(\varpi_r)\,
(\underline{m}^r,\underline{u}_k)(\underline{m}^r,\underline{u}_{k'}),
\end{equation}
where the stochastic pump spectral density is defined analogously
to Eq.~(\ref{rhom=Am}):
\begin{equation}
\nu_{\underline{m}^r}(\omega)=\int\limits_{-\infty}^\infty
\left\langle{e}^{i[\phidif(t_1)-\phidif(t_2)]
(\underline{m}^r,\underline{m}^g)/|\underline{m}^g|^2}\right\rangle_t
e^{i\omega(t_1-t_2)}\,d(t_1-t_2).
\end{equation}
The interference terms with $r\neq{r}'$ give a contribution
which does not grow with time and thus are neglected.
Using Eq.~(\ref{dp+dH=0}), we can also write
\begin{equation}\label{dWdpk=}
\frac{\langle\delta{p}_k\delta\Hdif_g\rangle_t}{t_2-t_1}=-\sum_{k'=2}^N
\frac{\langle\delta{p}_k\delta{p}_{k'}\rangle_t}{t_2-t_1}\,
(\underline{u}_k,\underline{\tilde\omega}),\quad
\frac{\langle(\delta\Hdif_g)^2\rangle_t}{t_2-t_1}=\sum_{k,k'=2}^N
\frac{\langle\delta{p}_k\delta{p}_{k'}\rangle_t}{t_2-t_1}\,
(\underline{u}_k,\underline{\tilde\omega})
(\underline{u}_{k'},\underline{\tilde\omega}).
\end{equation}

To write the diffusion equation, it is convenient to pass from the
pendulum energy $\Hdif_g$, Eq.~(\ref{Hpendulumg=}), to the pendulum
action~$W$ using the standard expression in terms of the complete
elliptic integrals $K(\kappa)$ and $E(\kappa)$,
\begin{equation}\begin{split}
&W=\frac{8}{\pi}\sqrt{\frac{2V_g}{g|\underline{m}^g|^2}}
\left\{\begin{array}{ll}
E(\kappa)-(1-\kappa^2)K(\kappa)&\kappa<1,\\
(\kappa/2)E(1/\kappa)+1/2&\kappa>1,
\end{array}\right.\quad
\kappa=\sqrt{\frac{\Hdif_g+2V_g}{4V_g}}.
\end{split}\end{equation}
Since transformation~(\ref{ptoI=}) is canonical, the volume element
in the phase space is preserved:
\begin{equation}
\frac{dW\,d\phidif}{2\pi}\prod_{k=2}^N\frac{dp_k\,d\phi_k'}{2\pi}
=\prod_{n=1}^N\frac{dI_n\,d\phi_n}{2\pi}.
\end{equation}
Not being interested in the oscillating phases, we introduce the
distribution function~$\hat{f}$ in the action space, so that the probability
to find the system in an infinitesimal volume element is given by
\begin{equation}\label{dPW=}
dP=\hat{f}(W,p_2,\ldots,p_n)\,dW\,dp_2\ldots{d}p_N.
\end{equation}
It is the flatness of the phase space measure appearing here that was the main
purpose of passing from $\Hdif_g$ to $W$.

Now, from Eqs.~(\ref{dpkdpk=}) and~(\ref{dWdpk=}) we can immediately
pass to the diffusion equation
\begin{subequations}\begin{eqnarray}
&&\frac{\partial\hat{f}}{\partial{t}}=
\frac\partial{\partial{W}}
\left[\hat{D}_{WW}\frac{\partial\hat{f}}{\partial{W}}
+\hat{D}_{Wk}\frac{\partial\hat{f}}{\partial{p}_k}\right]
+\frac{\partial}{\partial{p}_k}
\left[\hat{D}_{kW}\frac{\partial\hat{f}}{\partial{W}}
+\hat{D}_{kk'}\frac{\partial\hat{f}}{\partial{p}_{k'}}\right],\\
&&\hat{D}_{kk'}=\sum_rV_r^2
(\underline{m}^r,\underline{u}_k)(\underline{m}^r,\underline{u}_{k'})\,
\nu_{-2(\underline{m}^r,\underline{m}^g)/|\underline{m}^g|^2}(\varpi_r),\\
&&\hat{D}_{kW}=\hat{D}_{Wk}=
-\hat{D}_{kk'}(\underline{u}_{k'},\underline{\tilde\omega})\,
\frac{dW}{d\Hdif_g},\\
&&\hat{D}_{WW}=\hat{D}_{kk'}
(\underline{u}_k,\underline{\tilde\omega})
(\underline{u}_{k'},\underline{\tilde\omega})
\left(\frac{dW}{d\Hdif_g}\right)^2,
\end{eqnarray}\end{subequations}
where summation over repeating indices $k,k'=2,\ldots,N$
is assumed.

Diffusion in $W$ is confined to the narrow stochastic layer of the
width~$W_s$, given by Eq.~(\ref{Wsgl=}). At times longer than the
diffusion time across the layer we can assume the distribution
function to be independent on $W$, and write the probability
(\ref{dPW=}) as
\begin{equation}\label{fdef=}
dP=f(p_2,\ldots,p_n)\,W_s\,dp_2\ldots{d}p_N.
\end{equation}
As discussed in detail in \ref{app:diffusion}, projection
of the full diffusion equation on the layer leads to a closed
equation for $f$:
\begin{equation}\label{Arnolddif=}
W_s\,\frac{\partial{f}}{\partial{t}}
=\sum_{k,k'=2}^N\frac{\partial}{\partial{p}_k}\,W_sD_{kk'}\,
\frac{\partial{f}}{\partial{p}_k},
\end{equation}
which describes the Arnold diffusion in the space of the
variables $p_2,\ldots,p_N$.
The diffusion coefficient is given by
\begin{subequations}\begin{eqnarray}\label{ArnoldD=}
&&D_{kk'}=D_{k'k}
=\int\limits_\mathrm{layer}\frac{dW}{W_s}\left[
(\underline{u}_k,\underline{\underline{D}},\underline{u}_{k'})
-\frac{(\underline{u}_k,\underline{\underline{D}},\underline{\tilde\omega})
(\underline{\tilde\omega},\underline{\underline{D}},\underline{u}_{k'})}%
{(\underline{\tilde\omega},\underline{\underline{D}},
\underline{\tilde\omega})}\right],\\
&&\underline{\underline{D}}=\sum_r
[\underline{m}^r\otimes\underline{m}^r]V_r^2
\nu_{\underline{m}^r}(\varpi_r),
\end{eqnarray}\end{subequations}
and we use obvious tensor notations,
\begin{equation}
(\underline{u},[\underline{m}^r\otimes\underline{m}^r],\underline{u}')
\equiv(\underline{u},\underline{m}^r)(\underline{m}^r,\underline{u}'),
\quad
([\underline{m}^r\otimes\underline{m}^r],\underline{u}')\equiv
\underline{m}^r(\underline{m}^r,\underline{u}').
\end{equation}
As seen from Eq.~(\ref{ArnoldD=}), the diffusion coefficient vanishes
if the tensor $\underline{\underline{D}}$ is separable. This would be
the case if the sum over~$r$ involved only one term. If the sum involves
terms of very different strength, the contribution of the strongest one
is eliminated. This corresponds to the situation discussed in
Ref.~\cite{Chirikov1979}: the strongest perturbation of the guiding
resonance effectively determines the stochastic layer width~$W_s$,
but does not contribute to the Arnold diffusion. This strongest term
is called the layer resonance, $r=\ell$.
Also, the diffusion coefficient~(\ref{ArnoldD=}) vanishes identically
for the directions corresponding to the change of the total action
and of the total energy:
\begin{equation}\label{ArnoldDconstr=}
\sum_{k'=2}^ND_{kk'}(\underline{u}_{k'},\underline{u}^I)=0,\quad
\sum_{k'=2}^ND_{kk'}(\underline{u}_{k'},\underline{\tilde\omega})=0.
\end{equation}
Thus, the minimal number of oscillators, necessary to produce the
Arnold diffusion, is $N=4$ in a system with two conserved quantities.

Constraints (\ref{ArnoldDconstr=}) ensure that any distribution
function~$f$ which depends on~$p_k$ only via the combinations
\begin{equation}
I_{tot}(p_2,\ldots,p_N)=\sum_{k=2}^Np_k
(\underline{u}_k,\underline{u}^I),\quad
H_{tot}(p_2,\ldots,p_N)=\sum_{k=2}^N
\left[(\underline{\omega},\underline{u}_k)p_k
+\frac{g}2\,p_k^2\right],
\end{equation}
is a stationary solution of Eq.~(\ref{Arnolddif=}). Two specific
distributions are particularly important.
One is the microcanonical distribution, determined by two
parameters $I_{tot}^0,H_{tot}^0$:
\begin{subequations}
\begin{equation}
f_{I^0_{tot},H^0_{tot}}(p_2,\ldots,p_{N})=e^{-S}\,
\delta\left(H_{tot}(p_2,\ldots,p_{N})-H^0_{tot}\right)
\delta\left({I}_{tot}(p_2,\ldots,p_{N})-I^0_{tot}\right),
\label{microcanonical=}
\end{equation}
where $e^{-S}$ is the normalization factor, determined from
\begin{equation}
e^S=\int
\delta\left(H_{tot}(p_2,\ldots,p_{N})-H^0_{tot}\right)
\delta\left({I}_{tot}(p_2,\ldots,p_{N})-I^0_{tot}\right)
W_s\,dp_2\ldots{d}p_{N}.
\end{equation}
\end{subequations}
Thus defined
$S=S(I^0_{tot},H^0_{tot})$ is nothing else than the entropy.
The microcanonical distribution is the one to which
$N$~oscillators relax after having started from an initial
condition characterized by some values of
$I_{tot}(p_2,\ldots,p_{N})=I^0_{tot}$ and
$H_{tot}(p_2,\ldots,p_{N})=H^0_{tot}$, provided that these
$N$~oscillators are \emph{isolated} from the rest of the chain.
Indeed, the microcanonical distribution~(\ref{microcanonical=})
is the only one which (i)~is a stationary solution of the
diffusion equation (\ref{Arnolddif=}), and (ii)~has definite
values of $I_{tot}$ and $H_{tot}$. In other words, the diffusion
equation makes any initial distribution evolve towards the
flat one on the allowed manifold, which is precisely the
microcanonical distribution~(\ref{microcanonical=}).

The other important distribution is the grand canonical
distribution, determined by two parameters $\beta\equiv{1}/T$
(the inverse temperature) and $\mu$~(the chemical potential):
\begin{subequations}
\begin{equation}\label{grandcanonical=}
f_{\beta,\mu}(p_2,\ldots,p_N)=
e^{\beta{F}-\beta[H_{tot}(p_2,\ldots,p_N)
-\mu{I}_{tot}(p_2,\ldots,p_N)]},
\end{equation}
where $e^{\beta{F}}$ is the normalization factor, determined from 
\begin{equation}
e^{-\beta{F}}=\int
{e}^{-\beta[H_{tot}(p_2,\ldots,p_N)-\mu{I}_{tot}(p_2,\ldots,p_N)]}
W_s\,dp_2\ldots{d}p_N=1.
\end{equation}
\end{subequations}
Thus defined $F=F(\beta,\mu)$~is the free energy (more precisely,
the grand canonical thermodynamic potential).
The grand canonical distribution~(\ref{grandcanonical=})
is established on a finite segment of the chain, considered
to be a part of a larger segment, upon relaxation of this
larger segment to equilibrium. This larger segment may be
considered isolated, then the grand canonical distribution
for the smaller segment can be derived from the microcanonical
distribution for the larger segment using the standard
argumentation of statistical physics~\cite{LL5}.

In the region of the phase space where the guiding and the layer
resonances cannot be easily distinguished,
the above considerations should be slightly modified. Let us consider
a resonant triple occupying the sites $n=1,2,3$, and other oscillators
on the sites $n=4,\ldots,N$. As discussed in Sec.~\ref{sec:3oscchaotic},
the internal chaotic dynamics of the triple cannot be described in the
diffusion approximation. The relevant information about this dynamics
is contained in the spectral density $\nu_{\underline{m}^r}(\omega)$,
defined in Eq.~(\ref{xixi=}). Besides, the isolated triple can be
characterized by the values of its total action $\Itot_\mathrm{tr}$
and the total energy, which are conserved. The Hamiltonian of the triple
is given by Eq.~(\ref{H3expand=}), which we write as
$H_\mathrm{tr}(\Itot_\mathrm{tr})+\Hdif$.

Under the action of the perturbation, $I_n,\Itot_\mathrm{tr},\Hdif$
acquire the increments
\begin{subequations}\begin{eqnarray}
&&\delta{I}_{n>3}=-i\sum_rV_rm_n^r\int\limits_0^\infty
dt'\,e^{i\varpi_rt'}\Xi_{\underline{m}^r}(t') + \mbox{c. c.},\\
&&\sum_{n=4}^N\delta{I}_n+\delta\Itot_\mathrm{tr}=0,\\
&&\sum_{n=4}^N\tilde\omega_n\delta{I}_n
+\tilde\omega_\mathrm{tr}\delta\Itot_\mathrm{tr}
+\delta\Hdif=V(0)-V(t),
\end{eqnarray}\end{subequations}
where $\tilde\omega_\mathrm{tr}=
\partial{H}_\mathrm{tr}/\partial\Itot_\mathrm{tr}$, and
$\Xi_{\underline{m}^r}(t)$ was defined in Eq.~(\ref{xixi=}). As
discussed in Sec.~\ref{sec:3oscchaotic}, the excursion in $\Hdif$
is limited to the chaotic region of a finite phase volume, so the
Arnold diffusion occurs in the space of
$(\Itot_\mathrm{tr},I_4,I_5,\ldots,I_N)$. Thus, we introduce the
distribution function~$f$, such that the probability for the system
to be in an elementary volume of this space is
\begin{equation}
dP=f(\Itot_\mathrm{tr},I_4,\ldots,I_N)\,W_{ss}\,
d\Itot_\mathrm{tr}\,dI_4\ldots{d}I_N,
\end{equation}
where $W_{ss}$ is the chaotic phase volume in the space $(I_1,I_2,I_3)$
at fixed $\Itot_\mathrm{tr}=I_1+I_2+I_3$, as defined in
Eq.~(\ref{Wssdef=}). If we let the index~$k$ assume the values
$\mathrm{tr},4,5,\ldots{N}$ and define the vectors $\underline{u}_k$ as
\begin{equation}
u_{kn}=\left\{\begin{array}{ll}
\delta_{1n}+\delta_{2n}+\delta_{3n},&k=\mathrm{tr},\\
\delta_{kn},& k=4,\ldots,N,
\end{array}\right.
\end{equation}
then the diffusion equation for distribution equation has the same
form as Eq.~(\ref{Arnolddif=}) with the only replacement
$W_s\to{W}_{ss}$,
and the diffusion coefficient is again given by Eq.~(\ref{ArnoldD=}).

\subsection{Relaxation of remote oscillators}
\label{sec:relaxremote}

Here we apply the diffusion equation from Sec.~\ref{sec:Pumpdiffeq},
Eq.~(\ref{Arnolddif=}),
to study thermalization of an oscillator which is sufficiently far
away from the chaotic spot. In particular, we find the thermalization
time.

Let the $N$th oscillator have the weakest
coupling to the chaotic spot, i.~e., $D_{kN},D_{NN}\ll{D}_{kk'}$ for
$k,k'<N$. Then the first $N-1$ oscillators equilibrate among themselves
much faster than with the $N$th one. In this situation, equilibration
of the $N$th oscillator can be described by seeking the distribution
function in the form
\begin{equation}
f(\vec{p},I_N;t)=f_{\beta,\mu}^\perp(\vec{p})\,f^\|(I_N;t)
+\delta{f}(\vec{p},I_N;t),
\end{equation}
where we have denoted $\vec{p}\equiv(p_2,\ldots,p_{N-1})$. Being
sufficiently far away, the $N$th oscillator is not involved in the
guiding or layer resonances, so $p_N=I_N$, and $W_s$ is assumed
to be independent of~$I_N$. In the first term, $f_{\beta,\mu}^\perp$
is the grand canonical distribution function,
Eq.~(\ref{grandcanonical=}), for the first $N-1$ oscillators.
Finally, $\delta{f}$ is the small correction, $O(D_{NN}/D_{kk'})$,
whose component along $f_{\beta,\mu}^\perp$ in the space of functions
of~$\vec{p}$ is fixed by
\begin{equation}
\int\delta{f}\,{W}_s\,{d}p_2\ldots{d}p_{N-1}=0.
\end{equation}
Neglecting terms of the second order in $D_{NN}$ and using
constraints~(\ref{ArnoldDconstr=}), we write Eq.~(\ref{Arnolddif=})
as
\begin{equation}\begin{split}
W_sf_{\beta,\mu}^\perp\,\frac{\partial{f}^\|}{\partial{t}}={}&
f_{\beta,\mu}^\perp{b}_N\left[\frac{\partial}{\partial_{I_N}}
+\beta(\tilde\omega_N-\mu)\right]f^\|
+f_{\beta,\mu}^\perp{W}_s{D}_{NN}\left[\frac{\partial}{\partial_{I_N}}
+\beta(\tilde\omega_N-\mu)\right]^2f^\|+\\
&+\sum_{k,k'=2}^{N-1}\frac{\partial}{\partial{p}_k}\,
W_s{D}_{kk'}\,\frac{\partial\delta{f}}{\partial{p}_{k'}},
\label{ArnolddifkN=}
\end{split}\end{equation}
where we denoted
\begin{equation}
b_N\equiv\sum_{k=2}^{N-1}\frac{\partial({W}_sD_{kN})}{\partial{p}_k}.
\end{equation}
Note that the last term in Eq.~(\ref{ArnolddifkN=}) is of the same
order as others; its role is to compensate the dependence of
$W_s,b_N,D_{NN}$ on $\vec{p}$. However, being a total derivative,
it vanishes when integrated over $p_2,\ldots,p_{N-1}$. Thus,
integrating Eq.~(\ref{ArnolddifkN=}) over $p_2,\ldots,p_{N-1}$,
denoting
\begin{equation}
\langle\ldots\rangle_\perp=
\int(\ldots)\,f_{\beta,\mu}^\perp\,W_s\,{d}p_2\ldots{d}p_{N-1},
\end{equation}
and noting that
\begin{equation}
\int{b}_N\,f_{\beta,\mu}^\perp\,{d}p_2\ldots{d}p_{N-1}
=\frac{\partial\langle{D}_{NN}\rangle_\perp}{\partial{I}_N}
-\beta(\tilde\omega_N-\mu)\langle{D}_{NN}\rangle_\perp,
\end{equation}
we obtain a closed equation for $f^\|(I_N)$:
\begin{equation}\label{ArnolddifN=}
\frac{\partial{f}^\|}{\partial{t}}=
\frac\partial{\partial{I}_N}\,\langle{D}_{NN}\rangle_\perp
\left[\frac{\partial}{\partial{I_N}}+\beta(\tilde\omega_N-\mu)\right]
f^\|.
\end{equation}
Solution of this equation is not trivial since both
$\langle{D}_{NN}\rangle_\perp$ and $\tilde\omega_N$ depend
on $I_N$: $\tilde\omega_N=\omega_N+gI_N$, and
$\langle{D}_{NN}\rangle_\perp\propto{I}_N$, as discussed in
Sec.~\ref{sec:couplingremote}. Still, its stationary solution
is readily found to be given by $f_{\beta,\mu}(I_N)$.

An analytical solution of Eq.~(\ref{ArnolddifN=}) describing the
relaxation dynamics can be obtained when $gI_{N}\ll\omega_N-\mu$,
so we replace $\tilde\omega_N\to\omega_N$. After rescaling the
action and time variables, we can rewrite Eq.~(\ref{ArnolddifN=})
as
\begin{equation}\begin{split}
&\partial_{t'}u=\partial_x\left[x(\partial_x+1)u\right],\\
&x\equiv\beta(\omega_N-\mu)I_N,\quad
t'\equiv\beta(\omega_N-\mu)
\frac{\partial\langle{D}_{NN}\rangle_\perp}{\partial{I_N}}\,t
\equiv\Gamma{t}.
\end{split}\end{equation}
Seeking the solutions $\propto{e}^{-\lambda{t}'}$, and performing
the Laplace transform,
\begin{equation}
\tilde{u}(s)=\int\limits_0^\infty{u}(x)\,e^{-sx}\,dx,
\end{equation}
we obtain
\begin{equation}
s\,\frac{d}{ds}\left[(s+1)\tilde{u}\right]=\lambda\tilde{u}
\;\;\;\Rightarrow\;\;\;
\tilde{u}_\lambda(s)=\frac{s^\lambda}{(s+1)^{\lambda+1}}.
\end{equation}
The inverse Laplace transform takes especially simple form
when $\lambda$ is integer:
\begin{equation}
u_n(x)=\frac{1}{n!}\frac{d^n(x^ne^{-x})}{dx^n}
=\mathrm{L}_n(x)\,e^{-x},
\end{equation}
where $\mathrm{L}_n(x)$ are the Laguerre polynomials, which form
a complete and orthonormal set of functions on $0<x<\infty$ with
the weight~$e^{-x}$:
\begin{equation}\label{Laguerre=}
\int\limits_0^\infty\mathrm{L}_n(x)\,\mathrm{L}_{n'}(x)\,
{e}^{-x}\,dx=\delta_{nn'},\quad
\sum_{n=0}^\infty\mathrm{L}_n(x_0)\,\mathrm{L}_n(x)\,{e}^{-x}
=\delta(x-x_0),\quad x,x_0>0.
\end{equation}
The eigenfunction $u_0(x)=e^{-x}$ with zero relaxation rate
corresponds to the equilibrium distribution,
\begin{equation}
f^\|_{\beta,\mu}(I_N)=
\beta(\omega_N-\mu)\,e^{-\beta(\omega_N-\mu)I_N}.
\end{equation}
The eigenfunction $u_1(x)=(1-x)e^{-x}$ corresponds to a small
mismatch in the temperature [or in the chemical potential, as
they enter in one combination, $\beta(\omega_N-\mu)$]:
\begin{equation}
\frac{\partial{f}_{\beta,\mu}^\|(I_N)}{\partial\beta}\,\delta\beta
=\delta\beta\,(\omega_N-\mu)\left[1-\beta(\omega_N-\mu)I_{N}\right]
e^{-\beta(\omega_N-\mu)I_{N}}.
\end{equation}
Since the components of the distribution function, proportional
to eigenfunctions $u_{n>0}$, decay as $e^{-n\Gamma{t}}$, the
typical action (or energy) relaxation time of the oscillator can
be estimated as
$\Gamma^{-1}\sim(T/|\mu|)^2\langle{D}_{NN}\rangle_\perp^{-1}$.

From Eq.~(\ref{ArnolddifN=}) one can easily obtain the
fluctuation-dissipation theorem -- a relation between dynamical
fluctuations of the system in equilibrium and the system's
dynamical response to an external perturbing force~\cite{Callen1951}.
First, let us find the power spectrum of the fluctuations of the
action,
\begin{equation}
S_N(\omega)\equiv\int\limits_{-\infty}^\infty
\left[\langle{I}_N(t)\,I_N(0)\rangle-\langle{I}_N\rangle^2\right]
e^{i\omega{t}}\,dt.
\end{equation}
The average can be written as
\begin{equation}
\langle{I}_N(t)\,I_N(0)\rangle=
\int{I}_N^0{f}^\|_{\beta,\mu}(I_N^0)\,dI_N^0
\int{I}_N\,G(I_N,t|I_N^0)\,dI_N,
\end{equation}
where the first integral represents the averaging over the initial
conditions $I_N(0)=I_N^0$, and in the second integral we introduced
the conditional probability $G(I_N,t|I_N^0)$ -- the probability
distribution of $I_N$ at time~$t$, given that at $t=0$ it had a
fixed value $I_N=I_N^0$. Thus, $G(I_N,t|I_N^0)$ is given by the
solution of Eq.~(\ref{ArnolddifN=}) with the initial condtition
$f^\|(I_N)=\delta(I_N-I_N^0)$. This solution is easily found using
the second of the equations~(\ref{Laguerre=}):
\begin{equation}
G(I_N,t|I_N^0)=\beta(\omega_N-\mu)\sum_{n=0}^\infty
\mathrm{L}_n(\beta(\omega_N-\mu)I_N^0)\,
\mathrm{L}_n(\beta(\omega_N-\mu)I_N)\,
e^{-\beta(\omega_N-\mu)I_N-n\Gamma{t}}.
\end{equation}
This immediately gives
\begin{equation}\label{SNw=}
\langle{I}_N(t)\,I_N(0)\rangle=
\frac{1+e^{-\Gamma|t|}}{\beta^2(\omega_N-\mu)^2},\quad
S_N(\omega)=\frac{1}{\beta^2(\omega_N-\mu)^2}\,
\frac{2}{1+\omega^2/\Gamma^2}.
\end{equation}

Let us now find the susceptibility $\chi_N(\omega)$ which
determines the response of $I_N$ to a periodic perturbation
of the system's Hamiltonian,
\begin{equation}
\delta{H}=-\left(F_\omega{e}^{-i\omega{t}}
+F_\omega^*e^{i\omega{t}}\right)I_N,
\end{equation}
where $F_\omega$ is the generalized force corresponding to~$I_N$.
It has the dimensionality of a frequency, so it can be viewed as
a periodic modulation of the frequency~$\omega_N$. This modulation
produces a periodic deviation of the action $I_N$ from its
equilibrium average value; by definition,
\begin{equation}
\langle{I}_N\rangle=
\left.\langle{I_N}\rangle\right|_{F_\omega=0}+
\chi_N(\omega)\,F_\omega{e}^{-i\omega{t}}+
\chi_N^*(\omega)\,F_\omega{e}^{i\omega{t}}+O(F_\omega^2).
\end{equation}
The average of~$I_N$ on the left-hand side is performed over
the perturbed distribution function,
\begin{equation}
f^\|(I_N)=f^\|_{\beta,\mu}(I_N)
+\delta{f}^\|_\omega(I_N)\,e^{-i\omega{t}}
+[\delta{f}^\|_\omega(I_N)]^*\,e^{i\omega{t}}
+O(F_\omega^2),
\end{equation}
which should be found from the perturbed Eq.~(\ref{ArnolddifN=}).
The frequency~$\omega_N$ enters Eq.~(\ref{ArnolddifN=}) in two
places: first, explicitly in the square brackets; second, through
$\langle{D}_{NN}\rangle_\perp$. Note, however, that the modulation
of $\langle{D}_{NN}\rangle_\perp$ does not produce any correction
to the equilibrium distribution function:
$e^{-\beta(\omega_N-\mu)I_N}$ is a stationary solution of the
equation no matter what happens to $\langle{D}_{NN}\rangle_\perp$.
Thus, to find the linear in $F_\omega$ correction to~$f\|(I_N)$,
we should only include the modulation of $\omega_N$ in the square
brackets Eq.~(\ref{ArnolddifN=}). As a result,
$\delta{f}^\|_\omega(I_N)$ can be found from the equation
\begin{equation}
\left\{i\omega
+\frac\partial{\partial{I}_N}\,\langle{D}_{NN}\rangle_\perp
\left[\frac{\partial}{\partial{I_N}}+\beta(\omega_N-\mu)\right]
\right\}\delta{f}^\|_\omega(I_N)
=\beta{F}_\omega\,
\frac\partial{\partial{I}_N}\,
\frac{\langle{D}_{NN}\rangle_\perp{e}^{-\beta(\omega_N-\mu)I_N}}%
{\beta(\omega_N-\mu)}.
\end{equation}
Noting that the right-hand side of this equation is proportional
to the eigenfunction~$u_1(x)$, we easily find $\delta{f}^\|_\omega$
and the susceptibility:
\begin{equation}
\chi_N(\omega)=\frac{1}{1-i\omega/\Gamma}\,
\frac{1}{\beta(\omega_N-\mu)^2}.
\end{equation}
Comparing this with the expression for $S_N(\omega)$,
Eq.~(\ref{SNw=}), we obtain the relation
\begin{equation}
S_N(\omega)=\frac{2}{\beta\omega}\,\Im\chi_N(\omega),
\end{equation}
which is precisely the classical form of the fluctuation-dissipation
theorem~\cite{LL5}.

\subsection{Coupling to remote oscillators}
\label{sec:couplingremote}

Here we analyze the perturbation theory terms which determine
the diffusion in action of an oscillator which is sufficiently
far away from the chaotic spot. Let the rightmost oscillator of
the guiding resonance be on the site $n=0$, and let us focus on
the oscillator at $n=L$, assuming it to be further away than
the oscillators of the layer resonance, i.~e. $L\gg{L}_\ell$.
As we have seen in Sec.~\ref{sec:relaxremote},
relaxation of this oscillator is determined by the diagonal
element $D_{LL}$ of
the diffusion coefficient. The latter, in turn, is given by
Eq.~(\ref{ArnoldD=}) where the second term can be neglected as
its denominator is determined by oscillators with
$|n|\sim{L}_\ell\ll{L}$, and thus is large.

\begin{figure}
\begin{center}
\includegraphics[width=12cm]{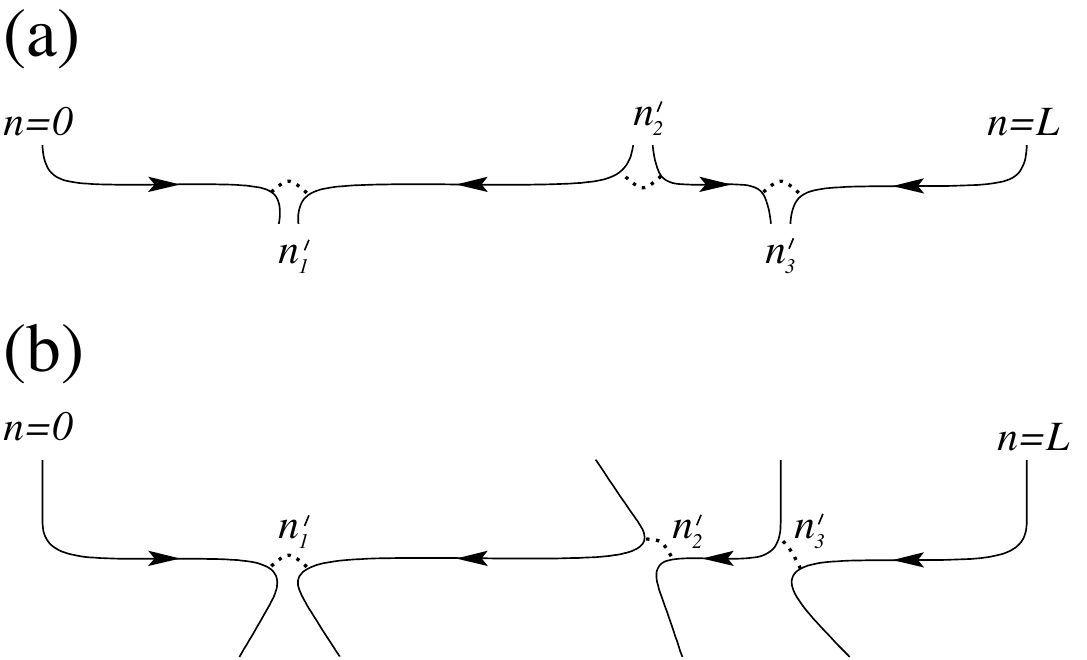}
\end{center}
\caption{\label{fig:diagremote}
Examples of diagrams describing the coupling between the chaotic
spot and a remote oscillator with $N_r=4$.}
\end{figure}

To describe the coupling between the oscillators at $n=0$ and $n=L$
for large enough $L$, it is sufficient to consider perturbation terms
whose spatial structure is described by diagrams of the kind shown in
Fig.~\ref{fig:diagremote}. First, consider Fig.~\ref{fig:diagremote}(a).
Namely, $N_r-1$ nonlinear vertices are placed on some sites
$n_1',\ldots,n_{N_r-1}'$ between $0$ and $L$, and each of these sites
has $m_{n_k'}^r=2(-1)^k$, while $m_0^r=1$, $m_L^r=(-1)^{N_r}$, and the
rest of $m_n^r=0$.
Indeed, $V_r\propto\tau^L$ which is clearly the lowest possible order
in~$\tau$. The number of resonances grows as
\begin{subequations}
\begin{equation}\label{NresLr=L=}
\mathcal{R}=\frac{(L-1)!}{(N_r-1)!\,(L-N_r)!}\approx
\sqrt{\frac{N_r}{2\pi{L}^2}}
\left(\frac{e^{1+O(N_r/L)}L}{N_r}\right)^{N_r},\quad
N_r\ll{L}.
\end{equation}
Next order in~$\tau$ is obtained by putting two oscillators
at neighboring sites near one of the $N_r-1$ nonlinear vertices,
so it introduces a factor $\sim{N}_r$ in the number of resonances.
Higher orders in~$\tau$ can be obtained if one separates the
two ends of the solid lines near the $k$th nonlinear vertex by
a distance~$l_k$. Assuming $l_k\ll{L}/N_r$, which is equivalent
to $L_r-L\ll{L}$, we can represent the number of resonances
corresponding to such configurations as
\begin{equation}\begin{split}
\label{nresLr>L=}
\mathcal{R}{}&=
\frac{(L-1)!}{(N_r-1)!\,(L-N_r)!}\sum_{l_1,\ldots,l_{N_r-1}\geqslant{0}}
\delta_{l_1+\ldots+l_{N_r-1},L_r-L}=\\
&=\frac{(L-1)!}{(N_r-1)!\,(L-N_r)!}
\frac{(L_r-L+N_r-2)!}{(L_r-L)!\,(N_r-2)!}.
\end{split}\end{equation}
\end{subequations}
This expression does not take into account changing the direction
of the solid lines, shown in Fig.~\ref{fig:diagremote}(b), which
can be done at those nonlinear vertices where $l_k>{0}$. Thus,
Eq.~(\ref{nresLr>L=}) gives a lower bound for the number of
resonances, while the upper bound is obtained by multiplying it
by $2^{N_r}$ for the two possible directions of each solid line.
The number of terms of the perturbation theory which have the same
order in $\tau,\rho$ and contribute to the same resonance, is
given by the product $\mathcal{N}=(l_1+1)\ldots(l_{N_r-1}+1)$.
Note that Eqs.~(\ref{NresLr=L=}), (\ref{nresLr>L=}) do not fit
in the discussion of Sec.~\ref{sec:numres}, as an extra constraint
is imposed (two sites of the resonance, $n=0$ and $n=L$, are fixed).

It is not always necessary to invoke terms shown in
Fig.~\ref{fig:diagremote}(b), as they bring an additional smallness
$\tau^{2(L_r-L)}$ to $\langle{D}_{LL}\rangle$, while the resonances
counted in Eq.~(\ref{NresLr=L=}) which correspond to
Fig.~\ref{fig:diagremote}(a) may already be sufficient to remove the
smallness of the Melnikov-Arnold exponential, especially for very
large~$L$. Thus, at ${L}\gtrsim{L}_\ell$ one can write the following
bounds:
\begin{equation}\label{Dremote=}
\frac{\Delta^3}{g^2}\,
w\,e^{-\mathcal{C}_3\ln^2(1/\tau^p\rho)}
\tau^{2(L-L_\ell)}
\left(\zeta_1^2+\ldots+\zeta_\mathcal{N}^2\right)
<\langle{D}_{LL}\rangle<
\frac{\Delta^3}{g^2}\,\tau^{2L}
\left(\zeta_1^2+\ldots+\zeta_\mathcal{N}^2\right).
\end{equation}
Here each $\zeta_k$ is a product of random denominators,
discussed in \ref{app:productdist}.
The upper bound is obtained by simply setting $L_r=L$,
$N_r=0$, and the Melnikov-Arnold exponential to unity.
The lower bound is written by noting that if the optimal
resonance is sought not on the full segment of the length~$L$,
but on a shorter one of the length $L_\ell$, it will be similar
to the layer resonance, so the corresponding factors are the same
as those determining the chaotic fraction,~$w$, except that
instead of the layer resonance one should take the next best
driving resonance, hence the factor
$e^{-\mathcal{C}_3\ln^2(1/\tau^p\rho)}$ with $\mathcal{C}_3\sim{1}$,
as discussed in the end of Sec.~\ref{sec:DensitySummary}.
Also, $w$~contains the thermal exponential $e^{-E_g/T}$,
while $D_{LL}$ does not.
Again, the disorder average of $\langle{D}_{LL}\rangle$ diverges,
as well as all higher moments, but
as we will see in Sec.~\ref{sec:Breaks}, it is the average of
$\langle{D}_{LL}\rangle^{-1}$ that enters the final result, so
the sum $\zeta_1^2+\ldots+\zeta_\mathcal{N}^2$ can be effectively
replaced by $(\mathrm{const}\sim{1})^L$. Moreover, even keeping
this factor along with $\tau^L,\rho^L$ would be beyond our
precision, so we set $\zeta_1^2+\ldots+\zeta_\mathcal{N}^2\to{1}$.

%
%

\subsection{Migration of chaotic spots}

The discussion of Sec.~\ref{sec:Pumpdiffeq} regarded a segment of
the chain containing one chaotic spot ``living'' on a given guiding
resonance $(\underline{m}^g,\underline{\tilde\omega})=0$. The system
dynamics in the space of the actions $\{I_n\}$ corresponded to the
diffusion on the hypersurface determined by the constraints
\begin{subequations}\begin{eqnarray}
&&I_{tot}=\sum_nI_n=\mathrm{const},\\
&&H_{tot}=\sum_n\left(\omega_nI_n+\frac{g}2\,I_n^2\right)=\mathrm{const},\\
&&(\underline{m}^g,\underline{\tilde{\omega}})=
\sum_nm^g_n(\omega_n+gI_n)=0.
\end{eqnarray}\end{subequations}
The last constraint is due to the fact that the system cannot leave
the stochastic layer around the guiding resonance, and the variables
$p_k$, introduced in Sec.~\ref{sec:Pumpdiffeq} are simply the
coordinates on the corresponding hyperplane. In the course of
diffusion the system may cross another hypersurface, determined by
the same action and energy conservation constraints
$I_{tot}=\mathrm{const}$, $H_{tot}=\mathrm{const}$, but with a
different $(\underline{m}^{g\prime},\underline{\tilde\omega})=0$.
The time required for this to happen can be estimated as
$\sim(T/|\mu|)^2\langle{D}_{LL}\rangle^{-1}e^{E_{g\prime}/T}$.
The prefactor corresponds to the time needed to cross the
typical thermal region of the action space, and ${D}_{LL}$ is
the diffusion coefficient of the most remote oscillator with
$m_n^{g\prime}\neq{0}$. This time coincides with the thermalization
time of this oscillator, obtained in Sec.~\ref{sec:relaxremote}.
The Arrhenius exponential corresponds to the number of attempts
needed to overcome the activation barrier for the oscillators
of the $g'$ resonance. Thus estimate is valid for lengths
$L\gg{E}_{g'}/T$; in the opposite limiting case the amount of action
needed to put the system on the $g'$~resonance is greater than the
typical action, $\sim{L}T/|\mu|$, stored in a segment of the
length~$L$, so the activation time is determined by the diffusion
coefficient at lengths $\sim{E}_g/T$.

\begin{figure}
\begin{center}
\includegraphics[width=6cm]{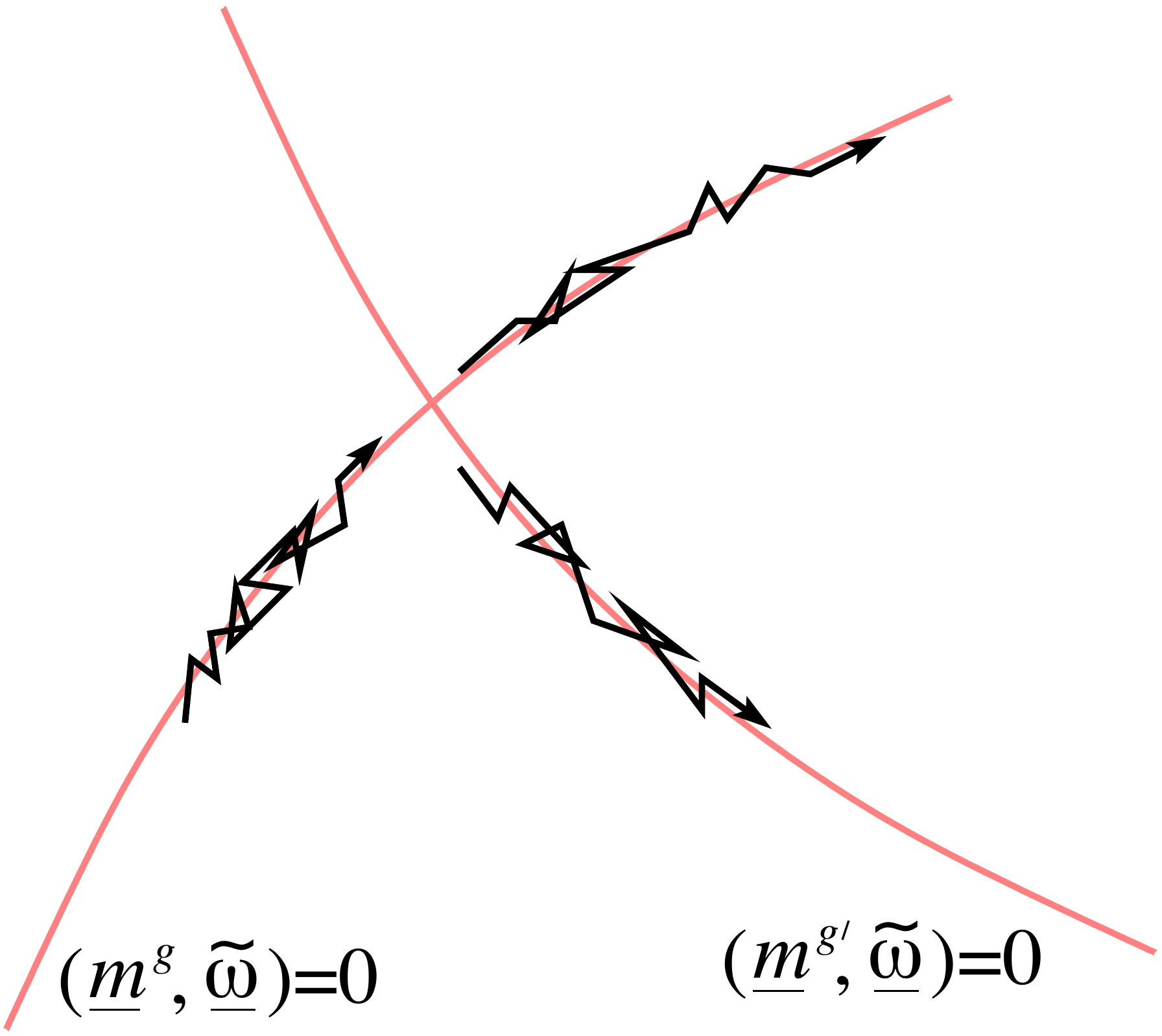}
\end{center}
\caption{\label{fig:resonance}
A schematic representation of an intersection between two resonant
hypersurfaces $(\underline{m}^g,\underline{\tilde\omega})=0$,
$(\underline{m}^{g\prime},\underline{\tilde\omega})=0$, shown
by the smooth curves (the plane of the figure corresponds to
the curved manifold $I_{tot}=\mathrm{const}$,
$H_{tot}=\mathrm{const}$). The wiggly trajectories represent the
system diffusing along the resonances. First, the system diffuses
along the $g$~resonance and arrives at the intersection point.
After some time it leaves the intersection point randomly choosing
whether to continue along the $g$~resonance or to switch to the
$g'$~resonance.
}
\end{figure}
Once the system variables fall in the stochastic layer of the
second resonance, we have two chaotic spots not far from each
other, each one acting as a stochastic pump.\footnote{
In connection with the problem of system sticking to various
structures inside the stochastic layer, discussed in the end
of Sec.~\ref{sec:2oscstoch}, we note that when the system is
changing the guiding resonance, sticking should be less probable
than in the situation of the diffusion along a single guiding
resonance. Indeed, when two chaotic spots are present relatively
close to each other, each of them effectively represents an
external source of   noise for the other one, preventing it
from sticking.}
Now nothing
prevents either of them to leave its stochastic layer under the
action of the other. Moreover, this happens in a relatively
short time as the system only has to diffuse across the thin
stochastic layer. Once this happens, one of the chaotic spots
is turned off, the other one remaining alone, and thus confined
to its own stochastic layer. If
it is the $g'$ resonance that is turned off, there is no
overall effect on the system. However, if it is the original
$g$~resonance, the system has effectively switched from one
resonance to the other, as shown schematically in
Fig.~\ref{fig:resonance}. As the two resonances are, generally
speaking, located in different regions of the chain, this
process corresponds to migration of the chaotic spot along the
chain.

In connection with migration of the chaotic spots it is worth
mentioning a recent mathematical work~\cite{Kaloshin2010}. For
a one-dimensional chain of pendula with weak local coupling,
energy above that of the unstable equilibrium was shown to be
able to perform an arbitrary random walk along the chain.

\section{Current through breaks}
\label{sec:Breaks}
\setcounter{equation}{0}

\subsection{General remarks on the procedure}
\label{sec:Breaksgeneral}

Global transport properties of random one-dimensional systems
tend to be determined by rare strong obstacles~\cite{Kurkijarvi1973}.
The reason for this is that in one dimension such obstacles cannot
be bypassed, in contrast to higher dimensions. Such rare obstacles
were called ``blockades''~\cite{Shante1973}, ``weak
links''~\cite{Lee1984}, or ``breaks''~\cite{Raikh1989,Rodin2009},
and here the latter term will be used. In the present problem,
a break is a region of the chain where the chaotic fraction~$w$
assumes anomalously small values on many consecutive sites,
i.~e., a region of the chain rarely visited by chaotic spots.
Equilibration to the left and to the right of the break is assumed
to be much faster than across the break, so the regions to the
left and to the right can be assumed to be equilibrated with
different values of the chemical potential and temperature,
$\mu_L,T_L\equiv{1}/\beta_L$ and $\mu_R,T_R\equiv{1}/\beta_R$,
respectively. The action and energy currents through the break,
$\curI,\curHwI$, are defined as the amounts of action and energy
transferred from left to right per unit time. In the linear
response approximation they can be written as
\begin{equation}
\left[\begin{array}{c} \curI \\ \curHwI \end{array}\right]
=R_b^{-1}
\left[\begin{array}{c} \beta_L\mu_L-\beta_R\mu_R
\\ \beta_R-\beta_L \end{array}\right].
\end{equation}
It is convenient to borrow the terminology from the electric
circuit theory, as discussed in Sec.~\ref{sec:QualitativeMacroscopic}:
the 2-column on the left-hand side will be simply
called the current, the 2-column on the right-hand side will be
called the voltage drop across the break. The $2\times{2}$ matrix
$R_b^{-1}$ corresponds to the break conductance, and its inverse,
$R_b$, to the resistance. It is determined by the realization of
disorder forming the break.

Quite generally, validity of the linear response theory and
finiteness of the coefficient $R_b^{-1}$ are immediate
consequences of the diffusive character of the system dynamics
in the space of actions. In the present work the origins of the
linear response theory are most easily seen in
Sec.~\ref{sec:relaxremote}, where a
perturbation of the distribution function of an oscillator from
its equilibrium value (say, putting it at a different temperature
or chemical potential) is shown to correspond to an eigenfunction
of the diffusion equation with a finite decay rate. The analysis
to be presented below corresponds to a more complicated situation,
but the diffusion equation which was presented in Sec.~\ref{sec:Pump}
will lead to a definite expression for $R_b^{-1}$.

In a stationary situation the current is constant along the chain,
while the voltage drop across a segment containing many breaks is
the sum of the voltage drops across each break. Thus, the macroscopic
conductivity~$\sigma$ entering Eq.~(\ref{currents=}) can be
expressed as
\begin{equation}\label{seriesresistors=}
\sigma=\left[\int{R}_b\,n_b\,db\right]^{-1},
\end{equation}
where $n_b\,db$ is the spatial density (or probability per
unit length) of breaks of a given type~$b$.
Eq.~(\ref{seriesresistors=}) is nothing but the usual addition
rule for resistors in series, as discussed in
Sec.~\ref{sec:QualitativeMacroscopic}. Analogously to the problem
of electronic hopping conduction in one-dimensional
samples~\cite{Kurkijarvi1973,Shante1973,Lee1984,Raikh1989,Rodin2009},
there are two strong competing factors: breaks with very large
resistances are extremely rare, while frequent breaks have too small
resistance.
As a result, the dominant contribution to the integral comes from
breaks which are close to a certain optimal configuration.
This also means that the macroscopic resistivity $\sigma^{-1}$ is
self-averaging on the length scale corresponding to the typical
distance between such optimal breaks. This section is dedicated
to finding such optimal breaks and to evaluation of
Eq.~(\ref{seriesresistors=}) analogously to Ref.~\cite{Raikh1989}.

In Eq.~(\ref{seriesresistors=}), $R_b$ is a random quantity,
determined by the frequencies of all oscillators in the break
region, while $n_b\,db$ is the probability measure in the space of
these frequencies.
However, evaluation of $R_b$ directly in terms of the frequencies
seems to be hopelessly complicated. Instead, a break will be
characterized by the corresponding spatial profile of the chaotic
fraction $\{w_n=e^{-\lambda_n}\}$, defined by Eq.~(\ref{wgdef=}).
Its probability distribution on a
given site is given by $p(\lambda)=d[1-e^{-S(\lambda)}]/d\lambda$,
where $S(\lambda)$ was calculated in
Sec.~\ref{sec:Density}. Assuming the chaotic fractions on different
sites to be uncorrelated, one can write the probability measure per
unit length as
\begin{equation}\label{nbdb=}
\int{n}_b\,db\to
\frac{1}L\int\prod_{|n|<L/2}d\left[1-e^{-S(\lambda_n)}\right],
\end{equation}
where $S(\lambda)$ is given by Eq.~(\ref{Slambda=}),
$S(\lambda)=\mathcal{C}_1\lambda^{p_1}\rho%
{e}^{\lambda/[\mathcal{C}\ln^2(1/\tau^p\rho)]}$,
and $L$~is a length which is much larger than the size of
a break, but much smaller than the distance between breaks.
The fact that the integral in right-hand side of Eq.~(\ref{nbdb=})
is proportional to the length follows from the translational
invariance of the probability distribution and is discussed
in detail in \ref{app:Jacobian}.

Strictly speaking, $R_b$~is not a functional of $\{\lambda_n\}$
only, since $\lambda$~is determined by the guiding and
layer resonances, while $R_b$ depends on the diffusion
coefficients, which are determined by the same guiding
resonance as~$\lambda$, but also by the driving resonance which
is different from the layer resonance. Still, it is quite
strongly correlated with~$\lambda$, as discussed in
Sec.~\ref{sec:couplingremote}, Eq.~(\ref{Dremote=}). In the
following, the relation between $R_b$ and $\{\lambda_n\}$ will
be approximated by a deterministic dependence $R_b=R(\{\lambda_n\})$,
and then the integral in Eq.~(\ref{seriesresistors=}) will be writen
as an integral over $\{\lambda_n\}$. Determination of the dependence
$R_b=R(\{\lambda_n\})$ is the main task of Secs.~\ref{sec:spotontop}
and~\ref{sec:spotonslope}.

Obviously, the profile $\lambda_n$ corresponding to the optimal
break must be symmetric with respect to its center, which can be
assumed to be either at $n=0$, or half-way between $n=0$ and
$n=1$, without loss of generality. Also, at the center the profile
$\lambda_n$ must have a maximum, $\lambda_0\gg{1}$.
Two mechanisms for the current flow through the break can be
identified, which require different treatment.
First, a chaotic spot may find itself near the center of the break,
at $|n|\sim{L}_\ell^\mathrm{max}\sim\lambda_0/\ln(1/\tau^p\rho)$
(the typical size of the layer resonance at the center of the break,
as estimated in Sec.~\ref{sec:Density}),
which occurs very rarely, once in a time $\propto{e}^{\lambda_0}$.
Second, a chaotic spot may reside somewhere on the slope of the
break, $|n|\gtrsim\lambda_0/\ln(1/\tau^p\rho)$ which occurs more
frequently. However, the current will still be determined by the
diffusion of oscillators near the center of the break, $\sim{D}_{00}$,
which has an additional smallness
$\sim\tau^{-2|n|}$. For the optimal break all these contributions
must have the same order; indeed, if one of the contributions to
the current is too small, this means that in the corresponding
region of the break $\lambda_n$ can be decreased with no effect
on the overall current, but with a gain in the probability.

\begin{figure}
\begin{center}
\includegraphics[width=12cm]{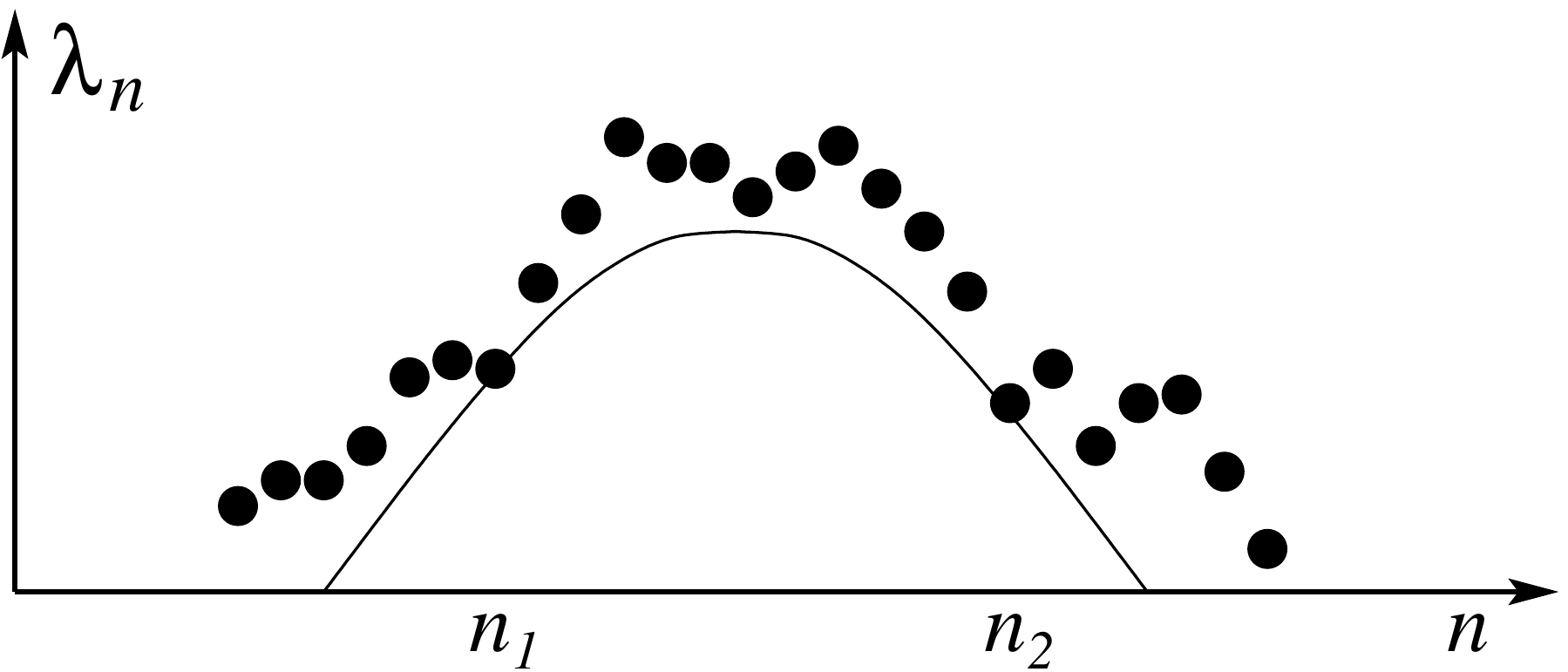}
\end{center}
\caption{\label{fig:breaks}
The profile $\{\lambda_n\}$ corresponding to some arbitrary break
(shown by circles at integer positions). The solid curve shows the
optimal break $b_\lambda(x)$, corresponding to this profile
$\{\lambda_n\}$, which touches $\{\lambda_n\}$ at two points
(denoted here by $n_1$ and $n_2$), and has the same resistance as
the break with the profile $\{\lambda_n\}$, up to power-law prefactors.}
\end{figure}

Putting the abovesaid in more formal terms, we introduce a
one-parametric family of functions
$b_\lambda(x)$, concave and symmetric, $b_\lambda(x)=b_\lambda(-x)$,
which we will call optimal breaks, parametrized by their value in
the maximum, $b_\lambda(0)=\lambda$. The resistance of a break with
an arbitrary profile $\{\lambda_n\}$ is the same (up to power-law
prefactors) as that of a break from the optimal family,
$\lambda_n^{opt}=b_\lambda(n)$,
with the value of~$\lambda$ determined as the largest $\lambda$ for
which exists a shift $x_0$ such that
$\lambda_n\geqslant{b}_\lambda(n-x_0)$ for all~$n$.
Indeed, this choice of $\lambda$ implies that there are at least
two sites, $n=n_1,n_2$, at which actually
$\lambda_{n_1}=b_\lambda(n_1)$, $\lambda_{n_2}=b_\lambda(n_2)$, and
which lie on different sides of the break, as illustrated by
Fig.~\ref{fig:breaks}. Then the transport through such a break will
be dominated by the chaotic spot when it resides on the sites
$n=n_1,n_2$. The resistance of the break becomes a function of
a single number $\lambda$ only, so that the integral in
Eq.~(\ref{seriesresistors=}) can be written as
\begin{equation}\label{Rintergal=}
\sigma^{-1}=\int\limits_0^\infty{d}\lambda\,R(\lambda)\,
\mathcal{J}(\lambda)\prod_ne^{-S(b_\lambda(n))},
\end{equation}
where the Jacobian $\mathcal{J}(\lambda)$ is analyzed in
\ref{app:Jacobian}.
The considerations of the previous paragraph already allow
one to guess that the optimal breaks are likely to be triangular,
$b_\lambda(x)\sim\lambda-2|x|\ln(1/\tau)$, at least at large enough~$x$,
and thus the size of the break $L_b\sim\lambda/\ln(1/\tau)$.

\subsection{Chaotic spot on the top of the break}
\label{sec:spotontop}

The definition of the currents given in Sec.~\ref{sec:Breaksgeneral}
leads to the following expression:
\begin{equation}
\left[\begin{array}{c} \curI \\ \curHwI \end{array}\right]
=\sum_{n>{0}}
\left[\begin{array}{c} 1 \\ \omega_n \end{array}\right]
\frac{d\langle{I}_n\rangle_f}{dt}
+\sum_{n>{0}}
\left[\begin{array}{c} 0 \\ g/2 \end{array}\right]
\frac{d\langle{I}_n^2\rangle_f}{dt}
+\left[\begin{array}{c} 0 \\ O(\tau) \end{array}\right].
\end{equation}
Here $\langle\ldots\rangle_f$ denotes the average with respect to
the distribution function~$f$, whose time dependence is governed
by the Arnold diffusion equation, Eq.~(\ref{Arnolddif=}).
Integrating by parts and introducing the vectors $\underline{u}^R$
such that $u^R_{n>{0}}=1$, $u^R_{n\leqslant{0}}=0$, 
and $\underline{\tilde\omega}^R$ such that
$\tilde\omega_{n>{0}}^R=\tilde\omega_n$,
$\tilde\omega_{n\leqslant{0}}^R=0$, we can write the current as
\begin{equation}\label{curf=}
\left[\begin{array}{c} \curI \\ \curHwI \end{array}\right]
=-\sum_{k,k'}\int
\left[\begin{array}{c} (\underline{u}^R,\underline{u}_k) \\ 
(\underline{\tilde\omega}^R,\underline{u}_k) \end{array}\right]
{D}_{kk'}\,\frac{\partial{f}}{\partial{p}_{k'}}\,W_s\prod_kdp_k.
\end{equation}
where the diffusion coefficient ${D}_{kk'}$ is given by
Eq.~(\ref{ArnoldD=}),
the distribution function~$f$ is defined in Eq.~(\ref{fdef=}), and
the stochastic layer width~$W_s$ is given by Eq.~(\ref{Wsgl=}).
Note that it would not be correct to determine the distribution
function itself from the stationary solution of Eq.~(\ref{Arnolddif=}).
Indeed, the latter was written for a given fixed guiding resonance,
while the position of the chaotic spot on the top of the break,
considered here, is in fact very rarely occupied for short periods
of time. The time required for the chaotic spot to jump off the top
is of the same order as the equilibration time, so the distribution
function on the left and on the right of the break is determined by
a certain average over different guiding resonances [see
Eq.~(\ref{Dsumspots=}) below].

Here we set
$f\propto\exp[-\beta_{L/R}(\omega_n-\mu_{L/R})I_n-\beta_{L/R}{g}I_n^2/2]$
for the oscillators to the left/right of the top of the break which do
not participate in the guiding resonance and for which $p_n=I_n$,
while the contribution of the oscillators participating in the guiding
resonance will be simply neglected. The relative error introduced by
neglecting these oscillators is small, since the number of the
oscillators in the guiding resonance, $2N_g$, is much smaller than that
for the layer resonance, $2N_\ell$, and other resonances driving the
diffusion (for these we expect $N_r\sim{N}_\ell$). More important is the
error introduced by setting the distribution function for other oscillators
to its equilibrium value ``by hands'': it is likely to overestimate the
current by a factor $\sim{1}$, which is not so important as our main
concern is about exponential factors. Using
constraints~(\ref{ArnoldDconstr=}), we express the currents as
\begin{equation}\label{curtop=}
\left[\begin{array}{c} \curI \\ \curHwI \end{array}\right]
=\sum_{{0}\leqslant{n}_1,n_2\notin{g}}\left\langle
\left[\begin{array}{c} 1 \\ \tilde\omega_{n_1} \end{array}\right]
{D}_{n_1n_2}
\left[\beta_R(\tilde\omega_{n_2}-\mu_R)-\beta_L(\tilde\omega_{n_2}-\mu_L)
\right]\right\rangle.
\end{equation}
Here the thermal average $\langle\ldots\rangle$ can already be
performed over the equilibrium distribution where the difference
between $\beta_R,\mu_R$ and $\beta_L,\mu_L$ is neglected, as we
are working in the linear response regime. Moreover, the diffusion
coefficient is always determined by the chaotic spot which has a
specific position, so contributions from different possible
positions~$n$ of the chaotic spots should be added with the
corresponding weights, which we write as
\begin{equation}\label{Dsumspots=}
{D}_{n_1n_2}=\sum_ne^{-\lambda_n}D_{n_1n_2}^{(n)}.
\end{equation}
As discussed in Sec.~\ref{sec:Breaksgeneral}, for the optimal
break all terms of this sum should be of the same order.
Since the dependence of $D_{n_1n_2}^{(n)}$ on $n$ is weak when
$|n|,|n_1|,|n_2|\sim{L}_\ell$, the top of the break is smooth:
$b_\lambda(x)=\lambda_0+O(1)$ for
$|x|\sim{L}_\ell^\mathrm{max}\sim\lambda_0/\ln(1/\tau^p\rho)$. 

As mentioned in Sec.~\ref{sec:Breaksgeneral},
the diffusion coefficient $D_{n_1n_2}^{(n)}$ is strongly
correlated with the chaotic fraction~$w_n$. Indeed, both are
determined by the same guiding resonance. The difference is
that $w_n$~contains the thermal exponential $e^{-E_g/T}$,
while $D_{n_1n_2}^{(n)}$ does not. Another difference is
that while the chaotic fraction is determined by the best
separatrix-destroying resonance, the diffusion coefficient is
determined by the second best one, the criteria for the
optimization being exactly the same. Indeed, the power spectrum
of the stochastic pump, Eq.~(\ref{rhom=Am}) contains the same
Melnikov-Arnold exponentials as the stochastic layer width,
Eq.~(\ref{Wsgl=}). The difference between contributions of the
best layer resonance and the second best resonance was discussed
in Sec.~\ref{sec:DensitySummary}, see Eq.~(\ref{secondlayer=}).
Thus, we adopt the following expression:
\begin{equation}
D_{n_1n_2}^{(n)}\sim{m}^r_{n_1}m^r_{n_2}\frac{\Delta^3}{g^2}\,
e^{-\mathcal{C}_2\lambda_n-\mathcal{C}_3\ln^2(1/\tau^p\rho)}.
\end{equation}
with $0<\mathcal{C}_2<1$, $\mathcal{C}_3\sim{1}$. Combining this
with Eqs.~(\ref{curtop=}),~(\ref{Dsumspots=}), we obtain
\begin{equation}\label{Rtop=mr}
\left.R^{-1}\right|_\mathrm{top}
\sim\frac{\Delta^3}{g^2}\sum_n
e^{-(1+\mathcal{C}_2)\lambda_n-\mathcal{C}_3\ln^2(1/\tau^p\rho)}
\left\langle
\left[\begin{array}{c} (\underline{u}^R,\underline{m}^r) \\
(\underline{\tilde{\omega}}^R,\underline{m}^r) \end{array}\right]
\left[\begin{array}{cc} (\underline{m}^r,\underline{u}^R) &
(\underline{m}^r,\underline{\tilde{\omega}}^R) \end{array}\right]
\right\rangle.
\end{equation}
It will be seen in Sec.~\ref{sec:optimization} that for a typical
break, contributing to the integral in Eq.~(\ref{seriesresistors=}),
the chaotic fractions $w_n$ are of the same order as for the optimal
break, $w_n\sim{e}^{-b_{\lambda_*}(n)}$. This means that the sum in
Eq.~(\ref{Rtop=mr}) contains many ($\sim{L}_\ell^\mathrm{max}$)
terms. Since $\underline{m}^r$'s are different for different~$n$,
the resulting matrix is not degenerate. Moreover, for each $n$ the
diagonal elements of the matrix can be estimated as
$(\underline{u}^R,\underline{m}^r)^2\sim{L}_\ell^\mathrm{max}$,
$\langle(\underline{m}^r,\underline{\tilde{\omega}}^R)\rangle\sim%
{L}_\ell^\mathrm{max}\Delta^2$, and they are positive. The
off-diagonal elements,
$\langle(\underline{\tilde{\omega}}^R,\underline{m}^r)%
(\underline{m}^r,\underline{u}^R)\rangle%
\sim{L}_\ell^\mathrm{max}\Delta$,
have random signs. Thus, the summation over~$n$ effectively
multiplies the diagonal elements by the number of the terms in the
sum, $\sim{L}_\ell^\mathrm{max}$, while the off-diagonal elements
are multiplied by $\sim\sqrt{{L}_\ell^\mathrm{max}}$. As a result,
we arrive at the following expression for the contribution from
top of the break:
\begin{equation}\label{Rtop=}
\left.R^{-1}\right|_\mathrm{top}
\sim\frac{\Delta^3}{g^2}\,
e^{-(1+\mathcal{C}_2)\lambda-\mathcal{C}_3\ln^2(1/\tau^p\rho)}
\left[\begin{array}{cc} 1 & \sim({L}_\ell^\mathrm{max})^{-1/2}\Delta \\
\sim({L}_\ell^\mathrm{max})^{-1/2}\Delta & \sim\Delta^2 \end{array}\right],
\end{equation}
where the overall prefactor $\sim({L}_\ell^\mathrm{max})^2$ is omitted,
as it is beyond our precision. In contrast, the factor
$\sim({L}_\ell^\mathrm{max})^{-1/2}$ determining the relative magnitude
of different matrix elements, is kept.

\subsection{Chaotic spot on the slopes of the break}
\label{sec:spotonslope}

When the chaotic spot resides on sites with
$|n|\gtrsim{L}_\ell^\mathrm{max}\sim\lambda_0/\ln(1/\tau^p\rho)$,
its coupling to oscillators which are near the center of the
break falls in the situation discussed in
Sec.~\ref{sec:couplingremote}.
The crucial point is that even for two neighboring oscillators
the diffusion coefficient of the one which is closer to the
chaotic spot is much larger than the diffusion coefficient of
the more remote oscillator. This holds even if one takes into
account that the diffusion coefficient is a strongly fluctuating
random quantity whose moments are divergent.%
\footnote{It is quite easy to check that for two independent
random variables $1<\zeta_1,\zeta_2<\infty$, distributed
according to a power law,
$p(\zeta)=\alpha\,\theta(\zeta-1)/\zeta^{1+\alpha}$,
the probability of $\tau\zeta_2>\zeta_1$ is given by
$\tau^\alpha/2$.}
%

Let us first consider a break which is symmetric with respect
to the bond between the sites $n=0$ and $n=1$. Then the oscillator
at $n=0$ is stronger coupled to the left part of the chain than
to the right, while the oscillator at $n=1$ is in the opposite
situation. Then their distribution functions can be approximated
by
$f_0\propto\exp[-\beta_L(\omega_0-\mu_L)I_0-\beta_L{g}I_0^2/2]$,
$f_1\propto\exp[-\beta_R(\omega_0-\mu_R)I_1-\beta_R{g}I_1^2/2]$,
and calculation of the current according to Eq.~(\ref{curtop=})
gives
\begin{equation}\begin{split}\label{Rbondsymm=}
\left.R^{-1}\right|_\mathrm{slope}
={}&\sum_{n\lesssim-L_\ell^\mathrm{max}}
e^{-\lambda_n}\left\langle
\left[\begin{array}{cc} 1 & \tilde\omega_1 \\
\tilde\omega_1 & \tilde\omega_1^2
\end{array}\right]D_{11}^{(n)}\right\rangle+\\
&+\sum_{n\gtrsim{L}_\ell^\mathrm{max}}
e^{-\lambda_n}\left\langle
\left[\begin{array}{cc} 1 & \tilde\omega_0 \\
\tilde\omega_0 & \tilde\omega_0^2
\end{array}\right]D_{00}^{(n)}\right\rangle.
\end{split}\end{equation}
We have neglected the off-diagonal components $D_{10},D_{01}$
of the diffusion coefficient, since the probability that an
optimal driving resonance involves both these sites is low
for $|n|\gg{L}_\ell^\mathrm{max}$.

Let us now consider a break which is symmetric with respect
to the site $n=0$. Then the oscillator at $n=0$ is coupled
equally well to the left and the right parts of the chain,
so its distribution function $f_0(I_0)$ should be found
from the diffusion equation which has the same form as
Eq.~(\ref{ArnolddifN=}), but with two terms on the right-hand
side, corresponding to the contributions from the left and
from the right. In contrast to Sec.~\ref{sec:spotontop},
here this procedure is justified since the dynamics of the
chaotic spots on the slopes of the break is faster than the
dynamics of the oscillator in the center, so we can introduce
the average diffusion coefficients
\begin{equation}
D_L=\sum_{n\lesssim-{L}_\ell^\mathrm{max}}
e^{-\lambda_n}\langle{D}_{00}^{(n)}\rangle_L,\quad
D_R=\sum_{n\gtrsim{L}_\ell^\mathrm{max}}
e^{-\lambda_n}\langle{D}_{00}^{(n)}\rangle_R.
\end{equation}
Here the thermal averaging is performed with respect to
the variables of oscillators with $n<0$ and $n>0$, so that
$D_L$ and $D_R$ depend on $I_0$ (namely,
$D_{L,R}\propto{I}_0$). For the distribution function we
obtain
\begin{equation}\begin{split}
f_0\propto\exp\left[-\frac{D_L\beta_L+D_R\beta_R}{D_L+D_R}\,
\left(\omega_0I_0+\frac{g}2\,I_0^2\right)
+\frac{D_L\beta_L\mu_L+D_R\beta_R\mu_R}{D_L+D_R}\,I_0\right].
\end{split}\end{equation}
Denoting the average with respect to this distribution
function by $\langle\ldots\rangle_0$, and calculating the
current according to Eq.~(\ref{curf=}), we write the
conductance as
\begin{equation}\begin{split}\label{Rsitesymm=}
\left.R^{-1}\right|_\mathrm{slope}={}&\left\langle
\left[\begin{array}{cc} 1 & \tilde\omega_0 \\
\tilde\omega_0 & \tilde\omega_0^2
\end{array}\right]\frac{D_LD_R}{D_L+D_R}\right\rangle_0+\\
&+\sum_{n\lesssim-L_\ell^\mathrm{max}}
e^{-\lambda_n}\left\langle
\left[\begin{array}{cc} 1 & \tilde\omega_1 \\
\tilde\omega_1 & \tilde\omega_1^2
\end{array}\right]D_{11}^{(n)}\right\rangle+\\
&+\sum_{n\gtrsim{L}_\ell^\mathrm{max}}
e^{-\lambda_n}\left\langle
\left[\begin{array}{cc} 1 & \tilde\omega_{-1} \\
\tilde\omega_{-1} & \tilde\omega_{-1}^2
\end{array}\right]D_{-1,-1}^{(n)}\right\rangle.
\end{split}\end{equation}
The two terms corresponding to the diffusion of the oscillators
at $n=\pm{1}$ are fully analogous to Eq.~(\ref{Rbondsymm=}).
They are of higher order in $\tau$ than the first term, but they
are kept since the matrix appearing in the first term is almost
degenerate. Indeed, for $D_{L,R}\propto{I}_0$ we can explicitly
calculate the average:
\begin{equation}\begin{split}
&\left\langle
\left[\begin{array}{cc} 1 & \tilde\omega_0 \\
\tilde\omega_0 & \tilde\omega_0
\end{array}\right]I_0\right\rangle_0=
\left[\begin{array}{cc} 1 & \omega_0' \\
\omega_0' & (\omega_0')^2+2g^2T^2/(\omega_0-\mu)^2+O(g^3)
\end{array}\right]
\left\langle{I}_0\right\rangle_0,\\
&\omega_0'=\omega_0+\frac{2gT}{\omega_0-\mu}
-\frac{6g^2T^2}{(\omega_0-\mu)^2}+O(g^3).
\end{split}\end{equation}
Note that the same kind of average actually appears in all
terms in Eqs.~(\ref{Rbondsymm=}),~(\ref{Rsitesymm=}).

If the size of the slopes is of the same order as the size of the
top, $\sim{L}_\ell^\mathrm{max}$, when the contribution from
Eq.~(\ref{Rbondsymm=}) or Eq.~(\ref{Rsitesymm=}) is added to that
from Eq.~(\ref{Rtop=}), they are of the same order, so the
resulting matrix is not degenerate, and can be inverted without
problem to obtain~$R$. In the opposite case, when the contribution
from Eq.~(\ref{Rbondsymm=}) or Eq.~(\ref{Rsitesymm=}) dominates,
matrix inversion should be done with care. 

Let us first consider Eq.~(\ref{Rbondsymm=}). The determinant
of the resulting matrix is
\[
\langle{D}_{00}\rangle\langle{D}_{11}\rangle(\omega_0'-\omega_1')^2
+2g^2T^2
\left(\langle{D}_{00}\rangle+\langle{D}_{11}\rangle\right)
\left[\frac{\langle{D}_{00}\rangle}{(\omega_0-\mu)^2}
+\frac{\langle{D}_{11}\rangle}{(\omega_1-\mu)^2}\right],
\]
where we denoted
\begin{equation}\label{D00avright=}
\langle{D}_{00}\rangle=\sum_{n\gtrsim{L}_\ell^\mathrm{max}}
e^{-\lambda_n}\left\langle{D}_{{00}}^{(n)}\right\rangle,\quad
\langle{D}_{11}\rangle=\sum_{n\lesssim-{L}_\ell^\mathrm{max}}
e^{-\lambda_n}\left\langle{D}_{11}^{(n)}\right\rangle.
\end{equation}
At first glance, the determinant can become small when
$\omega_0\approx\omega_1$. Note, however, that both
$\langle{D}_{00}\rangle$ and $\langle{D}_{11}\rangle$ contain
factors $1/(\omega_1-\omega_0)^2$, since coupling of the
oscillator at $n=1$ to the oscillators on the left slope
necessarily involves tunneling through the oscillator at $n=0$,
and vice versa. Thus, already the first term in the determinant
is regular, so the second one, which is small as $O(\rho^2)$,
can be safely neglected.

For Eq.~(\ref{Rsitesymm=}) the situation is somewhat different.
The determinant is given by
\[
\langle{D}_{00}\rangle\langle{D}_{11}\rangle(\omega_0'-\omega_1')^2
+\langle{D}_{00}\rangle\langle{D}_{-1,-1}\rangle(\omega_0'-\omega_{-1}')^2
+\langle{D}_{00}\rangle^2
\frac{2g^2T^2}{(\omega_0-\mu)^2},
\]
where we defined
$\langle{D}_{00}\rangle=\langle{D}_LD_R/(D_L+D_R)\rangle_0$,
and $\langle{D}_{11}\rangle,\langle{D}_{-1,-1}\rangle$ are
defined analogously to Eq.~(\ref{D00avright=}).
The first two terms have a smallness $O(\tau^2)$ coming from
$\langle{D}_{11}\rangle$, $\langle{D}_{-1,-1}\rangle$, while
the last term has a smallness $O(\rho^2)$. Still, the smallness
of the determinant amounts to an overall power-law prefactor
which is beyond our accuracy.

\subsection{Optimization}
\label{sec:optimization}

For the moment, let us ignore the matrix structure of the
break conductance $R^{-1}$. Indeed, as we have seen in
Secs.~\ref{sec:spotontop},~\ref{sec:spotonslope}, the matrix
elements differ by a factor which is at most a power of
$\tau,\rho$, or $\lambda$, and possible problems with the
matrix inversion amount to the same kind of factors. The
optimization of the break shape involves exponentials,
so it can be performed without paying attention to the matrix
structure.

\begin{figure}
\begin{center}
\includegraphics[width=12cm]{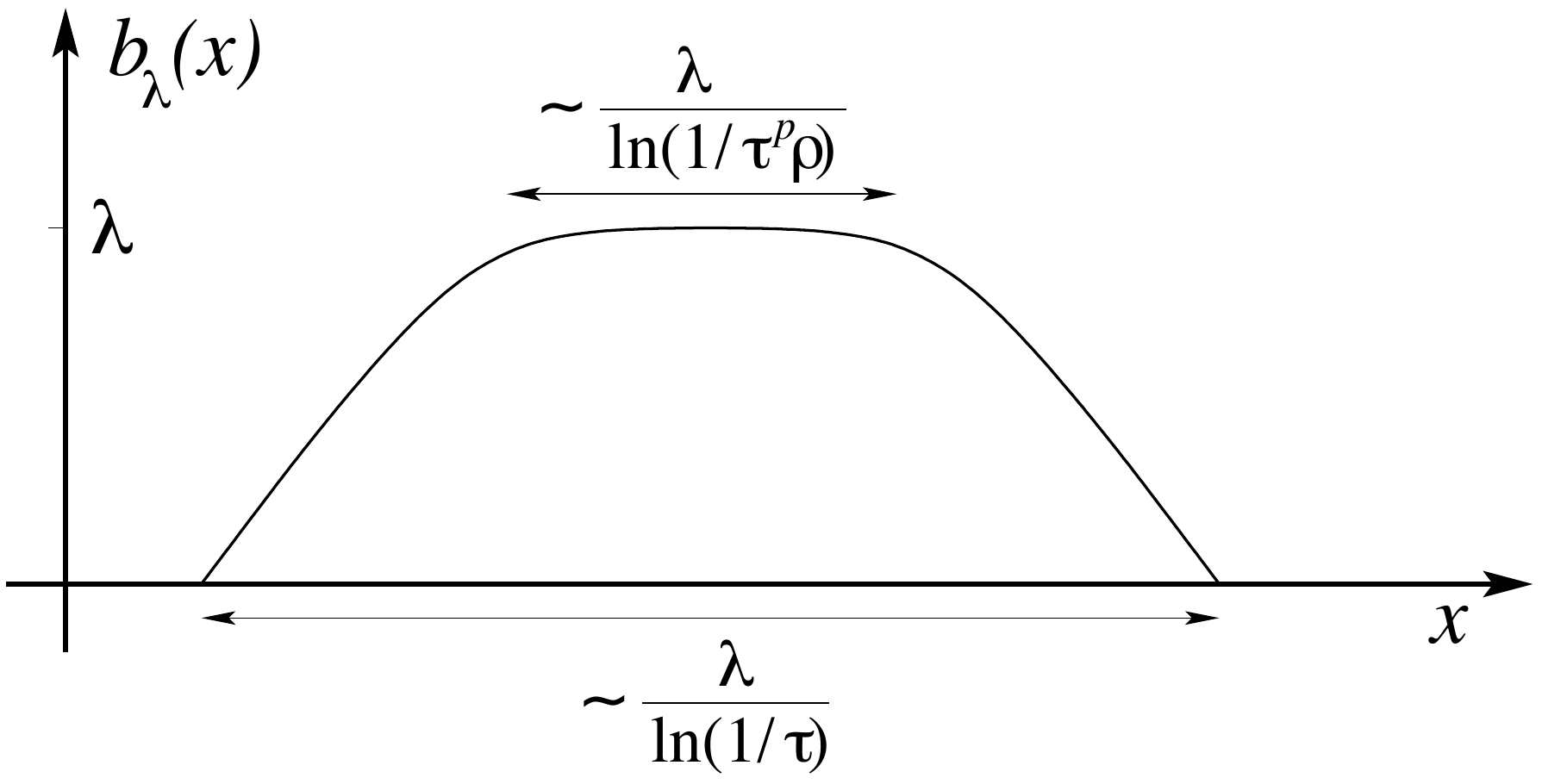}
\end{center}
\caption{\label{fig:optbreak}
The optimal break $b_\lambda(x)$ at~$\lambda\gg{1}$.
$b_\lambda(x)$ is defined as the family of functions, parametrized
by $\lambda$, such that the resistance of a break with an arbitrary
profile $\{\lambda_n\}$ is the same (up to power-law prefactors) as
that of an optimal break, $\lambda_n^{opt}=b_\lambda(n)$, with the
value of~$\lambda$ determined as the largest $\lambda$ for which
exists a shift $x_0$ such that $\lambda_n\geqslant{b}_\lambda(n-x_0)$
for all~$n$ (see Sec.~\ref{sec:Breaksgeneral} of the main text for
details).}
\end{figure}

From the discussion of Secs.~\ref{sec:spotontop}
and~\ref{sec:spotonslope} and the estimate (\ref{Dremote=})
one can conclude that the optimal break functions $b_\lambda(x)$,
defined in the end of Sec.~\ref{sec:Breaksgeneral},
should have the form, shown in Fig.~\ref{fig:optbreak}:
\begin{equation}
b_\lambda(x)=\left\{\begin{array}{ll}
\lambda+O(1),& |x|\sim\lambda/\ln(1/\tau^p\rho),\\
\lambda-2|x|\ln(1/\tau), & |x|\gg\lambda/\ln(1/\tau^p\rho).
\end{array}\right.
\end{equation}
Then the product in Eq.~(\ref{Rintergal=}) is expressed in terms of
[see Eq.~(\ref{Slambda=})]
\begin{equation}
\sum_nS(b_\lambda(n))=\mathcal{C}_1\rho\sum_n\left[b_\lambda(n)\right]^{p_1}
\exp\left[\frac{b_\lambda(n)}{\mathcal{C}\ln^2(1/\tau^p\rho)}\right]
=\mathcal{C}_1'\rho\lambda^{p_1}
\exp\left[\frac{\lambda}{\mathcal{C}\ln^2(1/\tau^p\rho)}\right],
\end{equation}
where $\mathcal{C}_1'$ is another logarithmic function of
$\tau$ and~$\rho$. The main dependence of $R$ on $\lambda$ is
exponential, $R(\lambda)\propto{e}^{\mathcal{C}_2'\lambda}$,
$1<\mathcal{C}_2'<2$ [see Eqs.~(\ref{Dremote=}) and~(\ref{Rtop=})],
so the integral in Eq.~(\ref{Rintergal=}) takes the form
\begin{equation}
\sigma^{-1}\propto\int\limits_0^\infty{d}\lambda
\exp\left\{\mathcal{C}_2'\lambda+\ln\mathcal{J}(\lambda)
-\mathcal{C}_1'\rho\lambda^{p_1}
\exp\left[\frac{\lambda}{\mathcal{C}\ln^2(1/\tau^p\rho)}\right]
\right\}.
\end{equation}
The Jacobian is calculated in \ref{app:Jacobian}, and it turns out that
\begin{equation}
\ln\mathcal{J}(\lambda)\sim\ln{S}(\lambda)\sim
\frac\lambda{\mathcal{C}\ln^2(1/\tau^p\rho)},
\end{equation}
so it can be neglected in comparison with $\mathcal{C}_2'\lambda$.
Then the optimal value~$\lambda_*$, maximizing the argument of the
exponential, is found from the equation
\begin{equation}
\mathcal{C}_2'=\left(\frac{p_1}{\lambda_*}
+\frac{1}{\mathcal{C}\ln^2(1/\tau^p\rho)}\right)
\mathcal{C}_1'\rho\lambda_*^{p_1}
\exp\left[\frac{\lambda_*}{\mathcal{C}\ln^2(1/\tau^p\rho)}\right],
\end{equation}
so that with logarithmic precision
\begin{equation}
\lambda_*=\mathcal{C}\ln^2\frac{1}{\tau^p\rho}\ln\frac{1}\rho.
\end{equation}
This determines the leading exponential in the macroscopic
conductivity,
\begin{equation}
\ln\sigma_*^{-1}=\mathcal{C}_2'\lambda_*
-\frac{\mathcal{C}_2'\mathcal{C}\ln^2(1/\tau^p\rho)}%
{1+(p_1/\lambda_*)\mathcal{C}\ln^2(1/\tau^p\rho)}\approx
\mathcal{C}_2'\mathcal{C}\ln^2\frac{1}{\tau^p\rho}\ln\frac{1}\rho,
\end{equation}
as well as the typical distance between such breaks,
\begin{equation}
\ln{L}_*=\frac{\mathcal{C}_2'\mathcal{C}\ln^2(1/\tau^p\rho)}%
{1+(p_1/\lambda_*)\mathcal{C}\ln^2(1/\tau^p\rho)}\approx
\mathcal{C}_2'\mathcal{C}\ln^2\frac{1}{\tau^p\rho}.
\end{equation}
This distance diverges as $\tau\to{0}$ or $\rho\to{0}$.
Segments of the chain shorter than $L_*$ cannot be
characterized by a conductivity, and the macroscopic
equations given in Sec.~\ref{sec:Results} are not valid. For
such segments only the total conductance of the segment can
be defined as the response of the currents to the difference
of potentials $\mu/T$, $1/T$, applied at the ends of the
segment. This conductance is determined by the strongest
break on the segment, and thus is strongly dependent on the
realization of disorder on the segment. Its probability
distribution can be calculated analogously to
Ref.~\cite{Raikh1989}, but is beyond the scope of the present
work.

The integration over $\lambda$ around the optimal value
$\lambda_*$ can be performed in the Gaussian approximation,
and the values of~$\lambda$ effectively contributing to the
integral are
\begin{equation}
|\lambda-\lambda_*|\sim\ln\frac{1}{\tau^p\rho}.
\end{equation}
It is easy to check that the cubic terms in the expansion
of the argument of the exponential in $\lambda-\lambda_*$
are small as $1/\ln(1/\tau^p\rho)$, so the Gaussian approximation
is valid.
Going back to the integrals over each individual~$\lambda_n$,
we find that the integral converges at
\begin{equation}
|\lambda_n-b_{\lambda_*}(n)|\sim{1}.
\end{equation}
This means that for a typical break, contributing to the integral
in Eq.~(\ref{seriesresistors=}),
the chaotic fractions $w_n$ are of the same order as for the optimal
break, $w_n\sim{e}^{-b_{\lambda_*}(n)}$. Hence, the conductance of
such a break is contributed by many possible positions of the chaotic
spot.  

Now one can turn to the study of the matrix structure of the
macroscopic conductivity~$\sigma$. It also becomes clear when each of
the two situation mentioned in the end of Sec.~\ref{sec:spotonslope}
is realized: the size of the top of the break,
$\sim\lambda_*/\ln(1/\tau^p\rho)$, and the size of the slopes,
$\sim\lambda_*/\ln(1/\tau)$, are of the same order unless
$\ln(1/\rho)\gg\ln(1/\tau)$, in which case the slopes are much longer.
The inverse of a linear combination of matrices appearing in
Secs.~\ref{sec:spotontop} and~\ref{sec:spotonslope} always has positive
quantities on the diagonal ($\sim\Delta^2$ and $\sim{1}$), while the
off-diagonal elements contain terms proportional to $\omega_n$
plus a thermal correction, $O(gT/(\omega_n-\mu))$. This inverse should
be averaged over the frequencies, as the macroscopic conductivity is
determined by many breaks. As a result, the diagonal elements remain
of the same order of magnitude as before the averaging, while in the
off-diagonal ones, $\omega_n$, which has a random sign,
averages to zero, and only the thermal correction survives. Thus,
the matrix structure of the conductivity is
\begin{equation}
\sigma\propto\left[\begin{array}{cc} 1 & O(gT/|\mu|) \\
O(gT/|\mu|) & O(\Delta^2) \end{array}\right].
\end{equation}
We remind the reader that the matrix elements of $\sigma$ have
the following meaning [see Eq.~(\ref{currents=})]:
$\sigma_{11}$ determines the response of the action current to
the gradient of $\mu/T$ when $T$~is constant, $\sigma_{12}$
determines the response of the action current to the gradient
of $1/T$ when $\mu/T$ is constant, while $\sigma_{21}$ and
$\sigma_{22}$ correspond to the same for the energy current.

\section*{Acknowledgements}
The author is grateful to I. L. Aleiner, B. L. Altshuler,
S.~Flach, and O.~M.~Yevtushenko for helpful discussions.

\appendix

\section{Thermodynamics of a strongly disordered nonlinear
Schr\"odinger chain}
\label{app:thermodynamics}
\setcounter{equation}{0}

Consider a chain of a long but finite length~$L$. Any microscopic
state of the chain (a set of complex numbers $\{\psi_n\}$) can be
characterized by definite values of the total energy~$H$, given by
Eq.~(\ref{Hpsipsi=}), and the total action (norm)
$I_{tot}=\sum_n|\psi_n|^2$ (or their densities,
$\mathcal{H}\equiv{H}/L$ and $\mathcal{I}\equiv{I}_{tot}/L$,
as introduced in Sec.~\ref{sec:Results}). In a thermal state,
characterized by two parameters, $\beta\equiv{1}/T$ and $\mu$, the
microscopic variables are distributed according to the grand
canonical distribution
$\mathcal{P}(\{I_n,\phi_n\})\propto{e}^{-\beta({H}-\mu{I}_{tot})}$.
The temperature~$T$ and the chemical potential~$\mu$ determine the
average action and energy per oscillator:
\begin{subequations}
\begin{eqnarray}
&&\mathcal{I}(T,\mu)=\frac{T}{L}\,\frac\partial{\partial\mu}
\ln\int
\exp\left[-\beta{H}(\{I_n,\phi_n\})+\beta\mu\sum_{n=1}^LI_n\right]
\prod_{n=1}^L\frac{dI_n\,d\phi_n}{2\pi},\label{ImuT=}\\
&&\mathcal{H}(T,\mu)=\mu\mathcal{I}(\mu,T)
-\frac{1}{L}\frac\partial{\partial\beta}
\ln\int
\exp\left[-\beta{H}(\{I_n,\phi_n\})+\beta\mu\sum_{n=1}^LI_n\right]
\prod_{n=1}^L\frac{dI_n\,d\phi_n}{2\pi}.\nonumber\\\label{HmuT=}
\end{eqnarray}
\end{subequations}
Here the integration limits are $0\leqslant{I}_n<\infty$,
$0\leqslant\phi_n<2\pi$.
In the limit of small~$\tau$ (strong disorder),
Eqs.~(\ref{ImuT=}), (\ref{HmuT=}) simplify:
\begin{subequations}
\begin{eqnarray}
&&\mathcal{I}(T,\mu)=\frac{1}{L}\sum_{n=1}^L
\int\limits_0^\infty{I}_n
{e}^{-\beta(\omega_n-\mu)I_n-\beta{g}I_n^2/2}\,
\frac{dI_n}{Z_n(T,\mu)}+O(\tau),\\
&&\mathcal{H}(T,\mu)=\frac{1}{L}\sum_{n=1}^L
\int\limits_0^\infty\left(\omega_nI_n+\frac{g}2\,I_n^2\right)
{e}^{-\beta(\omega_n-\mu)I_n-\beta{g}I_n^2/2}\,
\frac{dI_n}{Z_n(T,\mu)}+O(\tau),\\
&&Z_n(T,\mu)\equiv\int\limits_0^\infty
{e}^{-\beta(\omega_n-\mu)I_n-\beta{g}I_n^2/2}\,dI_n
\approx\left\{\begin{array}{ll}
T/(\omega_n-\mu),&\omega_n-\mu\gg{gT}\\
\sqrt{2\pi{T}/g}\,e^{(\mu-\omega_n)^2/(2gT)},&\mu-\omega_n\gg{g}T.
\end{array}\right.\nonumber\\
\end{eqnarray}
\end{subequations}
As usual, thermal fluctuations in the total energy and action are
proportional to $\sqrt{L}$, and are smaller than the average values
themselves, $L\mathcal{I}$ and $L\mathcal{H}$, for a sufficiently
large~$L$.
Moreover, even though each term in the sum is still a random
quantity due to the randomness of~$\omega_n$, summation over a large
number of sites performs the effective disorder average. Thus, in
the above equations one can replace
\begin{equation}
\frac{1}{L}\sum_{n=1}^L\to
\int\limits_{-\Delta/2}^{\Delta/2}
\frac{d\omega_n}\Delta.
\end{equation}
In other words, the total action and energy are self-averaging, and
one does not have to invoke many disorder realizations.

Performing the integrations, and keeping only the leading and
subleading terms in the small parameter $gT/(|\mu|\Delta)$
(i.~e., expanding the exponential in~$g$), we obtain
\begin{subequations}
\begin{eqnarray}
&&\mathcal{I}=\frac{T}{\Delta}\ln\frac{\Delta/2-\mu}{-\Delta/2-\mu}
-\frac{2(-\mu)gT^2}{(\mu^2-\Delta^2/4)^2}+O(g^2)+O(\tau),
\label{ImuTap=}\\
&&\mathcal{H}=T
\left(1+\frac{\mu}{\Delta}\ln\frac{\Delta/2-\mu}{-\Delta/2-\mu}\right)
+\frac{gT^2(\mu^2+\Delta^2/4)}{(\mu^2-\Delta^2/4)^2}+O(g^2)+O(\tau).
\label{HmuTap=}
\end{eqnarray}
\end{subequations}
These expressions work only for $\mu<-\Delta/2$,
$|\mu+\Delta/2|\gg\sqrt{gT}$ (this is the condition for smallness
of the subleading term with respect to the leading one).
The assumption $|\mathcal{H}|\lesssim(\Delta/2)\mathcal{I}$
corresponds to $\mu$~not being too close to $-\Delta/2$, and the
strong inequality $|\mathcal{H}|\ll\mathcal{I}\Delta$,
Eq.~(\ref{HllIDelta=}), translates into $|\mu|\gg\Delta$.
Expanding Eqs.~(\ref{ImuTap=}), (\ref{HmuTap=}) in the small
parameter $\Delta/|\mu|$, we obtain
\begin{subequations}
\begin{eqnarray}
&&\mathcal{I}=\frac{T}{|\mu|}
\left(1+\frac{\Delta^2}{12\mu^2}-\frac{2gT}{\mu^2}\right)
+O(g^2T^2/\mu^2\Delta^2)+O(\Delta^4/\mu^4)+O(\tau),\\
&&\mathcal{H}=T\,\frac{12gT-\Delta^2}{12\mu^2}
+O(g^2/\mu^2\Delta^2)+O(\Delta^4/\mu^4)+O(\tau),
\end{eqnarray}
\end{subequations}
which leads to Eq.~(\ref{rhoh=Tmu}).
For $\mu>-\Delta/2$ it is essential to keep $gI_n^2/2$ in the
exponential. In this case instead of Eqs.~(\ref{ImuTap=}),
(\ref{HmuTap=}) one obtains
\begin{subequations}
\begin{eqnarray}
&&\mathcal{I}=\left\{\begin{array}{ll}
(\mu+\Delta/2)^2/(2g\Delta),&-\Delta/2<\mu\leqslant\Delta/2,\\
\mu/g,&\mu\geqslant\Delta/2,
\end{array}\right.\label{IzeroT=}\\
&&\mathcal{H}=\left\{\begin{array}{ll}
(2\mu^3+3\mu^2\Delta/2-\Delta^3/8)^2/(6g\Delta),&
-\Delta/2<\mu\leqslant\Delta/2,\\
(\mu^2-\Delta^2/12)/2g,&\mu\geqslant\Delta/2.
\end{array}\right.\label{HzeroT=}
\end{eqnarray}
\end{subequations}

So far we were concerned the mapping
$(T,\mu)\to(\mathcal{I},\mathcal{H})$.
It turns out that the inverse correspondence,
$(\mathcal{I},\mathcal{H})\to(T,\mu)$, is not always well-defined,
as was noted in Ref.~\cite{Rasmussen2000} for the case without
disorder.
Below we study this problem for the strongly disordered case.

First, let us determine the allowed values of $\mathcal{I}$
and $\mathcal{H}$. For each $\mathcal{I}>0$ the Hamiltonian~$H$
has a minimum. Minimization of $H$ under the constraint
$\sum{I}_n=\mathcal{I}L$ is performed by introducing the Lagrange
multiplier~$\mu$, and is equivalent to elimination of $\mu$ from
Eqs.~(\ref{IzeroT=}), (\ref{HzeroT=}):
\begin{equation}
\mathcal{H}\geqslant\left\{\begin{array}{ll}
(\mathcal{I}/6)(\sqrt{32g\mathcal{I}\Delta}-3\Delta),&
0\leqslant\mathcal{I}\leqslant\Delta/(2g),\\
g\mathcal{I}^2/2-\Delta^2(24g),&
\mathcal{I}\geqslant\Delta/(2g).
\end{array}\right.
\end{equation}
Clearly, there is no upper bound on~$\mathcal{H}$ for a
given~$\mathcal{I}$. Indeed, one can consider a state
$I_1=\mathcal{I}L$, $I_2=\ldots=I_L=0$, for which
$H=g\mathcal{I}^2L^2/2+O(L)$, that is for $L\to\infty$ the average
energy density $\mathcal{H}$ can be made arbitrarily large.

\begin{figure}
\begin{center}
\includegraphics[width=7cm]{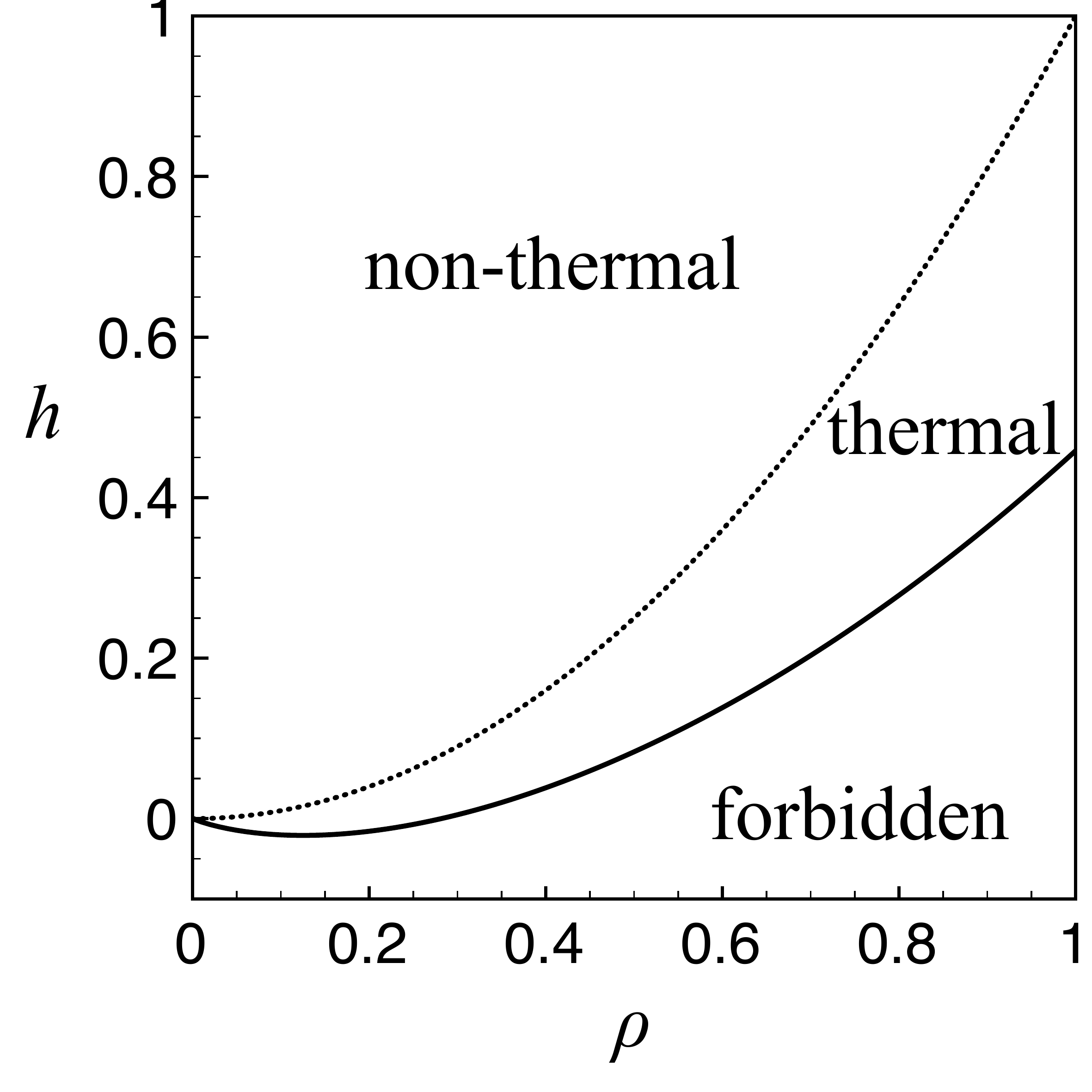}
\end{center}
\caption{\label{fig:nonthermal}
The $(\mathcal{I}>0,\mathcal{H})$ half-plane in the dimensionless
variables $\rho=g\mathcal{I}/\Delta$, $h=g\mathcal{H}/\Delta^2$.
The solid line corresponds to $T=0$; the region below this line
is forbidden. The dashed parabola corresponds to $T\to\infty$;
the region above the parabola corresponds to non-thermal states
which cannot be described by the grand canonical distribution.}
\end{figure}

However, for thermal states the largest value of~$\mathcal{H}$
is obtained when $T\to\infty$. In order for the partition function
to remain finite one should keep $\beta\mu=\mathrm{const}<0$,
then all $I_n=\mathcal{I}=-1/(\beta\mu)$, and we obtain the
$T\to\infty$ line to be $\mathcal{H}=g\mathcal{I}^2$ (in fact,
it is the same as in Ref.~\cite{Rasmussen2000}; indeed, $\beta=0$
effectively removes the disorder from the partition function).
Thus, the whole region $\mathcal{H}>g\mathcal{I}^2$ of the right
$(\mathcal{I},\mathcal{H})$ half-plane corresponds to non-thermal
states (i.~e., no $T$ and $\mu$ corresponding to given
$\mathcal{I},\mathcal{H}$ can be found in this region); see the
discussion in Ref.~\cite{Rasmussen2000} for details and
Fig.~\ref{fig:nonthermal} for the graphical representation of
the right $(\mathcal{I},\mathcal{H})$ half-plane.
The assumptions Eqs.~(\ref{HllIDelta=}), (\ref{Hless0}) ensure
the fact that the considerations of the present work are
restricted to thermal states only.

\section{On the spreading of a finite-norm wave packet}
\label{app:spreading}
\setcounter{equation}{0}

Here we are going to see what the macroscopic equations of
Sec.~\ref{sec:Results} predict when an initial condition in the
form of a finite-size wave packet is taken for $\mathcal{I}(x)$,
$\mathcal{H}(x)$.
Let us use the variables $\rho,h$, introduced in Eq.~(\ref{rho=h=}),
and assume
\begin{equation}\label{hllrho=}
h<0,\quad\rho^2\ll|h|\ll\rho.
\end{equation}
Then we can rewrite the transport equations in a closed compact form:
\begin{equation}
\frac{\partial}{\partial(t\Delta)}
\left[\begin{array}{c} \rho \\ h \end{array}\right]
=\frac\partial{\partial{x}}\,
e^{-\mathcal{C}'\ln^2(1/\tau^p\rho)\ln(1/\rho)}
\left[\begin{array}{cc} 1 & C_1\rho \\ C_1\rho & C_2
\end{array}\right]
\frac\partial{\partial{x}}
\left[\begin{array}{c} -1/\rho \\ h/(12\rho^2) \end{array}\right],
\quad C_1,C_2\sim{1}.
\end{equation}
Now one can estimate the growth of the typical size of the cloud,
$L(t)$, by writing $\rho\sim\rho_0L_0/L(t)$ and approximating the
spatial derivatives as $\partial_x\to{1}/L(t)\to\rho/(\rho_0L_0)$.
Let us also assume $\ln(1/\tau)\ll\ln(1/\rho)$. Then one obtains
the following estimate:
\begin{equation}\label{L(t)=}
\frac{d\rho^{-1}}{d\tilde{t}}
\sim{e}^{-\mathcal{C}'\ln^3\rho^{-1}}
\;\;\;\Rightarrow\;\;\;
L(t)\sim{e}^{[(\mathcal{C}')^{-1}\ln{t}]^{1/3}},
\end{equation}
which is Eq.~(\ref{Ltsim}). Here we rescaled the time as
$\tilde{t}\sim{t}\Delta/(\rho_0^2L_0^2)$
and approximated the integral
\[
\int{e}^{\mathcal{C}'\ln^3\rho^{-1}}d\rho^{-1}
\sim{e}^{\mathcal{C}'\ln^3\rho^{-1}},
\]
as power-law prefactors are beyond our precision.

It is important to check that if inequalities~(\ref{hllrho=}) are
satisfied at the initial moment, they remain satisfied for all times.
This is certainly the case if the $\rho$~cloud and the $h$~cloud have
the same length scale~$L(t)$ during the expansion:
$\rho\sim\rho_0L_0/L(t)$, $h\sim{h}_0L_0/L(t)$. Let us check that in
the course of the expansion the two length scales $L_\rho,L_h$ do not
separate. Clearly, $L_h\gg{L}_\rho$ is impossible since the expansion
of the $h$~cloud outside the $\rho$~cloud is blocked by the exponential.
To see that $L_h\ll{L}_\rho$ is also impossible, let us write the
currents as (the terms proportional to~$C_1$ are small and can be
neglected)
\begin{equation}
\left[\begin{array}{c} J^\rho \\ J^h \end{array}\right]\propto
\left[\begin{array}{c} \partial_x\rho \\
(C_2/12)\partial_xh-(C_2/6)(h/\rho)\partial_x\rho \end{array}\right].
\end{equation}
Thus, the $\rho$~cloud expands on its own,
so one can study the expansion focusing on $\rho$~only. If at
some instant $L_h\ll{L}_\rho$, then $\partial{h}/\partial{t}$
will be larger than $\partial\rho/\partial{t}$ by a factor
$L_\rho^2/L_h^2$, so the expansion of the $h$~cloud will be faster,
and it will eventually catch up.

Note that the length scale~$L_*$, Eq.~(\ref{Lstar=}),
determining the validity of the macroscopic transport equations,
diverges much faster than~$L$. This means that, strictly speaking,
the spreading of a wave packet cannot be described by macroscopic
equations.
Still, as discussed in Sec.~\ref{sec:onMott}, the effect of
the fluctuations at short length scales is just to modify the
factor~$\mathcal{C}'$, so the functional form of the dependence
$L(t)$, Eq.~(\ref{L(t)=}), should not be affected.
However, the spreading of the wave packet is subject to strong
sample-dependent fluctuations.

For not sufficiently large~$L$ another point is worth checking.
One of the key assumptions of the present work is that the
system can overcome an arbitrarily high activation barrier~$E$,
in a sufficient time, $\propto{e}^{E/T}$, i.~e., an arbitrarily
large amount of action and energy can be concentrated at a given
point at some instant of time. However, the wave packet contains
only a fixed finite amount of action and energy, so sufficiently
large fluctuations are forbidden. The condition for applicability 
of the Gibbs distribution can be written as
\begin{equation}
\frac{H-\mu{I}_{tot}}{T}\gg
\mathcal{C}'\ln^2\frac{1}{\tau^p\rho}\,\ln\frac{1}{\rho}.
\end{equation}
The right-hand side of this inequality represents the typical
value of the activation barrier corresponding to the optimal
break, as discussed in Sec.~\ref{sec:optimization}.
The left-hand side is just the size of the
cloud,~$L$. Since $\rho\propto{1}/L$, this inequality is
satisfied after a sufficiently long time of the expansion.

\section{Probability distribution of a product of denominators}
\setcounter{equation}{0}
\label{app:productdist}

Here we study the properties of the probability distribution
\begin{equation}\label{pnz=}
p_n(\zeta)=\int\limits_0^1
\delta\left(\zeta-\frac{1}{x_1\ldots{x}_n}\right)dx_1\ldots{d}x_n
=\frac{\theta(\zeta-1)}{\zeta^2}\,\frac{\ln^{n-1}\zeta}{(n-1)!}.
\end{equation}
If one adopts Stirling's formula for $(n-1)!$, Eq.~(\ref{pnz=})
can be rewritten as
\begin{equation}\label{lognormal=}
p_n(\zeta)=\frac{\theta(\zeta-1)}{\zeta\sqrt{2\pi(n-1)}}\,
\exp\left[-\frac{\ln^2(\zeta/e^{n-1})}{2(n-1)}
+\frac{\ln^3(\zeta/e^{n-1})}{3(n-1)^2}-\ldots\right].
\end{equation}
When all terms in the exponent except the first one are much
smaller than unity, which is the case if
\begin{equation}\label{lognormalcond=}
\left|\ln\frac\zeta{e^{n-1}}\right|\ll{3}^{1/3}(n-1)^{2/3},
\end{equation}
they can be neglected, and Eq.~(\ref{lognormal=}) corresponds
to the log-normal distribution. The total probability contained
in the log-normal region is determined by the error function
$\mathop{\mathrm{erf}}([(9/8)(n-1)]^{1/6})$. Thus, the rest of
the probability decays as $e^{-n^{1/3}}$, i.~e., slower than
exponentially. The average of $\zeta^2$, which diverges
for the actual distribution, for the log-normal one is finite
and equal to $e^{4(n-1)}$.

Another interesting region is the tail of the distribution, where
the probability integral can be calculated by the steepest
descent:
\begin{equation}\label{pzetasteepest=}
\int\limits_\zeta^\infty{p}_n(\zeta')\,d\zeta'=
\int\limits_{\ln\zeta}^\infty{e}^{-l+(n-1)\ln{l}}\frac{dl}{(n-1)!}
=\frac{1}{\zeta\,(n-1)!}\frac{\ln^n\zeta}{\ln(\zeta/e^{n-1})}
\left[1-\frac{n-1}{\ln^2(\zeta/e^{n-1})}+\ldots\right],
\end{equation}
where we have linearized the argument of the exponential with
respect to~$l$. The correction is small when
\begin{equation}
\left|\ln\frac\zeta{e^{n-1}}\right|\gg\sqrt{n-1}.
\end{equation}
Comparing this with Eq.~(\ref{lognormalcond=}), we see that the
two regions actually overlap. The approximate form of the
distribution $p_n(\zeta)$ in the tail region can be obtained by
differentiating Eq.~(\ref{pzetasteepest=}) with respect to~$\zeta$.

Let us now study the sum of $\mathcal{N}$ independent random
variables~$\zeta$ with random signs,
$z=|\pm\zeta_1\pm\ldots\pm\zeta_\mathcal{N}|$.
As all moments of the distribution $p_n(\zeta)$ diverge, the
central limit theorem is not valid and the distribution of~$z$,
$p_z(z)$, does not tend to a Gaussian. 
In the tail region (large $z$), $p_z(z)$ coincides with the
distribution of $\max\{\zeta_1,\ldots,\zeta_\mathcal{N}\}$,
so the sum is simply determined by its largest term.
The maximum of $p_z(z)$ is at $z=0$, due to the possibility
of the cancellation between different terms in the sum, even
though each individual term is bounded from below [$p_n(\zeta)$
vanishes for $\zeta<1$]. These properties will be discussed in
more detail in the end of this appendix. Here we note that
the random variable $\max\{\zeta_1,\ldots,\zeta_\mathcal{N}\}$
represents an upper bound on~$z$ since their distributions
coincide at large~$z$, while the probability of small values
of $\max\{\zeta_1,\ldots,\zeta_\mathcal{N}\}$ is suppressed,
in contrast to small values of~$z$, as discussed above.

The probability for $\max\{\zeta_1,\ldots,\zeta_\mathcal{N}\}$
to be smaller than some value $z$ is given by the product of
the probabilities,
\begin{equation}\label{Pzeta1zetan=}
\mathcal{P}\{\zeta_1,\ldots,\zeta_\mathcal{N}<z\}
\approx\left(1-\int\limits_{z}^\infty{p}_n(\zeta)\,d\zeta\right)^\mathcal{N}
\approx\exp\left[-\frac{\mathcal{N}}{z\,(n-1)!}\frac{\ln^nz}{\ln(z/e^{n-1})}\right].
\end{equation}
Most of the probability is contained in the region of~$z$ where
the argument of the exponential is of the order of unity, i.~e.
\begin{equation}\label{z=calN}
\frac{1}{z\,(n-1)!}\frac{\ln^nz}{\ln(z/e^{n-1})}
=\frac{1}{\mathcal{N}}.
\end{equation}
At large $n$ the solution of this equation is given by
\begin{equation}
z=\left[\sqrt{2\pi{n}}\,\ln{f}(\mathcal{N}^{1/n})\right]%
^{-1-1/\ln{f}(\mathcal{N}^{1/n})}
\left[e\,f(\mathcal{N}^{1/n})\right]^n,
\end{equation}
where the function $f(x)$ is defined as the solution of the equation
\begin{equation}
\frac{f}{1+\ln{f}}=x.
\end{equation}
In particular, if $\mathcal{N}\sim{c}^n$, then $z\sim(ef(c))^n$.
To give an idea of the behavior of $f(x)$, we note that
$f(1)=1$, and $x\ln(ex)\leqslant{f}(x)<2x\ln(ex)$ for $x>1$.
In the region around the solution of Eq.~(\ref{z=calN}) the
probability~(\ref{Pzeta1zetan=}) can be approximated by linearizing
the logarithm of the argument of the exponential in
Eq.~(\ref{Pzeta1zetan=}) with respect to $\ln{z}$, as in
Eq.~(\ref{pzetasteepest=}):
\begin{equation}
\mathcal{P}\{\zeta_1,\ldots,\zeta_\mathcal{N}<e^{-\lambda}\}
=\exp\left[-\exp\left[
\frac{\ln{f}}{1+\ln{f}}\left(\lambda+n\ln(ef)\right)\right]\right].
\end{equation}

A lower bound on~$z$ can be obtained if one replaces $p_z(z)$ by a
box distribution $p_z(0)\,\theta(1/p_z(0)-z)$ thus suppressing
the probability of large~$z$. Unable to calculate $p_z(0)$
explicitly, we obtain the lower bound in three steps:
(i)~replace $p_n(\zeta)$ by the corresponding log-normal
distribution, which already suppresses the probability of
large~$z$; (ii)~apply the central limit theorem to the sum
of log-normally distributed random variables; (iii)~replace
the resulting Gaussian by a box of the same height.
This procedure results in a box distribution with a height
larger than $p_z(0)$. To account for random signs of $\zeta_k$,
we redistribute the log-normal symmetrically between the positive
and negative~$\zeta$, then the second moment of this distribution
is given by the average of $\zeta^2$ over the log-normal,
$e^{4(n-1)}$, so the central limit theorem immediately gives
\begin{equation}
p_z(0)<\frac{2}{\sqrt{2\pi\mathcal{N}}\,e^{2(n-1)}}.
\end{equation}

To conclude this appendix, let us return to the distribution of
the sum of random variables not satisfying the central limit theorem.
Consider a random variable $\zeta$ whose probability distribution
$p_\zeta(\zeta)$ is symmetric, $p_\zeta(\zeta)=p_\zeta(-\zeta)$,
and for $|\zeta|\gtrsim\zeta_\infty$ can be approximated by
\begin{equation}\label{pzetazeta=}
p_\zeta(\zeta)\approx\frac{1}{2}\frac{\zeta_0^{\alpha-1}}{|\zeta|^\alpha},
\quad |\zeta|\gtrsim\zeta_\infty,
\end{equation}
where $\zeta_0\lesssim\zeta_\infty$ so that the total probability
contained in the region $|x|\gtrsim\zeta_\infty$ does not exceed
unity. Let $1<\alpha\leqslant{3}$ so that the total probability
converges, but the second moment diverges. Then the central
limit theorem is not applicable for the sum
$z=\zeta_1+\ldots+\zeta_\mathcal{N}$. The probability distribution
for this sum can be expressed as
\begin{equation}
p_z(z)=\int\limits_{-\infty}^\infty\frac{dk}{2\pi}\,e^{ikz}
\left[\int\limits_{-\infty}^\infty{e}^{-ik\zeta}\,
p_\zeta(\zeta)\,d\zeta\right]^\mathcal{N}
\approx
\int\limits_{-\infty}^\infty\frac{dk}{2\pi}\,e^{ikz}
\exp\left[-\mathcal{N}\int\limits_{-\infty}^\infty(1-e^{-ik\zeta})\,
p_\zeta(\zeta)\,d\zeta\right].
\end{equation}
At small $k\ll{1}/\zeta_\infty$ the $\zeta$ integral can be approximated as
\begin{equation}\begin{split}\label{Fourierpzetazeta=}
{}&\int\limits_{-\infty}^\infty{e}^{-ik\zeta}\,p_\zeta(\zeta)\,d\zeta
=1-\zeta_0^{\alpha-1}\int\limits_{\zeta_\infty}^\infty
\frac{1-\cos{k\zeta}}{\zeta^\alpha}\,d\zeta
+O(k^2\zeta_\infty^2)=\\
&=1-(k\zeta_0)^{\alpha-1}\left[\int\limits_0^\infty\frac{1-\cos{u}}{u^\alpha}\,du
+O(k^{3-\alpha}\zeta_\infty^{3-\alpha})\right]+O(k^2\zeta_\infty^2).
\end{split}\end{equation}
The integral over $u$ is equal to $-\Gamma(1-\alpha)\sin(\pi\alpha/2)$.
The $O(k^2\zeta_\infty^2)$ term can be neglected when
$\mathcal{N}\gg(\zeta_\infty/\zeta_0)^{(3-\alpha)/(\alpha-1)}$, so we can write
\begin{equation}\label{pzzabskz=}
p_z(z)\approx
\int\limits_{-\infty}^\infty\frac{dk}{2\pi}\,e^{ikz-|kz_\mathcal{N}|^{\alpha-1}}
=\left\{\begin{array}{ll} [\pi(\alpha-1)z_\mathcal{N}]^{-1}\,
\Gamma\left(\frac{1}{\alpha-1}\right),&|z|\ll{z}_\mathcal{N},\\
(\mathcal{N}/2)\zeta_0^{\alpha-1}/z^\alpha,&|z|\gg{z}_\mathcal{N},
\end{array}\right.
\end{equation}
where $z_\mathcal{N}$ represents the typical value of $|z|$:
\begin{equation}
z_\mathcal{N}=\zeta_0
\left[-\mathcal{N}\,\Gamma(1-\alpha)\sin\frac{\pi\alpha}2\right]^{1/(\alpha-1)}.
\end{equation}
The expression for $p_z(z)$ at large $|z|\gg{z}_\mathcal{N}$ was obtained
by expanding the exponential as
$e^{-|kz_\mathcal{N}|^{\alpha-1}}=1-|kz_\mathcal{N}|^{\alpha-1}+\ldots$,
performing analytical continuation from the positive imaginary semiaxis
of~$z$, and using the relation
$\Gamma(\alpha)\,\Gamma(1-\alpha)\,\sin\pi\alpha=\pi$.
It turns out to coincide with the tail of the probability
distribution of $\max\{|\zeta_1|,\ldots,|\zeta_\mathcal{N}|\}$
(up to a factor of 2, which accounts for the difference between the
distributions of~$z$ and~$|z|$):
\begin{equation}
\frac{d}{dz}\left(\mathcal{P}\{|\zeta|<z\}\right)^\mathcal{N}\approx
\frac{d}{dz}\exp\left(-\frac{\mathcal{N}}{\alpha-1}
\frac{\zeta_0^{\alpha-1}}{z^{\alpha-1}}\right)
\approx\frac{\mathcal{N}\zeta_0^{\alpha-1}}{z^{\alpha-1}}.
\end{equation}
Let us now consider the sum of positive random variables, assuming
$p_\zeta(\zeta)$ to vanish at $\zeta<0$ and to have the
form~(\ref{pzetazeta=}) without the factor 1/2 at large~$\zeta$.
For $1<\alpha<2$ the integration is analogous to
Eq.~(\ref{Fourierpzetazeta=}), but the result must be analytical
in the lower complex half-plane of~$k$. Then, instead of 
Eq.~(\ref{pzzabskz=}) we obtain
\begin{equation}
p_z(z)\approx
\int\limits_{-\infty}^\infty\frac{dk}{2\pi}\,
e^{ikz+\mathcal{N}\,\Gamma(1-\alpha)\,(ik\zeta_0)^{\alpha-1}}
=-\int\limits_0^\infty\frac{d\kappa}\pi\,e^{-\kappa{z}}
\Im\exp\left[-\mathcal{N}\,\Gamma(1-\alpha)\,(\kappa\zeta_0)^{\alpha-1}
e^{i\pi\alpha}\right],
\end{equation}
which at large $z$ again gives $\mathcal{N}\zeta_0^{\alpha-1}/z^\alpha$.
For $2<\alpha<3$ the first moment, $\bar\zeta$, of $p_\zeta(\zeta)$,
is finite, so it should be subtracted, and at large~$z$ we obtain
$\mathcal{N}\zeta_0^{\alpha-1}/(z-\mathcal{N}\bar\zeta)^\alpha$.

\section{Change of variables in the vicinity of a resonance}
\setcounter{equation}{0}
\label{app:change}

Consider $N$ oscillators, $n=1,\ldots,N$, and suppose that
the condition
\begin{equation}
\sum_{n=1}^Nm_n\,\frac{\partial{H}(I_1,\ldots,I_N)}{\partial{I}_n}
\equiv(\underline{m},\underline{\tilde{\omega}})=0.
\end{equation}
is satisfied for some values of the actions $\underline{I}=\underline{I}^r$.
Then the slow phase is introduced as
\begin{subequations}\begin{equation}\label{A:phidifN=}
\phidif=\sum_{n=1}^Nm_n\phi_n.
\end{equation}
To complete the transformation to new phases $\phi_k'$, we write
\begin{equation}\label{A:phikN=}
\phi_1'=\phidif,\quad
\phi_{k>1}'=\sum_{n=1}^Nu_{kn}\phi_n,
\end{equation}\end{subequations}
where the matrix $u_{kn}$ is chosen so that the vectors
$\underline{u}_k\equiv(u_{k1},\ldots,u_{kN})$ satisfy the
orthogonality relations
\begin{equation}
(\underline{u}_{k>1},\underline{u}_{k'>1})=\delta_{kk'},\quad
(\underline{u}_{k>1},\underline{m})=0.
\end{equation}
Then, defining $\underline{u}_1=\underline{m}$, we can easily
invert the matrix:
\begin{equation}
(u^{-1})_{nk}=\left\{\begin{array}{ll}
m_n/|\underline{m}|^2,&k=1,\\ u_{kn},&k>1,
\end{array}\right.\quad
|\underline{m}|^2\equiv(\underline{u}_1,\underline{u}_1)
=\sum_{n=1}^Nm_n^2,
\end{equation}
and write the completeness relation as
\begin{equation}
\frac{m_nm_n'}{|\underline{m}^2|}+\sum_{k=2}^Nu_{kn}u_{kn'}=\delta_{nn'}.
\end{equation}

The transformation for phases fixes the transformation to the
new actions $p_k$ in the vicinity of the resonance point (so
that the whole transformation is canonical):
\begin{equation}
\underline{I}=\sum_{k=1}^Np_k\underline{u}_k.
\end{equation}
This gives
\begin{equation}
\frac{\partial{H}}{\partial{p}_k}=
\sum_{n=1}^N\frac{\partial{H}}{\partial{I}_n}\frac{\partial{I}_n}{\partial{p}_k}
=\sum_{n=1}^Nu_{kn}\,\frac{\partial{H}}{\partial{I}_n},
\end{equation}
so that $\partial{H}/\partial{p}_1=0$, and $p_1$~is the effective
pendulum momentum. The kinetic part of the pendulum Hamiltonian is
then determined by $\partial^2H/\partial{p}_1^2=g|\underline{m}|^2$,
and this is what appears in Eq.~(\ref{Hpendulumg=}).

Let us now approximate
\begin{equation}
\phi_k'(t)=\frac{\partial{H}}{\partial{p}_k}\,t
+\left(\frac{\partial^2H}{\partial{p}_1^2}\right)^{-1}
\left(\frac{\partial^2H}{\partial{p}_k\partial{p}_1}\right)\phi_1'(t),
\end{equation}
where the derivatives are taken at the point $\underline{I}^r$.
The second term appears when the equations of motion are
written to the first order in $p_1$ which gives
$d\phi_1'/dt=(\partial^2H/\partial{p}_1^2)p_1$.
Inverting Eqs.~(\ref{A:phidifN=}),~(\ref{A:phikN=}), we obtain
\begin{equation}\label{phintapp=}
\phi_n(t)=\sum_{k=1}^N\phi_k'(t)\,u_{kn}
=\frac{\partial{H}}{\partial{I}_{n}}\,t
+\left(\underline{m},
\frac{\partial^2{H}}{\partial\underline{I}\partial\underline{I}},
\underline{m}\right)^{-1}
\frac{\partial(\underline{m},\underline{\tilde\omega})}{\partial{I}_n}\,\phidif(t),
\end{equation}
which is Eq.~(\ref{phint=}).

Let us now see how the stochastic layer thickness translates into
an effective integration measure on the resonant manifold. Namely,
we consider the integral of a function $F(I_1,\ldots,I_n)$, smooth
on the scale of the stochastic layer width:
\begin{equation}
\int\prod_{k=2}^N\frac{dp_k\,d\phi_k'}{2\pi}
\int\limits_\mathrm{layer}\frac{dp_1\,d\phi_1'}{2\pi}\,F
=\int{F}\mathcal{M}\,\delta(m_1\tilde\omega_1+\ldots+m_N\tilde\omega_N)
\prod_{n=1}^NdI_n,
\end{equation}
where $\mathcal{M}$ is the measure to be determined. Since the
transformation is canonical,
\begin{equation}
\prod_{k=1}^N\frac{dp_k\,d\phi_k'}{2\pi}=\prod_{n=1}^N\frac{dI_n\,d\phi_n}{2\pi},
\end{equation}
$\mathcal{M}$ is determined from the condition
\begin{equation}
W_s\equiv\int\limits_\mathrm{layer}\frac{dp_1\,d\phi_1'}{2\pi}
=\int{d}p_1\,\mathcal{M}\,\delta(m_1\tilde\omega_1+\ldots+m_N\tilde\omega_N),
\end{equation}
where $W_s$ is the volume of the stochastic layer in the
pendulum phase space. 
In order to resolve the $\delta$-function we write
\begin{equation}\begin{split}
d(\underline{m},\underline{\tilde\omega}){}&=
\sum_{n,n'=1}^Nm_n\,\frac{\partial^2H}{\partial{I}_n\partial{I}_{n'}}\,dI_{n'}=
\left(\underline{m},
\frac{\partial^2{H}}{\partial\underline{I}\partial\underline{I}},
\underline{m}\right)dp_1
+\sum_{k=2}^N\left(\underline{u}_k,
\frac{\partial^2{H}}{\partial\underline{I}\partial\underline{I}},
\underline{m}\right)dp_k
\end{split}\end{equation}
The last term vanishes, and the integration over $p_1$ gives
\begin{equation}
\mathcal{M}=W_s
\sum_{n,n'=1}^Nm_n\,\frac{\partial^2H}{\partial{I}_n\partial{I}_{n'}}\,m_{n'}
=g|\underline{m}|^2W_s.
\end{equation}
It is this measure that appears in Eq.~(\ref{wgdef=}).

\section{Integrals used in thermal averaging}
\setcounter{equation}{0}
\label{app:intGibbs}

Let us consider the following integral:
\begin{equation}
J_{\alpha_1\ldots\alpha_n}(A)=
\int\limits_0^\infty 
\exp\left(-x_1-\ldots-x_n-\frac{A}{x_1^{\alpha_1}\ldots{x}_n^{\alpha_n}}\right)
dx_1\ldots{d}x_n.
\end{equation}
At $A\gg{1}$ the leading exponential asymptotics is easily
found by noticing that the maximum of the exponent in the
integrand is reached at
\begin{equation}
x_k^{max}=\alpha_k
\left(\frac{A}{\alpha_1^{\alpha_1}\ldots\alpha_n^{\alpha_n}}\right)%
^{1/(\alpha_1+\ldots+\alpha_n+1)},
\end{equation}
so we obtain
\begin{equation}
\ln{J}_{\alpha_1\ldots\alpha_n}(A\gg{1})=
-(\alpha_1+\ldots+\alpha_n+1)\left(\frac{A}%
{\alpha_1^{\alpha_1}\ldots\alpha_n^{\alpha_n}}\right)%
^{1/(\alpha_1+\ldots+\alpha_n+1)}.
\end{equation}
Let us note that
\begin{equation}
(\alpha_1+\ldots+\alpha_n)\ln\frac{\alpha_1+\ldots+\alpha_n}{n}
\leqslant\sum_{k=1}^n\alpha_k\ln\alpha_k\leqslant
(\alpha_1+\ldots+\alpha_n)\ln(\alpha_1+\ldots+\alpha_n).
\end{equation}
The left inequality is the Jensen's inequality following
from the convexity of the function $x\ln{x}$ (the equality
holds when all $\alpha_k=\alpha$), while the right inequality
obviously follows from the monotony of $\ln{x}$ (the equality
holds when all $\alpha_k$ except one are zero).
Denoting $\alpha_1+\ldots+\alpha_n=n\alpha$, we obtain an
estimate
\begin{equation}\label{appintestim=}
\exp\left[-\frac{n\alpha+1}{\alpha^{n\alpha/(n\alpha+1)}}\,
A^{1/(n\alpha+1)}\right]
\leqslant{J}_{\alpha_1\ldots\alpha_n}(A\gg{1})\leqslant
\exp\left[-\frac{n\alpha+1}{(n\alpha)^{n\alpha/(n\alpha+1)}}\,
A^{1/(n\alpha+1)}\right].
\end{equation}
Even though the saddle point approximation is not supposed
to be valid when some of the $\alpha_k$'s are zero, the
right inequality is still correct since the integration
over the corresponding $n-1$ coordinates simply gives
$\int{e}^{-x_k}dx_k=1$, while the integration over the
remaining coordinate in the saddle point approximation
gives the same result, as can be checked directly.

Let us now focus on the case when when all $\alpha_k=\alpha$,
\begin{equation}
J_\alpha(A)=\int\limits_0^\infty 
\exp\left[-x_1-\ldots-x_n-\frac{A}{(x_1\ldots{x}_n)^\alpha}\right]
dx_1\ldots{d}x_n,
\end{equation}
and improve this estimate by calculating the prefactors
in front of the exponentials in the Gaussian approximation:
Let us perform an orthogonal change of variables
\begin{equation}
x_k=\frac{y_\|}{\sqrt{n}}+\sum_{m=1}^{n-1}e_k^my_m,
\end{equation}
where the unit basis vectors $e^m$ are orthogonal between
themselves, as well as to the direction of $y_\|$:
\begin{equation}
\sum_{k=1}^ne^m_ke^{m'}_k=\delta_{mm'},\quad
\sum_{k=1}^ne_k^m=0
\end{equation}
To the second order in the perpendicular deviations,
\begin{equation}\begin{split}
x_1\ldots{x}_n{}&=\left(\frac{y_\|}{\sqrt{n}}\right)^n
+\frac{1}{2}\left(\frac{y_\|}{\sqrt{n}}\right)^{n-2}
\sum_{m,m'=1}^{n-1}
\left(\sum_{k\neq{k}'}e_k^me_{k'}^{m'}\right)y_my_{m'}=\\
&=\left(\frac{y_\|}{\sqrt{n}}\right)^n
-\frac{1}2\left(\frac{y_\|}{\sqrt{n}}\right)^{n-2}
\sum_{m=1}^{n-1}y_m^2,
\end{split}\end{equation}
where we took advantage of the orthogonality:
\[
\sum_{k\neq{k}'}e_k^me_{k'}^{m'}
=\left(\sum_{k=1}^me_k^m\right)
\left(\sum_{k'=1}^me_{k'}^{m'}\right)
-\sum_{k=1}^ne^m_ke^{m'}_k
=-\delta_{mm'}
\]
Then the integral is easily calculated as ($Y=y_\|/\sqrt{n}$)
\begin{equation}\begin{split}
J_\alpha(A\gg{1}){}&=\int\limits_0^\infty \sqrt{n}\,dY
\int{d}^{n-1}\vec{y}_\perp
\exp\left[-nY-\frac{A}{Y^{n\alpha}}
-\frac\alpha{2}\frac{A}{Y^{n\alpha}}\frac{|\vec{y}_\perp|^2}{Y^2}\right]=\\
&=\sqrt{\frac{(2\pi)^{n-1}{n}}{(\alpha{A})^{n-1}}}\int\limits_0^\infty 
e^{-nY-A/Y^{n\alpha}}Y^{(n\alpha+2)(n-1)/2}\,dY=\\
&=\frac{n(\alpha{n}+2-\alpha)}2
\sqrt{\frac{(2\pi)^n}{n\alpha+1}\,(\alpha{A})^{n/(n\alpha+1)}}\,
\exp\left[-\frac{n\alpha+1}\alpha\,(\alpha{A})^{1/(n\alpha+1)}\right].
\label{JalphaA=}
\end{split}\end{equation}
The third line is obtained from the second one by denoting
$u=Y^{(n\alpha+2)(n-1)/2+1}$, and integrating over~$u$ in the
Gaussian approximation. The prefactor is large, so to obtain
a lower bound we can simply omit it. Then, setting
$n=2N_g-m_{n^g_*}$ and $\alpha=1/4$,  we arrive at Eq.~(\ref{lowerestint=}).

To obtain the upper bound, we note that
$x^{n/2}e^{-nx/2}\leqslant{e}^{-n/2}$, which gives
\begin{subequations}
\begin{equation}
J_\alpha(A\gg{1})\leqslant{n}(n\alpha+2)(2\pi/e)^{n/2}
\exp\left[-\frac{n}{2}\,(\alpha{A})^{1/(n\alpha+1)}\right].
\end{equation}
Note that the integral over $\vec{y}_\perp$ can be calculated in the
Gaussian approximation only when $\alpha{A}/Y^{n\alpha}\gg{1}$,
otherwise it is of the order of unity. Thus, to obtain an upper
bound, we just set the factor $(Y^{n\alpha}/\alpha{A})^{n-1}$
in the second line of Eq.~(\ref{JalphaA=}) to unity and integrate
over $Y^n=u$ in the Gaussian approximation:
\begin{equation}\begin{split}
J_\alpha(A\gg{1}){}&\leqslant\sqrt{\frac{(2\pi)^nn^2}{n\alpha+1}\,
(\alpha{A})^{(2n-1)/(n\alpha+1)}}\,
\exp\left[-\frac{n\alpha+1}\alpha\,(\alpha{A})^{1/(n\alpha+1)}\right]
\leqslant\\
&\leqslant{n}(8\pi/e^2)^{n/2}
\exp\left[-\frac{n}{2}\,(\alpha{A})^{1/(n\alpha+1)}\right],
\end{split}\end{equation}\end{subequations}
where the last inequality was written using
$(xe^{-x/2})^n\leqslant(2/e)^n$. Since $8\pi/e^2=3.4\ldots>2\pi/e=2.3\ldots$,
we adopt the latter estimate. This is precisely Eq.~(\ref{uglyineq=}).


\section{Effective diffusion coefficient along a thin layer}
\label{app:diffusion}
\setcounter{equation}{0}

Consider the diffusion equation for the distribution function
$\hat{f}(W,p_2,\ldots,p_N)\equiv{f}(W,\vec{p})$
in a flat space (the volume element being
${d}W\,dp_2\ldots{d}p_N$):
\begin{subequations}
\begin{equation}\begin{split}\label{appdiffeq=}
\frac{\partial\hat{f}}{\partial{t}}=-\partial_WJ_W-\partial_kJ_k,\quad
J_W=-\hat{D}_{WW}\partial_W\hat{f}
-\hat{D}_{Wk}\partial_k\hat{f},\quad
J_k=-\hat{D}_{kW}\partial_W\hat{f}
-\hat{D}_{kk'}\partial_{k'}\hat{f}.
\end{split}\end{equation}
Here $\partial_W=\partial/\partial{W}$,
$\partial_k=\partial/\partial{p}_k$, and the summation over
repeating indices $k=2,\ldots,N$ is assumed. The diffusion is
confined to a thin layer $W_1(\vec{p})<W<W_2(\vec{p})$,
$W_2(\vec{p})-W_1(\vec{p})\equiv{W}_s(\vec{p})$,
so that the normal component of the current vanishes at the
boundaries:
\begin{equation}
\left.\left(J_W-\frac{\partial{W}_{1,2}}{\partial{p}_k}\,
J_k\right)\right|_{W=W_{1,2}}=0.
\end{equation}
\end{subequations}
All dependencies on $p_k$ are assumed to be slow on the scale~$W_s$:
$W_s\partial_k\hat{f}\ll\hat{f}$. Thus, one can seek the distribution
function in the form
\begin{equation}
\hat{f}(W,\vec{p})={f}(\vec{p})+\tilde{f}(W,\vec{p}),\quad
\int\limits_{W_1(\vec{p})}^{W_2(\vec{p})}
\tilde{f}(W,\vec{p})\,dW=0.
\end{equation} 
Then our goal is to eliminate $\tilde{f}$ and find the effective
equation for $f(\vec{p})$.

Integrating Eq.~(\ref{appdiffeq=}) over $W$ and using the boundary
conditions, we obtain
\begin{equation}\label{appdiffeqav=}
W_s(\vec{p})\,\frac{\partial{f}(\vec{p})}{\partial{t}}=
-\partial_k\int\limits_{W_1(\vec{p})}^{W_2(\vec{p})}J_k(W,\vec{p})\,dW.
\end{equation}
Let us perform the gradient expansion,
$\tilde{f}=\tilde{f}^{(1)}+\tilde{f}^{(2)}+O(\partial_k^3)$.
To deal with the diffusion equation and the boundary condition
we need $J_k$ to $O(\partial_k)$ and $J_W$ to $O(\partial_k^2)$:
\begin{subequations}\begin{eqnarray}
&&-J_k=\hat{D}_{kk'}\partial_{k'}{f}
+\hat{D}_{kW}\partial_W\tilde{f}^{(1)}+O(\partial_k^2),\\
&&-J_W=\hat{D}_{Wk}\partial_k{f}+\hat{D}_{WW}\partial_W\tilde{f}^{(1)}
+\hat{D}_{Wk}\partial_k\tilde{f}^{(1)}+\hat{D}_{WW}\partial_W\tilde{f}^{(2)}
+O(\partial_k^3).
\end{eqnarray}\end{subequations}
Let us multiply Eq.~(\ref{appdiffeq=}) by $W_s$ and subtract from
it Eq.~(\ref{appdiffeqav=}). Since $\partial_t=O(\partial_k^2)$,
we can neglect $\partial_{t}\tilde{f}$ and write
\begin{equation}\begin{split}\label{tildef=}
0={}&W_s\partial_W
\left[\hat{D}_{Wk}\partial_k{f}+\hat{D}_{Wk}\partial_k\tilde{f}
+\hat{D}_{WW}\partial_W\tilde{f}\right]+\\
&{}+W_s\partial_k\hat{D}_{kk'}\partial_{k'}{f}
-\partial_k\left(\int\limits_{W_1}^{W_2}\hat{D}_{kk'}\,dW\right)
\partial_{k'}{f}+\\
&{}+W_s\partial_k\hat{D}_{kW}\partial_W\tilde{f}
-\partial_k\int\limits_{W_1}^{W_2}\hat{D}_{kW}\partial_W\tilde{f}\,dW
+O(\partial_k^3).
\end{split}\end{equation}
There are only two terms of the order $O(\partial_k)$, from which we
obtain
\begin{equation}
\tilde{f}^{(1)}=-\left(\int\limits^W\frac{\hat{D}_{Wk}}{\hat{D}_{WW}}\,dW\right)
\partial_k{f},
\quad
-J_k=\left(D_{kk'}-\frac{D_{kW}D_{Wk'}}{D_{WW}}\right)\partial_{k'}{f}
\equiv{D}_{kk'}^\|\partial_{k'}{f}.
\end{equation}
Substituting $J_k$ into Eq.~(\ref{appdiffeqav=}), we obtain the desired
effective equation for ${f}(\vec{p})$:
\begin{equation}
\frac{\partial{f}(\vec{p})}{\partial{t}}=
\frac{1}{W_s(\vec{p})}\,\frac{\partial}{\partial{p}_k}
\int\limits_{W_1(\vec{p})}^{W_2(\vec{p})}dW
\left[\hat{D}_{kk'}(W,\vec{p})
-\frac{\hat{D}_{kW}(W,\vec{p})\,\hat{D}_{Wk'}(W,\vec{p})}%
{\hat{D}_{WW}(W,\vec{p})}\right]
\frac{\partial{f}(\vec{p})}{\partial{p}_{k'}}.
\end{equation}
Finally, we have to make sure that the correction $\tilde{f}^{(2)}$
is regular (its explicit form is not even needed). Collecting the
$O(\partial_k^2)$ terms in Eq.~(\ref{tildef=}), we obtain
\begin{equation}\begin{split}
\label{tildef2=}
\partial_W\hat{D}_{WW}\partial_W\tilde{f}^{(2)}=&{}
\partial_W\hat{D}_{Wk}\partial_k
\left(\int\limits^W\frac{\hat{D}_{Wk'}}{\hat{D}_{WW}}\,dW\right)
\partial_{k'}{f}
-\partial_k{D}_{kk'}^\|\partial_{k'}{f}+\\
&{}+\frac{1}{W_s}\,\partial_k
\left(\int\limits_{W_1}^{W_2}{D}_{kk'}^\|\,dW\right)
\partial_{k'}{f}.
\end{split}\end{equation}
The boundary conditions translate into
\begin{equation}
\left.\hat{D}_{WW}\partial_W\tilde{f}^{(2)}\right|_{W=W_{1,2}}
=\hat{D}_{Wk}\partial_k\left(\int\limits^{W_{1,2}}
\frac{\hat{D}_{Wk'}}{\hat{D}_{WW}}\,dW\right)
\partial_{k'}{f}
+\frac{\partial{W}_{1,2}}{\partial{p}_k}
{D}_{kk'}^\|\partial_{k'}{f}.
\end{equation}
Eq.~(\ref{tildef2=}) is compatible with these boundary conditions
and can be straightforwardly integrated, indeed giving $\tilde{f}^{(2)}$
of the second order of smallness.

\section{Jacobian for integration over breaks}
\label{app:Jacobian}

Our goal is to find the Jacobian $\mathcal{J}(\lambda)$ in the
transformation
\begin{equation}
\int\prod_np(\lambda_n)\,d\lambda_n\to\int{d}\lambda\,
\mathcal{J}(\lambda)\prod_ne^{-S(b_\lambda(n))},
\end{equation}
where the probability distribution $p(\lambda_n)$ is represented as
\begin{equation}
p(\lambda_n)=\frac{d[1-e^{-S(\lambda_n)}]}{d\lambda_n}
\equiv{S}'(\lambda_n)\,e^{-S(\lambda_n)}.
\end{equation}
The calculation is based on three main observations.

First, as discussed in Sec.~\ref{sec:Breaksgeneral}, for each
break profile $\{\lambda_n\}$ one can specify
the two sites, $n_1,n_2$, where it touches the corresponding
optimal break,
$\lambda_{n_{1,2}}=b_\lambda(n_{1,2}-x_0)$,
for some values of $\lambda,x_0$, which are determined by
the profile $\{\lambda_n\}$. Thus, the whole integration
manifold of $\{\lambda_n\}$ can be split into submanifolds
corresponding to given values of $n_1,n_2$:
\begin{equation}
\int{d}\vec\lambda\equiv\int\prod_nd\lambda_n
\to\sum_{n_1\leqslant{n}_2}\int{d}\vec\lambda^{(n_1,n_2)}
\end{equation}
Second, at fixed $n_1,n_2$ there is one-to-one correspondence
between $\lambda_{n_1},\lambda_{n_2}$ and the parameters
$\lambda,x_0$ of the optimal break $b_\lambda(n-x_0)$.
That is, for an arbitrary profile $\{\lambda_n\}$ the
parameters $\lambda,x_0$ are in fact determined by four numbers
$n_1,n_2,\lambda_{n_1},\lambda_{n_2}$. The transformation
$(\lambda_{n_1},\lambda_{n_2})\leftrightarrow(\lambda,x_0)$
is determined by
\begin{equation}
d\lambda_{n_{1,2}}=\partial_\lambda{b}_\lambda(n_{1,2}-x_0)\,d\lambda
-b_\lambda'(n_{1,2}-x_0)\,dx_0,
\end{equation}
where we denoted
$b_\lambda'(x)\equiv\partial{b}_\lambda(x)/\partial{x}$. The
integration measure transforms as
\begin{equation}
d\lambda_{n_1}d\lambda_{n_2}=
\left[b_\lambda'(n_1-x_0)\,\partial_\lambda{b}_\lambda(n_2-x_0)
-b_\lambda'(n_2-x_0)\,\partial_\lambda{b}_\lambda(n_1-x_0)\right]
dx_0\,d\lambda.
\end{equation}
Third, at fixed $n_1,n_2,\lambda_{n_1},\lambda_{n_2}$, or,
equivalently, at fixed $n_1,n_2,\lambda,x_0$, the integration range
for $\lambda_n$ with $n\neq{n}_1,n_2$ is simply
$b_\lambda(n-x_0)<\lambda_n<\infty$, so the integration is
straightforward. Thus, one can write
\begin{equation}\begin{split}
\int{d}\vec\lambda\prod_np(\lambda_n){}&=
\sum_{n_1\leqslant{n}_2}\int{d}\lambda_{n_1}d\lambda_{n_2}\,
p(\lambda_{n_1})\,p(\lambda_{n_2})
\prod_{n\neq{n_1,n_2}}\exp\left[-S(b_\lambda(n-x_0))\right]=\\
{}&=\sum_{n_1\leqslant{n}_2}\int{d}\lambda_{n_1}d\lambda_{n_2}\,
S'(\lambda_{n_1})\,S'(\lambda_{n_2})\,
e^{-\Xi(\lambda)}.
\end{split}\end{equation}
Here we introduced
\begin{equation}
e^{-\Xi(\lambda)}=\prod_n\exp\left[-S(b_\lambda(n-x_0))\right],
\end{equation}
assuming it to be independent of $x_0$. This is not strictly
true, since $n$~is integer and $x_0$ is continuous, so it is
actually a periodic function of~$x_0$ with unit period, but we
neglect this modulation.

Let us now pass from $\lambda_{n_1},\lambda_{n_2}$ to
$\lambda,x_0$ keeping $\lambda$ fixed, and
change the summation variables as $n_1=n$, $n_2=n+l$
as well as the integration variable as $x_0=n+l-x$:
\begin{equation}\begin{split}
&\sum_{n_1\leqslant{n}_2}
\int\limits_{n_1}^{n_2}dx_0\,
S'(b_\lambda(n_1-x_0))\,S'(b_\lambda(n_2-x_0)){}\times{}\\
&\qquad\qquad\times\left[b_\lambda'(n_1-x_0)\,\partial_\lambda{b}_\lambda(n_2-x_0)
-b_\lambda'(n_2-x_0)\,\partial_\lambda{b}_\lambda(n_1-x_0)\right]=\\
&=\sum_{n=-\infty}^{\infty}\sum_{l=0}^{l_{max}}
\int\limits_{x_{min}}^{x_{max}}dx
\left[\frac{\partial{S}(b_\lambda(x-l))}{\partial{x}}\,
\frac{\partial{S}(b_\lambda(x))}{\partial\lambda}
-\frac{\partial{S}(b_\lambda(x))}{\partial{x}}\,
\frac{\partial{S}(b_\lambda(x-l))}{\partial\lambda}\right]=\\
&=L\,\mathcal{J}(\lambda),
\label{Jacobian=}
\end{split}\end{equation}
which gives precisely the sought Jacobian.
Due to translational invariance, the sum over $n$ produces the
total length~$L$, which appears in Eq.~(\ref{nbdb=}).
The value of $l_{max}=l_{max}(\lambda)$ is determined from the
equation $b_{\lambda}(l_{max}/2)=0$, and the range of $x$~integration
is given by
\begin{equation}
\begin{split}
&0\leqslant{x}\leqslant{l},\quad 0\leqslant{l}\leqslant\frac{l_{max}}2,\\
&l-\frac{l_{max}}2\leqslant{x}\leqslant{l_{max}},\quad
\frac{l_{max}}2\leqslant{l}\leqslant{l_{max}}.
\end{split}
\end{equation}
Replacing the summation over~$l$ by integration,
we can rewrite the Jacobian as
\begin{equation}\begin{split}
\mathcal{J}(\lambda){}&=-\int\limits_0^{l_{max}/2}{d}x\,dy
\left[\frac{\partial{S}(b_\lambda(-y))}{\partial{y}}
\,\frac{\partial{S}(b_\lambda(x))}{\partial\lambda}
+\frac{\partial{S}(b_\lambda(x))}{\partial{x}}
\,\frac{\partial{S}(b_\lambda(-y))}{\partial\lambda}\right]=\\
&=S(\lambda)\int\limits_{-l_{max}/2}^{l_{max}/2}
\frac{\partial{S}(b_\lambda(x))}{\partial\lambda}\,dx.
\end{split}
\end{equation}



\section*{References}


\end{document}